\documentstyle[graphics]{sxthesis}
\title{Gravitational Properties of Quantum Bosonic Strings}
\author{Alfredo V\'{a}zquez-Cruz}
\qualification{D. Phil.}
\date{September, 1996\\Revised version: June, 1997}
\input epsf
\let\include\input

\begin{document}
\maketitle
\begin{declaration}\addcontentsline{toc}{chapter}{DECLARATION}
I hereby declare that this thesis has not been submitted, 
either in the same or different form, to this or any other University for a 
degree. However, the thesis incorporates in chapter \ref{ChapTmunu} 
work done 
in collaboration with my supervisor Dr. E. J. Copeland and Prof. H. J. de Vega 
from the University of Paris and submitted for publication to 
the editors of ``Physical Review D'' and appears also in the list of preprints 
of Los Alamos National Laboratory under the preprint number:
\\
\begin{center}
hep-th-9601012
\end{center}
\vspace*{5cm}
Alfredo V\'{a}zquez-Cruz.
\end{declaration}

\begin{acknowledgments}
I will always remember with great pleasure the years I have spent at Sussex
University.\ These have been exciting years of personal growth for me as a 
scientist and as an individual.\ With much
pleasure, I take this opportunity to thank the people who helped me
through this important and most challenging period of my life.

I have to start by thanking the people who worked with me 
in the development of this thesis: 
Ed Copeland, 
my teacher and supervisor, who was always supportive during
my entire graduate career. He always gave me encouragement and useful
advice. Through his permanent questioning of our methods I have been 
able to learn the hidden `marvels' of physics and mathematics. 
Hector de Vega, my friend and a remarkable scientist, whose impressive
knowledge on the subject of String Theory made my studies an exciting and
enjoyable experience.\ His many contributions to the subject of String Theory 
certainly made this work possible. 
His works have certainly been an inspiration to me.

I must now thank  my teachers at Sussex University, who created  an always
stimulating environment for me and one that
I will greatly miss.\ In particular, I am grateful to Prof. David Bailin for
broadening my understanding of String theory, for his advice and many
useful comments. I am also grateful to Dr. Mark Hindmarsh 
who taught me how to use some `mysterious' latex commands. 
I also wish to thank Prof.\ Norman Dombey for all the
support I received from him. Without his help it would not have  been
possible for me to attend many of the workshops I attended during my
stay in Sussex. 

Special thanks go also to Prof.\ Norma S\'{a}nchez for her support for my 
attendance at the 1995 Erice School in String Gravity and
Physics at the Planck Scale and also for her organisation of such a remarkable
School; and to the ICTP organisers for all the support I received from them 
in the two ocassions I participated in their Summer School 
in High Energy Physics and Cosmology.  

My friends at Brighton, Araceli, Duncan, Dominique, Martha and Vicente, 
Paty Morris and Paty Cook 
shared beautiful moments of friendship with me and kept me from being overly
dedicated to my research. I am specially grateful to Duncan Shoebridge, 
who was kind enough to read the whole manuscript of this thesis and help me 
to correct and improve it, through his vast and thorough knowledge of the 
English language. (Any mistakes in grammatical 
structure/phrasing and the like still featuring in this work 
are entirely my fault and not his.)

I must also thank  Mr. and Mrs. Walker, who opened the doors of their house 
to me during the last months of my studies and always treated me as a member of 
their family. I will not forget their hospitality. In the same way I 
have to thank also Ed and Natalie Copeland for their generous hospitality 
during the correction time of this thesis I certainly will not forget their 
kindness and support.

My friends in M\'{e}xico, Carolina, Isabel, Genaro and Luis
Carlos Chavira, always believed in my ability to succeed and encouraged me 
in all my endeavours.

The Physics Division staff was always supportive and helpful.\ Sally
Church, Anthea Clark and Sue Bullock helped me navigate the complex 
`maze' of the Sussex University bureaucracy.

Special thanks are due to my ever-supportive family. I wish to
thank my parents, Jos\'{e} and Esperanza, and my brothers, Jos\'{e} Luis,
Oscar, Carlos Alejandro and Juan Antonio. 

The great support I received from my mother and my brother, 
Juan Antonio, made this work possible. To them I dedicate this work.

Finally, I wish to thank God for the love He has always shown me
during my entire life.

\end{acknowledgments}

\begin{abstract}\addcontentsline{toc}{chapter}{ABSTRACT}
In this thesis we are interested in the study of the gravitational 
properties of quantum bosonic strings. We start by computing the quantum 
energy-momentum tensor  ${\hat T}^{\mu\nu}(x)$
for strings in
Minkowski space-time. We perform the calculation of its 
expectation value for different
physical string states both for open and closed bosonic strings.
The states we consider are described by normalizable wave-packets in  
the centre of mass coordinates.
Amongst our results, we find in particular that ${\hat
T}^{\mu\nu}(x)$ becomes a non-local operator at the quantum level, its
position appears to be smeared out by quantum fluctuations. We find that
the expectation value acquires a non-zero value for both massive and 
massless string
states. 

After computing $\langle {\hat T}^{\mu\nu}(x)\rangle$ we proceed to
calculate the gravitational field due to a quantum massless bosonic string 
in the framework of a weak-field approximation to Einstein's
equations. We obtain a multipole expansion for the weak-field metric
$h^{\mu\nu}(x)$ and present its gravitational properties, including the
gravitational radiation produced by such a string. Our results are then
compared to those found for classical (cosmic) strings.

\end{abstract}

\tableofcontents  
\listoffigures
\pagebreak
\thispagestyle{empty}

\starttext
\starttext
\chapter{Introduction}
One of the greatest challenges facing physics nowadays is the
construction of a consistent theory of quantum gravity and thence a
consistent theory for the unification of the fundamental forces in
nature.

The last decades have witnessed a remarkable advance in the construction
of a unified theory of the known fundamental interactions, namely: the
electro-magnetic, the weak, the strong, and the gravitational forces.\
First, the electro-magnetic and weak forces were unified in the
description given by the Weinberg-Salam theory  \cite{UFT1,UFT2}
and subsequently there have been important successes in incorporating 
the strong
interactions, described by quantum chromodynamics, into a larger gauge
theory \cite{UFT3}.\ However,
despite all these successes gravity has remained the odd one out in this
scheme of grand unification.\ Furthermore, here we are confronted with
two problems: first, there is the problem of the actual unification of
gravity with the other fundamental forces in a single grand unified
theory and second the quantisation of gravity itself.\ We will see
in section \ref{QGprob} that both problems 
are really encompassed in the second one.

\section{Thesis}

\subsection{Gravitational properties}
This thesis focusses mainly on the quantum aspects of fundamental bosonic
strings in the framework of gravity.\ That is, we want to study the
gravitational properties emerging from quantum strings.

The classical aspects of String Theory in this framework, {\em cosmic
strings}, have been studied in numerous papers (see for example Vilenkin \& 
Shellard \cite{shellard} and references within).\ We would like to see
how the quantum nature of strings affects some of the main results for
cosmic strings. For example: is there an equivalent to the deficit angle 
obtained in
the classical theory?. The theory of
fundamental strings, although mathematically equivalent to that of cosmic
strings at the classical level, is very different in concept.\
Cosmic strings emerge when we break a gauge symmetry.\ Fundamental
strings, on the other hand, are supposed to 
describe all the known (and unknown) particle fields in nature since
String Theory can be seen as the most promising candidate for 
a quantum theory for the gravitational field.\ So how different
is the gravitational field due to fundamental strings 
compared to the one due to a Cosmic string?. 
We will see that because of the quantum nature of the 
strings considered in this thesis, we will also be  making contact with  
what is called a {\bf semi-classical approach to quantum gravity}.

\subsection{The expectation value of the string energy-momentum tensor}
One very important aspect of our work is the computation of the string
energy-momentum tensor in a truly stringy way, by this we mean that we will 
keep the extended nature of the string as opposed to other works where 
the string is integrated over a spatial volume and treated like a point 
particle \cite{HJV}. Such a computation leads to the ordinary results 
for a point particle because 
the authors considered the string precisely to
behave like a point particle; that is, they integrated over a spatial
volume completely surrounding  the string.\ In this way the calculation
turns out to be very simple; however, if we do not integrate over a volume 
around the
string, we keep all the string features and, as we will show in chapter 
\ref{ChapTmunu}, neither
the calculation nor the results are simple anymore.\ The 
consequences and differences that arise in this approach will be presented in
chapter \ref{ChapTmunu}.\ At the quantum level the string
energy-momentum tensor in Minkowski space-time seems to emulate a vertex
operator. This occurs not only because we are considering the string to
be a quantum object but also because we are maintaining its string
nature.
As we will see, the gravitational field produced by fundamental 
strings gives very different
results if we keep the string nature compared 
to that given by integrating over a 
volume embedding the string. \ A similar situation, although much more
complicated, occurs for the calculation of the
expectation value of the string energy-momentum tensor in curved
space-times (for example in gravitational shock-wave space-times).  

We have mentioned that our calculations also correspond to a semi-classical
theory of quantum gravity.\ So it is convenient at this point to talk
about what a quantum theory of gravity
should be. 

\section{Theories of quantum gravity}
\label{QGprob}
What is a theory of quantum gravity? A quantum theory of gravity 
(assuming it exists) is a
theory in which general relativity can be unified with quantum field
theory. A quantum theory of gravity must necessarily be:

\begin{enumerate}
\item{A finite theory (not in a renormalizable way but it has to be exactly
finite).}
\item{A theory of everything.}
\end{enumerate}

The reason for this is as follows \cite{HJVQG}: 
first let us recall what 
the meaning of a renormalizable quantum field theory (QFT) is.\ It is a
theory that has a domain of validity characterised by energies $E$ such
that these energies remain below the scale of energies relevant to the
model under consideration.\ That is
$$E<\Lambda,$$
where $\Lambda$ is of the order of 1 GeV for QED, 100 GeV for the
standard model of strong and electro-weak interactions, etc.\ One always
applies QFT up to an infinite energy (equivalent to zero distances) for
virtual processes and what we find in most cases is that the theory has
ultra-violet divergences.\ These divergences reflect the fact that our
model is unphysical for energies much larger than $\Lambda$.\ In a
renormalizable QFT, these infinities can be absorbed in a finite number
of parameters (like coupling constants and mass ratios).\ Since these
parameters are not predicted by the model in question, they have to be
fixed by their experimental values.\ This means that we need a more
general theory valid at energies beyond $\Lambda$ in order to compute
these renormalized parameters.\ For example $\frac{M_W}{M_Z}$ can be
computed in a {\em grand unified theory} whereas it must be fitted to 
its experimental value in the standard electro-weak model.

Now, let us analyse the consequences of Heisenberg's principle
in quantum mechanics when it is combined with the notion of
gravitational (Schwarzschild) radius in general relativity.

If we can make measurements at a very small distance $\Delta x$, then
\begin{equation}
\Delta p\sim\Delta E\sim\frac{1}{\Delta x},\label{one}
\end{equation}
where we have set c=$\hbar$=1.\ For sufficiently large $\Delta E$, particles
with masses $m\sim 1/\Delta x$ will be produced.\ The gravitational radius
of these particles will be of the order
\begin{equation}
r_G\sim Gm\sim\frac{(l_{Planck})^2}{\Delta x}
\label{two}
\end{equation}
where $l_{Planck}\sim 10^{-33}cm$.\ General relativity allows us to make
measurements at distances $\Delta x$ provided
\begin{eqnarray}
\Delta x > r_G & \rightarrow & \Delta x >\frac{(l_{Planck})^2}{\Delta
x},\label{three}
\end{eqnarray}
which implies
\begin{eqnarray}
\Delta x>l_{Planck} & or & m<M_{Planck}.\label{four}
\end{eqnarray}
From the discussion above we can see that no measurements can be made
at distances smaller than the Planck length and that there are no
particles heavier than the Planck mass.\ This is a consequence of
combining relativistic quantum mechanics with general relativity.

Since the Planck mass is the heaviest possible particle state, a theory
valid in this domain has to be valid at any other lower energy scale.\
One may ignore higher energy phenomena in a low energy theory; but the
opposite is not true.\ Therefore, we conclude that a theory valid at the
energy scale of the Planck mass will be a {\em theory of everything}.\ It
has to describe all known particle physics as well as the {\em desert}
which exists beyond the standard model.\ A theoretical prediction for
graviton-graviton scattering at energies of $M_{Planck}$ must include
all particles produced in a real experiment, which in practice means all
existing particles in nature since all matter is coupled to gravity.
\ It should be noted that the conclusion of a quantum theory of gravity
being a theory of everything is independent of any model we want to
study.

Once we have arrived at the conclusion above, it is clear that we must
now conclude the following: a consistent theory of quantum gravity has
to be a finite theory.\ In a quantum theory of gravity we have
$\Lambda=M_{Planck}$ and there cannot be any theory of particles beyond
it.\ Therefore if ultraviolet divergences appear in quantum gravity,
there is no way to interpret them as coming  from a higher energy scale
as is done in QFT.\ No physical understanding can be given to such
ultraviolet infinities.\ Therefore, the theory of quantum gravity has to
be exactly finite and not renormalizable finite. 
Of course, in the discussion above we are assuming 
that both general relativity and quantum mechanics will hold 
at energies in the order of the Planck scale, something 
which it is far from obvious, and that we are dealing with a 3+1 
space-time geometry.

The general theory of relativity is completely compatible with all other
classical theories; however, the only observable classical fields are
the electro-magnetic field and the gravitational field.\ The many other
interactions between the fundamental fields of nature can only be
described properly with the aid of quantum mechanics.\ Hence the
need to have a theory which includes all the interaction on an equal
footing, a {\em grand unified theory}.\ Currently some of the most fruitful 
approaches to quantise the gravitational field are supergravity
theories in which the graviton is regarded as one member of a multiplet
of gauge particles including bosons and fermions 
\cite{SG1}-\cite{SG2}.\ And of course,
the most serious candidate for not only quantising gravity but
for actually being a theory of everything is {\em String Theory}.
\subsection{The problems quantising the gravitation field}
\begin{figure}
\centerline{\epsfxsize=10cm\epsfbox{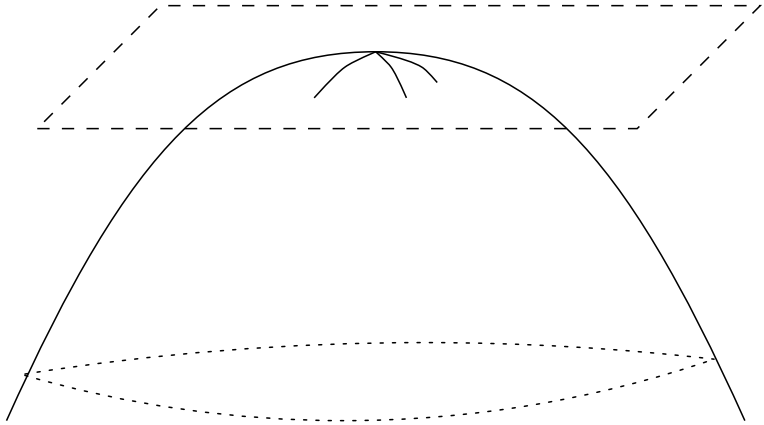}}
\caption{The space-time is locally flat, the postulates of quantum 
mechanics are valid 
as long as we do not deviate much from this picture.}
\label{loc}
\end{figure}

Let us first say something about why gravity seems not to be  compatible
with quantum mechanics. One of the main postulates of relativity 
is that a locally
geodesic coordinate system can be introduced at every region of
space-time (see fig.(\ref{loc})) so the action of the gravitational 
field becomes locally
ineffective and the space is approximately flat Minkowski space.\
Therefore, to say that in our neighbourhood, with its small
curvature, the space is flat seems to be natural.\ 
This essential postulate is
what in quantum mechanics is often taken for granted.\ This postulate
also shows us the limitations of the coexistence between quantum theory
and relativity theory: when we want to study physical processes in
regions of space dominated by a strong curvature regime (close to
singularities) and when one considers the behaviour of a far extended
physical system, quantum mechanics and general relativity are no longer
compatible with each other, the reason being that they start from
different space structures.
Quantum mechanics always presupposes a flat Minkowski space-time of
infinite extent both in its fundamental commutator rules, which are
formulated explicitly by the Lorentz group and in technical issues like
expansion in plane waves, asymptotic behaviour at infinity or the
formulation of conservation laws.\ Relativity theory tell us, however,
that the space is Riemannian. Quantum mechanics is valid if we do not 
deviate too much from the Minkowski space-time. However in strong curvature 
regimes, the validity of quantum mechanics is far from obvious.
Another contradiction arises from the idea of relativity theory that the
propagators of space are the propagators of the interactions of the
matter and can be measured out by material test bodies, including
measurements over very small distances and with the metric in very
small regions of space.\ If the dimensions of these regions are so small
that atoms or elementary particles should be taken as objects, then
their location is no longer so precisely defined and therefore we cannot
speak of making any measurement.

Despite these observations, we know that there exist macroscopic bodies
(e.\ g.\ stars) which consist of elementary particles and we know that
the motion of these objects obey the laws of gravitation.\ Therefore, a
consistent theory involving the merging of quantum mechanics and
relativity should be possible, although it is expected that at
least one of the two theories must be modified.\ In this direction,
there are at least three possibilities towards unifying gravity with
quantum mechanics.\ These will be presented in the following subsections. 
Before going into them, let us review briefly the `standard' description 
of our Universe.

\subsection{The standard Big-Bang model}
The most cherished model that aims to explain the evolution of the   
Universe is the so called Big-Bang scenario.\ The Universe seems once to
have been  a great expanding {\it fire ball} of quarks, leptons and
gluons existing at an enormous temperature.

The standard Big-Bang cosmology rests on three fundamental theoretical
pillars: the cosmological principle, the theory of general relativity
and a perfect fluid description of matter \cite{ext2}.
\subsubsection{The cosmological principle}
The cosmological principle states that on large distance scales the
Universe is homogeneous \cite{ext1}.\ 
From the observational point of view, this is a
nontrivial statement.\ On small scales the Universe looks extremely
inhomogeneous: we can see with our naked eye that stars are not
randomly, distributed.\ They are clustered in galaxies.\ Further observations
tell us that galaxies are not randomly distributed either: galaxies also 
cluster into clusters of galaxies.\ Until recently every new
survey has showed that there are new structures on the scale of the sample 
volume.\ In terms of the visible distribution of matter there seems to be 
no evidence 
for large scale homogeneity \cite{piet}. Recently 
there has been the discovery of 
considerable large-scale disturbances in the Hubble expansion in the 
neighbourhood of our Local Group of galaxies \cite{GA0,GA3,GA4,GA6,GA8}. 
This has been claimed to be one 
of the most important results of observational cosmology in recent years 
\cite{GA7}. It is not known in the present status of the theory, when most 
of the structure of a given scale formed and what most of this mass is. 
The large coherent flow of galaxies (including the Milky Way) suggests the 
presence of an extensive over density of matter that is now known 
with the name 
of {\em great attractor} \cite{GA3}. Recent observations have 
shown that there seems to exist 
a more distant and greater attractor than the great attractor, the entire 
Pisces region is moving in the general direction of their local supercluster 
and in the direction of the great attractor; however, 
its peculiar velocity of
around 400 Km/s is too fast even for the combined action of the gravitational 
pull from the local supercluster and the great attractor suggesting the 
presence of a larger mega-structure \cite{GA5}. These observations are 
difficult to reconcile with the homogeneous picture of the Universe 
\cite{piet, Barys}. In fact nowadays there is an ongoing debate regarding whether 
the Universe is homogeneous or not on large scale \cite{Mark, piet}. 
However, the most compelling evidence for the homogeneity of 
the universe is given by the smoothness of the {\em cosmic microwave background 
radiation} (CMBR). Data from COBE indicates that the Universe is extremely 
homogeneous. Work done by \cite{Martinez, Ellis} strongly suggests that given 
the smallness of the anisotropies in the CMBR the cosmological principle is 
indeed valid.

\subsubsection{The general theory of relativity}
The second theoretical pillar, the theory of general relativity, is the
theory which determines the dynamics of the Universe.\ According to
general relativity, space-time is a smooth manifold.\ Together with the
cosmological principle, this tells us that it is possible to choose a  
family of hypersurfaces with maximal symmetry.\ These are the
homogeneous constant time hypersurfaces.\ The metric of these surfaces
is (in spherical coordinates) \cite{Wei}
\begin{equation}
ds^2=a^2(t)\left[\frac{dr^2}{1-kr^2}+r^2(d\theta^2+\sin^2\theta
d\varphi^2)\right].
\end{equation}
The constant $k$ is $+1$, $0$ or $-1$ for closed, flat or open surfaces
respectively.\ The function $a(t)$ is the scale factor of the
Universe.\
By a coordinate choice, it could be set equal to 1 at any given time.\
However, the time dependence of $a(t)$ indicates how the spatial sections
evolve as a function of time.\ The full space-time metric is given by
\begin{equation}
ds^2=dt^2-a^2(t)\left[\frac{dr^2}{1-kr^2}+r^2(d\theta^2+\sin^2\theta
d\varphi^2)\right].
\end{equation}
The most important consequence of general relativity for the history of
the Universe is that it relates the expansion rate to the matter
content in the following way: Hubble in 1929 found that the velocity of 
celestial bodies was somehow proportional to their relative distances, the 
relation between the velocity and the distance given by
$$v=\frac{\dot{a}(t)}{a(t)}\;d=H\;d$$
this is the so called `Hubble's law'. In this relation $H$ is 
`Hubble's constant' (which as we can see, is not constant but varies 
as the Universe evolves in time). The value of this constant today is:
$$H_0=100\;h\, Km\;s^{-1}\;Mpc^{-1}$$
and $h$ is believe to be in the range of $0.4\leq h\leq 1$.

The relation between the expansion of the Universe and its matter content is 
given by the following equation of motion:
$$\left(\frac{\dot{a}(t)}{a(t)}\right)^2-
\frac{8\pi G}{3}\rho=-\frac{k}{a^2(t)}.
$$
Here the first term in the LHS is the kinetic energy term and the second one 
is the potential term. $\rho$ is the matter density and $k$ is a constant 
that can be positive, negative or zero depending on the matter density. If 
$k<0$, the expansion dominates over the matter gravitational pull and the 
Universe will expand forever, this is called an {\em open 
Universe}; if $k>0$, the matter term dominates over the expansion one and 
eventually the expansion ceases giving place then to a `contraction' era, 
the Universe then ends in a `big crunch', 
this Universe is called a {\em closed 
Universe}; finally, if $k=0$ neither of the two terms dominates and we call 
this Universe a {\em flat Universe}. So far the bulk of observational evidence 
suggests that we live in an {\em open Universe}, although with the large 
percent of `missing' matter (dark matter) still to find this view may 
change in the future.
\subsubsection{Observational evidence}
The three observational pillars of the standard Big-Bang
cosmology are Hubble's redshift-distance relation, the existence of a   
black body spectrum of the cosmic microwave background, and the
concordance between observed and theoretically determined light element
abundances. The cosmic microwave background radiation observational data 
suggests that the Universe was extremely uniform at its infancy; therefore, 
in order to explain great attractor-like structures 
mentioned above, we may want to postulate that 
there were also primordial density variations which in time grew. Their 
gravitational pull then acted like a `gravity amplifier' drawing more 
and more matter forming over dense regions 
(at the expense of under dense regions) of matter. However, this 
`gravity amplifier' would not have been enough to turn the quantum fluctuations 
expected in the Big-Bang into today's galaxies and galaxy clusters let alone 
mega-structures like the great attractor (given the age of the 
Universe between 
10 and 20 billion years old). However, if we consider inflation, the situation 
may improve.
\subsubsection{The problems facing the standard Big-Bang cosmology}
Standard Big-Bang cosmology is faced with several important and
fundamental problems: the age, dark matter, homogeneity, flatness, and 
formation of structure problems.\ In addition the model does not explain the
small value of the cosmological constant nor the
very fundamental problem: What triggered the `bang'?.
\ Can we go backwards in time to times before the
Big-Bang explosion? An answer to these questions is not possible within
the current status of physical theories, although present data is
consistent with the standard `Big-Bang' model (if we consider that an epoch 
of inflation at the very early stages of the Universe indeed occurred).

\subsubsection{Inflation}
Because inflation is nowadays one of the most studied models for the 
description of the Universe we live in, it is worth 
to present here a few words about it.

Inflation is an epoch of near exponential expansion in the very early history 
of the Universe, in this way all the presently observable Universe comes from 
a tiny initial region of space.
Models of inflation generally rely on the dynamics of a scalar field usually 
called the {\em inflaton}. At a certain `early' time it is assumed that the 
inflaton is displaced from the absolute minimum of its potential, the 
potential energy density dominates all other sources of energy density and leads 
to a period of exponential expansion 
in the Universe as time evolves until it no longer dominates as the inflaton 
reaches its absolute minimum \cite{Inf4,LKolb}.

Inflation besides solving the horizon 
and entropy (flatness) problems, can predict an almost  
scale invariant initial spectrum of fluctuations consistent with the 
COBE data \cite{cobe,cobe2}. These observations 
have been considered a great success for 
inflation. However, it is also possible to produce a similar spectrum 
of fluctuations from topological defects emerging from phase transitions in 
the early Universe (in particular from cosmic strings). 

Another unresolved issue for inflation is that in most attempts to incorporate 
inflation into specific models of particle physics 
there exists at least two important problems: parameters such as coupling 
constants must be fine tuned to extremely small values in order to avoid 
overproduction of density fluctuations \cite{Inf1,Inf3}.
\subsection{The quantisation of space-time: The wave-function of the
Universe}
Let us go back now to the possibilities we were considering before regarding 
how it may be possible to unify gravity with quantum mechanics.

First, the physics in our classical Universe may well 
be very different from
that of a quantum Universe. One possible way to unify gravity with the notions 
of quantum mechanics is to consider a Universe which is a mixture
of states, each with an a priori probability of occurrence.\ Each state
corresponds to a possible three geometry, including its topological
properties, and can be described by a point in superspace.
This approach involves the so called `wave-function' of the Universe: 
$\Psi$. It is possible to define this wave-function by fixing the metric and 
other fields that may be present on a hypersurface $\Sigma$ and then performing 
a path integral over bounded metrics and other fields.
In this approach, the constants of nature can take on
different values in different Universes resulting from: 
different choices of vacuum
states, different compactifications or to more complicated phenomena
such as worm-hole effects. It is to be also noticed that this approach 
is particularly sensitive to the initial
conditions of the theory.\ However, we know that $\Psi$ has to be
unique; therefore, we need a law of boundary conditions  
\cite{Alex2}.\ The
following are some of the boundary conditions that have been proposed in
the literature \cite{Alex2}:

\begin{enumerate}
\item{The Hartle-Hawking boundary condition \cite{HH1}: 
they proposed that $\Psi(h,\phi)$ 
should be given by a path integral over 
Euclidean 4-geometries $g^{\mu\nu}(\vec{x},t)$ and bounded by 
the 3-geometry $h_{ij}(\vec{x})$ 
in the following way:
$$\Psi(h,\phi)=\int[d g][d\phi]e^{-S_{E}}.$$
The wave-function describes basically a semi-classical tunnelling from 
`nothingness' \cite{Car2} to a Universe.

It should be noticed that in most attempts to compute 
the Hartle-Hawking wave-function, two 
important approximations are made \cite{Car1}: 
1) the path integral is evaluated by means 
of a saddle point approximation; 2) only the leading `least' action extremum 
 is taken into account.
 
 With this proposal, Grishchuk and Rozhansky \cite{GrisA} computed the 
 `most likely' values of the scalar field $\phi$ that is predicted in this 
 framework at the beginning of the Lorentzian stage in the evolution of the 
 Universe. Their results show that the most probable values of $\phi$ are 
 smaller than those which provide the minimally sufficient duration of 
 inflation, concluding that such a wave-function does not 
 describe in an adequate way 
 the Universe we live in. However, the values they found are only smaller 
 by a factor of 3 or 4 than the values needed, which still can make this 
 wave-function approach consistent with inflation and the large scale structure 
 of the observable Universe \cite{Car3,AOB}.}

\item{Lorentz path integral (a proposal by Alex Vilenkin, see for 
example \cite{AlexA}): the wave-function of the Universe 
should be obtained by integrating over Lorentzian histories interpolating 
between a vanishing 3-geometry and a finite 3-geometry $h_{ij}(\vec{x})$ 
\begin{eqnarray}
\Psi(h,\phi)&=&\int[d g][d\phi]e^{iS_{E}}\nonumber\\
 &= & K(h,\phi|0)\nonumber
\end{eqnarray}
where $K(h,\phi|0)$ is a causal propagator. 
In this scenario Vilenkin is able to obtain just the right initial conditions 
for inflation \cite{AlexA}.}
\item{Linde's proposal \cite{Linde}: He suggested that the Wick's rotation 
in the path integral should be done in the opposite direction 
$t\rightarrow +i\tau $
$$\Psi(h,\phi)=\int[d g][d\phi]e^{S_{E}}.
$$}
\end{enumerate}
Here $S_{E}$ is the Euclidean action given by \cite{HH1}:
$$S_{E}=\int d^4 x\sqrt{-g}\left(-R+\Lambda-\frac{1}{2}(\nabla\phi)^2-
\frac{R}{12}\phi^2\right).$$
$R$ is the scalar curvature and $\Lambda$ is a cosmological constant.

Another approach is given in \cite{Moss1} where the authors studied the 
wave-function of superstring theories in curved space-time. In doing that, 
some insight can be gained about the origin and evolution of the very early 
Universe. In particular they investigated the influences of the string vacuum 
fluctuations on processes occurring in the early Universe.

From a philosophical point of view, this approach of quantising the
gravitational field via a wave-function may run into some problems. 
Since we are talking
about a probability wave-function for the Universe, a randomly picked
Universe may not be suitable for life.
\subsection{Semi-classical gravity}
A less radical approach to that of quantising 
gravity is to treat the gravitational field classically, but quantise
all other fields. (This line of work is the one we will follow throughout 
this dissertation.) In a semi-classical theory, the coupling of gravity to
the quantised field depends on the one hand on the fact that the field
equations for gravity can be formulated covariantly, and therefore can
be made to depend on the gravity field, and on the other hand on the fact that 
the gravitational field can be seen to have the quantum fields as its
source.\ These fields occur, however, in the source of Einstein's field
equations (the energy-momentum tensor) not as operators but as
expectation values:
\begin{equation}
R_{\mu\nu}-\frac{1}{2}Rg_{\mu\nu}=\kappa\langle{\hat T}^{\mu\nu}\rangle.
\label{five}
\end{equation}
And we need the further requirement that the expectation value of the
energy-momentum tensor be divergence free in order for the field equations to
be integrable.\ That is:
\begin{equation}
\langle{\hat T}^{\mu\nu}\rangle_{;\nu}=0.
\label{sixx}
\end{equation}
Here, as usual `;' means covariant differentiation. 
We can see that this expression is not a simple consequence of the
equations governing the quantum fields, but rather a constraint on these
quantities; for example, the states which are used to form the
expectation values. 
\subsection{Quantum origins of the Universe}
Another possibility for the unification of gravity with quantum mechanics 
is to consider how quantum mechanics actually 
allowed for the possibility for the existence of the Universe as we know 
it today.
\subsubsection{The bouncing Universe}
Can the Universe have begun with a nonsingular although very violent
event?
One possibility is considered in \cite{Nariai,Park}, 
the Universe 'bounces'; there are
terms in the effective action for gravity which are induced by quantum
effects that reverse the collapse.\ If the matter density is enough, so
that the Universe is closed, the question of origin need not arise; in
some models these universes have always existed, eternally expanding
and contracting.
\subsubsection{Another possibility: a tunnelling effect}
Another possibility \cite{Atk} is that the Universe originated as a tunnelling 
effect from a classically stable, static space-time configuration.\ The Big
Bang is analogous to a single radioactive decay, on a huge scale.

Studies in which people considered a quantum origin of the Universe began
with the work of Tryon \cite{Tryon}. He suggested that 
the Universe might be a
vacuum fluctuation; it began as nothing at all.\ If this is so, then the
net quantum numbers of the Universe must be zero.\ In this respect, the
total electric charge of the Universe is consistent with zero.\ The   
total baryon number is not consistent with zero.\ However, this is not 
so troubling since the grand unified theories of strong, weak, and
electro-magnetic interaction imply proton instability.\ Tryon also
adopts the view that the total energy must be strictly conserved in the
creation process.\ This means that a Universe which originated as a   
vacuum fluctuation must have zero total energy.
\subsubsection{Vacuum origin of the Universe}
Brout, Englert, and Gunzig \cite{Nar, Gun} have further developed the idea of a 
vacuum origin of the Universe.\ They consider that the Universe was
initiated by a {\it local} quantum fluctuation of the space-time metric.
This results in particle creation.\ This creation of matter causes a
further change in the metric, and a cooperative process is set up.\
During this fire ball stage of particle creation, which is characterised
phenomenologically by negative pressure, the Universe is an open de
Sitter space-time which will develop a singularity, a future event
horizon, within a finite proper time.\ Before this horizon is reached,
however, the authors postulate that the cooperative process stops, and
particle creation ends.\ Then begins the second stage of the evolution of 
the Universe: adiabatic free expansion with positive pressure, the usual
post Big-Bang expansion.\ The authors examine in detail the particle
creation mechanism and the joining of the fire ball and Big-Bang stages,
but the origin of the quantum fluctuation is not examined.
\subsection{String Theory}
String Theory is the best candidate we have of a quantum theory of
gravity. The way String Theory unifies the various interactions of nature is, 
roughly speaking, similar to the way a violin string gives a
`fundamental' description of musical sound. The musical notes are
not the `fundamental' entities; it is the violin string which is the
`fundamental' object. This object can give the description not only of
the different tones that exist in music but also can give a description
of full harmonies, which are constructed of different musical tones.
In the case of String Theory, each `note' may be interpreted as the
different particles and forces that exist in nature, the string here being, 
as in the case of the violin, the fundamental object.

Superstring theory has no anomalies. The huge symmetry of the theory
makes it possible to cancel all the potential anomalies the theory
could have. Furthermore, it is hoped that the theory will remain finite 
to all orders
in perturbation theory. An important point to stress here is that 
all these nice properties of String Theory are
very sensitive to the string background we choose; therefore, there is very 
limited freedom in the theory. A consequence of this is that there are 
not so many free parameters involved in String Theory as we find in the GUT's.
However, String Theory is not free of problems, the main problems 
that String Theory faces are the following:

\begin{enumerate}
\item{First of all, the high energy region of the 
theory seems to be untestable since the
energies involved are those found at the Planck scale ($10^{19}$ GeV.).\
Obviously, a theory that cannot be tested is not an acceptable physical
theory. Of course, there is hope that somehow low energy effects may arise 
from String Theory. There have been many studies on superstring phenomenology 
(see for example \cite{Bai1}-\cite{Bai3}) but its predictions are not well 
understood yet, this is in part due to the many ways we can break the theory 
to low energies. Recently further advances have been made in order to test 
the low energy sector of superstring theory by investigating the 
possibility of spontaneous breaking of CP symmetry \cite{NuevoBailin}. 
In the future we may hope to have indeed a way to test 
the theory.}

\item{The bosonic string theory has problems because of the appearance of a
tachyon in the theory and because it does not include fermions.\ We can
incorporate fermions into the theory by introducing supersymmetry.\
However, we still do not have direct evidence to confirm the existence of
supersymmetry.}

\item{The theory seems to contain the general theory of relativity;
however, it does not explain why the cosmological constant is zero.}

\item{The theory has thousands of ways to break-down to low energy.\ We
do not know, therefore, which one is the correct vacuum for the theory. 
However, with the advent of $M$-theory the realisation of a $united$ 
superstring picture seems to be closer. $M$-theory tells us that what 
was believe to be different superstring theories may actually be 
related to each other \cite{MTheory}. 
This has been one of the most important results in String Theory lately.}

\item{We do not know how to break the 26 dimensional bosonic string theory 
(10 dimensional in the case of superstrings) down to 4 dimensions 
in a dynamical way (in some analogy to the way of how we break the 
symmetry spontaneously in field theory), although 
there exists many ways to actually reduce the theory to 4 dimensions (see 
for example \cite{Bai1}, \cite{Bai4}-\cite{Bai9}). This difficulty 
is the most fundamental one and therefore of paramount importance in 
String Theory.}
\end{enumerate}

\chapter{A brief introduction to String Theory}
In this chapter, I will try to give a brief introduction to some 
of the most important issues regarding String Theory.\ The material
covered here, however, has been selected on the basis of the work I will
be developing in this thesis, mainly
to try to present this work in a self-contained fashion.
\section{Background}
Today's physics is based on two fundamental theories. On the one hand we
have the theory of general relativity, which has proved to be very
successful in explaining not only the behaviour of the cosmos on large
scales but also in leading to a sensible description of the behaviour 
of the Universe
itself; on the other hand we have quantum field theory, which explains
the physics of the microcosmos. Quantum field theory has been
particularly successful in describing the weak and electro-magnetic
interactions and some remarkable advances have been made in the quantum
field description of strong interactions (Quantum Chromodynamics). 
Between the poles of these two theories all the present knowledge of physics is
covered; that is, we have a description of nature for over 40 orders of
magnitude.

Probably the most important goal in the minds of
physicists these days yet to be achieved is to present a 
unified picture of all forces
known in nature: the short-range forces, {\em strong} and {\em weak}; and
the long-range forces, {\em electro-magnetic} and {\em gravitational}. The
strong, weak and electro-magnetic forces can be described fairly well in
the realm of quantum field theory whilst the gravitational one can be
described by means of general relativity. However, up to this day
the theory of general relativity has proved to be incompatible with
quantum field theory; therefore, the ultimate goal of incorporating
gravity together with the other interactions in a single theory has not
been realised.
String theories emerged in the late sixties, when G. Veneziano postulated
his {\em Beta}-function amplitude for strong interactions. In the 1960's
the main problem facing physicists was the enormous proliferation of
strongly interacting particles. One characteristic of these particles is
that they seemed to have a spin proportional to the $mass^2$ of the
particle. A theory of quantum fields consistent with higher spins is
still lacking to this day. Another intriguing characteristic was that the
scattering amplitudes seemed to have a duality in the $s$ and
$t$-channels. With this in mind, Veneziano constructed an amplitude
which had precisely this dual behaviour \cite{str3}. Later on, it was discovered 
that behind this amplitude was really a relativistic string \cite{str15}. Dual models
constructed from this idea give one way to incorporate particles of high
spin without having ultraviolet divergences \cite{str1}-\cite{str15}. However, 
crucial developments in the seventies showed that these theories were not the
right way of describing strong interactions. In particular the failure
of dual models to incorporate the parton-like behaviour of strong
interactions was the main reason for abandoning such models. Another problem
with the dual model is that they predicted massless particles which
experimentally are not observed in the strong interactions. One of these
massless particles had a spin equal to 2. The coupling of this particle
was similar to the ones of general relativity. This particle is now
interpreted as a {\em graviton}. So now the view regarding dual models
has changed: the possibility of treating these models as a theory for
quantum gravity has given rise to a re-emergence of String Theory, not
in the context of strong interactions but in the wider context of a
possible theory incorporating gravity along with the other interactions
of nature. At present the most promising hope for a truly unified and finite
description of quantum field theory and general relativity is {\em Superstring
Theory}. This may be due mainly to the huge set of gauge symmetries the
theory possesses. We have to recall that all the advances in unifying the
strong, weak and electro-magnetic interactions have been made thanks to
the discovery of gauge symmetries.

\section{Free point-particles}
Before committing ourselves to work with extended objects such as strings, 
let us recall some of the aspects of the physics of point-particles. The 
notation we will follow throughout the chapters of this thesis will be as 
follows: greek indexes ($\mu$, $\nu$\dots) will run from 0 to $D-1$ where 
$D$ are the number of space-time dimensions. Latin indexes ($i$, $j$\dots) 
will run from 1 to $D-1$, that is, they will represent the $spatial$ 
dimensions of our system. The metric will have signature (+,-,-,-,\dots).

Let us start now by considering a free spinless particle with 
mass $m$.\ It is clear
that such a particle will follow a trajectory with only one parameter.\ Let us 
denote this trajectory by $X^{\mu}(\tau)$.
\ This trajectory is usually known 
by the name of {\em world-line}.\ The classical action 
describing this particle must be independent of how the trajectory is 
parametrised and it is defined to be proportional to the arc-length
travelled by the particle
\begin{equation}
S=-m\int^{s_f}_{s_i}ds=-m\int^{\tau_f}_{\tau_i}d\tau\sqrt{\dot{X}^{2}(\tau)}.
\label{Actp1}
\end{equation} 
Where $\dot{X}=dX/d\tau$ and $\tau$ is an arbitrary parameter, that label 
points along the world-line.
As we have said above, this action must be reparametrization invariant so
let us show that that is precisely the case here. Let us make a change
of coordinates from $\tau$ to $\tilde{\tau}$:
$$\tau\rightarrow\tilde{\tau}(\tau)$$
With this transformation we obtain
$$d\tau=\frac{d\tau}{d\tilde{\tau}}d\tilde{\tau}$$
$$
\frac{dX}{d\tau}=\frac{dX}{d\tilde{\tau}}\frac{d\tilde{\tau}}{d\tau}$$
therefore,
$$m\int d\tau\left[\left(\frac{dX}{d\tau}\right)^2\right]^{1/2}=
m\int d\tilde{\tau}\left[\left(\frac{dX}
{d\tilde{\tau}}\right)^2\right]^{1/2}.$$
Thus the action is invariant under arbitrary reparametrizations of the
variable $\tau$.

Introducing canonical conjugates:
$$P^{\mu}=\frac{\partial{\cal{L}}}{\partial\dot{X}^{\mu}}=
-\frac{m\dot{X}^{\mu}}{\sqrt{\dot{X}^2}}$$
we see that not all the canonical momenta are independent. (Here the 
Lagrangian is given by $L=
-m\sqrt{\dot{X}^{\mu}(\tau)\dot{X}_{\mu}(\tau)}$).
There is the constraint
\begin{equation}
P^{2}-m^2\equiv 0.
\end{equation}
This is a mass shell condition.

Now let us introduce the following Lagrangian with the help of a
Lagrange multiplier $e$:
\begin{equation}
L=P_{\mu}\dot{X}^{\mu}-\frac{1}{2}e(P^2-m^2). 
\end{equation}
By varying this Lagrangian with respect to $e$, 
we recover our constraint on the momenta. Let us now perform a path
integral integrating functionally over the variable $P$
\begin{eqnarray}
\int DP \exp\left\{i\int
d\tau[P\dot{X}-\frac{1}{2}e(P^2-m^2)]\right\}\nonumber\\ 
\sim\exp\left\{i\int d\tau\frac{1}{2}(e^{-1}\dot{X}^2-em^2)\right\}.
\end{eqnarray}
If in this expression we regard $X^{\mu}(\tau)$ as a set of scalar 
fields in one dimension, we can couple them to a metric $g\equiv g_{\tau\tau}$ 
and we can then write an action in a way similar to that in which 
we can couple scalar fields to gravity in four dimensions
\begin{equation}
S=-\frac{1}{2}\int d\tau\sqrt{-g}(g^{-1}\dot{X^{2}}-m^2).
\label{Actp2}
\end{equation}
Here we can see that the mass term behaves like a cosmological constant.\ We 
can then vary this action with respect to the metric to obtain the equation 
of motion for $g$.\ Solving this equation of motion and substituting the solution 
in eq.(\ref{Actp2}), we obtain our previous action eq.(\ref{Actp1}).
We can conclude that both actions 
are equivalent at least at the classical 
level.\ One important difference between these two actions is the fact that 
in eq.(\ref{Actp2}) we can actually take the limit of $m=0$.
Furthermore, we have eliminated the square root of the original action
so now we have a linearised action which is easier to handle than the
original one.
\section{Free bosonic strings}
\subsection{The Nambu-Goto string action}
The simplest extended object we can think of, built up from a single 
material point, is a one dimensional object: {\em a string}.\ 
Just as we associate to a point of matter some mass $m$ (which may be zero),
we can associate to each of the points 
of a string a $tension$ $T=1/2\pi\alpha'$ 
where $\alpha'$ is called the {\em Regge slope} parameter.
\ In this sense, we may regard the string as a distinguishable 
collection of points in space-time, distinguishable
precisely due to the tension associated to each of its points \cite{strmex}.

The string action is now proportional to the area the string generates
as it moves. This area is usually called the {\em world-sheet} of the
string. This action is given by \cite{str16}
\begin{equation}
S=-\frac{1}{2\pi\alpha'}\int d\sigma
d\tau\sqrt{(\dot{X}^{\mu}X'_{\mu})^2-\dot{X}^2 X'^2}
\end{equation}
and is called the Nambu-Goto action of the string. 
Here $\sigma$ and $\tau$ are arbitrary parameters which label 
points of the world-sheet. The prime denotes differentiation 
with respect to $\sigma$ and the dot differentiations with respect to $\tau$.

In the case of open strings it is necessary but not sufficient to ensure
that the action is invariant under a general
transformation 
$$X^{\mu}\rightarrow X^{\mu}+\delta X^{\mu}.$$
The variation of the action under this transformation involves a term
proportional to the wave equation and also the following surface term:
$$-\frac{1}{2\pi\alpha'}\int d\tau\left[X'_{\mu}\,\delta
X^{\mu}\left|_{\sigma=\pi}-X'_{\mu}
\,\delta X^{\mu}\right|_{\sigma=0}\right]=0.$$ 
It is the vanishing of this surface term which gives the open string
boundary condition.\ For closed strings the wave equation and
periodicity of $X$ is necessary and sufficient to ensure that the action
is stationary, as we can see from the calculation below.

Let us consider an arbitrary change in the configuration of the string
$$X^{\mu}(\sigma,\tau)\rightarrow X^{\mu}(\sigma,\tau)+\delta
X^{\mu}(\sigma,\tau).$$
We can now calculate the change in the action from the change in the 
Lagrangian \cite{gsw,strmex},
\begin{eqnarray}
\delta{\cal L}(\dot{X}^{\mu},X'^{\mu})&=&\frac{\partial{\cal
L}}{\partial \dot{X}^{\mu}}\delta \dot{X}^{\mu}+ \frac{\partial{\cal
L}}{\partial X'^{\mu}}\delta X'^{\mu}=\frac{\partial{\cal
L}}{\partial \dot{X}^{\mu}}\frac{\partial}{\partial\tau}\delta X^{\mu}
+\frac{\partial{\cal
L}}{\partial X'^{\mu}}\frac{\partial}{\partial\sigma}\delta
X^{\mu}\nonumber\\ 
& & \nonumber\\
&=&\frac{\partial}{\partial\tau}\left(\frac{\partial{\cal
L}}{\partial \dot{X}^{\mu}}\delta
X^{\mu}\right)+\frac{\partial}{\partial\sigma}\left(\frac{\partial{\cal
L}}{\partial X'^{\mu}}\delta X^{\mu}\right)-\delta X^{\mu}
\left[\frac{\partial}{\partial\tau}\left(\frac{\partial{\cal
L}}{\partial \dot{X}^{\mu}}\right)+\frac{\partial}{\partial\sigma}
\left(\frac{\partial{\cal L}}{\partial X'^{\mu}}\right)\right]\cr
& &
\end{eqnarray}
\begin{eqnarray}
\delta S &=&-\frac{1}{2\pi\alpha'}\int^{\tau_f}_{\tau_i}d\tau\int^{\pi}_{0}
d\sigma
\delta X^{\mu}(\sigma,\tau)
\left[\frac{\partial}{\partial\tau}\left(\frac{\partial{\cal
L}}{\partial \dot{X}^{\mu}}\right)+\frac{\partial}{\partial\sigma}
\left(\frac{\partial{\cal L}}{\partial X'^{\mu}}\right)\right]\nonumber\\
& & \nonumber\\ 
& & +\frac{1}{2\pi\alpha'}\int^{\pi}_{0}d\sigma\left.
\frac{\partial{\cal L}}{\partial
\dot{X}^{\mu}}\delta X^{\mu}\right|^{\tau_f}_{\tau_{i}}+
\frac{1}{2\pi\alpha'}\int^{\tau_f}_{\tau_i}d\tau\left.
\frac{\partial{\cal L}}{\partial
X'^{\mu}}\delta X^{\mu}\right|^{\pi}_{0}.\label{EOM}
\end{eqnarray}
From expression (\ref{EOM})we can obtain the boundary conditions mentioned
above as well as the equations of motion for the string. (The open string is 
conventially described by the parameter $\sigma$ running from $0$ to $\pi$; 
for closed strings the same convention is used.) The equations
of motion are given by
\begin{equation}
\frac{\partial}{\partial\tau}\left(\frac{\partial{\cal
L}}{\partial \dot{X}^{\mu}}\right)+\frac{\partial}{\partial\sigma}
\left(\frac{\partial{\cal L}}{\partial X'^{\mu}}\right)=0,
\label{EOM2}
\end{equation}
and the boundary conditions are given by
\begin{equation}
\int^{\tau_f}_{\tau_i}d\tau\left.\frac{\partial{\cal L}}{\partial
X'^{\mu}}\delta X^{\mu}\right|_{\pi}-\int^{\tau_f}_{\tau_i}
d\tau\left.\frac{\partial{\cal L}}{\partial X'^{\mu}}
\delta X^{\mu}\right|_0=0.
\label{BC1}
\end{equation}
That is, for open strings
\begin{equation}
\left.\frac{\partial{\cal L}}{\partial X'^{\mu}}\delta X^{\mu}\right|_{\pi}=
\left.\frac{\partial{\cal L}}{\partial X'^{\mu}}\delta X^{\mu}\right|_0=0
\end{equation}
and
\begin{equation}
\left.\frac{\partial{\cal L}}{\partial X'^{\mu}}\delta X^{\mu}\right|_{\pi}=
\left.\frac{\partial{\cal L}}{\partial X'^{\mu}}\delta X^{\mu}\right|_0
\end{equation}
for closed strings.
For open strings this is clear from eq.(\ref{BC1}). For closed strings 
this is due to the periodicity 
$\delta X^{\mu}(\pi,\tau)=\delta X^{\mu}(0,\tau)$.

Now we can define the conjugate momenta to $\sigma$ and $\tau$ as 
\cite{str16,strmex}
\begin{equation}
P^{\mu}_{\tau}(\sigma,\tau)\equiv -\frac{\partial{\cal
L}}{\partial \dot{X}^{\mu}}=
\frac{1}{2\pi\alpha'}\frac{(\dot{X}^{\rho}X'_{\rho})
X'^{\mu}-X'^2\dot{X}^{\mu}}
{\sqrt{(\dot{X}^{\rho}X'_{\rho})^2-X'^2\dot{X}^2}}
\end{equation}
and
\begin{equation}
P^{\mu}_{\sigma}(\sigma,\tau)\equiv -\frac{\partial{\cal
L}}{\partial X'^{\mu}}=
\frac{1}{2\pi\alpha'}\frac{(\dot{X}^{\rho}X'_{\rho})
\dot{X}^{\mu}-\dot{X}^2 X'^{\mu}}
{\sqrt{(\dot{X}^{\rho}X'_{\rho})^2-X'^2\dot{X}^2}}.
\end{equation}
Thus the equations of motions (\ref{EOM2}) become
$$\frac{\partial P^{\mu}_{\tau}}{\partial\tau}+\frac{\partial
P^{\mu}_{\sigma}}{\partial\sigma}=0,$$
that is
\begin{equation}
\frac{\partial P^{\mu}_{\alpha}}{\partial\sigma^{\alpha}}=0
\end{equation}
where ($\sigma^{\alpha}=(\tau,\sigma)$).
Notice that the equations of motions obtained here are very complicated. 
This is due to the fact that we have not chosen any gauge yet. 
\subsection{The Polyakov string action}
\begin{figure}
\centerline{\epsfxsize=10cm\epsfbox{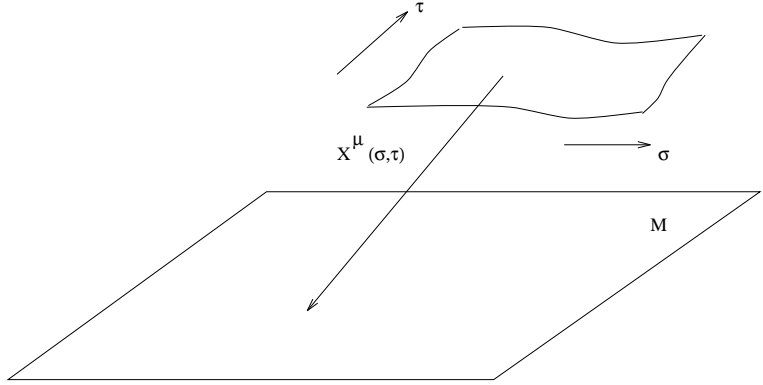}}
\caption{String world-sheet.}\label{world-sheet}
\end{figure}

As in the point-particle case, we can express this action in a more
convenient way, which is totally equivalent to the Nambu-Goto action of the 
previous section. The Polyakov form of the action is given by \cite{str5,gsw}
\begin{equation}
S=-\frac{1}{2\pi\alpha'}\int d\sigma
d\tau\sqrt{h}h^{\alpha\beta}(\sigma)g_{\mu\nu}(X)\partial_{\alpha}
X^{\mu}\partial_{\beta}X^{\nu},
\end{equation}
where $h^{\alpha\beta}$ ($\alpha$, $\beta=0,1$) is the inverse of 
$h_{\alpha\beta}$ and $h$ is
the absolute value of the determinant of $h_{\alpha\beta}$.\ The metric
$h_{\alpha\beta}$ is a Minkowskian metric in $1+1$ dimensions.\ The
string coordinates $X^{\mu}(\sigma,\tau)$ give a map of the world-sheet
manifold into physical space-time (fig.\ref{world-sheet}).
The reparametrization invariance of this action allows us to make a covariant 
gauge choice, namely:
\begin{equation}
\sqrt{h}h^{\alpha\beta}=\eta^{\alpha\beta},
\label{CG}
\end{equation}
where $\eta^{\alpha\beta}$ is of course a two-dimensional Minkowski metric. 
This gauge gives the following action:
\begin{equation}
S=-\frac{1}{2\pi\alpha'}\int d\sigma
d\tau g_{\mu\nu}(X)\partial_{\alpha}
X^{\mu}\partial^{\alpha}X^{\nu},
\end{equation}
and the constraints:
\begin{equation}
\left(\partial_{\tau}X^{\mu}\pm\partial_{\sigma}X^{\mu}\right)^2=0.
\label{GCO}
\end{equation}
This is equivalent to saying that the energy-momentum tensor on 
the world-sheet must vanish in this gauge. The gauge choice we have made is 
known as the {\em conformal gauge} \cite{SSSS}.
In the conformal gauge we find the wave equation
\begin{equation}
\Box X^{\mu}=\left(\frac{\partial^2}{\partial
\sigma^2}-\frac{\partial^2}{\partial\tau^2}\right)\;X^{\mu}=0.\label{onda}
\end{equation}
We can
write the solution as a sum of two arbitrary left and right 
moving wave functions
\begin{equation}
X^{\mu}(\sigma,\tau)=X^{\mu}_{R}(\tau-\sigma)+X^{\mu}_{L}(\tau+\sigma).
\end{equation}
$X^{\mu}_{R}$ here describes the right-moving modes of the string
whilst $X^{\mu}_{L}$ describes the left-moving ones.\ Thus we can see
that the string coordinates are given by
\begin{equation}
X^{\mu}(\sigma,\tau)=q^{\mu}+2\alpha' p^{\mu}\tau +
i\sqrt{\alpha'}\sum^{\infty}_{n\neq 0}\frac{e^{-in\tau}}{n}\alpha^{\mu}_{n}
\cos n\sigma
\end{equation}
for open strings and
\begin{equation}
X^{\mu}(\sigma,\tau)=q^{\mu}+2\alpha' p^{\mu}\tau+
i\sqrt{\alpha'}\sum^{\infty}_{n\neq 0}
\frac{e^{-in\tau}}{n}\left(\alpha^{\mu}_{n}
e^{in\sigma}+\tilde{\alpha}^{\mu}_{n}e^{-in\sigma}\right)
\end{equation}
for closed strings.

Here, $q^{\mu}+2\alpha' p^{\mu}\tau$ describes the centre of mass coordinates 
and the $\alpha^{\mu}_{n}$ and $\tilde{\alpha}^{\mu}_{n}$ describe the 
right and left oscillation
modes of the string respectively. In order to simplify the notation, we will 
from here onwards substitute $\alpha'$ for its value $1/2$\footnote{
There are two conventions for the value of $\alpha'$. For open strings $\alpha'
=1$ whilst for closed strings $\alpha'=1/2$. Many of the results in String 
Theory are independent of whether we are working with closed strings or open 
strings so we will indicate which convention 
is followed only when the result is related 
to the open or closed string.}. 

Covariant quantisation gives the following
commutation relations:
\begin{eqnarray}
\left[X^{\mu}(\sigma,\tau),P^{\nu}_{\tau}(\sigma',\tau)\right]
&=&-i\eta^{\mu\nu}\delta
(\sigma-\sigma'),\nonumber\\
\left[X^{\mu}(\sigma,\tau),X^{\nu}(\sigma',\tau)\right]&=&0,\nonumber\\
\left[P^{\mu}_{\tau}(\sigma,\tau),P^{\nu}_{\tau}(\sigma',\tau)\right]&=&0,
\end{eqnarray}
which in turn yield the commutation relations for the operators 
$\alpha^{\mu}$ and 
$\tilde{\alpha}^{\mu}$.
\begin{eqnarray}
\left[\alpha^{\mu}_{m},\alpha^{\nu}_{-n}\right] =
 &  -m\delta_{m,n}\eta^{\mu\nu},& n,\;m>0\nonumber\\
\left[\alpha^{\mu}_{m},\alpha^{\nu}_{n}\right]=
&\left[\alpha^{\mu}_{-m},\alpha^{\nu}_{-n}\right]=0,&n,\;m>0\nonumber\\
\left[\alpha^{\mu}_{m},\tilde{\alpha}^{\nu}_{n}\right]=& 0 &\nonumber\\
\left[\tilde{\alpha}^{\mu}_{m},\tilde{\alpha}^{\nu}_{-n}\right]=&
-m\delta_{m,n}\eta^{\mu\nu},&n\;m>0,\nonumber\\
\left[\tilde{\alpha}^{\mu}_{m},\tilde{\alpha}^{\nu}_{n}\right]=&
\left[\tilde{\alpha}^{\mu}_{-m},\tilde{\alpha}^{\nu}_{-n}\right]=
0,&n,\;m>0.
\end{eqnarray} 
At the quantum level, the $\alpha_{n}$ are related to 
the familiar normalised harmonic oscillator operators by
\begin{eqnarray}
\alpha^{\mu}_{n}=&\sqrt{n}a^{\mu}_{n},& n>0\nonumber\\ 
\alpha^{\mu}_{-n}=&\sqrt{n}a^{\mu\dag}_{n},& n>0,
\end{eqnarray}
 with similar expressions for the $\tilde{\alpha}^{\mu}$. 
 Here the $\alpha^{\mu}_{n}$ and $\tilde{\alpha}^{\mu}_{n}$ are annihilation 
 operators whilst the $\alpha^{\mu}_{-n}$ and $\tilde{\alpha}^{\mu}_{-n}$ 
 are creation operators.
\section{The Virasoro algebra}
The Fock space is built up by applying the raising operators
$\alpha^{\mu}_{-n}$ to the ground state $|0\rangle$.\ This Fock space is
not positive definite since the time components of these operators have a
minus sign in their commutation relations,
$[\alpha^{0}_{n},\alpha^{0}_{-n}]=-n$, and therefore a state of the form
$\alpha^{0}_{-n}|0\rangle $ has negative norm since $\langle
0|\alpha^{0}_{n}\alpha^{0}_{-n}|0\rangle=-n$.\ From this discussion we 
may see that the physical space of
allowed string states consists of a subspace of the Fock space.\ This
subspace is specified by certain subsidiary conditions.\ For instance, in the
classical theory the vanishing of the world-sheet energy-momentum tensor 
represents the subsidiary conditions.\ The Fourier modes
of the world-sheet energy-momentum components give the Virasoro generators:
\begin{equation}
L_{m}=-\frac{1}{4\pi}\int^{\pi}_{-\pi}e^{im\sigma}
\left(\dot{X}+X'\right)^2 d\sigma
=-\frac{1}{2}\sum^{\infty}_{-\infty}\alpha_{n}\cdot\alpha_{m-n},
\label{VIR}
\end{equation}
as well as a similar expression $\tilde{L}_{m}$ in the case of closed strings.

Because of a normal ordering ambiguity arising from the expression for
$m=0$, we include an undetermined
constant $a$ and then we say that a physical state $|\phi\rangle$ must satisfy
\begin{equation}
(L_0-a)|\phi\rangle=0.
\label{consta}
\end{equation}
For closed string we have in addition the following condition:
\begin{equation}
(L_m-\tilde{L}_m)|\phi\rangle=0.
\label{consta2}
\end{equation}
Equations (\ref{consta}) and (\ref{consta2}) determine the mass of 
a string state in terms of its
internal state of oscillation. We can see this from eq.(\ref{VIR}) as follows:
\begin{equation}
-\frac{1}{2}\left(\alpha^{\mu}_0\right)^2-\sum^{\infty}_{n=1}
\alpha_{-n}\cdot\alpha_{n}-a=0,
\label{MA1}
\end{equation}
Here $\alpha^{\mu}_0$ and $\tilde{\alpha}^{\mu}_0$ are identified with the 
string momentum $p^{\mu}$ in the 
following way:
$$\left(\alpha^{\mu}_0\right)^2=p^{\mu}p_{\mu}$$
for open string, and
$$\left(\alpha^{\mu}_0\right)^2=\left(\tilde{\alpha}^{\mu}_0\right)^2=
\frac{1}{4}p^{\mu}p_{\mu}$$
for closed string. So substituting back in
expression (\ref{MA1}) we obtain:
\begin{equation}
\frac{1}{2}p^{\mu}p_{\mu}+\sum^{\infty}_{n=1}
\alpha_{-n}\cdot\alpha_{n}+a=0,
\label{MA2}
\end{equation}
for open strings, and
\begin{equation}
\frac{1}{8}p^{\mu}p_{\mu}+\sum^{\infty}_{n=1}
\alpha_{-n}\cdot\alpha_{n}+a=0,
\label{MA3}
\end{equation}
for closed strings.
With these expressions we obtain the mass shell condition for open 
and closed strings:
\begin{equation}
M^2=-2a-2\sum^{\infty}_{n=1}
\alpha_{-n}\cdot\alpha_{n},
\label{MASO}
\end{equation}
in the open string case; whilst for the closed string case we have:
\begin{equation}
M^2=-8a-8\sum^{\infty}_{n=1}
\alpha_{-n}\cdot\alpha_{n}=-8a-8
\sum^{\infty}_{n=1}
\tilde{\alpha}_{-n}\cdot\tilde{\alpha}_{n}.
\label{MASC}
\end{equation}
Expressions (\ref{MASO}) and (\ref{MASC}) show that the bosonic string ground 
state has a negative mass squared; that is, the ground state of a bosonic string 
is a {\em tachyon}.

Physical states also need to satisfy the following condition:
\begin{equation}
L_n|\phi\rangle=0
\end{equation}
for $n>0$, and an identical equation for $\tilde{L}_m$.
The algebra generated by these operators is:
\begin{equation}
\left[L_m,L_n\right]=\left(m-n\right)L_{m+n}+\frac{D}{12}\left(m^3-m\right)
\delta_{m,-n}\label{anomal}
\end{equation}
where $D$ is the dimension of the space-time. 
\subsection{Derivation of the Virasoro algebra}
As we can see from expression eq.(\ref{anomal}), the commutator of the 
Virasoro operator consists of two parts. The second term is a quantum anomaly 
not present in the classical theory.

Notice that the anomalous commutation relations of the 
Virasoro operators make it
impossible to find states annihilated by all of them.
\ A ghost-free spectrum is only possible
for certain values of the constant $a$ and the space-time dimension $D$.
If there are no ghosts among the allowed states, then Lorentz invariance
of the covariant formalism ensures that there are no ghosts in the
physical Hilbert space, 
which only makes physical sense when the space-time has the critical 
dimension of 26. In 26 dimensions there are 
no negative norm physical states.

In order to appreciate the statements above let us derive them explicitly. First 
let us obtain the classic commutator relations for the Virasoro operators.

From the definition of the Virasoro operator we have: 
\begin{equation}
\left[L_m,L_n\right]=\frac{1}{4}\sum^{\infty}_{k,l=-\infty}\left[
\alpha_{m-k}\cdot\alpha_k,\alpha_{n-l}\cdot\alpha_l\right],
\end{equation}
where the dot means a scalar product. 
We can then use the identity $\left[AB,CD\right]=A\left[B,C\right]D+
AC\left[B,D\right]+\left[A,C\right]DB+C\left[A,D\right]B$. So the 
commutator becomes:
\begin{eqnarray}
\left[L_m,L_n\right]&=&\frac{1}{4}\sum^{\infty}_{k,l=-\infty}
\left(k\alpha_{m-k}\cdot\alpha_l\delta_{k+n-l}+k\alpha_{m-k}\cdot
\alpha_{n-l}\delta_{k+l}\right.\nonumber\\
& & \left.+\left(m-k\right)\alpha_l\cdot\alpha_k\delta_{m-k+n-l}+
\left(m-k\right)\alpha_{n-l}\cdot\alpha_k\delta_{m-k+l}\right).
\end{eqnarray}
We can simplify the expression above obtaining the following:
\begin{equation}
\left[L_m,L_n\right]=\frac{1}{2}\sum^{\infty}_{k=-\infty}
\left(k\alpha_{m-k}\cdot\alpha_{k+n}+\left(m-k\right)\alpha_{m-k+n}\cdot
\alpha_k\right).
\end{equation}
Changing variables in the first term ($k\rightarrow k'=k+n$), we arrive at the 
classical version of the Virasoro algebra
\begin{equation}
\left[L_m,L_n\right]=\left(m-n\right)L_{m+n}.
\end{equation}
Let us proceed now to compute the quantum counterpart of the Virasoro 
algebra \cite{str16,strmex,green}. We start from
$$
L_m=-\frac{1}{4\pi}\int^{\pi}_{-\pi}d\sigma 
e^{im\sigma}\left(\dot{X}+X'\right)^2.
$$
Making the following change of variable: $z=e^{i\sigma}$ we have
\begin{equation}
L_m=i\frac{1}{4\pi}\int_{c(z)}dz z^{m-1}{\cal P}^{\mu}(z){\cal P}_{\mu}(z),
\end{equation}
where
\begin{equation}
{\cal P}^{\mu}(z)=\dot{X}^{\mu}+X'^{\mu}.
\label{PAR}
\end{equation}
From here we see that
\begin{eqnarray}
\left[L_m,L_n\right]&=&-\frac{1}{16\pi^2}
\int_{c(z)}\int_{c(w)}dz\,dw\, 
z^{m-1}w^{n-1}
{\cal P}^{\mu}(z){\cal P}_{\mu}(z){\cal P}^{\nu}(w){\cal P}_{\nu}(w)
\nonumber\\ & & +
\frac{1}{16\pi^2}\int_{c(z)}\int_{c(w)}dz\,dw\,z^{m-1}w^{n-1}
{\cal P}^{\mu}(w){\cal P}_{\mu}(w){\cal P}^{\nu}(z){\cal P}_{\nu}(z).
\end{eqnarray}
Time ordering this expression, we obtain:
\begin{eqnarray}
\left[L_m,L_n\right]&=&-\frac{1}{16\pi^2}
\int_{c(z)}\int_{c(w)}z^{m-1}w^{n-1}
\left[:{\cal P}^{\mu}(z){\cal P}_{\mu}(z){\cal P}^{\nu}(w){\cal P}_{\nu}(w):
\right.\nonumber\\ & &
\left.+4\underline{{\cal P}^{\mu}(z){\cal P}^{\nu}(w)}:{\cal P}_{\mu}(z)
{\cal P}_{\nu}(w):+
2\underline{{\cal P}^{\mu}(z){\cal P}^{\nu}(w)}\,
\underline{{\cal P}_{\mu}(z){\cal P}_{\nu}(w)}\right]+
\nonumber\\
& & \frac{1}{16\pi^2}\int_{c(z)}\int_{c(w)}z^{m-1}w^{n-1}
\left[:{\cal P}^{\mu}(w){\cal P}_{\mu}(w){\cal P}^{\nu}(z){\cal P}_{\nu}(z):
\right.\nonumber\\ & &
\left.+4\underline{{\cal P}^{\mu}(w){\cal P}^{\nu}(z)}:
{\cal P}_{\mu}(w){\cal P}_{\nu}(z):+
2\underline{{\cal P}^{\mu}(w){\cal P}^{\nu}(z)}\,
\underline{{\cal P}_{\mu}(w){\cal P}_{\nu}(z)}\right]
\label{Wick}
\end{eqnarray}
and $::$ means normal order and 
$\underline{\hspace*{1cm}}$ means a Dyson-Wick contraction \cite{shaw}:
$$\underline{AB}=AB-:AB:.$$
Using the explicit form of ${\cal P}^{\mu}(z)$ (\ref{PAR}) we find that
\begin{equation}
\underline{{\cal P}^{\mu}(z){\cal P}^{\nu}(w)}=
2\sum^{\infty}_{m=1}m\eta^{\mu\nu}\left(\frac{w}{z}\right)^m,
\label{Dyson}
\end{equation}
if and only if $w<z$. Substituting eq.(\ref{Dyson}) back into 
expression (\ref{Wick}) we obtain
\begin{eqnarray}
\left[L_m,L_n\right]&=&-\frac{1}{16\pi^2}
\int_{c(z)}\int_{c(w)}dz\;dw\;z^{m-1}w^{n-1}
\left[:{\cal P}^{\mu}(z){\cal P}_{\mu}(z){\cal P}^{\nu}(w){\cal P}_{\nu}(w):
\right.\nonumber\\ & &
\left.+\frac{8wz}{(z-w)^2}:{\cal P}^{\nu}(z){\cal P}_{\nu}(w):+
\frac{8w^2 z^2}{(z-w)^4}D
\right]\;\; w<z\nonumber\\
& &+\frac{1}{16\pi^2}\int_{c(z)}\int_{c(w)}dz\;dw\;z^{m-1}w^{n-1}
\left[:{\cal P}^{\mu}(w){\cal P}_{\mu}(w){\cal P}^{\nu}(z){\cal P}_{\nu}(z):
\right.\nonumber\\ & &
\left.+\frac{8wz}{(z-w)^2}:{\cal P}^{\nu}(w){\cal P}_{\nu}(z):+
\frac{8w^2 z^2}{(z-w)^4}D
\right]\;\; z<w.\label{comm}
\end{eqnarray}
\begin{figure}
\centerline{\epsfxsize=10cm\epsfbox{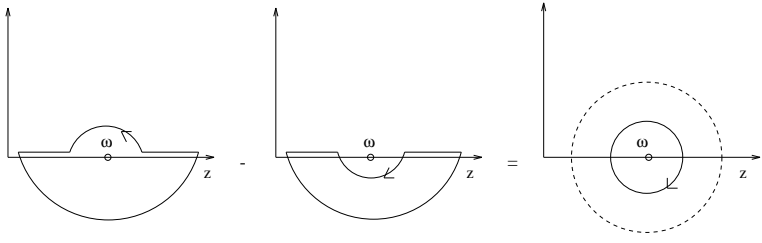}}
\caption{The z-contour for each w.}
\label{count}
\end{figure}
Using the contours of fig.(\ref{count}) we obtain the following results: the 
first integral in eq.(\ref{comm}) identically vanishes. For the second integral 
we have: 
\begin{eqnarray}
\int_{c(z)}\int_{c(w)}dz\;dw\;\frac{z^{m}w^{n}}{\left(z-w\right)^2}
:{\cal P}^{\nu}(z){\cal P}_{\nu}(w):&=&
2\pi i\; m\int_{c(w)}dw\frac{w^{n+m}}{w}:{\cal P}^{\nu}(w){\cal P}_{\nu}(w):
\nonumber\\
& &+2\pi i\int_{c(w)}dw\;w^{m+n}:{\cal P}'^{\nu}(w){\cal P}_{\nu}(w):
\nonumber\\ & &
\label{inttot}
\end{eqnarray}
where ${\cal P}'^{\nu}(w)=d{\cal P}(w)/dw$. The second integral in 
(\ref{inttot}) can be written as follows:
\begin{eqnarray}
2\pi i\int_{c(w)}dw\;w^{m+n}:{\cal P}'^{\nu}(w){\cal P}_{\nu}(w):&=&
\pi\; i\int_{c(w)}dw\frac{d}{dw}\left(w^{n+m}:{\cal P}^{\nu}(w)
{\cal P}_{\nu}(w):\right)
\nonumber\\
& &-\pi\; i\left(n+m\right)\int_{c(w)}dw\frac{w^{n+m}}{w}
:{\cal P}^{\nu}(w){\cal P}_{\nu}(w):\nonumber\\ & &
\end{eqnarray}
and from this expression we obtain:
\begin{eqnarray}
2\pi i\int_{c(w)}dw\;w^{m+n}:{\cal P}'^{\nu}(w){\cal P}_{\nu}(w):&=&
4\left(m+n\right)L_{m+n}.
\label{res1}
\end{eqnarray}
Looking now at the first integral in eq.(\ref{inttot}) we see that
\begin{equation}
2\pi i\; m\int_{c(w)}dw\frac{w^{n+m}}{w}:{\cal P}^{\nu}(w){\cal P}_{\nu}(w):=
-8m L_{m+n}.
\label{res2}
\end{equation}
Finally, the third integral in eq.(\ref{comm}) gives:
\begin{equation}
\int_{c(z)}\int_{c(w)}dz\;dw\;\frac{z^{m+1}w^{n+1}}{(z-w)^4}D=
-\frac{2\pi^2}{3}mD\left(m^2-1\right)\delta_{m,-n}.
\label{res3}
\end{equation}
Substituting eqs.(\ref{res1})-(\ref{res3}) back into eq.(\ref{comm}), 
we obtain our final result:
\begin{equation}
\left[L_m,L_n\right]=\left(m-n\right)L_{m+n}+\frac{D}{12}\left(m^3-m\right)
\delta_{m,-n}
\end{equation}
Now, let us derive the critical dimension of the space-time 
and the constant $a$ appearing in eq.(\ref{consta}). We need to vary the 
parameters $a$ and $D$ in order to find the regions where there are no negative 
norm states in the physical Hilbert space. For simplicity, let us work with the
open string case. If we denote the ground state of the open string with momentum 
$k^{\mu}$ as $|0;k\rangle$, we find that the mass-shell condition $L_0=a$ 
means that $k^2=-2a$. If we go to the first excited level, we will need to 
introduce a polarisation vector with $D$ independent components (before gauging 
away some of them). The states in the first excitation level are given by
$$\zeta\cdot\alpha_{-1}|0;k\rangle.$$
The mass-shell condition now implies $k^2=-2(a-1)$, and the $L_1$ subsidiary
condition implies that $\zeta\cdot k=0$. This means that we only have $D-1$ 
allowed polarisations. The norm of these states is given by $\zeta\cdot\zeta$. 
If the vector $k^{\mu}$ lies in the $(0,\vec{1})$ plane, then the $D-2$ states with
polarisation normal to that plane have positive norms. If 
the first excited state 
is a tachyon, then $k^2<0$ and $k^\mu$ can be chosen to have no time component. 
This state has a negative norm. If $k^2>0$, $k$ can be chosen to have only 
a time component and the norm will be positive. Finally if $k^2=0$, we find that
the norm is zero. With these results we find that one of the conditions 
for the absence of unphysical states is
$$a\leq 1.$$  

Now let us define a {\em spurious} state. A spurious state satisfies the following 
conditions:
$$\left(L_0-a\right)|\psi\rangle=0$$
and
$$\langle\phi|\psi\rangle=0,$$
for all physical states $|\phi\rangle$. These states can be written in the form 
$$|\psi\rangle=\sum_{n>0}L_{-n}|\chi_{n}\rangle,$$
where $|\chi_n\rangle$ is a state that satisfies
$$\left(L_0-a+n\right)|\chi_n\rangle=0.$$

Now, let us now construct a state that is both spurious and physical. 
With this in mind, let us define a spurious state of the following form:
$$|\psi\rangle=L_{-1}|\chi\rangle,$$
where $\chi$ is an arbitrary state which satisfies:  
$L_m|\chi\rangle=0$ for $m>0$ and $\left(L_0-a+1\right)|\chi\rangle=0.$ 
The state $|\psi\rangle$ is not a physical state as it stands. Let us apply 
the operator $L_1$ to this state thus:
$$L_1|\psi\rangle=L_1L_{-1}|\chi\rangle=2L_0|\chi\rangle,$$
which does not vanish unless $a=1$. Now let us construct another spurious 
state as follows:
$$|\psi\rangle=\left(L_{_2}+\gamma L^2_{-1}\right)|\chi\rangle.$$
This state needs to satisfy $L_1|\psi\rangle=L_2|\psi\rangle=0$ in order to be 
a physical state as well. From these conditions we find two equations:
$$\left(L_1L_{-2}+\gamma L_1 L^2_{-1}\right)|\psi\rangle=0$$ and
$$\left(L_2L_{-2}+\gamma L_2 L^2_{-1}\right)|\psi\rangle=0.$$
From these equations we find now that $\gamma=3/2$ and that $D=26$. 
Notice that from this discussion about spurious states we have found 
a set of states which are physical and whose norm is zero. That is, they are 
at the {\em border} between states with negative norm and states with positive 
norm. Therefore, we have found that $D=26$ is a critical dimension.
Of course, this derivation of the critical dimension $D$ of the space-time and 
of the value of the parameter $a$ is far from rigourous; however, it gives a 
physical insight as to how they emerge in String Theory. It can be proved that 
the bosonic string spectrum has no ghosts and the theory is unitary for 
precisely these values of the space-time dimension $D$ and the parameter $a$
\cite{gsw}. 

\section{Vertex operators}
\begin{figure}
\centerline{\epsfxsize=10cm\epsfbox{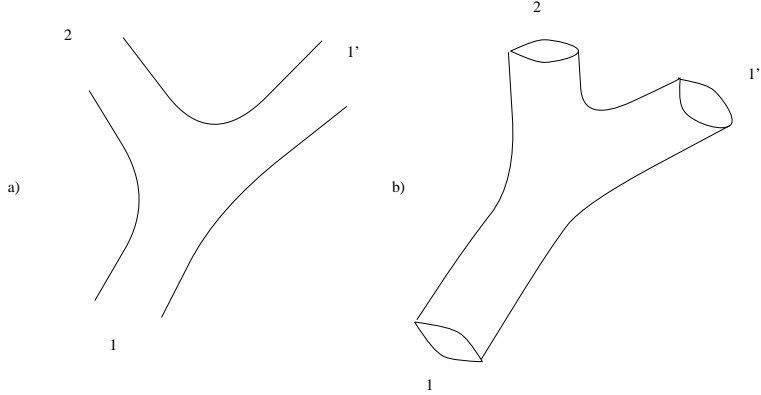}}
\caption{a) Splitting of open strings and b) closed strings.}
\label{inter}
\end{figure}

Interactions can be seen in the open String Theory as a process in
which a single string splits into two or one in which two strings join
together to form a single one (see for example fig.(\ref{inter})). In
String Theory one is naturally led to introduce an operator by means of
which a second string can be obtained from the original string. This
operator, in the open string case, is introduced at the ends of the
string where string 2 is emitted. 
\subsection{Conformal dimension}
Let us start by considering open strings only. Let us now introduce a local  
operator $A(\tau)$. This operator is in reality a function of both $\sigma$ and 
$\tau$ but because we want to study this operator at the end-points of the 
string \cite{gsw} (we are working with open strings) we have chosen 
$\sigma=0$ for simplicity, although we could have chosen the other 
end-point: $\sigma=2\pi$. $A(\tau)$ is defined to have conformal dimension 
$J$ {\bf if and only if} under reparametrization it transforms to 
$$
A'(\tau')=\left(\frac{d\tau}{d\tau'}\right)^J A(\tau).
$$
If we now consider an infinitesimal transformation:
$\tau\rightarrow\tau'=\tau+\epsilon(\tau),$
then the transformation law of an operator of conformal dimension $J$ reads:
\begin{equation}
\delta A(\tau)=-\epsilon\frac{d A}{d\tau}-JA\frac{d\epsilon}{d\tau}.
\end{equation}
If we look now at the commutation relations between the Virasoro operators and 
the local operator $A(\tau)$, we find that the condition for $A(\tau)$ to have 
conformal dimension $J$ is
\begin{equation}
\left[L_n,A(\tau)\right]=e^{in\tau}\left(-i\frac{d}{d\tau}+mJ\right)A(\tau).
\label{ConD}
\end{equation}
If an expansion in Fourier modes of the operator $A(\tau)$ is possible then 
eq.(\ref{ConD}) becomes:
\begin{equation}
\left[L_n,A_m\right]=\left[n\left(J-1\right)-m\right]A_{n+m}.
\end{equation}
Using this expression and after some algebra we find 
that the string coordinate $X^{\mu}(\tau)$ has 
conformal dimension $J=0$ and $\dot{X}^{\mu}(\tau)$ has conformal dimension 
$J=1$.
The operators that satisfy eq.(\ref{ConD}) are said to have a {\em definite} 
conformal dimension. These operators are useful because with them we can build 
physical states from others. For example, if we have a physical state $|\phi
\rangle$ and $A(\tau)$ has conformal dimension $J=1$, we can easily see that 
$$\left[L_n,A_0\right]=0$$ and therefore that 
$$|\phi'\rangle=A_0|\phi\rangle$$
is also a physical state. 

Looking back to the issue of constructing a vertex 
operator, we see that vertex operators need to map an initial physical state 
to a final physical state. From the discussion above it is expected that 
the open string operator should have conformal dimension $J=1$ ($J=2$ in the
case of closed strings). Vertex operators need to be conformally 
invariant in order for the theory to
make sense (to be anomaly free), which means that it needs to 
be invariant under
reparametrization of the world-sheet. The vertex operator $V(k,\tau)$
for the emission at time $\tau$ and $\sigma=0$ (or $\pi$) of a physical
state of momentum $k^\mu$ or absorption of a physical state of momentum
$-k^\mu$ has the following general structure:
\begin{equation}
V=\int d\tau\, W_v(X^{\mu},\dot{X}^{\mu},\ddot{X},\dots) 
e^{ik\cdot X(0,\tau)}
\end{equation}
for open strings, and
\begin{equation}
V=\int d\sigma d\tau\, W_v(X^{\mu},\dot{X}^{\mu},\ddot{X},\dots) 
e^{ik\cdot X(\sigma,\tau)}
\end{equation}
for closed strings.

It must change, amongst several other things, the momentum of the
states it acts on by an amount $k^\mu$. This is achieved by the 
introduction of the factors
$ e^{ik\cdot X(0,\tau)}$ and $ e^{ik\cdot X(\sigma,\tau)}$
which appear in the vertex operator for open and closed strings 
respectively. Notice that these factors by 
themselves require normal
ordering. 

We can see that the main difference between open and closed string vertex 
operators is that in 
the closed String Theory the vertex operator is introduced in the world-sheet
instead of the ends of the string.
Another difference is that the conformal dimension of the vertex
operator differs by a factor of 2 from that for the open string.
\section{The bosonic string spectrum}
\begin{figure}
\centerline{\epsfxsize=10cm\epsfbox{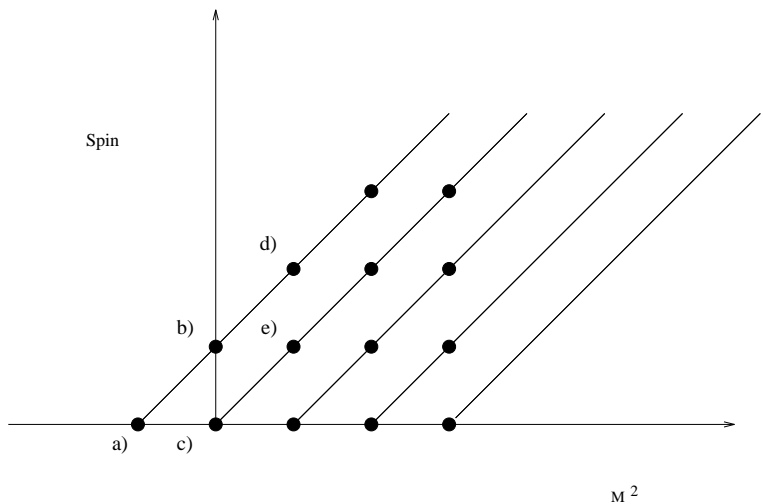}}
\caption{Regge trajectories for open strings.}
\label{specopen}
\end{figure}

\begin{figure}
\centerline{\epsfxsize=10cm\epsfbox{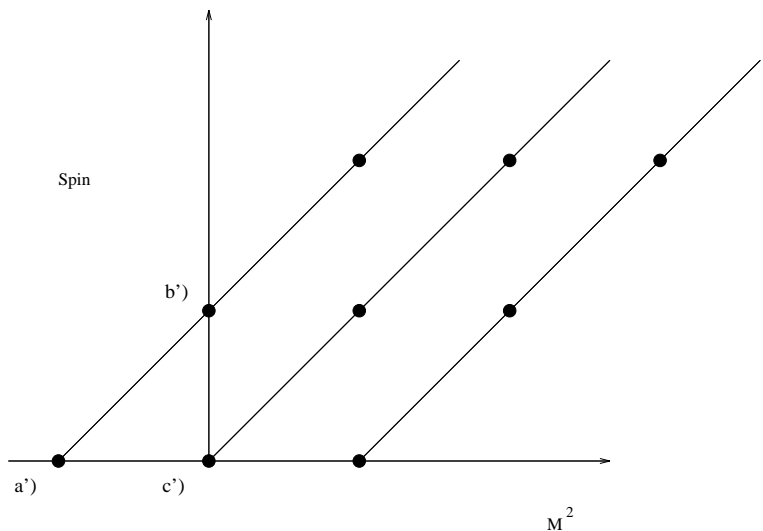}}
\caption{Regge trajectories for closed strings.}
\label{specclose}
\end{figure}

It was in the early 1960's that it was realised that 
hadronic resonances possessed rather high 
spins which increased proportionally to the mass squared of the 
state following 
roughly the prescription: $m^2=J/\alpha'$ where $J$ is the spin of the state, 
$m$ the mass of the state and $\alpha'$ again the Regge slope. If we 
plot this 
with the energy squared on the $x-$axis and the spin of the state on 
the $y-$axis, then we obtain a curve called the {\em Regge trajectory}.\
As is well known today, String Theory started as an attempt to explain 
the physics of the strong interaction including this behaviour of 
the mass spectrum. Indeed the theory possesses a similar behaviour for the mass 
spectrum of its states. At that time the setbacks of the theory (as a theory 
for strong interactions) were 
(amongst others) that the string spectrum also 
has massless states which do not belong to the hadronic world of strong 
interactions and it has the wrong Regge slope. However, nowadays 
we know that String Theory is not the correct 
theory for studying strong interactions but rather a theory which 
may provide us with a unified picture of all known interactions. Having said 
this, let us see what the string spectrum looks like. The spectrum of the 
lower-lying states in the open string case 
can be categorised as (see fig.\ref{specopen}):
\begin{center}
a) Tachyon$\rightarrow |0\rangle$\\
b) Massless vector$\rightarrow a^{\dag\mu}_{1}|0\rangle$\\
c) Massless scalar$\rightarrow k_{\mu}a^{\dag\mu}_{1}|0\rangle$\\
d) Massive spin two$\rightarrow a^{\dag\mu}_{1}a^{\dag\nu}_{1}|0\rangle$\\
e) Massive vector$\rightarrow a^{\dag\mu}_{2}|0\rangle$
\end{center}
and for closed string it can be categorised as\footnote{The masses of these 
particles may be obtained quickly from the conformal 
dimension of the appropriate vertex operator as discussed in the previous 
section. In short, knowing that the vertex operators for open 
string states have to have conformal dimension $J=1$ and the ones for closed string 
states have to have conformal dimension $J=2$, we can calculate the mass squared of 
the state in question: $J_v=-k^2/2$ for open strings and $J_v=-k^2/4$ for closed 
strings. Here $J_v$ is the conformal dimension of the factor 
$e^{ik\cdot X}$ appearing in all vertex operators of String Theory.}
(see fig.\ref{specclose})
\begin{center}
a') Tachyon$\rightarrow |0\rangle$\\
b') Massless spin two$\rightarrow a^{\dag\mu}_{1}a^{\dag\nu}_{1}|0\rangle$\\
c') Massless scalar$\rightarrow
k_{\mu}k_{\nu}a^{\dag\mu}_{1}\tilde{a}^{\dag\nu}_{1}|0\rangle$
\end{center}

It is the massless spin two particle appearing in the mass spectrum of
the closed string that made people think of String Theory as a possible
theory of quantum gravity.\ This particle is the graviton.\ As we can
see, the graviton enters the theory as a member of a multiplet.\ The
other important massless particle appearing in the spectrum for closed
string is the scalar massless particle.\ This particle is the dilaton.

So we can see that the description of free bosonic strings gives a theory 
which is much more constrained than any other corresponding theory 
for point-particles.
\section{Fixing the gauge}
There are three formalisms in which to fix the gauge of the theory. This means 
that there exist at least three ways of first quantising the string \cite{kak}:
\begin{enumerate}
\item{The Gupta-Bleuler is perhaps the simplest of the three
formalisms.\ We allow ghosts to appear in the action, which permits us
to maintain manifest Lorentz invariance.\ The price we must pay,
however, is that we must impose ghost-killing constraints on the Hilbert
space.\ Projection operators must be inserted in all propagators.\ At
tree level, but for higher loops, this is extremely
difficult.

The Gupta-Bleuler formalism will maintain Lorentz invariance by imposing
the Virasoro constraints on the state vectors of the theory:
\begin{equation}
\langle\phi|L_{f}|\psi\rangle=0
\end{equation}
where $\langle\phi|$ and $|\psi\rangle$ represent states of the theory.\
These constraints will eliminate the ghost which appears in the state
vectors.\ Therefore, we are allowed to maintain the non-physical states of
the theory in the action.}
\item{The light cone gauge formalism has the advantage that it is
explicitly ghost-free in the action as well as in the Hilbert space.\
There exist no complications when we go to higher loops.\ However, the
formalism is slightly awkward and we have to check Lorentz invariance at
each step.}
\item{The BRST formalism combines the best features of the two formalisms
mentioned above.\ It is manifestly covariant, like the Gupta-Bleuler
formalism, and it is unitary, like the light cone formalism.\ This is
because the negative norm ghosts cancel against the Faddeev-Popov
ghosts} which need to be introduced in the string action.
\end{enumerate}
We will be using mainly the Gupta-Bleuler formalism. Notice that 
most of the results presented in this chapter were derived in this formalism.
\section{String compactification}
One of the most important problems of String Theory is the one regarding
the extra dimensions of the theory. As we have seen, bosonic string
theory is only consistent in a 26-dimensional space-time whereas the
more realistic superstring counterpart only makes sense in
a 10-dimensional space-time. Therefore, we need somehow to compactify the
theory to a realistic 4-dimensional one (see for example 
\cite{Bai1}~-~\cite{Bai9}). Until such a reduction is made,
the theory will have a lack of contact with real physical observables. 

Because we do not have a quantum field theory framework on which this
dimensional breaking can possibly be done, the best we can do is to look
at classical solutions where spontaneous compactification of the
extra dimensions has already taken place.


\chapter{The quantum bosonic string energy-momentum tensor 
  in Minkowski space-time}
\label{ChapTmunu}  
In this chapter\footnote{This chapter is based upon  
  work with E.\@ J.\@ Copeland and H.\@ J.\@ de Vega.} and the next, I 
  will present some of the main derivations and results of this 
thesis. Here we will compute the quantum energy-momentum tensor 
${\hat T}^{\mu\nu}(x)$ for bosonic strings in
Minkowski space-time \cite{CVV}. Its expectation value, for different
physical string states both for open and closed bosonic strings, will be 
computed. 
The states considered are described by normalizable wave-packets in
the centre of mass coordinates. We will find in particular that 
${\hat T}^{\mu\nu}(x)$ loses its locality, which could imply 
that the classical
divergence  
that occurs in String Theory as we approach the string position is removed at 
the quantum level as the string position is smeared out by quantum
fluctuations. 
\section{Introduction}
String Theory has emerged as the most promising
candidate to reconcile general relativity with quantum mechanics and
unify gravity with the other fundamental interactions. 
Because of this merging of quantum field theory with general 
relativity in String Theory, it makes sense to investigate the gravitational 
consequences of strings as we approach the Planck scale.
We must remember that when particles scatter at energies of the order of or 
larger than the
Planck mass, the interaction that dominates their collision is the
gravitational one. At these energies the picture of particle fields or
strings in flat space-time ceases to be valid, the curved space-time
geometry created by the particles has to be taken into account.\ This
has been the motivation here, to investigate the possible gravitational
effects arising from an isolated quantum bosonic string living in a flat
space-time background, so we may begin a study of the
scattering process of strings merely by the gravitational interaction
between them.\ 
A systematic and thorough study of quantum strings in physically relevant 
curved space-times has been started in \cite{dvs} and has been reviewed 
in ref.~\cite{ERI}.  

As we have said, in this chapter, we calculate the energy-momentum tensor 
of both closed and open
quantum bosonic strings in $3+1$ dimensions. Our target space is assumed to 
be the direct product of four dimensional Minkowski space-time times a compact
manifold taking care of  conformal anomalies.

As a starting point we recall that it has already been shown \cite{edal} 
that the 
back-reaction for a classical bosonic string in $3+1$ dimensions has a
logarithmic divergence when the space-time coordinate 
$x\rightarrow X(\sigma)$; that is, when we
approach the core of the string.\ This divergence is absorbed 
into a renormalization of the string tension.\ Copeland et al \cite{edal} 
showed that by demanding that both 
it and the divergence in the energy-momentum tensor
vanish, the string is forced to have the couplings of compactified $N=1$,
$D=10$ supergravity.\ In this calculation we will be able to see that when 
we take into
account the quantum nature of the strings, we lose all  
information regarding the position of the
string and therefore any divergences that may appear when one calculates
the back-reaction of quantum strings are not related at all to the classical 
position of the string.

In the present work, the energy-momentum tensor of the string,
${\hat T}^{\mu\nu}$, is a quantum operator and it may be regarded 
as a vertex operator for the
emission (absorption) of gravitons. We compute 
its expectation value in one-particle string
states, choosing for the string centre of mass wave function a wave-packet
centred at the origin.  
Notice that this computation will preserve all the {\em stringy} 
features, in 
contrast to similar computations presented in \cite{HJV}, 
where these features 
are lost since they integrate the energy-momentum tensor of 
the bosonic string 
over a spatial volume totally enclosing the string.

Our results will depend on the mass of the string state
chosen. We will concentrate mainly on the string massless states although 
some results for the string tachyonic state will also be shown.
In our computation, we consider spherically symmetric and cylindrically 
symmetric configurations for our string states.
The components for the string
energy density and energy flux behave like massless waves, with the string 
energy being radiated outwards as a massless lump peaked at $r=t$
(in the spherically symmetric case). We
provide integral representations for $\langle{\hat T}^{\mu\nu}(r,t)\rangle$
[eq.(\ref{tmassles})]. We can see that $\langle{\hat T}^{\mu\nu}(r,t)\rangle$ 
propagates as outgoing (plus ingoing ) spherical waves.
After exhibiting the tensor structure of  $\langle{\hat T}^{\mu\nu}\rangle$
[eqs.(\ref{rotaT})-(\ref{forfac})], the asymptotic behaviour of  
$\langle{\hat T}^{\mu\nu}(r,t)\rangle$ 
for $ r \to \infty $ and $ t $ fixed and for  $ t \to \infty $ with $
r $ fixed is computed. 
In the first regime the energy density and the stress
tensor decay as $ r^{-1} $ whereas the energy flux decays as  $ r^{-2} $.
For  $ t \to \infty $ with $ r $ fixed, the energy density tends to $ 0^- $
as $ e^{-t^2} $. That is, the spherical wave leaves behind a rapidly
vanishing negative energy density.

For cylindrically symmetric configurations,  
$\langle{\hat T}^{\mu\nu}(\rho,t)\rangle$
propagates as outgoing (plus ingoing ) cylindrical waves. For large $
\rho $ and fixed $ t $ the energy density  decays
as $1/\rho$ and the energy flux decays as $1/\rho^2$. For large   $ t
\to \infty $ with $ \rho $ fixed.

\section{The string energy-momentum tensor}
We recall that the energy-momentum tensor for a classical bosonic string 
with tension
$(\alpha')^{-1}$ (see for example \cite{HJV}) is given by 
\begin{equation}\label{six}
    T^{\mu\nu}(x)=\frac{1}{2\pi\alpha'}\int d\sigma d\tau   \;
(\dot{X}^{\mu}\dot{X}^{\nu}-X'^{\mu}X'^{\nu})\delta(x-X(\sigma,\tau))
\end{equation}
This expression is easily obtained from the Polyakov form of 
the string action 
presented in the previous chapter by varying the action with respect to the 
space-time metric. The delta function in the expression above comes from the 
fact that the string energy-momentum tensor needs to vanish everywhere except 
at the position of the string. Jumping a few steps ahead, 
we notice that at the 
quantum level the locality given by this delta function will disappear since 
quantum fluctuations will smear the localised behaviour of the 
classical string. 

The string coordinates are given in Minkowski space-time by
\begin{equation}\label{seven}
    X^{\mu}(\sigma,\tau)=q^{\mu}+2\alpha'p^{\mu}\tau
    +i\sqrt{\alpha'}\sum_{n\neq 0}\frac{1}{n}[\alpha^{\mu}_{n}\;
    e^{-in(\tau-\sigma)}
    +\tilde{\alpha}^{\mu}_{n}\;e^{-in(\tau+\sigma)}]
\end{equation}
for closed strings and
\begin{equation}\label{sevena}
    X^{\mu}(\sigma,\tau)=q^{\mu}+2\alpha'p^{\mu}\tau+
    i\sqrt{\alpha'}\sum_{n\neq 0}\frac{1}{n}\; 
    \alpha^{\mu}_{n}\;e^{-in\tau}\cos n\sigma
\end{equation}
for open strings, as was also stated in the previous chapter.

For closed strings we can set $\alpha'=1/2$; thus,  
inserting eq.(\ref{seven}) in eq.(\ref{six}) and rewriting the four 
dimensional delta function in integral form, we obtain for closed strings:
\begin{eqnarray}
T^{\mu\nu}(x)=\frac{1}{\pi}\int d\sigma
d\tau\frac{d^{4}\lambda}{(2\pi)^{4}}
\{p^{\mu}p^{\nu}+\frac{p^{\mu}}{\sqrt{2}}\sum_{n\neq0}
[\alpha^{\nu}_{n}e^{-in(\tau-\sigma)}+\tilde{\alpha}^{\nu}_{n}
e^{-in(\tau+\sigma)}]+\nonumber\\
 +\sum_{n\neq0}(\alpha^{\mu}_{n}
e^{-in(\tau-\sigma)}+\tilde{\alpha}^{\mu}_{n}
e^{-in(\tau+\sigma)})\frac{p^{\nu}}{\sqrt{2}}
+\nonumber\\+\sum_{n\neq0}\sum_{m\neq0}[\alpha^{\mu}_{n}
\tilde{\alpha}^{\nu}_{m}e^{-in(\tau-\sigma)}e^{-im(\tau+\sigma)}
+\tilde{\alpha}^{\mu}_{n}\alpha^{\nu}_{m}e^{-in(\tau+\sigma)}
e^{-im(\tau-\sigma)}]\}\;e^{i\lambda\cdot x}\;e^{-i\lambda\cdot
X(\sigma,\tau)}.\label{twelve}
\end{eqnarray}
We can now write
$$X(\sigma,\tau)=X_{cm}+X_{+}+X_{-},$$
where $X_{cm}=q+p\tau$ is the centre of mass coordinate and 
$X_{+}$ and $X_{-}$ refer to the terms with $\alpha_{n>0}$ and
$\alpha_{n<0}$ in $X(\sigma,\tau)$ respectively.
\ In this way we can 
now see that our energy-momentum tensor has the same form as that of a vertex
operator \cite{gsw, kak}. 

We should notice that eq.(\ref{six}) is meaningful 
at the classical level.\ However, at 
the quantum level one must be careful with the order of the operators 
since $\dot{X}^{\mu}$ and $\dot{X}^{\nu}$
do not commute with $X(\sigma,\tau)$. We shall define the quantum
operator $ {\hat T}^{\mu\nu}(x) $ by symmetric ordering. That is,
\begin{equation}
{\hat T}^{\mu\nu}(x)\equiv\frac{1}{3}\left[{\hat T}^{\mu\nu}_{a}(x)+
{\hat T}^{\mu\nu}_{b}(x)+{\hat T}^{\mu\nu}_{c}(x)\right]
\label{Tcua}
\end{equation}
where
\begin{eqnarray}\label{Tcuaa}
   {\hat T}^{\mu\nu}_{a}(x)&=& \frac{1}{2\pi\alpha'}\int d\sigma d\tau   
\left(\dot{X}^{\mu}\dot{X}^{\nu}-X'^{\mu}X'^{\nu}\right)
\delta(x-X(\sigma,\tau))\nonumber\\  
{\hat T}^{\mu\nu}_{b}(x)&=& \frac{1}{2\pi\alpha'}\int d\sigma d\tau 
\delta(x-X(\sigma,\tau))
\left(\dot{X}^{\mu}\dot{X}^{\nu}-X'^{\mu}X'^{\nu}\right)\nonumber\\ 
{\hat T}^{\mu\nu}_{c}(x)&=& \frac{1}{2\pi\alpha'}\int d\sigma d\tau 
\left[\frac{1}{2}\left(\dot{X}^{\mu}\delta(x-X(\sigma,\tau))\dot{X}^{\nu}-
X'^{\mu}\delta(x-X(\sigma,\tau))X'^{\nu}+\right.\right.\nonumber\\ & &
\left.\left.
\dot{X}^{\nu}\delta(x-X(\sigma,\tau))\dot{X}^{\mu}-
X'^{\nu}\delta(x-X(\sigma,\tau))X'^{\mu} 
\right)\right].
\end{eqnarray}
This definition ensures hermiticity:
$$
{\hat T}^{\mu\nu}(x)^{\dag}= {\hat T}^{\mu\nu}(x).
$$
Since we are interested in computing an 
expectation for the above quantum operator, 
we must use invariantly normalizable particle states as 
wave packets.

Let us consider a string on a mass and spin eigenstate with a centre
of mass wave function.\ An 
on-shell scalar string state is then 
\begin{equation}
 | \Psi\rangle = \int d^4 p \;
\frac{\varphi(\vec{p})}{E}\; 
\delta(p^{0}-\sqrt{\vec{p}^{2}+m^{2}})\;  | p \rangle
.
\end{equation}
Where 
$$E=p^0=\sqrt{\vec{p}^2+m^2}$$
and $\varphi(\vec{p})$ is a wave-packet that we are free to choose.

Here we assume the extra space-time dimensions (beyond four) to be
appropriately compactified and consider string states in the physical
(uncompactified) four dimensional Minkowski space-time. 

This string state needs to be correctly normalised. Evaluating the scalar 
product we obtain:
\begin{eqnarray}
\langle\Psi|\Psi\rangle &=&\int d^3p d^3p'\delta^{3}(\vec{p}-\vec{p}')
\delta(p'^{0}-\sqrt{p'^{2}+m^2})\delta(p^{0}-\sqrt{p^{2}+m^2})
\, dp^0 dp'^0\delta(p^0-p'^0)\times\nonumber\\ & & 
\frac{\varphi^{*}(\vec{p})}{E}\frac{\varphi(\vec{p}')}{E'}=
\int d^3p\frac{|\varphi(\vec{p})|^2}{E^2} \delta(0).
\label{escalar}
\end{eqnarray}
The last Dirac delta can be regularised by considering a large finite temporal 
box of size~$T$.
\begin{equation}
\delta(0)=\frac{1}{2\pi}\int^{T}_{0}dt=\frac{T}{2\pi}.
\end{equation}
Hence we have
\begin{equation}
\langle\Psi|\Psi\rangle =\frac{T}{2\pi}
\int d^3p\frac{|\varphi(\vec{p})|^2}{E^2}.
\label{promedio}
\end{equation}
 
Returning to ${\hat T}^{\mu\nu}(x)$, because it is a quantum operator, it 
requires normal ordering. For closed strings we obtain:  
\begin{eqnarray}
\langle\Psi|\hat{T}^{\mu\nu}_{a}(x)|\Psi\rangle &=& \frac{1}{\pi}
\int d\sigma d\tau
\frac{d^4\lambda}{(2\pi)^4}\;e^{i\lambda\cdot x}
\left\{\langle\Psi|e^{-i\lambda\cdot X_{-}}
p^{\mu}p^{\nu}e^{-i\lambda\cdot X_{+}}e^{-i\lambda\cdot X_{cm}}
|\Psi\rangle+\right.\nonumber\\ & & \left.
\langle \Psi|e^{-i\lambda\cdot X_{-}}\frac{p^{\mu}}{\sqrt{2}}
\sum_{n=1}\left[\alpha^{\nu}_{-n}e^{in(\tau-\sigma)}+
\tilde{\alpha}^{\nu}_{-n}e^{in(\tau+\sigma)}\right]e^{-i\lambda\cdot X_{+}}
e^{-i\lambda\cdot X_{cm}}
|\Psi\rangle +
\right.\nonumber\\ & & 
\left.  \langle \Psi|e^{-i\lambda\cdot X_{-}}
\sum_{n=1}\left[\alpha^{\nu}_{n}e^{-in(\tau-\sigma)}+
\tilde{\alpha}^{\nu}_{n}e^{-in(\tau+\sigma)}\right]\frac{p^{\mu}}{\sqrt{2}}
e^{-i\lambda\cdot X_{+}}
e^{-i\lambda\cdot X_{cm}}|\Psi\rangle+\right.\nonumber\\ & & \left. 
\sum_{n=1}\sum_{m=1}\langle\Psi|e^{-i\lambda\cdot X_{-}}
\alpha^{\mu}_n\tilde{\alpha}^{\nu}_m
e^{-in(\tau-\sigma)}e^{-im(\tau+\sigma)}e^{-i\lambda\cdot X_{+}}
e^{-i\lambda\cdot X_{cm}}|\Psi\rangle+\right.\nonumber\\ & & \left.
\sum_{n=1}\sum_{m=1}\langle\Psi|e^{-i\lambda\cdot X_{-}}
\alpha^{\mu}_{-n}\tilde{\alpha}^{\nu}_{-m}
e^{in(\tau-\sigma)}e^{im(\tau+\sigma)}e^{-i\lambda\cdot X_{+}}
e^{-i\lambda\cdot X_{cm}}
|\Psi\rangle+\right.\nonumber\\ & & 
\left.\sum_{n=1}\sum_{m=1}\langle\Psi|e^{-i\lambda\cdot X_{-}}
\alpha^{\mu}_{n}\tilde{\alpha}^{\nu}_{-m}
e^{-in(\tau-\sigma)}e^{im(\tau+\sigma)}e^{-i\lambda\cdot X_{+}}
e^{-i\lambda\cdot X_{cm}}|\Psi\rangle+\right.\nonumber\\ & & \left.
\sum_{n=1}\sum_{m=1}\langle\Psi|e^{-i\lambda\cdot X_{-}}
\alpha^{\mu}_{-n}\tilde{\alpha}^{\nu}_{m}
e^{in(\tau-\sigma)}e^{-im(\tau+\sigma)}e^{-i\lambda\cdot X_{+}}
e^{-i\lambda\cdot X_{cm}}
|\Psi\rangle +\right.\nonumber\\ & & 
\left.
\sum_{n=1}\sum_{m=1}\langle\Psi|e^{-i\lambda\cdot X_{-}}
\tilde{\alpha}^{\mu}_n\alpha^{\nu}_m
e^{-in(\tau+\sigma)}e^{-im(\tau-\sigma)}e^{-i\lambda\cdot X_{+}}
e^{-i\lambda\cdot X_{cm}}|\Psi\rangle+\right.\nonumber\\ & & \left.
\sum_{n=1}\sum_{m=1}\langle\Psi|e^{-i\lambda\cdot X_{-}}
\tilde{\alpha}^{\mu}_{-n}\alpha^{\nu}_{-m}
e^{in(\tau+\sigma)}e^{im(\tau-\sigma)}e^{-i\lambda\cdot X_{+}}
e^{-i\lambda\cdot X_{cm}}
|\Psi\rangle +\right.\nonumber\\ & & 
\left.\sum_{n=1}\sum_{m=1}\langle\Psi|e^{-i\lambda\cdot X_{-}}
\tilde{\alpha}^{\mu}_{n}\alpha^{\nu}_{-m}
e^{-in(\tau+\sigma)}e^{im(\tau-\sigma)}e^{-i\lambda\cdot X_{+}}
e^{-i\lambda\cdot X_{cm}}|\Psi\rangle+\right.\nonumber\\ & & \left.
\sum_{n=1}\sum_{m=1}\langle\Psi|e^{-i\lambda\cdot X_{-}}
\tilde{\alpha}^{\mu}_{-n}\alpha^{\nu}_{m}
e^{in(\tau+\sigma)}e^{-im(\tau-\sigma)}e^{-i\lambda\cdot X_{+}}
e^{-i\lambda\cdot X_{cm}}
|\Psi\rangle\right\}
\label{RR1}
\end{eqnarray}
and we obtain similar expressions for the other terms in eq.(\ref{Tcua}). 
It is clear that in the tachyonic case, since 
$\hat{T}^{\mu\nu}(x)$ is already normal ordered, $e^{-i\lambda\cdot X_{+}}$, 
will annihilate the ground state. Thus, we only need to worry about the 
action of $e^{-i\lambda\cdot X_{cm}}$ on the state $|p_2\rangle$ (Here, 
$|p_2\rangle$ is the initial momenta and $|p_1\rangle$ is the final momenta). 
We obtain the following result:
\begin{eqnarray}
\frac{\langle \Psi | {\hat T}^{\mu\nu}(x) | \Psi \rangle}
 {\langle\Psi |\Psi\rangle} \equiv  \langle
{\hat T}^{\mu\nu}(x)\rangle 
=  & & \nonumber\\ \frac{1}{3\pi\langle\Psi |\Psi\rangle}\int\frac{d^{4}\lambda}
{(2\pi)^{4}}\; d^4p_{1}d^4p_{2}\;d\sigma 
d\tau \; e^{i\lambda\cdot x}[p^{\mu}_{1}p^{\nu}_{1} + 
p^{\mu}_{2}p^{\nu}_{2} +
\frac{1}{2}\left(p^{\mu}_{1}p^{\nu}_{2}+p^{\nu}_{1}p^{\mu}_{2}\right)]
\times \nonumber\\  
\langle p_{1}|e^{-i\lambda\cdot X_{cm}}|p_{2}\rangle\;
\frac{\varphi^{*}(\vec{p}_{1})}{E_1}\frac{\varphi(\vec{p}_{2})}{E_2}
\; \delta(p^{0}_{1}-\sqrt{\vec{p}_{1}^{2}+m^{2}_{1}})
\delta(p^{0}_{2}-\sqrt{\vec{p}_{2}^{2}+m^{2}_{2}}). & &
\label{fifteen} 
\end{eqnarray} 
Writing
\begin{equation}
\langle p_{1}|e^{-i\lambda\cdot X_{cm}}|p_{2}\rangle =
e^{i\frac{\tau\lambda^{2}}{2}-i\lambda\cdot p_{2}\tau}
\delta^{4}(\lambda+p_{1}-p_{2}),
\label{lm}
\end{equation}
eq.(\ref{fifteen}) becomes
\begin{eqnarray}
\langle {\hat T}^{\mu\nu}(x)\rangle =\frac{2}{3\langle\Psi |\Psi\rangle}
\int\frac{d^{4}\lambda}{(2\pi)^{3}}\; 
d^4p_{1}d^4p_{2}\;d\tau \; e^{i\lambda\cdot x}\; 
[p^{\mu}_{1}p^{\nu}_{1}+p^{\mu}_{2}p^{\nu}_{2} +
\frac{1}{2}\left(p^{\mu}_{1}p^{\nu}_{2}+p^{\nu}_{1}p^{\mu}_{2}
\right)]\times\nonumber\\  
e^{i\frac{\tau\lambda^{2}}{2}-i\lambda p_{2}\tau}
\; \delta^{4}(\lambda+p_{1}-p_{2}) 
\frac{\varphi^{*}(\vec{p}_{1})}{E_1}\frac{\varphi(\vec{p}_{2})}{E_2}\; 
\delta(p^{0}_{1}-\sqrt{\vec{p}_{1}^{2}+m^{2}_{1}})
\delta(p^{0}_{2}-\sqrt{\vec{p}_{2}^{2}+m^{2}_{2}})\; .
\end{eqnarray}
Performing the $\lambda$, $p^0_1$ and $p^0_2$ integrals, we get:  
\begin{eqnarray}\label{taq}
\langle {\hat T}^{\mu\nu}(x)\rangle =
\frac{4}{3(2\pi)^{3}\langle\Psi |\Psi\rangle}\int d^3p_{1}d^3p_{2}
\; e^{i(p_{2}-p_{1})\cdot x}
\; [p^{\mu}_{1}p^{\nu}_{1}+p^{\mu}_{2}p^{\nu}_{2} +
\frac{1}{2}\left(p^{\mu}_{1}p^{\nu}_{2}+p^{\nu}_{1}p^{\mu}_{2}
\right)]\times\nonumber\\ 
\;\frac{\varphi^{*}(\vec{p}_{1})}{E_1}\frac{\varphi(\vec{p}_{2})}{E_2}\; 
\int^{\tau_1}_{\tau_2}d\tau\; e^{i\frac{\tau}{2}(p^2_1-p^2_ 2)}.
\end{eqnarray} 
The calculation for open strings is obtained by substituting
eq.(\ref{sevena}) into eq.(\ref{Tcua}), with the result 
(setting $\alpha'=1$):
\begin{eqnarray}
\langle {\hat T}^{\mu\nu}(x)\rangle =
\frac{2}{3(2\pi)^{3}\langle\Psi |\Psi\rangle}\int d^3p_{1}d^3p_{2}
\; e^{i(p_{2}-p_{1})\cdot x}\; 
[p^{\mu}_{1}p^{\nu}_{1}+p^{\mu}_{2}p^{\nu}_{2} +
\frac{1}{2}\left(p^{\mu}_{1}p^{\nu}_{2}+p^{\nu}_{1}p^{\mu}_{2}
\right)]\times\nonumber\\ 
\;\frac{\varphi^{*}(\vec{p}_{1})}{E_1}\frac{\varphi(\vec{p}_{2})}{E_2}\; 
\int^{\tau_1}_{\tau_2}d\tau\; e^{i\frac{\tau}{2}(p^2_1-p^2_2)}.
\end{eqnarray}
The limits in the $\tau$ integral, come from some very early time 
$\tau_2$ to some much later time $\tau_1$. We are thinking here of 
state two as the {\em in state} and state one as the {\em out state}. 

Now, we know from the fact we are on shell (i.e. the $p^0$ integrals 
we have performed) that $p^2_1=p^2_2=m^2$. Hence the $tau$ integral becomes:
$$\int^{\tau_1}_{\tau_2}d\tau=\tau_1-\tau_2.$$
This time is not the physical time $T$ and we 
would like to write it in terms of the physical time. We can 
do this from the earlier expression for the coordinate of the string. 
We expect at very early and very late times that 
the asymptotic behaviour of the string to be dominated by the zero modes. In 
particular for the physical time coordinate $X^0$ we can write it as
$$X^0=q^0+2\alpha'p^0\tau,$$
a result true in the limits $\tau\rightarrow\pm\infty$, 
$X^0\rightarrow\pm\infty$. Let us call the initial 
physical time $-T/2$ and the final physical time $+T/2$. 
Then we have the following two equations:
$$
-T/2=q^0+2\alpha' E_2\tau_2 
$$
and
$$
T/2=q^0+2\alpha' E_1\tau_1 
$$
Since $\tau_1,\tau_2\rightarrow\infty$ we can ignore the $q^0$ terms 
in the expressions above. Taking the 
difference we reach the following result:
$$\tau_1-\tau_2=\frac{T}{4\alpha'}\left(\frac{1}{E_1}+\frac{1}{E_2}\right).$$
We can now replace $\tau$ terms in the expressions for 
$\langle\hat{T}^{\mu\nu}(x)\rangle$ 
to obtain:
\begin{eqnarray}\label{taq2}
\langle {\hat T}^{\mu\nu}(x)\rangle =
\frac{2T}{3(2\pi)^{3}\langle\Psi |\Psi\rangle}\int d^3p_{1}d^3p_{2}
\; e^{i(p_{2}-p_{1})\cdot x}
\; [p^{\mu}_{1}p^{\nu}_{1}+p^{\mu}_{2}p^{\nu}_{2} +
\frac{1}{2}\left(p^{\mu}_{1}p^{\nu}_{2}+p^{\nu}_{1}p^{\mu}_{2}
\right)]\times\nonumber\\ 
\;\varphi^{*}(\vec{p}_{1})\varphi(\vec{p}_{2})\;
\left(\frac{E_1+E_2}{(E_1E_2)^2}\right)
\label{taqclosed}
\end{eqnarray} 
for closed strings, whereas for open strings we obtain
\begin{eqnarray}
\langle {\hat T}^{\mu\nu}(x)\rangle =
\frac{T}{6(2\pi)^{3}\langle\Psi |\Psi\rangle}\int d^3p_{1}d^3p_{2}
\; e^{i(p_{2}-p_{1})\cdot x}\; 
[p^{\mu}_{1}p^{\nu}_{1}+p^{\mu}_{2}p^{\nu}_{2} +
\frac{1}{2}\left(p^{\mu}_{1}p^{\nu}_{2}+p^{\nu}_{1}p^{\mu}_{2}
\right)]\times\nonumber\\ 
\;\varphi^{*}(\vec{p}_{1})\varphi(\vec{p}_{2})\;
\left(\frac{E_1+E_2}{(E_1E_2)^2}\right).
\label{taqopen}
\end{eqnarray}


Now, let us consider the massless string states. 
For the closed string the massless states are the graviton
\begin{equation}\label{graviton}
|p,1\rangle_s \equiv {P_{i l}}_s (n)\;
\tilde{\alpha}^l_{-1}\;\alpha^i_{-1}\;|p,0\rangle 
\end{equation}
and the dilaton
 \begin{equation}\label{dilaton}
|p,1  \rangle \equiv P_{i l}(n)\;
\tilde{\alpha}^l_{-1}\;\alpha^i_{-1}\;|p,0\rangle 
\end{equation}
(we will refer from now on to the states $|p, 0\rangle$ simply as $|p\rangle$ 
unless said otherwise) where $p^2 = 0$, $s=1,2$ labels the
graviton helicity, 
$ P^{i l}_s (n), \; 1\leq i, l \leq 3$ projects onto the spin 2 
graviton states and 
$ P^{i l}(n)$ onto the (scalar) dilaton state, and the projection 
operators satisfy \cite{Leon}
\begin{eqnarray}
 P^{i l}_s (n) &=&  P^{ l i}_s (n) \quad , \quad  n_i\, P^{i l}_s (n)= 0
\cr\cr
 P^{l}_{l\; s} (n) &=& 0  \quad , \quad  
P^{i l}_s (n) P_{il\; s'} (n) = 2\delta_{s s'} \cr \cr
 P^{i l}_s (n) &=&  P^{i l}_s (-n)  \quad , \quad \cr \cr
 P^{i l} (n) &=&  P^{ l i} (n) \quad , \quad  n_i\, P^{i l} (n)= 0
\cr\cr 
P^{i l} (n) P_{i l} (n) &=& 2 \quad , \quad
 P^{i l} (n) = P^{i l}(-n) \quad \; . \nonumber
\end{eqnarray}
A possible representation for these operators is as follows: first we
introduce the orthonormal unit vectors:
$$\frac{n_i}{n}=(\sin\gamma\cos\phi,\sin\gamma\sin\phi,\cos\gamma),$$
$$l_i=(\sin\phi,-\cos\phi,0),$$
$$m_i=\pm(\cos\phi\cos\gamma,\sin\phi\cos\gamma,-\sin\gamma),$$
$+$ for $\gamma<\pi/2$, $-$ for $\gamma>\pi/2$. Thus, for $P^{il}_s(n)$
we have:
$$\buildrel{1}\over{P^{il}}(n)=(l_il_j-m_im_j),$$
and
$$\buildrel{2}\over{P^{il}}(n)=(l_im_j+l_jm_i).$$
For $P^{il}(n)$ we have:
$$P^{il}(n)=\delta^{il}-\frac{n^in^l}{n^2}.$$
For the open string we have the massless vector states (photons), 
which are given by
$$
|p,1  \rangle \equiv P^{i l} (n) \; \alpha^l_{-1}\,|p,0\rangle \; .
$$

It is clear from eq.(\ref{RR1}), since it is (as we have said before) 
already normal ordered, that the only oscillation 
modes that need to be kept 
in $e^{-i\lambda\cdot X_{+}}$ and $e^{-i\lambda\cdot X_{-}}$ 
are the ones 
with $n=1$ since all the other modes commute with 
$\alpha_{-1}\;(\alpha_{1})$ 
and $\tilde{\alpha}_{-1}\;(\tilde{\alpha}_{1})$ and will annihilate 
the ground state. Thus, we can write for gravitons
$$\langle\Psi|e^{-i\lambda\cdot X_{-}}$$ 
and 
$$e^{-i\lambda\cdot X_{+}}|\Psi\rangle$$ 
as
\begin{equation}
\langle 1,p|e^{-i\lambda\cdot X_{-}}=
\langle p_1|\tilde{\alpha}_{1}^{l}\alpha^{i}_{1}\;{P_{il}}_{s}(1)
e^{-i\lambda\cdot X_{-}}
\end{equation}
and
\begin{equation}
e^{-i\lambda\cdot X_{+}}|\Psi\rangle=
e^{-i\lambda\cdot X_{+}}{P_{il}}_{s'}(2)
\tilde{\alpha}_{-1}^{j}\alpha^{m}_{-1}|p_2\rangle
\end{equation}
respectively. We can now compute the action of $e^{-i\lambda\cdot X_{+}}$ 
on $|\Psi\rangle$ and of $e^{-i\lambda\cdot X_{-}}$ on $\langle\Psi|$ 
with the help of the identities:
\begin{equation}
\left[\alpha^{\mu}_{n},e^{-\lambda\cdot X_{-}}\right]=
-\frac{\lambda^{\mu}}{\sqrt{2}}e^{i(\tau-\sigma)}\;e^{-\lambda\cdot X_{-}}
\end{equation}
\begin{equation}
\left[\tilde{\alpha}^{\mu}_{n},e^{-\lambda\cdot X_{-}}\right]=
-\frac{\lambda^{\mu}}{\sqrt{2}}e^{i(\tau+\sigma)}\;e^{-\lambda\cdot X_{-}}
\end{equation}
\begin{equation}
\left[e^{\lambda\cdot X_{+}},\alpha^{\mu}_{-n}\right]=
\frac{\lambda^{\mu}}{\sqrt{2}}e^{-i(\tau-\sigma)}\;e^{-\lambda\cdot X_{+}}
\end{equation}
\begin{equation}
\left[e^{\lambda\cdot X_{+}},\tilde{\alpha}^{\mu}_{-n}\right]=
\frac{\lambda^{\mu}}{\sqrt{2}}e^{-i(\tau+\sigma)}\;e^{-\lambda\cdot X_{+}}
\end{equation}
obtaining:
\begin{equation}
\langle\Psi|e^{-i\lambda\cdot X_{-}}=
\langle p_1|(\tilde{\alpha}^{l}_{1}\alpha^{i}_{1}-
\frac{\lambda^{l}}{\sqrt{2}}e^{i(\tau+\sigma)}\alpha^{i}_{1}-
\frac{\lambda^{i}}{\sqrt{2}}e^{i(\tau-\sigma)}\tilde{\alpha}^{l}_{1}+
\frac{\lambda^{l}\lambda^{i}}{2}e^{2i\tau})\;{P_{il}}_{s}(1)
\label{R1}
\end{equation}
and
\begin{equation}
e^{-i\lambda\cdot X_{+}}|\Psi\rangle=
{P_{il}}_{s'}(2)(\tilde{\alpha}^{j}_{-1}\alpha^{m}_{-1}+
\frac{\lambda^{j}}{\sqrt{2}}e^{-i(\tau+\sigma)}\alpha^{m}_{-1}+
\frac{\lambda^{m}}{\sqrt{2}}e^{-i(\tau-\sigma)}\tilde{\alpha}^{j}_{-1}+
\frac{\lambda^{j}\lambda^{m}}{2}e^{-2i\tau})|p_2\rangle.
\label{R2}
\end{equation}
${P_{il}}_{s'}(2)$ and ${P_{il}}_{s}(1)$ refer to the projection 
operator needed for the initial (2) and final (1) states respectively.
Substituting expressions (\ref{R1}) and (\ref{R2}) into eq.(\ref{RR1}) 
and the equivalent expressions for $\hat{T}^{\mu\nu}_{b}$ and 
$\hat{T}^{\mu\nu}_{c}$ and 
taking the matrix elements we obtain for the graviton:
\begin{eqnarray}
\frac{\langle p_1 | \tilde{\alpha}^l_{1}\alpha^i_{1}
{\hat T}^{\mu\nu}(x) \tilde{\alpha}^j_{-1}\alpha^m_{-1}
|p_2 \rangle}{\langle\Psi|\Psi\rangle}
&=&\frac{1}
{6\pi\langle\Psi|\Psi\rangle}\int\frac{d^{4}\lambda}{(2\pi)^{4}}
d^4p_{1}d^4p_{2}\; d\sigma 
d\tau \;{P_{il}}_{s}(1)\;{P_{jm}}_{s'}(2)\;A^{ljim}\times\nonumber\\ & &
[p^{\mu}_{1}p^{\nu}_{1}+p^{\mu}_{2}p^{\nu}_{2} +
\frac{1}{2}\left(p^{\mu}_{1}p^{\nu}_{2}+p^{\nu}_{1}p^{\mu}_{2}
\right)]\langle p_{1}|
e^{i\lambda\cdot x}e^{-i\lambda\cdot X_{cm}}|p_{2}
\rangle\times\nonumber\\ & & 
\frac{\varphi^{*}(\vec{p}_{1})}{E_1}\;\frac{\varphi(\vec{p}_{2})}{E_2} \; 
\delta(p^{0}_{1}-\sqrt{\vec{p}_{1}^{2}+m^{2}})\;
\delta(p^{0}_{2}-\sqrt{\vec{p}_{2}^{2}+m^{2}}) \; ,\nonumber\\ & &
\label{antproy}
\end{eqnarray}
where
$$
A^{ljim}=4\delta^{lj}\delta^{im}-2\delta^{lj}\lambda^{i}\lambda^{m}
-2\delta^{im}\lambda^{l}\lambda^{j} +
\lambda^{l}\lambda^{j}
\lambda^{i}\lambda^{m}. 
$$
A similar expression is found for the dilaton case. For the open 
string case we obtain through a similar calculation: 
\begin{eqnarray}
\frac{\langle p_1|\alpha^i_{1} {\hat T}^{\mu\nu}(x)\alpha^m_{-1}
|p_2 \rangle}{\langle\Psi|\Psi\rangle}
 &=&\frac{2}{3\pi\langle\Psi|\Psi\rangle}
\int d^4p_{1}d^4p_{2} \; d^4\lambda\;d\tau\;\frac{e^{i\lambda\cdot
x}}{(2\pi)^{3}} P^{i}_{l}(1)\;P^{l}_{i}(2)\times\nonumber\\ & &
 \left[\delta^{i m}\; \left(p^{\mu}_{1}p^{\nu}_{1}
+p^{\mu}_{2}p^{\nu}_{2} +\frac{1}{2}
\left(p^{\mu}_{1}p^{\nu}_{2}+p^{\nu}_{1}p^{\mu}_{2}
\right)\right)\right.\nonumber\\  & & \left. -\frac{3}{2}
\eta^{\mu\nu}\lambda^{i}\lambda^{m}\right] \;\langle p_{1}|
e^{i\lambda\cdot x}e^{-i\lambda\cdot X_{cm}}|p_{2}\rangle
\times\nonumber\\ & &
\frac{\varphi^{*}(\vec{p}_{1})}{E_1}\frac{\varphi(\vec{p}_{2})}{E_2} \; 
\delta(p^{0}_{1}-\sqrt{\vec{p}_{1}^{2}+m^{2}})
\delta(p^{0}_{2}-\sqrt{\vec{p}_{2}^{2}+m^{2}}),\nonumber\\ & &
\label{unop}
\end{eqnarray}
having performed the $\sigma$ integration.

Substituting eq.(\ref{lm}) into eqs.(\ref{antproy}) and (\ref{unop}) 
and integrating over $\lambda$ , $p ^{0}_{1}$ and  
$p ^{0}_{2}$, we find
\begin{eqnarray}
\frac{\langle \Psi |
{\hat T}^{\mu\nu}(x)_{{\cal A}}| \Psi \rangle}{\langle\Psi|\Psi\rangle}&=&
\frac{T}{6(2\pi)^{3}\langle\Psi|\Psi\rangle}
\int d^{3}p_{1}d^{3}p_{2}e^{i(p_{2}-p_{1})\cdot x}\;
{\cal B}^{\mu\nu}_{{\cal A}}
\;\varphi^{*}(\vec{p}_{1})\;\varphi(\vec{p}_{2}) 
\left(\frac{E_1+E_2}{\left(E_1E_2\right)^2}\right),\nonumber\\ & &
\label{ssmas}
\end{eqnarray}
with ${\cal A}=(g,d,ph)$ where $g,d,ph$ refer to the graviton, dilaton or 
photon case. With this notation we have that
\begin{eqnarray}
{\cal B}^{\mu\nu}_{g}
 &=&[p^{\mu}_{1}p^{\nu}_{1}+p^{\mu}_{2}p^{\nu}_{2} +
\frac{1}{2}\left(p^{\mu}_{1}p^{\nu}_{2}+p^{\nu}_{1}p^{\mu}_{2}
\right)]\times\nonumber\\ 
& & \left[
4\buildrel{s}\over{P_{jm}}(1)\buildrel{s'}\over{P^{jm}}(2)+
2\buildrel{s}\over{P^{il}}(1)\buildrel{s'}\over{P^{m}_{l}}(2)p_2^ip^m_1+
2\buildrel{s}\over{P^{il}}(1)\buildrel{s'}\over{P^{j}_{i}}(2)p_2^lp^j_1
\right.\nonumber\\
 & &\left.
+\buildrel{s}\over{P^{il}}(1)\buildrel{s'}\over{P^{jm}}(2)p_2^lp^j_1p_2^ip^m_1
\right],
\label{ss1}
\end{eqnarray}
\begin{eqnarray}
{\cal B}^{\mu\nu}_{d}
 &=&
[p^{\mu}_{1}p^{\nu}_{1}+p^{\mu}_{2}p^{\nu}_{2} +
\frac{1}{2}\left(p^{\mu}_{1}p^{\nu}_{2}+p^{\nu}_{1}p^{\mu}_{2}
\right)]\times\nonumber\\ 
& & \left[4P_{jm}(1)P^{jm}(2)+
2P^{il}(1)P^{m}_{l}(2)p_2^ip^m_1+
2P^{il}(1)P^{j}_{i}(2)p_2^lp^j_1\right.\nonumber\\ & &\left.
+P^{il}(1)P^{jm}(2)p_2^lp^j_1p_2^ip^m_1\right]
\label{ss2}
\end{eqnarray}
and
\begin{eqnarray}
{\cal B}^{\mu\nu}_{ph}
 &=&
 \left[P^{i}_{l}(1)P^{l}_i(2)
\left(p^{\mu}_{1}p^{\nu}_{1}+p^{\mu}_{2}p^{\nu}_{2} +
\frac{1}{2}\left(p^{\mu}_{1}p^{\nu}_{2}+p^{\nu}_{1}p^{\mu}_{2}
\right)\right)\right.\nonumber\\  
& & \left.-\frac{3}{2}
\eta^{\mu\nu}P^{i}_l(1)P^{lm}(2)(p_2^i-p^i_1)(p_2^m-p^m_1)\right].
\label{ss3}
\end{eqnarray}
It is convenient to extend at this point the definition 
of ${\cal B}^{\mu\nu}_{{\cal A}}$ to include also the tachyon case:
\begin{eqnarray}
{\cal B}^{\mu\nu}_{ty1}
 &=&4[p^{\mu}_{1}p^{\nu}_{1}+p^{\mu}_{2}p^{\nu}_{2} +
\frac{1}{2}\left(p^{\mu}_{1}p^{\nu}_{2}+p^{\nu}_{1}p^{\mu}_{2}
\right)]
\label{tac1}
\end{eqnarray}
for closed strings, whreas for open strings we obtain
\begin{eqnarray}
{\cal B}^{\mu\nu}_{ty2}
 &=&[p^{\mu}_{1}p^{\nu}_{1}+p^{\mu}_{2}p^{\nu}_{2} +
\frac{1}{2}\left(p^{\mu}_{1}p^{\nu}_{2}+p^{\nu}_{1}p^{\mu}_{2}
\right)].
\label{tac2}
\end{eqnarray}

There is an important restriction on the allowed combination of the momenta 
$p_1$ and $p_2$ that emerges when we demand the physical condition of 
energy-momentum conservation. This is equivalent to
\begin{eqnarray}
\frac{\langle \Psi |
{\hat T}^{\mu\nu}(x)_{{\cal A}}| \Psi \rangle_{,\nu}}{\langle\Psi|\Psi\rangle}
 &=&\frac{iT}{6(2\pi)^{3}\langle\Psi|\Psi\rangle}
\int d^{3}p_{1}d^{3}p_{2}e^{i(p_{2}-p_{1})\cdot x}\;(p_{2}-p_{1})_{\nu}
{\cal B}^{\mu\nu}_{{\cal A}}
\;\varphi^{*}(\vec{p}_{1})\;\varphi(\vec{p}_{2}) \times\nonumber\\ & &
\left(\frac{E_1+E_2}{\left(E_1E_2\right)^2}\right)=0.
\label{ssmas2}
\end{eqnarray}
Thus energy-momentum conservation tells us that only a subset of the 
states we are considering are physical. The constraint we have obtained 
in order to satisfy energy-momentum conservation is consistent with
\begin{equation}
(p_{2}-p_{1})_{\nu}{\cal B}^{\mu\nu}_{{\cal A}}=0,
\end{equation}
which reduces in all cases to
\begin{equation}
(p^{\nu}_2-p^{\nu}_1)(p_{2\; \mu}p^{\mu}_{1}-m^2)=0.
\label{A6}
\end{equation}
This expression tell us that the  4-vectors $p_1$ and  
$p_2$ must satisfy either:
\begin{equation}
p^{\mu}_1=p^{\mu}_2
\label{A6a}
\end{equation}
which is a trivial solution and which we will not be considering, or 
\begin{equation}
p_{2\mu}p^{\mu}_{1}=m^2.
\label{A6b}
\end{equation}
If $p^{\mu}_1$ and 
$p^{\mu}_2$ satisfy either eq.(\ref{A6a}) or eq.(\ref{A6b}) 
then energy-momentum conservation is 
ensured. It should be noted that the imposition of eq.(\ref{A6b}), rather 
than eq.(\ref{A6a}) is rather speculative. One point we should also notice is 
the fact that the imposition of eq.(\ref{A6a}) would lead us to the classical 
energy-momentum tensor for point particles following a Gaussian distribution.

Let us consider now the $\langle {\hat T}^{\mu\nu}(x)\rangle$ for the massless 
string states. In order to satisfy eq.(\ref{A6b}), $\vec{p}_1$ and $\vec{p}_2$ 
must be parallel to each other. Thus, we can now write expression 
eq.(\ref{ssmas}) in terms of eqs.(\ref{ss1})-(\ref{ss3}) as:
\begin{eqnarray}\label{taq3}
\langle {\hat T}^{\mu\nu}(x)\rangle =
\frac{2T}{3(2\pi)^{3}\langle\Psi|\Psi\rangle}
\int d^3p_{1}d^3p_{2}\; e^{i(p_{2}-p_{1})\cdot x}
\; [p^{\mu}_{1}p^{\nu}_{1}+p^{\mu}_{2}p^{\nu}_{2} +
\frac{1}{2}\left(p^{\mu}_{1}p^{\nu}_{2}+p^{\nu}_{1}p^{\mu}_{2}
\right)]\times\nonumber\\ 
\;\varphi^{*}(\vec{p}_{1})\varphi(\vec{p}_{2})\;
\left(\frac{E_1+E_2}{(E_1E_2)^2}\right)
\end{eqnarray} 
where, since $\vec{p}_1$ is parallel to $\vec{p}_2$, all the 
terms involving $P(n)\cdot \vec{p}$ have vanished. The integrals above 
are now restricted integrals, we are now considering only $\vec{p}_1$ and 
$\vec{p}_2$ which are parallel to each other.


\subsection{Conformal Invariance}
Closely related to the issue of energy-momentum conservation is that 
of conformal invariance. We now proceed to establish 
the conformal invariance of our results.

Conformal transformations are simpler to describe in the coordinates
$$
x^{\pm} \equiv \frac1{\sqrt2}(\sigma \pm \tau)\; .
$$
The classical energy-momentum tensor eq.(\ref{six}) then takes the form
\begin{equation}\label{sixb}
    T^{\mu\nu}(x)=\frac{1}{2\pi\alpha'}\int dx^+ dx^-   \;
\int \frac{d^{4}\lambda}{(2\pi)^{4}} \; e^{i\lambda\cdot x}\;
\partial_{+} X^{\mu} \partial_{-} X^{\nu} \;
e^{-i\lambda\cdot X(\sigma,\tau) }\;  .
\end{equation}
This expression is clearly invariant under the conformal transformations
$$
x^+ \to f^+(x^+) \quad , \quad x^- \to f^-(x^-)
$$
for arbitrary $f^{\pm}$. In other words, conformal invariance holds if
the integrand $ \partial_{+} X^{\mu} \partial_{-} X^{\nu} $ has
$(1,1)$ as conformal weights.  

At the quantum level it is enough to require conformal invariance
on-shell. That is, the matrix elements of  $ {\hat T}^{\mu\nu}(x) $ 
[eq. (\ref{Tcua})] on physical states must be conformally invariant. 
We must then look to eq.(\ref{taq3})
and show that the  integrands have conformal weights $(1,1)$.
Actually this expression is close to the matrix elements of the
graviton vertex operator 
$$
\epsilon_{\mu \nu}(\lambda)\;  \partial_+X^{\mu}\partial_-X^{\nu}
\; e^{i\lambda \cdot x}\; .
$$
This operator has conformal weights $(1,1)$ provided $ \lambda^2 = 0 $
and $ \lambda^{\mu}\epsilon_{\mu \nu}(\lambda)= 0 =
\lambda^{\nu}\epsilon_{\mu \nu}(\lambda) $. 

We see in our calculation that the fact that we are working with states 
with $p^{\mu}_1p_{\mu\; 2}=0$ in eq. (\ref{taq3}),
set $ \lambda = p_2 - p_1 $ to be
light-like :  $ \lambda^2 = 0 $. However, we used the full operator 
$ \partial_{+} X^{\mu} \partial_{-} X^{\nu} $, which, generally
speaking, may contain in Fourier space a longitudinal part of the form
$$
\gamma_+^{\mu}\; \lambda^{\nu} + \gamma_-^{\nu}\; \lambda^{\mu}, 
$$
where $\gamma_{\pm}^{\mu}$ are some vectors,
plus the transverse part. We can find  longitudinal pieces by contracting 
the integrand in eq.(\ref{taq3}) with
$ \lambda^{\nu} $ and with $ \lambda^{\mu} $ separately. It is not difficult 
to see that the result is identically zero. Hence,  the whole 
integrands come from the transverse part of  $ \partial_{+} X^{\mu}
\partial_{-} X^{\nu} $ with conformal weights  $(1,1)$.
This completes the proof of the conformal invariance.

It must be noticed that the same transversality property of the
integrand ensures the energy-momentum conservation $ \partial_{\mu}
{\hat T}^{\mu\nu}(x) = 0$. This reflects the connection between
world-sheet conformal invariance and space-time invariances in String Theory.
\section{The classical energy-momentum tensor for point-particles} 
It is to be noticed that when we choose the wave-packets $\varphi(\vec{p})$ 
to have zero width, that is, when the state is completely localised in 
momentum space: 
$\varphi(\vec{p})=A\delta(\vec{p}-\vec{p}_{a})$, we recover the usual results 
found in \cite{HJV} as a short calculation shows. This adds confidence to 
the calculations we are about to perform in the next sections. 
From eq.(\ref{taq3}) we have:
\begin{eqnarray}
\langle {\hat T}^{\mu\nu}(x)\rangle =
\frac{2T|A|^2}{3(2\pi)^{3}\langle\Psi|\Psi\rangle}
\int d^3p_{1}d^3p_{2}\; e^{i(p_{2}-p_{1})\cdot x}
\; [p^{\mu}_{1}p^{\nu}_{1}+p^{\mu}_{2}p^{\nu}_{2} +
\frac{1}{2}\left(p^{\mu}_{1}p^{\nu}_{2}+p^{\nu}_{1}p^{\mu}_{2}
\right)]\times\nonumber\\ 
\;\delta(\vec{p}_{1}-\vec{p}_a)\delta(\vec{p}_{2}-\vec{p}_a)\;
\left(\frac{E_1+E_2}{(E_1E_2)^2}\right)
\label{sinmasa}
\end{eqnarray} 
which becomes
\begin{equation}
\langle {\hat T}^{\mu\nu}(x)\rangle =
\frac{2TA^2}{3(2\pi)^{3}\langle\Psi|\Psi\rangle}
\left(\frac{6p^{\mu}p^{\nu}}{E^3}\right),
\end{equation}
and
\begin{eqnarray}
\langle \Psi|\Psi\rangle &=& A^2\int \frac{d^4p_1}{E_1}\frac{d^4p_2}{E_2}\;
P(\vec{p}_1)\cdot P(\vec{p}_2)\;\delta(\vec{p}_{1}-\vec{p}_a)\;
\delta(\vec{p}_{2}-\vec{p}_a)\;\delta(\vec{p}_{1}-\vec{p}_2)\times
\nonumber\\ & &
\delta(p^{0}_{1}-p^{0}_2)\;
\delta(p^{0}_{1}-\sqrt{\vec{p}^{2}_{1}+m^2})\;
\delta(p^{0}_{2}-\sqrt{\vec{p}^{2}_{2}+m^2}),
\end{eqnarray}
\begin{equation}
\langle \Psi|\Psi\rangle=A^2\int \frac{d^3p_1 d^3p_2}{E_1 E_2}\;
P(\vec{p}_1)\cdot P(\vec{p}_2)\;\delta(\vec{p}_{1}-\vec{p}_a)\;
\delta(\vec{p}_{2}-\vec{p}_a)\;\delta(\vec{p}_{1}-\vec{p}_2)\;
\delta(E_1-E_2),
\end{equation}
\begin{equation}
\langle \Psi|\Psi\rangle=2A^2\int \frac{d^3p_2}{E_a E_2}
\delta(\vec{p}_{2}-\vec{p}_a)\;
\delta(\vec{p}_{2}-\vec{p}_a)\;
\delta(E_2-E_a),
\end{equation}
$$
\langle \Psi|\Psi\rangle=\frac{2A^2}{E^2_a}\delta(\vec{p}_{a}-\vec{p}_a)\;
\delta(E_a-E_a).
$$
We can regularise the Dirac deltas obtaining the following result
$$
\langle \Psi|\Psi\rangle=\frac{A^2}{(2\pi)^{4}E^2_a}\;V\;T.
$$
Putting everything together we find:
\begin{equation}
\langle{\hat T}^{\mu\nu}(x)\rangle =
4\pi\;
\frac{p^{\mu}p^{\nu}}{V\;E}.
\end{equation}
This result is nothing more than the energy-momentum tensor of a point-like 
particle. Looking now at the energy we have:
\begin{equation}
\int\frac{dV}{(2\pi)^3}\langle{\hat T}^{00}(x) \rangle
\rightarrow
E
\end{equation}
just as it should be. A similar calculation can be performed for the tachyonic 
case by replacing eq.(\ref{taq3}) in eq.(\ref{sinmasa}) by 
either eq.(\ref{taqclosed}) or eq.(\ref{taqopen}) resulting 
also in
$$\int\frac{dV}{(2\pi)^3}\langle{\hat T}^{00}(x) \rangle
\rightarrow
E.$$
\section{${\hat T}^{\mu\nu}(x)$ for massless string states  
in spherically symmetric configurations}
Let us now consider the expectation value of the energy-momentum tensor 
for the massless closed string state given by eq.(\ref{taq3}).  
Notice that such an expectation value is independent of the value of the
particle spin (zero, one or two).

For such a case it is convenient to use the parametrisation:
$$
p_1 = E_1 (1,  {\hat u_1}) \quad , \quad
p_2 = E_2 (1, {\hat u_2})
$$
with $ E_1 ,  E_2 \geq 0$ and
$${\hat u_1}=(\cos\phi\sin\gamma,\sin\phi\sin\gamma,\cos\gamma)$$
$${\hat u_2}=(\cos\beta\sin\delta,\sin\beta\sin\delta,\cos\delta).$$
But the constraint $p_1\cdot p_2=0$ [eq.(\ref{A6b})] 
implies ${\hat u}_1\cdot{\hat u}_2=1$. Thus, 
eq.(\ref{taq3}) becomes
\begin{eqnarray}
\langle {\hat T}^{\mu\nu}(x)\rangle &=& 
\frac{2T}{3(2\pi)^{3}\langle\Psi|\Psi\rangle} \int_0^{\infty}dE_1 
\int_0^{\infty} dE_2 \;\int d{\hat u}_1
\varphi^{*}(E_1,{\hat u_{1}})\; \varphi(E_2,{\hat u_{1}})\;
e^{i(E_2-E_1)(t-\vec{x}\cdot{\hat u_{1}})} \nonumber\\ 
& & \left[p^{\mu}_{1}p^{\nu}_{1}
+p^{\mu}_{2}p^{\nu}_{2}+
\frac{1}{2}\left(p^{\mu}_{1}p^{\nu}_{2}+p^{\nu}_{1}p^{\mu}_{2}
\right)\right](E_1+E_2)
\label{tmassles}
\end{eqnarray}
where $d{\hat u_{1}} \equiv \sin\gamma \, d\gamma \, d\phi $ and we have 
taken into account the constraint ${\hat u}_1\cdot{\hat u}_2=1$.
Let us consider for simplicity 
spherically symmetric wave packets
$ \varphi(E,{\hat u}) = \varphi(E)$. If $\vec{x}=(0,0,r)$ we can integrate 
over the angles with the result:
\begin{eqnarray}\label{esfsim0}
 \langle {\hat T}^{00}(t,r)\rangle&=&
  \frac{8T}{3(2\pi)^3\langle\Psi|\Psi\rangle}  
\int_0^{\infty} dE_1 \int_0^{\infty}  dE_2 \;
\varphi^{*}(E_1)\varphi(E_2)\; e^{i(E_2-E_1)t} \nonumber\\  
& & \frac{\sin(E_2-E_1)r}{(E_2-E_1)r} \;
\left[ E^{2}_{1} + E^{2}_{2} +  E_1 E_2 \right](E_1+E_2)
\end{eqnarray}
We can relate the result for
arbitrary $x=(t, {\vec x})$ with the special case   $ x = (t,0,0,z) $
using rotational invariance as follows,
\begin{eqnarray}\label{rotaT}
\langle {\hat T}^{0 i}(x)\rangle &=& {\hat x}^i \; C(t,r) \quad ,
\quad i=1,2,3 ,\cr \cr
\langle {\hat T}^{i j}(x)\rangle &=& \delta^{ij} \; A(t,r) +  {\hat
x}^i \, {\hat x}^j  \; B(t,r) \; \; , \; i,j = 1,2,3.
\end{eqnarray}
Here
\begin{eqnarray}\label{defABC}
 C(t,r)&=& \langle {\hat T}^{03}(t,r=z)\rangle \cr \cr
 A(t,r)=  \langle {\hat T}^{22}(t,r=z)\rangle \quad &,& \quad
 B(t,r)= \langle {\hat T}^{33}(t,r=z)\rangle -  \langle {\hat
T}^{11}(t,r=z)\rangle
\end{eqnarray}
with ${\hat x}^{i}=\frac{x^i}{r}$ the unit vector, and 
\begin{eqnarray}\label{esfsimr}
\langle {\hat T}^{11}(t,r=z)\rangle&=& -  
\frac{8T}{3(2\pi)^3\langle\Psi|\Psi\rangle}  \;
\int_0^{\infty} dE_1 \int_0^{\infty}  dE_2 \;
\varphi^{*}(E_1)\varphi(E_2)\;
\frac{e^{i(E_2-E_1)t}}{(E_2-E_1)^{2}r^{2}} \nonumber\\ 
& & \left[\cos(E_2-E_1)r - \frac{\sin(E_2-E_1)r}{(E_2-E_1)r}\right]
 \left[  E^{2}_{1} + E^{2}_{2} + E_1 E_2 \right](E_1+E_2), \cr
\langle {\hat T}^{33}(t,r=z)\rangle&=&  
\frac{8T}{3(2\pi)^3\langle\Psi|\Psi\rangle}\,  
\int_0^{\infty} dE_1 \int_0^{\infty} dE_2 \;
 \left[  E^{2}_{1} + E^{2}_{2} + E_1 E_2 \right](E_1+E_2)\nonumber\\
& &  \varphi^{*}(E_1)\varphi(E_2)\;
\frac{e^{i(E_2-E_1)t}}{(E_2-E_1)r} \;
\left[ \sin(E_2-E_1)r+2\frac{\cos(E_2-E_1)r}{(E_2-E_1)r} \right.\nonumber\\ 
& & \left. - 2\frac{\sin(E_2-E_1)r}{(E_2-E_1)^{2}r^{2}}\right], \nonumber\\
 \langle {\hat T}^{03}(t,r=z)\rangle&=& 
  \frac{i8T}{3(2\pi)^3\langle\Psi|\Psi\rangle} \,  
\int_0^{\infty} dE_1 \int_0^{\infty}dE_2  \; \left[ 
E^{2}_{1} + E^{2}_{2} +  E_1 E_2 \right](E_1+E_2) \nonumber\\
& &  \varphi^{*}(E_1)\varphi(E_2)\;
\frac{e^{i(E_2-E_1)t}}{(E_2-E_1)r} \;
\left[\cos(E_2-E_1)r -
\frac{\sin(E_2-E_1)r}{(E_2-E_1)r}\right].
\end{eqnarray}
The other components satisfy 
$\langle {\hat T}^{22}(t,r=z)\rangle=\langle {\hat T}^{11}(t,r=z)\rangle$,
 $\langle {\hat T}^{01}(t,r=z)\rangle=\langle {\hat T}^{02}(t,r=z)\rangle=
\langle {\hat T}^{12}(t,r=z)\rangle=
\langle {\hat T}^{13}(t,r=z)\rangle=
\langle {\hat T}^{23}(t,r=z)\rangle=0$, as they must from rotational
invariance.

As we can see the trace of the expectation value of the string
energy-momentum 
tensor vanishes.
Notice that 
$$
3 A(t,r) + B(t,r) =  \langle {\hat T}^{00}(t,r)\rangle 
$$ 
due to the tracelessness of the energy-momentum tensor. 

The invariant functions $  \langle {\hat
T}^{00}(t,r)\rangle,  A(t,r),\;
B(t,r)$ and $C(t,r)$ in eqs.(\ref{esfsim0}), (\ref{defABC}) and
(\ref{esfsimr}) can be written in terms of outgoing and incoming waves as
\begin{eqnarray}\label{forfac}
\langle {\hat T}^{00}(t,r)\rangle&=&\frac{1}{r}\; \left[ F(t+r) -
F(t-r)\right],\cr\cr 
A(t,r) &=&        -\frac{1}{r^2} \left[ H(t+r) + H(t-r)\right] 
        +\frac{1}{r^3} \left[ E(t+r) - E(t-r)\right],\cr\cr
B(t,r)&=&\frac{1}{r}\; \left[ F(t+r) - F(t-r)\right]
         +\frac{3}{r^2} \left[ H(t+r) + H(t-r)\right]
 -\frac{3}{r^3} \left[ E(t+r) - E(t-r)\right],\cr\cr
C(t,r)&=&-\frac{1}{r} \left[ F(t+r) + F(t-r)\right]-
\frac{1}{r^2} \left[ H(t+r) - H(t-r)\right]\; .  
\end{eqnarray}
where
\begin{eqnarray}\label{fint}
F(x) &=& \frac{4T}{3(2\pi)^3\langle\Psi|\Psi\rangle}\; 
\int_0^{\infty} dE_1 \int_0^{\infty}  dE_2 \;
\varphi(E_1)\varphi(E_2)\nonumber\\  
& & \frac{\sin(E_2-E_1)x}{(E_2-E_1)} \;
\left[ E^{2}_{1} + E^{2}_{2} +  E_1 E_2 \right](E_1+E_2)\; .
\end{eqnarray}
Notice that $ F(x) = - H'(x) $ and $ H(x) = E'(x) $.
Let us choose now a real wave-packet $\varphi(E)$, typically peaked at $E=0$ 
and decreasing rapidly away from there. 
For example consider a Gaussian wave-packet $\varphi(E)$:
\begin{equation}\label{pagau}
\varphi(E) = N\;e^{-\alpha E^2}
\end{equation}
where $N$ is an arbitrary dimensionless constant and $\alpha\equiv 1/\sigma$ 
$\sigma$ being the width of the wave-packet. An expression for $\alpha$ can be 
obtained by demanding 
the wave-functions to be correctly normalised. In particular when we consider
\begin{equation}
\langle\Psi|\Psi\rangle=2T\int^{\infty}_0 
dp \frac{p^2}{E^2}|\varphi(\vec{p})|^2=1,
\label{promedio2}
\end{equation}
\begin{equation}
\frac{1}{2T}=\int^{\infty}_0 dE\;|\varphi(E)|^2=
N^2\int^{\infty}_0 dE\;e^{-2\alpha E^2}
\end{equation}
from this expression we arrive at 
\begin{equation}
\alpha=\frac{\pi\;N^4\;T^2}{2}.
\end{equation}
We see from eq.(\ref{forfac}) that the energy density 
$\langle {\hat T}^{00}(r,t)\rangle$ and the energy flux $\langle
{\hat T}^{0i}(r,t)\rangle$ behave like spherical waves describing 
the way a massless string 
state spreads out starting from the initial wave-packet we choose. 

$F(x)$ is difficult to calculate exactly, but its asymptotic behaviour 
can be obtained. Changing the integration variables in eq.(\ref{fint}) to  
$$
E_2-E_1 = v \tau / x \quad , \quad E_2+E_1 = v\; ,
$$
we find
\begin{eqnarray}
F(x) &=& \frac{T}{3(2\pi)^3\langle\Psi|\Psi\rangle}\; 
\int_0^{\infty} v^3 \, dv \; \int_0^{x} {{d\tau}\over \tau}\;
\varphi(\frac{v}{2}[1
+\frac{\tau}{x}])\varphi(\frac{v}{2}[1
-\frac{\tau}{x}])\nonumber\\   
& & \sin(v\tau) \;
\left[3 +  {{\tau^2}\over {x^2}}\right]\; .
\label{Y3}
\end{eqnarray}
Now, we can let $x \to \infty$ obtaining the following integrals: 
\begin{equation}
\int^{\infty}_0\frac{d\tau}{\tau}\;\sin v\tau\;e^{-\frac{\alpha v^2\tau^2}{2x^2}}
=\frac{\pi}{2}\Phi(\frac{x}{\sqrt{2\alpha}})
\label{Y1}
\end{equation}
and
\begin{equation}
\int^{\infty}_0\frac{d\tau}{x^2}\;\tau\;
\sin v\tau\;e^{-\frac{\alpha v^2\tau^2}{2x^2}}
=\frac{x}{2\alpha}\;\sqrt{\frac{2\pi}{\alpha}}\;e^{-\frac{x^2}{2\alpha}}.
\label{Y2}
\end{equation}
Substituting eqs.(\ref{Y1}) and (\ref{Y2}) into eq.(\ref{Y3}) and taking 
the limit $x \to \infty$ we arrive to the result
\begin{equation}
F(x)  \buildrel{x\to \pm\infty}\over = \pm 
\;\frac{1}{(2\pi)^3\alpha^2}\sqrt{\frac{2\alpha}{\pi}}
+ O(e^{-x^2}).
\end{equation}
We find through similar calculations:
\begin{equation}
H(x)  \buildrel{x\to \pm\infty}\over = 
-\frac{1}{(2\pi)^3\alpha^2}\sqrt{\frac{2\alpha}{\pi}}\;|x|
+ O(e^{-x^2}),
\end{equation}
\begin{equation}
E(x)  \buildrel{x\to \pm\infty}\over =
-\frac{1}{2(2\pi)^3\alpha^2}\sqrt{\frac{2\alpha}{\pi}}\;x^2\;sign(x)
+ O(e^{-x^2}),
\end{equation}
To gain an insight into the behaviour of 
$\langle\hat{T}^{\mu\nu}\rangle$, we consider
the limiting cases: $ r\rightarrow\infty , \;  t $ fixed 
and $ t\rightarrow\infty , \;  r $ fixed
with the following results:
\\\\
a) $ r\rightarrow\infty , \; t$ fixed 
\begin{equation}
\langle {\hat T}^{00}(r,t)\rangle \buildrel{r\to \infty}\over = 
\frac{2}{(2\pi)^3\alpha^2}\sqrt{\frac{2\alpha}{\pi}}\frac{1}{r} ,
 \label{M1}
 \end{equation}
 \begin{equation}
 A(r,t) \buildrel{r\to \infty}\over =
\frac{1}{(2\pi)^3\alpha^2}\sqrt{\frac{2\alpha}{\pi}}\frac{1}{r} ,
\end{equation}
\begin{equation}
B(r,t)\buildrel{r\to \infty}\over=
-\frac{1}{(2\pi)^3\alpha^2}\sqrt{\frac{2\alpha}{\pi}}\frac{1}{r} , 
\end{equation}
and
\begin{equation}
C(r,t) \buildrel{r\to \infty}\over=
\frac{2}{(2\pi)^3\alpha^2}\sqrt{\frac{2\alpha}{\pi}}\frac{t}{r^2} ,.
\end{equation}
\\
b) $t\rightarrow\infty, r$ fixed
\begin{equation}
\langle {\hat T}^{00}(r,t)\rangle \buildrel{t\to \infty}\over=
-\frac{2}{3\alpha^2(2\pi)^3}\; 
e^{-\frac{t^2}{2\alpha}}\;\sinh\frac{tr}{\alpha}\;\frac{t}{r},
\end{equation}
\begin{equation}
A(r,t)\buildrel{t\to \infty}\over=\frac{2}{\alpha(2\pi)^3}\; 
e^{-\frac{t^2}{2\alpha}}\;\sinh\frac{tr}{\alpha}\;\frac{t}{r^2} ,
\end{equation}
\begin{equation}
B(r,t)\buildrel{t\to \infty}\over= -\frac{2}{3\alpha^2(2\pi)^3}\; 
e^{-\frac{t^2}{2\alpha}}\;\sinh\frac{tr}{\alpha}\;\frac{t}{r}
\end{equation}
and
\begin{equation}
C(r,t)\buildrel{t\to \infty}\over= -\frac{2}{3\alpha^2(2\pi)^3}\; 
e^{-\frac{t^2}{2\alpha}}\;\sinh\frac{tr}{\alpha}\;\frac{t}{r}.
\label{MFINAL}
\end{equation}
Thus, as we mentioned earlier the  energy density, the energy flux  and the
components of the stress tensor propagate 
as spherical outgoing waves. For  $t$ fixed, the  energy density
decays as $ r^{-1} $ while the  energy flux decays as  $ r^{-2} $.

For $r$ fixed and large $t$, the energy-momentum components decay 
exponentially fast as
$O(e^{-t^2})$. In this regime we find a negative energy density. 
The spherical wave seems to leave behind a small but negative
energy density. Notice that $ T^{00} $ is not a positive definite
quantity for strings [see eq.(\ref{six})]. We will 
discuss this result in the next chapter.

The results in this section have been derived using a Gaussian 
shape for the wave
packet.\ It is clear that the results will be qualitatively similar
for any fast-decaying wave function $\varphi(E)$.

It should be noted that analogous but not identical results follow for
massless scalar waves in the point particle case. 
A spherically symmetric solution of the wave
equation in $D=3+1$ dimensions 
$$
\partial^2 \phi(r,t) = 0
$$
takes the form,
\begin{equation}\label{solfi}
\phi(r,t) = {1 \over r} \left[ f(t-r) + g(t+r) \right]
\end{equation}
where $f(x)$ and $g(x)$ are arbitrary functions.\  
The energy-momentum tensor for such a massless scalar field can be
written as:
\begin{equation}\label{tmnfi}
 {\hat T}^{\mu\nu}(x)_{\phi}= \partial^{\mu}\phi \partial^{\nu}\phi-
{{\eta^{\mu\nu}}\over 2} (\partial\phi)^2.
\end{equation}
Inserting eq.(\ref{solfi}) into  eq.(\ref{tmnfi}) yields,
\begin{eqnarray}
 {\hat T}^{00}(r,t)_{\phi} &=& {1 \over {r^2}}\left\{ f'^2 +g'^2 + {{f+g}\over
{2r}}\left[ 2(f'-g')+ {{f+g}\over r} \right] \right\}, \cr \cr 
 {\hat T}^{0i}(r,t)_{\phi} &=& {{x^i}\over {r^3}}\;\left[ g'^2 -  f'^2
- \frac1r(f+g)(f' - g') \right],\cr \cr
 {\hat T}^{ij}(r,t)_{\phi} &=& \frac1{2r^2}\;\left[ \delta^{ij} +
{{2x^ix^j}\over {r^2}} \right]\;\left[f'-g'+\frac1r(f+g)\right]^2.
\nonumber
\end{eqnarray}
We see that the string $ {\hat T}^{\mu\nu}$ scales as $1/r$ [eq.(\ref{forfac})]
whereas the field $ {\hat T}^{\mu\nu}_{\phi}$ scales as $1/r^2$.
Perhaps the slower $ {\hat T}^{\mu\nu}$ decay for strings can be related to the
fact that they are extended objects.

\subsection{The total energy for a quantum bosonic string}
The asymptotic behaviour of $ {\hat T}^{00}$ which scales as $1/r$ can make 
us suspect that the total energy may diverge, since the total 
energy is given by the integral over the volume enclosing $ {\hat T}^{00}$ and 
such an integral involves a power of $r^2$ in the numerator. Let us 
then estimate the total energy for the quantum string. 

There are basically two regions where we would like to compute the total energy:
the region where $r<T$ and the region where $r>T$ (remembering that $T$ is 
the size of a finite temporal box and that $\alpha\sim T^2$). \\ \\
a)$r>T$, $T$ large but fixed. From eq.(\ref{M1}) we have:
\begin{equation}
E_{T}\sim\int^{R} \frac{dV}{\alpha^{3/2}\;r},
\end{equation}
\begin{equation}
E_{T}\sim\int^{R} \frac{dV}{r\;T^{3}},
\end{equation}
\begin{equation}
E_{T}\sim \left(\frac{R}{T}\right)^2\frac{1}{T}.
\label{energyI}
\end{equation}
b)$r<T$, $T$ large but fixed. From eq.(\ref{esfsim0}) we have:
\begin{eqnarray}
E_{T}&\sim &\int^{R}_{0}r^2\langle {\hat T}^{00}(r,t)\rangle\; dr\nonumber\\ 
&\sim &T\int^{\infty}_0  dE_1\int^{\infty}_0 \frac{ dE_2}{(E_1-E_2)^3}
\left[\sin(E_1-E_2)R-(E_1-E_2)R\cos(E_1-E_2)R\right]\times\nonumber\\ & &
\left(E^{2}_{1}+E^{2}_{1}+E_{1}E_{2}\right)(E_1+E_2)\varphi(E_1)\varphi(E_2).
\end{eqnarray}
Since $\varphi(E)\sim e^{-T^2E^2}$ we see that these integrals are dominated 
by $E\ll 1/T$ and since $R<T$ we have
\begin{equation}
E_{T}\sim T\int^{_{\ll}\;1/T}_{0}dE_1\int^{_{\ll}\;1/T}_{0} dE_2\; 
R^3\;
\left(E^{2}_{1}+E^{2}_{1}+E_{1}E_{2}\right)(E_1+E_2)\varphi(E_1)\varphi(E_2).
\end{equation}
Solving this last set of integrals we obtain the following result: 
\begin{equation}
E_{T}\sim \frac{R^3}{T^4}\sim\left(\frac{R}{T}\right)^3\frac{1}{T}.
\label{energyF}
\end{equation}
What we can see from these expressions is that the total energy diverges only 
when we work in a temporal box which is very small compared to the radius of 
the volume of space, but this is an acausal situation, which is not physically 
realisable.

In regions where $R\sim T$ (that is in the 
border of our causal horizon) the total energy is finite and small. 
When our temporal box 
is larger than the radius of the space volume the total 
energy converges to a finite small value. 
In the expressions above $R$ acts like a cut-off for the 
size of the string source. There is another reason why we can expect finite 
energies to arise. The calculation is for a single string, and in reality 
we expect other strings to be present as well and these will provide a 
natural cut-off scale, as in the case of global cosmic 
strings.
\section{${\hat T}^{\mu\nu}$ for massless string states in 
cylindrically symmetric configurations.}
Cosmic strings can be considered as essentially very long straight
strings in
(almost) cylindrically symmetric configurations.\ Although they
behave as fundamental strings only classically and in the
Nambu approximation (that is, zero string thickness), it
is important to study the expectation value of  $ {\hat T}^{\mu\nu}(x)$
for a cylindrically symmetric configuration.\ It would be interesting
to see if there is an equivalent quantum version of the deficit angle
found for cosmic strings \cite{shellard}.\ Consider
a cylindrically symmetric wave packet

\begin{equation}\label{paqcil}
\varphi(E,{\hat u}) = \varphi(E , \gamma).
\end{equation}
We can re-write the integral eq.(\ref{tmassles}) in a more convenient way to
analyse $\langle {\hat T}^{\mu\nu}(x)\rangle$ in a cylindrical
configuration as follows:
\begin{eqnarray}
\langle {\hat T}^{\mu\nu}(x)\rangle &=& 
\frac{4T}{3(2\pi)^4\langle\Psi|\Psi\rangle} \int_0^{\infty} dE_1 
\int_{0}^{\infty} dE_2\int_0^{\pi}
d\gamma\sin\gamma\int_0^{2\pi}d\phi 
\;\varphi^{*}(E_1,\gamma)\; \varphi(E_2,\gamma)\nonumber\\
& & e^{i(E_2-E_1)t}\;e^{i(E_2-E_1)\rho\cos\phi\sin\gamma} 
\left[p^{\mu}_{1}p^{\nu}_{1}
+p^{\mu}_{2}p^{\nu}_{2}+
\frac{1}{2}\left(p^{\mu}_{1}p^{\nu}_{2}+p^{\nu}_{1}p^{\mu}_{2}
\right)\right](E_1+E_2)\nonumber\\ & &
\label{tmassles2}
\end{eqnarray}
where we have chosen ${\vec x} = (\rho,0,0)$ , ${\hat u}_1.{\vec x} = \rho
\cos\phi\,\sin\gamma$.
Integrating over $ \phi $ in eq.(\ref{tmassles2}) we find
\begin{eqnarray}\label{cilsim}
\langle {\hat T}^{00}(t,\rho)\rangle &=& 
\frac{4T}{3(2\pi)^3\langle\Psi|\Psi\rangle}  
\int_0^{\infty} dE_1 \int_0^{\infty} 
dE_2\int_0^{\pi}d\gamma\sin\gamma\;
\varphi^{*}(E_1,\gamma)\varphi(E_2,\gamma) \nonumber\\
& & \left[  E^{2}_{1} + E^{2}_{2} +  E_1 E_2 \right](E_1+E_2)
 J_0([E_1-E_2]\rho\,\sin\gamma)\; e^{i(E_2-E_1)t} \; ,\cr \cr
\langle {\hat T}^{33}(t,\rho)\rangle &=& 
\frac{4T}{3(2\pi)^3\langle\Psi|\Psi\rangle}  
\int_0^{\infty}  dE_1 \int_0^{\infty}   dE_2 
\int_0^{\pi}d\gamma\sin\gamma\cos^{2}\gamma\;
\varphi^{*}(E_1,\gamma)\varphi(E_2,\gamma)\;  \nonumber\\
& &  \left[  E^{2}_{1} + E^{2}_{2} + E_1 E_2 \right](E_1+E_2) 
J_0([E_1-E_2]\rho\,\sin\gamma)\; e^{i(E_2-E_1)t} \;,\cr \cr 
\langle {\hat T}^{03}(t,\rho)\rangle &=&
\frac{2T}{3(2\pi)^3\langle\Psi|\Psi\rangle}
\int_0^{\infty}  dE_1 \int_0^{\infty}   dE_2 
\int_0^{\pi}d\gamma\sin\gamma\cos\gamma\;
\varphi^{*}(E_1,\gamma)\varphi(E_2,\gamma) \nonumber\\
& &  \left[  E^{2}_{1} + E^{2}_{2} + E_1 E_2 \right](E_1+E_2)
\; J_0([E_1-E_2]\rho\,\sin\gamma)\; e^{i(E_2-E_1)t} \; .
\end{eqnarray}

Using rotational invariance around the $z$-axis we can express 
${\hat T}^{\alpha\beta}(x)$, ${\hat T}^{3 \alpha}(x)$ and $ {\hat T}^{0
\alpha}(x)$, $\alpha, \beta=1,2 $ as follows
\begin{eqnarray}
{\hat T}^{\alpha\beta}(x) &=& \delta^{\alpha\beta} \; a(t,\rho) +
e^{\alpha} \, e^{\beta}\; b(t,\rho) \cr \cr
 {\hat T}^{0\alpha}(x) &=& e^{\alpha} \,c(t,\rho) \quad , \quad 
 {\hat T}^{3 \alpha}(x)= e^{\alpha} \,d(t,\rho) \nonumber
\end{eqnarray}
where $ e^{\alpha} = (\cos\phi, \sin\phi)$. The coefficients $
a(t,\rho),  b(t,\rho), c(t,\rho)$ and $d(t,\rho)$ follow from the
calculation for $\rho = x$ (that is $\phi = 0$):
\begin{eqnarray}
a(t,\rho) &=& \langle {\hat T}^{22}(t,\rho=x)\rangle \quad , \quad 
  b(t,\rho) =  \langle {\hat T}^{11}(t,\rho=x)\rangle -  \langle {\hat
T}^{22}(t,\rho=x)\rangle \; ,\cr \cr
c(t,\rho) &=& \langle {\hat T}^{01}(t,\rho=x)\rangle \quad , \quad 
d(t,\rho)  = \langle {\hat T}^{13}(t,\rho=x)\rangle  \nonumber
\end{eqnarray}
We find from eqs.(\ref{paqcil}, \ref{tmassles2}) at $\rho = x$,
\begin{eqnarray}\label{cilsim2}
\langle {\hat T}^{11}(t,\rho=x)\rangle &=& 
\frac{2T}{3(2\pi)^3\langle\Psi|\Psi\rangle}
\int_0^{\infty} dE_1 \int_0^{\infty}  dE_2 
\int_0^{\pi}d\gamma\sin^{3}\gamma\;
\varphi^{*}(E_1,\gamma)\varphi(E_2,\gamma)\;  
e^{i(E_2-E_1)t} \nonumber\\
& &  \left[  E^{2}_{1} + E^{2}_{2} + E_1 E_2 \right](E_1+E_2)
\;\left[J_0([E_1-E_2]\rho\,\sin\gamma)-
J_2([E_1-E_2]\rho\,\sin\gamma)\right], \cr \cr
\langle {\hat T}^{22}(t,\rho=x)\rangle &=&
\frac{4T}{3(2\pi)^3\langle\Psi|\Psi\rangle}
\int_0^{\infty}  dE_1 \int_0^{\infty}  dE_2 
\int_0^{\pi}d\gamma\sin^{2}\gamma\;
\varphi^{*}(E_1,\gamma)\varphi(E_2,\gamma) \nonumber\\
& &  \left[  E^{2}_{1} + E^{2}_{2} + E_1 E_2 \right](E_1+E_2)
\;J_1([E_1-E_2]\rho\,\sin\gamma)\;
\frac{e^{i(E_2-E_1)t}}{(E_1-E_2)\rho}\; , \cr \cr 
\langle {\hat T}^{01}(t,\rho=x)\rangle &=& 
\frac{i4T}{3(2\pi)^3\langle\Psi|\Psi\rangle}
\int_0^{\infty}  dE_1 \int_0^{\infty}   dE_2 
\int_0^{\pi}d\gamma\sin^{2}\gamma\;
\varphi^{*}(E_1,\gamma)\varphi(E_2,\gamma) \nonumber\\
& &  \left[  E^{2}_{1} + E^{2}_{2} + E_1 E_2 \right](E_1+E_2)
\;J_1([E_1-E_2]\rho\,\sin\gamma)\; e^{i(E_2-E_1)t}\; , \cr \cr
 \langle {\hat T}^{13}(t,\rho=x)\rangle & = &
 \frac{i4T }{3(2\pi)^3\langle\Psi|\Psi\rangle}  
\int_0^{\infty} dE_1 \int_0^{\infty}  dE_2 
\int_0^{\pi}d\gamma\sin^{2}\gamma\cos\gamma\;
\varphi^{*}(E_1,\gamma)\varphi(E_2,\gamma)  \nonumber\\
& & \left[  E^{2}_{1} + E^{2}_{2} + E_1 E_2 \right](E_1+E_2)\;
J_1([E_1-E_2]\rho\,\sin\gamma)\; e^{i(E_2-E_1)t} \; ,
\end{eqnarray}
where $J_{n}(z)$ is a Bessel function of integer order. Cylindrical 
symmetry also results in
$$
\langle {\hat T}^{02}(t,\rho=x)\rangle = \langle {\hat
T}^{12}(t,\rho=x)\rangle =\langle {\hat T}^{23}(t,\rho=x)\rangle = 0.
$$

For a wave packet  symmetric  with respect to the $ x y $ plane,
$$
\varphi(E,\gamma) = \varphi(E,\pi - \gamma) \; ,
$$
and we find
$$
d(t,\rho)  = \langle {\hat T}^{13}(t,\rho=x)\rangle = 0 \; , \; 
 \langle {\hat T}^{03}(t,\rho)\rangle = 0 \; .
$$

Notice that 
$$
2 a(t,\rho) + b(t,\rho) +\langle {\hat T}^{33}(t,\rho)\rangle =
\langle {\hat T}^{00}(t,\rho)\rangle  
$$ 
due to the tracelessness of the energy-momentum tensor. 
In order to compute the asymptotic behaviour $\rho \to \infty$, 
we change the integration variables in eq.(\ref{cilsim})-(\ref{cilsim2}) to 
$$
E_2-E_1 = v \tau / \rho \quad , \quad E_2+E_1 = v\; .
$$
We find for the energy-density for a real $ \varphi(E, \gamma) $,
\begin{eqnarray}
\langle {\hat T}^{00}(t,\rho)\rangle &=&
 \frac{T}{6(2\pi)^3\langle\Psi|\Psi\rangle \; \rho}  \; 
\int_0^{\infty} v^4 \, dv \; \int_0^{\rho} d\tau \; \int_0^{\pi}
\sin\gamma \; d\gamma \; \varphi(\frac{v}{2}[1+\frac{\tau}{\rho}],\gamma )
\varphi(\frac{v}{2}[1-\frac{\tau}{\rho}],\gamma)\nonumber\\   
& & \cos(v\tau t/\rho ) \;
\left[3 - \frac{\tau^2}{\rho^2} \right]\; J_0(v\tau\sin\gamma)\; .
\nonumber
\end{eqnarray}
This representation is appropriate to compute the  limit
$\rho\rightarrow\infty$ with $t$ fixed. We find in such a limit,
\begin{eqnarray}
\langle {\hat T}^{00}(t,\rho)\rangle & \buildrel{\rho\to \infty}\over=&
\displaystyle{ 
\frac{T}{(2\pi)^3\langle\Psi|\Psi\rangle\; \rho}  \; \int_0^{\infty} 
E^3 dE \;  \int_{\sin\gamma > t/\rho} \; d\gamma \; 
{{\varphi(E, \gamma)^2}\over{\sqrt{1 - \left({t
\over{\rho\sin\gamma}}\right)^2}}}} \; ,\nonumber
\end{eqnarray}
where we have used the formula
$$
\int_0^{\infty} J_0(ax)\; \cos bx \; dx = {{\theta(a^2 - b^2)}\over
{\sqrt{a^2 - b^2}}} \; .
$$
That is, the energy decays as $ \rho^{-1} $ for large $  \rho $ and
fixed $ t $. 

In an analogous manner we compute the $\rho\rightarrow\infty, \; t  $ fixed
behaviour of the other components of $\langle {\hat
T}^{\mu\nu}(x)\rangle $  with the following results:
\begin{eqnarray}
a(t,\rho) & \buildrel{\rho\to \infty}\over=&
\frac{8T}{(2\pi)^3\langle\Psi|\Psi\rangle \; \rho}  \; 
\int_0^{\infty} E^3 dE \;  \int_{\sin\gamma >
t/\rho} \; d\gamma \; \varphi(E, \gamma)^2\nonumber\\
& &\sin^2\gamma \;\sqrt{1 - \left({t\over{\rho\sin\gamma}}\right)^2} 
\; , \cr \cr
a(t,\rho) + b(t,\rho) & \buildrel{\rho\to \infty}\over=& 
\frac{8T\;t^2}{(2\pi)^3\langle\Psi|\Psi\rangle \; \rho^3}  
\; \int_0^{\infty} E^3 dE \;  \int_{\sin\gamma >
t/\rho} \; d\gamma \; {{\varphi(E, \gamma)^2}\over{\sqrt{1 - \left({t
\over{\rho\sin\gamma}}\right)^2}}}  \cr \cr
 & \buildrel{\rho\to \infty}\over=& {{t^2}\over{ \rho^2}}\; \langle
{\hat T}^{00}(t,\rho)\rangle \; , \cr \cr
\langle {\hat T}^{33}(t,\rho)\rangle  &\buildrel{\rho\to \infty}\over=&
\frac{8T}{(2\pi)^3\langle\Psi|\Psi\rangle\; \rho}  \; 
\int_0^{\infty} E^3 dE \;  \int_{\sin\gamma >
t/\rho} \; d\gamma \cos^2\gamma\; {{\varphi(E, \gamma)^2}\over{\sqrt{1
- \left({t \over{\rho\sin\gamma}}\right)^2}}} \; , \cr \cr
c(t,\rho) &\buildrel{\rho\to \infty}\over=&
\frac{8T\; t}{(2\pi)^3\langle\Psi|\Psi\rangle \; \rho^2}  \; 
\int_0^{\infty} E^3 dE \;  \int_{\sin\gamma >
t/\rho} \; d\gamma \; {{\varphi(E, \gamma)^2}\over{\sqrt{1 - \left({t
\over{\rho\sin\gamma}}\right)^2}}} \cr \cr
 &\buildrel{\rho\to \infty}\over=&
 {{t}\over{ \rho}}\; \langle {\hat T}^{00}(t,\rho)\rangle \; .\nonumber
\end{eqnarray}

\bigskip

The calculation of the $ t \to  \infty$ behaviour for  $\rho$ fixed can
be easily obtained from  eq.(\ref{tmassles2}). That is,  
\begin{eqnarray}
\langle {\hat T}^{00}(t,\rho)\rangle &=& 
\frac{4T}{3(2\pi)^4\langle\Psi|\Psi\rangle } \; 
\int_0^{\pi} d\gamma\sin\gamma\int_0^{2\pi}d\phi 
\left[ f_3(\gamma, \phi) f^{*}_0(\gamma, \phi) + 2 f_1(\gamma,
\phi) f^{*}_2(\gamma, \phi) \right.\nonumber\\ & & \left.  +2 f_2(\gamma,
\phi) f^{*}_1(\gamma, \phi)
f_0(\gamma, \phi) f^{*}_3(\gamma, \phi)\right]
\label{asiEt}
\end{eqnarray}
where
\begin{equation}\label{fnc}
 f_n(\gamma, \phi)\equiv \int_0^{\infty} E^n \; dE \; \varphi(E, \gamma)
\; e^{iE(\rho \sin\gamma\sin\phi - t)} \; .
\end{equation}
We explicitly see here $ \langle {\hat T}^{00}(t,\rho)\rangle $ as a
superposition of  outgoing and ingoing  cylindrical waves.
As before, a typical form of the wave function $\varphi(E,\gamma)$
is  a Gaussian Ansatz 
\begin{equation}\label{cilgau}
\varphi(E,\gamma) = N\; e^{- E^2 (\alpha^2 \sin^2\gamma + \beta^2 \cos^2\gamma)}
\end{equation}
where $N$ is an arbitrary dimensionless constant. An expression 
for $\alpha$ and $\beta$ can be 
obtained by demanding 
the wave-functions to be correctly normalised. In particular when we consider
$\langle\Psi|\Psi\rangle=1$, 
we obtain (analogously to eq.(\ref{promedio2}))
\begin{equation}
\langle\Psi|\Psi\rangle=T\int^{\infty}_0 
dp \;\sin\gamma\;d\gamma\;\frac{p^2}{E^2}|\varphi(\vec{p})|^2=1,
\end{equation}
hence from eq.(\ref{cilgau}),
\begin{equation}
\langle\Psi|\Psi\rangle=TN^2\int^{\infty}_0 
dE\;\sin\gamma\;d\gamma\;e^{-2E^2(\alpha^2\sin^2\gamma+\beta^2\cos^2\gamma)}.
\end{equation}
Making the following change of variable:
$$u=\cos\gamma,$$ we obtain
$$\frac{1}{N^2\;T}=\int^{\infty}_0dE\int^{1}_{-1}du\;e^{-2E^2[\alpha^2+
u^2(\beta^2-\alpha^2)]},$$
$$\frac{1}{N^2\;T}=\frac{1}{2A}\sqrt{\frac{\pi}{2}}\int^{1}_{-1}
\frac{du}{\sqrt{u^2+\alpha^2/A^2}},$$
where $A=(\beta^2-\alpha^2)^{1/2}$, therefore $\beta>\alpha$. Thus, we find 
$$\frac{1}{N^2\;T}=\sqrt{\frac{\pi}{2(\beta^2-\alpha^2)}}
\sinh^{-1}\left(\frac{\sqrt{\beta^2-\alpha^2}}{\alpha}\right),$$
$$\sinh\left(\frac{\sqrt{2/\pi(\beta^2-\alpha^2)}}{N^2\;T}\right)=
\frac{\sqrt{\beta^2-\alpha^2}}{\alpha}.$$
If $\beta^2-\alpha^2\ll T^2$ we obtain
$$\alpha\simeq\sqrt{\frac{\pi}{2}}N^2\;T$$
whilst $\beta$ must satisfy $\beta^2>\alpha^2$ and 
$(\beta^2-\alpha^2)\ll T^2$. 

Returning to the integral (\ref{fnc}), we can see that this integral is 
dominated in the $ t \to \infty $ limit by
its lower bound. We find,
$$
 f_n(\gamma, \phi) \buildrel{t\to \infty}\over= n! (-i)^{n+1} \;
\varphi(0, \gamma) \; t^{-n-1} \left[ 1 + O(t^{-1})\right]+O(e^{-t^2}) \; .
$$
Inserting this result in eq.(\ref{asiEt}) yields
$$
\langle {\hat T}^{00}(t,\rho)\rangle \buildrel{t\to \infty}\over=
0-O(e^{-t^2}) \; .
$$

Using the same technique we find  the $ t \to  \infty$ behaviour
$\rho$ fixed, for the other $ \langle {\hat T}^{\mu \nu}(t,\rho)\rangle $
components,
\begin{eqnarray}
\langle {\hat T}^{33}(t,\rho)\rangle &\buildrel{t\to \infty}\over=&
0-O(e^{-t^2}),\nonumber\\
a(t,\rho) &\buildrel{t\to \infty}\over=&
0-O(e^{-t^2}),\nonumber\\
b(t,\rho) &\buildrel{t\to \infty}\over=& 0 + O(e^{-t^2}), \nonumber\\
c(t,\rho) &\buildrel{t\to \infty}\over=& 0 - O(e^{-t^2}). \nonumber
\end{eqnarray}

{F}rom the expressions above we can see that 
for large $t$ and $\rho$ fixed the energy density decays 
exponentially fast as $O(e^{-t^2})$ leaving behind a rapidly vanishing 
negative contribution just as in the spherical case.\ 
For $\rho$ large and fixed $t$ the 
situation is similar to the spherical one as well, the energy density  decays
as $1/\rho$ and the energy flux decays as $1/\rho^2$.

\section{Final remarks.}

An important point we need to stress, 
is the fact that, in contrast to what happens in the
classical theory where there is a logarithmic divergence as we approach
the core of the string \cite{edal}, in the quantum theory we can
obtain a metric  where, if any
divergences are present these are not related at all to the position of
the string.\ We showed, that the localised behaviour of the 
string given
by $\delta(x-X(\sigma,\tau))$ disappears when we take into account the
quantum nature of the strings.\ 
 In the {\em classical theory of radiating strings},
divergences coincide with the source of the gravitational field. However,
as we have seen, quantum mechanically this is not the case. 
The source position is smeared by the quantum fluctuations. The probability 
amplitude for the centre of mass string position
is given by the Fourier transform of $\varphi ({\vec p})$ and is a
smooth function peaked at the origin.


\input{epsf}
\chapter{Another derivation for $\langle {\hat T}^{\mu\nu}(r,t)\rangle$}
\label{Alpha}
In the previous chapter we computed the expectation value for the
energy-momentum tensor both in spherically and cylindrically symmetric
configurations. We presented exact asymptotic results for 
$\langle \hat{T}^{\mu\nu}(r,t)\rangle$ in several
interesting regions of space-time. We are interested in 
calculating also the gravitational field arising from the quantum string, 
so the need for an explicit 
expression for the $\langle\hat{T}^{\mu\nu}(r,t)\rangle$ components is apparent. 
In this chapter, we will try to obtain this by computing 
$\langle \hat{T}^{\mu\nu}(r,t)\rangle$, for the spherically symmetric case, 
in an approximate way which will allow us to work with
this expectation value for all $r$ and $t$.

\section{The $\alpha-$large approximation}
Our approximation will be based on the following observations: 
first, in the
spherical case we know that the functions $F(z)$, $H(z)$ and
$E(z)$ defined in the last chapter do not depend on the angles 
$\phi$ and $\gamma$ which defined
our spherical configuration; second, the wave-packets 
$\varphi(\vec{p})$
have a Gaussian ansatz; in other words, they are fast decaying
functions, the width of the packet being given by $1/\alpha$. (This
$\alpha$ is not to be confused with $\alpha'$ in the string tension.) If
we consider that most of this distribution has a width which is very
small in momentum space i.e. an almost monochromatic wave-function
(monochromatic in the sense that because we are working in a massless case,
any distribution in momentum is basically a distribution in energy),
then $1/\alpha$ has to be very small ($\alpha$ is large). We believe that
working in this approximation makes physical sense.

\begin{figure}
\centerline{\epsfxsize=15cm\epsfbox{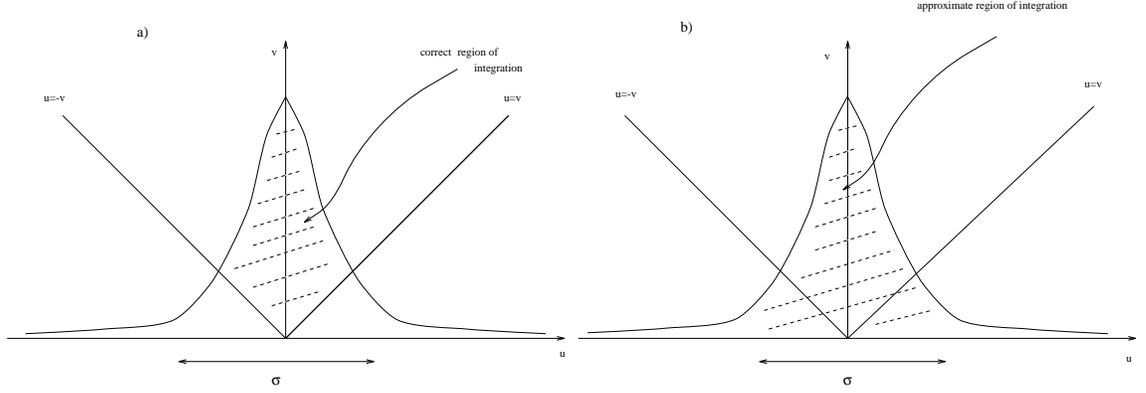}}
\caption{a) The integration region for the integral $F(z)$.  
It is exceedingly difficult to integrate explicitly over such a region. In 
contrast we have in  
b) that if the width of the wave function $\sigma$ is sufficiently small 
we can integrate over the whole range: 
$-\infty\leq u\leq\infty$ explicitly introducing small errors. To have 
a wave-packet $\varphi(E)$ 
with small $\sigma$ means that most particles in the particle ensemble 
posses the same energy.}
\label{packet}
\end{figure}

With this approximation it is convenient to make the following change of 
variables 
$$u=E_2-E_1 \quad, v=E_2+E_1$$
Eq.(\ref{fint}) then becomes
\begin{equation}
F(z)=\frac{1}{6(2\pi)^3}\;\sqrt{\frac{2\alpha}{\pi}}\;\int^{\infty}_{0}dv
\int^{v}_{-v}du\;e^{-\frac{\alpha}{2}(u^2+v^2)}\frac{\sin uz}{u}
\left[3v^3+vu^2\right]
\label{Forig}
\end{equation}
Now, in the approximation of $\alpha$ being large, we can change the 
limits of integration over $u$ (see fig.(\ref{packet})). Thus 
eq.(\ref{Forig}) can be written as
\begin{equation}
F(z)=\frac{1}{3(2\pi)^3}\;\sqrt{\frac{2\alpha}{\pi}}\;\int^{\infty}_{0}dv
\int^{\infty}_{0}du\;e^{-\frac{\alpha}{2}(u^2+v^2)}\frac{\sin uz}{u}
\left[3v^3+vu^2\right].
\label{Fmodif}
\end{equation}
We can now solve both integrals (a list of important integrals are presented in
appendix \ref{APEXPRIM}). After some work we obtain the following result
\begin{equation}
F(z)=\frac{1}{\alpha^2(2\pi)^3}\;\sqrt{\frac{2\alpha}{\pi}}\;
\left[
\pi\Phi(\frac{z}{\sqrt{2\alpha}})+
\frac{z}{6}\;\sqrt{\frac{2\pi}{\alpha}}\;e^{-\frac{z^2}{2\alpha}}\right].
\label{FZA}
\end{equation}
Where $\Phi(x)$ is the probability integral as defined in appendix \ref{APEXPRIM}. 
Integrating eq.(\ref{FZA}) over $z$ and multiplying by $-1$, 
we obtain:
\begin{equation}
H(z)=-\frac{\pi}{\alpha^2(2\pi)^3}\;\sqrt{\frac{2\alpha}{\pi}}\;
\left[z\Phi(\frac{z}{\sqrt{2\alpha}})+
\frac{5}{6}\;\sqrt{\frac{2\alpha}{\pi}}\; e^{-\frac{z^2}{2\alpha}}\right]
\label{HZA}
\end{equation}
and integrating this expression over $z$, we obtain
\begin{equation}
E(z)= -\frac{\pi}{2\alpha^2(2\pi)^3}\;\sqrt{\frac{2\alpha}{\pi}}\;
\left[\Phi(\frac{z}{\sqrt{2\alpha}})\left(z^2+\frac{2\alpha}{3}\right)+
z\sqrt{\frac{2\alpha}{\pi}}\;e^{-\frac{z^2}{2\alpha}}\right].
\label{EZA}
\end{equation}
In eqs.(\ref{HZA})-(\ref{EZA}) we have used the fact that we know the 
exact asymptotic behaviour of 
these functions in order to set the arbitrary integration constant to zero.
\section{Behaviour of $\langle \hat{T}^{\mu\nu}(r,t)\rangle$ when $r$ and 
$t\rightarrow\infty$}
Now let us compute once more the limit of $\langle \hat{T}^{\mu\nu}(r,t)
\rangle$ when $r$ and $t\rightarrow\infty$. The results we obtain in these 
limits are:
\\
a) $t$ fixed, $r\rightarrow\infty$
$$\langle \hat{T}^{00}(r,t)\rangle\stackrel{r\rightarrow\infty}{=}
\frac{2}{\alpha^2(2\pi)^3}\sqrt{\frac{2\alpha}{\pi}}\;\frac{1}{r},$$
$$A(r,t)\stackrel{r\rightarrow\infty}{=}
\frac{1}{\alpha^2(2\pi)^3}\sqrt{\frac{2\alpha}{\pi}}\;\frac{1}{r},$$
$$B(r,t)\stackrel{r\rightarrow\infty}{=}
-\frac{1}{\alpha^2(2\pi)^3}\sqrt{\frac{2\alpha}{\pi}}\;\frac{1}{r}$$
and
$$C(r,t)\stackrel{r\rightarrow\infty}{=}
\frac{2}{\alpha^2(2\pi)^3}\sqrt{\frac{2\alpha}{\pi}}\;\frac{t}{r^2}.$$
\\
b) $r$ fixed, $t\rightarrow\infty$
$$\langle \hat{T}^{00}(r,t)\rangle\stackrel{t\rightarrow\infty}{=}
-\frac{2}{3\alpha^2(2\pi)^3}\; 
e^{-\frac{t^2}{2\alpha}}\;\sinh\frac{tr}{\alpha}\;\frac{t}{r},
$$
$$
A(r,t)\stackrel{t\rightarrow\infty}{=}
\frac{2}{\alpha(2\pi)^3}\; 
e^{-\frac{t^2}{2\alpha}}\;\sinh\frac{tr}{\alpha}\;\frac{t}{r^2},
$$
$$
B(r,t)\stackrel{t\rightarrow\infty}{=}
-\frac{2}{3\alpha^2(2\pi)^3}\; 
e^{-\frac{t^2}{2\alpha}}\;\sinh\frac{tr}{\alpha}\;\frac{t}{r}
$$
and
$$C(r,t)\stackrel{t\rightarrow\infty}{=}
-\frac{2}{3\alpha^2(2\pi)^3}\; 
e^{-\frac{t^2}{2\alpha}}\;\sinh\frac{tr}{\alpha}\;\frac{t}{r}.
$$
As we can see, with this approximation we have recovered our previous 
results eqs.(\ref{M1})-(\ref{MFINAL}) (at leading order). 

Now let us examine more carefully the results presented above. For a better 
understanding of these results it is convenient to plot the energy-density and 
energy-flux for each of the regions we are studying. 

\begin{figure}
\centerline{\rotatebox{-90}{\resizebox{10cm}{!}{\includegraphics{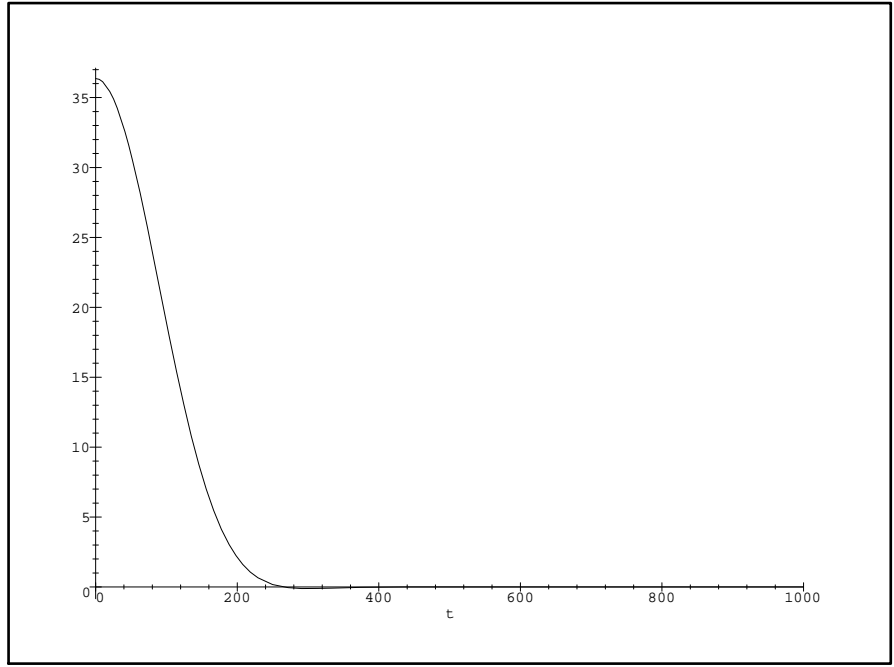}}}
}
\caption{$\langle\hat{T}^{00}(r,t)\rangle$ ($\times10^{-12}$) 
for $r$ fixed and equal to $0.5$, $\alpha$=10000.}
\label{T00rfixed} 
\end{figure}

\begin{figure}
\centerline{\rotatebox{-90}{\resizebox{10cm}{!}
{\includegraphics{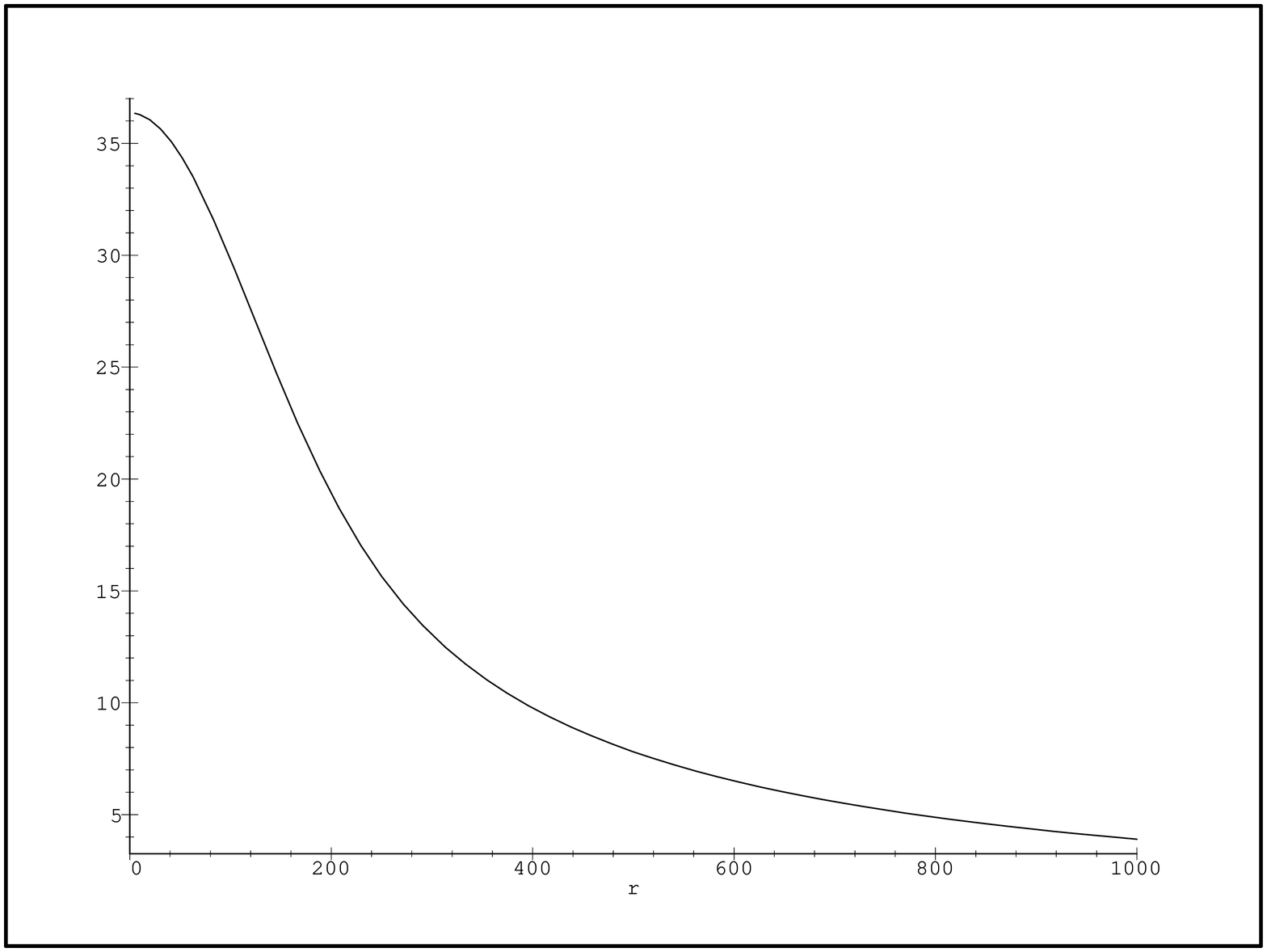}}}}
\caption{$\langle\hat{T}^{00}(r,t)\rangle$ ($\times10^{-12}$) 
for $t$ fixed and equal to $0.5$, $\alpha$=10000.}
\label{T00tfixed}
\end{figure}

Figures (\ref{T00rfixed}) and (\ref{T00tfixed}) show $\langle\hat{T}^{00}(r,t)
\rangle$ for $r$ fixed and $t$ fixed respectively. (In all these plots, 
we are taking a value of $\alpha=10000$.) We can see that 
$\langle\hat{T}^{00}(r,t)
\rangle$ is always finite. Notice that for large $t$ and $r$ fixed
$\langle\hat{T}^{00}(r,t)\rangle$  develops small but negative values 
(we will discuss this result in the next section). If we 
fix $t$, the energy-density is always positive 
and vanishes asymptotically. 

\begin{figure}
\centerline{\rotatebox{-90}{\resizebox{10cm}{!}
{\includegraphics{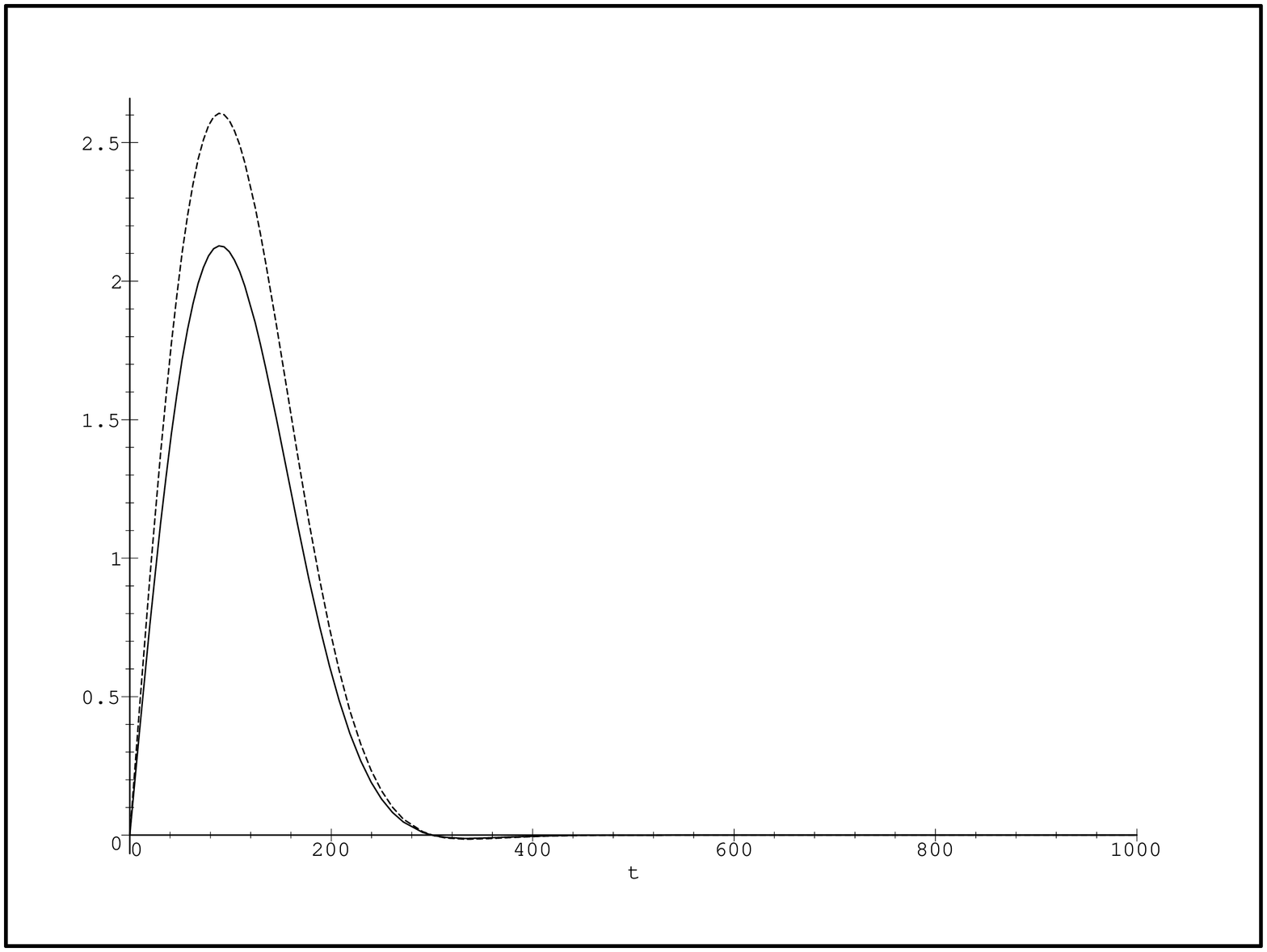}}}}
\caption{$\langle\hat{T}^{01}(r,t)\rangle$ ($\times10^{-14}$) 
for $r$ fixed and equal to $0.5$, $\alpha$=10000. Here the solid line 
shows $\langle\hat{T}^{01}(r,t)\rangle$ in the direction $\phi=\pi/4$, 
$\gamma=\pi/4$ whereas the dashed line shows $\langle\hat{T}^{01}(r,t)\rangle$ 
in the direction $\phi=\pi/4$, $\gamma=\pi/3$. The other 
2 components of $\langle\hat{T}^{0i}(r,t)\rangle$ behave in a similar fashion, 
different only from $\langle\hat{T}^{01}(r,t)\rangle$ by a factor introduced 
by ${\hat x}^i=(\cos\phi\sin\gamma,\sin\phi\sin\gamma,\cos\gamma)$}
\label{T0irfixed}
\end{figure}

\begin{figure}
\centerline{\rotatebox{-90}{\resizebox{10cm}{!}
{\includegraphics{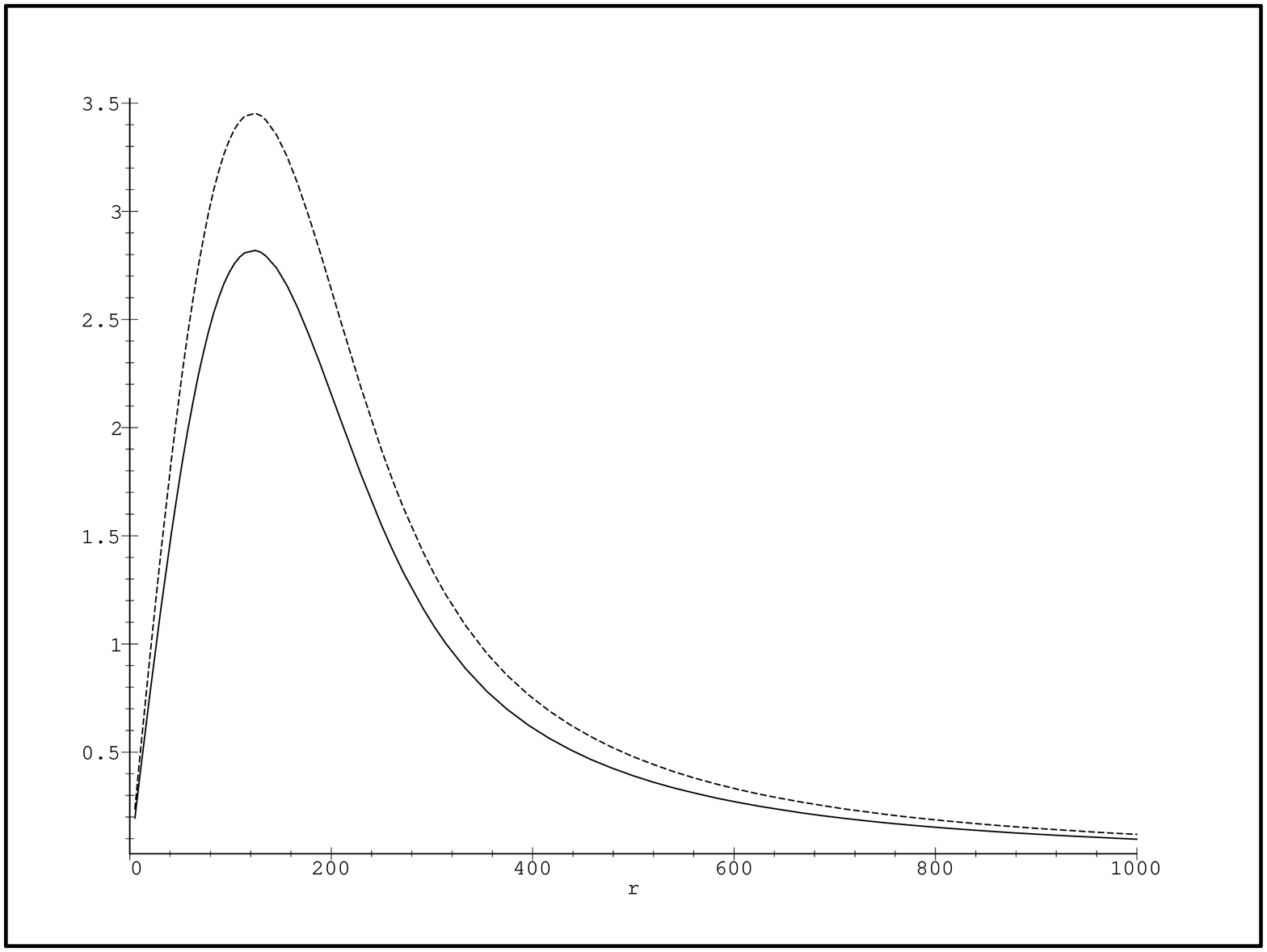}}}}
\caption{$\langle\hat{T}^{01}(r,t)\rangle$ ($\times10^{-14}$) 
for $t$ fixed and equal to $0.5$, $\alpha$=10000. Here the solid line 
shows $\langle\hat{T}^{01}(r,t)\rangle$ in the direction $\phi=\pi/4$, 
$\gamma=\pi/4$ whereas the dashed line shows $\langle\hat{T}^{01}(r,t)\rangle$ 
in the direction $\phi=\pi/4$, $\gamma=\pi/3$. The other 
2 components of $\langle\hat{T}^{0i}(r,t)\rangle$ behave in a similar fashion, 
different only from $\langle\hat{T}^{01}(r,t)\rangle$ by a factor introduced 
by ${\hat x}^i=(\cos\phi\sin\gamma,\sin\phi\sin\gamma,\cos\gamma)$}
\label{T0itfixed}
\end{figure}

The energy-flux is plotted in figures 
(\ref{T0irfixed})-(\ref{T0itfixed}). $\langle\hat{T}^{0i}(r,t)\rangle$ 
also develops small but negative values when $r$ remains fixed and time 
evolves. 
It is interesting to notice one aspect in the evolution of the energy-flux 
as time evolves namely that the energy-flux always presents very 
small values, in comparison to the corresponding values for 
the energy-density (also true for the $t-$fixed and case). 
\section{The total energy of the string}
Now, let us compute an analytic expression for the total energy of the string. 
The total energy is given by the following expression:
\begin{equation}
E_T=\int \langle\hat{T}^{00}(r,t)\rangle dV =
4\pi\int_0^R \langle\hat{T}^{00}(r,t)\rangle r^2\;dr
\end{equation}
where $R$ acts as a cut-off. From eqs.(\ref{forfac}) and 
(\ref{FZA}) we have:
\begin{eqnarray}
E_T&=&
\frac{2}{\alpha^2(2\pi)^2}\;\sqrt{\frac{2\alpha}{\pi}}
\int^R_0 r\;dr\left[\pi\left(\Phi(\frac{t+r}{\sqrt{2\alpha}})-
\Phi(\frac{t-r}{\sqrt{2\alpha}})\right)+\right.\nonumber\\ & & \left.
\frac{1}{6}\sqrt{\frac{2\pi}{\alpha}}\left((t+r)\;
e^{-\frac{(t+r)^2}{2\alpha}}-(t-r)\;
e^{-\frac{(t-r)^2}{2\alpha}}\right)\right].\nonumber\\ & &
\label{Energia0}
\end{eqnarray}
Making the following changes of variables
$$x=t+r,\,\,y=t-r$$
we can re-write eq.(\ref{Energia0}) as
\begin{eqnarray}
E_T&=&
\frac{2}{\alpha^2(2\pi)^2}\;\sqrt{\frac{2\alpha}{\pi}}\left[
-\int^{t+R}_t (t-x)\;dx\;\pi\;\Phi(\frac{x}{\sqrt{2\alpha}})+
\int^{t-R}_t (t-y)\;dy\;\pi\;\Phi(\frac{y}{\sqrt{2\alpha}})
\right.\nonumber\\ & & \left.
-\frac{1}{6}\sqrt{\frac{2\pi}{\alpha}}\int^{t+R}_t (t-x)\;dx\;x
\;e^{-\frac{x^2}{2\alpha}}+\frac{1}{6}
\sqrt{\frac{2\pi}{\alpha}}\int^{t-R}_t (t-y)\;dy\;y
\;e^{-\frac{y^2}{2\alpha}}
\right].\nonumber\\ & &
\label{EnergiaA}
\end{eqnarray}
Thus, the total energy can be written in the following form:
\begin{equation}
E_T=\frac{2}{\alpha^2(2\pi)^2}\sqrt{\frac{2\alpha}{\pi}}\left[
P(t+R)-P(t-R)+Q(t+R)-Q(t-R)\right]
\label{Energia1}
\end{equation}
where
\begin{equation}
P(z)=\pi\Phi(\frac{z}{\sqrt{2\alpha}})\left[\frac{1}{2}(z^2-\alpha)-tz
\right]-\sqrt{2\alpha\pi}\;e^{-\frac{z^2}{2\alpha}}\;
\left(t-\frac{z}{2}\right)
\end{equation}
and
\begin{equation}
Q(z)=\frac{\alpha\pi}{6}\left[\sqrt{\frac{2\alpha}{\pi}}\;\frac{t}{\alpha}
\;e^{-\frac{z^2}{2\alpha}}\left(t-z\right)+
\Phi(\frac{z}{\sqrt{2\alpha}})\right].
\end{equation}
We can now, take the limits we discussed in chapter \ref{ChapTmunu} but before 
doing that, let us address our result found in the previous section for the 
energy-density in the regime $t\rightarrow\infty$, $r$ fixed.

Although the notion of a state with negative energy is not known in 
classical physics this is not the case in quantum field theory. The existence of 
quantum states with negative energy density is proved to be inevitable as some 
authors have been able to show \cite{kuo,eps}. In quantum field theory, the 
energy density may be negative in a certain space-time region. In order to 
avoid disturbing effects arising from these negative energy-density regions, we 
need some constraint on the temporal or spatial extent of the negative energy density 
as well as on its magnitude. It has been argued \cite{ford, helf} that the 
negative energy $-\Delta E$ localisable in a time of order $\Delta t$ should 
satisfy the following quantum inequality:
\begin{equation}
\left|\Delta E\right|\Delta t\stackrel{<}{\sim}\hbar.
\label{ineq}
\end{equation}
The physical implication of this quantum inequality is that an observer cannot 
see unboundedly large negative energy densities which persist for arbitrarily 
long periods of time.
%
It is not difficult to see that in the region where the energy density we 
have computed takes on vanishingly small negative values, the quantum inequality 
eq.(\ref{ineq}) is indeed satisfied. Thus, the negative 
energy density we have found does not play any noticeable physical role as we 
may have expected.

Working now the asymptotic behaviour for the total energy we obtain\\

a) $R\gg T$, $T$ large but fixed (remembering that $\alpha\sim T^{2}$) 
\begin{equation}
E_T\sim\left(\frac{R}{T}\right)^2\;\frac{1}{T}
\label{EnergyI}
\end{equation}
and \\
b)$R<T$, $T$ large but fixed. From eq.(\ref{esfsim0}) we find again: 
\begin{equation}
E_T\sim \left(\frac{R}{T}\right)^3\frac{1}{T}.
\label{EnergyF}
\end{equation}
As we can see these results are in total agreement with the ones obtained 
in chapter \ref{ChapTmunu}, as they should be.


\section{Final Remarks}
As we can see from the results obtained throughout this chapter, we 
have an accurate method to obtain more complete 
results for the expectation value of the massless sector of the 
quantum bosonic string energy-momentum 
tensor in spherical configurations. We have been able to show that this method 
reproduces the asymptotic behaviour for $\langle\hat{T}^{\mu\nu}(r,t)\rangle$ 
computed in the previous chapter as well as the asymptotic behaviour of 
the total energy. We can now use these explicit expression 
for $\langle\hat{T}^{\mu\nu}(r,t)\rangle$ in order 
to investigate the weak field limit of Einstein's equations.

\chapter{The gravitational field of a quantum bosonic string}
\label{Back}
In this chapter and the next, we will compute and analyse the 
weak-field metric produced by a quantum bosonic 
string. We shall see that the gravitational field 
for a quantum string differs in many ways from that of a 
{\em classical string} (cosmic string)\footnote{
Strictly speaking, this is true in the Nambu approximation where 
the string is consider to have zero width.}. Cosmic strings 
(amongst other defects) could have been produced as the early Universe went 
through a phase transition. They have been proposed as a possible explanation 
for the density fluctuations in the early Universe that may have given rise 
to the origin of galaxies. 
Studies of how their gravitational field affects their motion have 
been presented by a number of authors (see for example: \cite{Moss4,edal,KIB}).
It has been shown, for example, that the metric  around a 
straight static string is that of a conical space-time; that is, although 
particles moving near the cosmic string are not affected by any gravitational 
force from the string, gravitational lensing occurs. This leads to double images 
with a characteristic separation of about a few seconds of arc \cite{Alex}. 
This in principle should lead to observational evidence for the existence 
of cosmic strings in the early Universe. However, current astronomical 
observations have not shown yet any evidence in that respect.

Another important effect due to the space-time geometry produced by a 
cosmic string is 
the appearance of wakes behind moving strings \cite{Hisc,Sil}. 
This causes particles 
passing near the string to conglomerate in a region and eventually 
this region will have a total mass comparable with that of the 
string itself. These wakes may provide us with an explanation for the 
observed large scale structure of the Universe.

In this calculation, we will be working 
in a {\em far field} approximation to Einstein's equations, and will show that 
the solution for 
the gravitational field of a quantum string will hold in regions far from 
the source. The metric of a quantum string approaches that of a flat Minkowski 
space-time as we go farther away from the source. This is in agreement with our 
weak-field approximation assumption. It should be noticed though, that in  
these calculations we are not considering the string solutions in a fully 
self-consistent way; we must not forget that, once we obtain the metric 
generated by the string, 
the string will depend now on this new metric. 
We must remember that String Theory is 
extremely sensitive to the background upon which the string is moving. So if 
our weak-field approximation of the metric starts 
to deviate noticeably 
from the Minkowski metric, our string solution used in the computation of 
the string energy-momentum tensor will no longer be valid. The results presented 
in this chapter, therefore, may be used as long as the string solutions can 
be considered as those for a flat Minkowski space-time which is the case here.
\section{The weak field approximation to Einstein's field equations}
The weak field approximation is a linearised version of Einstein's field 
equations. It is based upon the assumption that the metric tensor will 
deviate only slightly from the Minkowski metric \cite{Wei}:
$$g_{\mu\nu}=\eta_{\mu\nu}+h_{\mu\nu},$$
where $\left|h^{\mu\nu}\right|\ll 1$. In this way we may neglect 
all terms not linear in $h_{\mu\nu}$ or its derivatives.

In this approximation, Einstein's field equations read:
\begin{equation}
\Box h^{\mu\nu}-\frac{\partial^2 h^{\nu}_{\lambda}}{\partial x^\lambda 
\partial x^{\mu}}-\frac{\partial^2 h^{\mu}_{\lambda}}{\partial x^\lambda 
\partial x^{\nu}}+\frac{\partial^2 h^{\lambda}_{\lambda}}{\partial x^\mu 
\partial x^{\nu}}=-16\pi G S^{\mu\nu},
\label{WFE}
\end{equation}
where
$$ S^{\mu\nu}=T^{\mu\nu}-\frac{1}{2}\eta^{\mu\nu}T^{\lambda}_{\lambda}.$$
These equations cannot be solved in a unique way, the reason for this being 
that we can always perform the following general transformation:
$$x^{\mu}\rightarrow x'^{\mu}=x^\mu+\epsilon^{\mu}(x),$$
where $\frac{\partial\epsilon^{\mu}(x)}{\partial x^{\nu}}$ is at most of the 
same magnitude as $h^{\mu\nu}$; $\epsilon^{\mu}$ are small and arbitrary 
functions of $x$. 

In this new coordinate system, $x'^{\mu}$, the metric is given by
$$g'^{\mu\nu}=\frac{\partial x'^\mu}{\partial x^{\lambda}}
\frac{\partial x'^\nu}{\partial x^{\rho}}g^{\lambda\rho}
$$
or, since $g^{\mu\nu}\simeq\eta^{\mu\nu}-h^{\mu\nu}$:
$$h'^{\mu\nu}=h^{\mu\nu}-\frac{\partial\epsilon^{\mu}}{\partial x^{\lambda}}
\eta^{\lambda\nu}-\frac{\partial\epsilon^{\nu}}{\partial x^{\rho}}
\eta^{\rho\mu}.$$
Therefore, if $h^{\mu\nu}$ is a solution,
$$h'^{\mu\nu}=h^{\mu\nu}-\frac{\partial\epsilon^{\mu}}{\partial x_{\nu}}
-\frac{\partial\epsilon^{\nu}}{\partial x_{\mu}}
$$
is also a solution. 

We can now choose a gauge in order to work in a particular 
(convenient) coordinate system. Choosing the Harmonic gauge $g^{\mu\nu}
\Gamma^{\lambda}_{\mu\nu}=0$ (here $\Gamma^{\lambda}_{\mu\nu}$ are the 
Christoffel's symbols), we get to a first order:
\begin{equation}
\frac{\partial h^{\mu}_{\nu}}{\partial x^{\mu}}=\frac{1}{2}
\frac{\partial h^{\mu}_{\mu}}{\partial x^{\nu}}.
\label{HGC}
\end{equation}
Notice that if $h^{\mu\nu}$ does not satisfy this relation, we can always 
make it comply by performing a change of coordinates. 
In this gauge, the weak 
field approximation to Einstein's equations reads:
\begin{equation}
\Box h^{\mu\nu}(x)=-16\pi G\left(T^{\mu\nu}(x)-\frac{1}{2}\eta^{\mu\nu}
T^{\lambda}_{\lambda}(x)\right).
\label{EFE}
\end{equation}
Here it may be appropriate to mention briefly, as a reminder, 
that although working with the linearised version of Einstein's field equations 
presents clear advantages over working with its exact form, we always have to 
be very careful about the conclusions we reach from this approximation. The 
exact gravitational field corresponding to the exact solution may deviate 
significantly from that derived from the linearised theory if the sources of 
the field move in a different way from what was in the first place assumed. 
Therefore, the 
reliability of the results found in the linearised theory will be higher if 
we have a good knowledge of the behaviour of our source and if the source is 
not too massive.

Since in our case the source of the gravitational field is a quantum operator 
(${\hat T}^{\mu\nu}(x)$), we must modify eq.(\ref{EFE}) slightly replacing in its 
RHS $T^{\mu\nu}(x)$ by $\langle {\hat T}^{\mu\nu}(x)\rangle$:
\begin{equation}
\Box h^{\mu\nu}(x)=-16\pi G\left(\langle {\hat T}^{\mu\nu}(x)\rangle-
\frac{1}{2}\eta^{\mu\nu}
\langle {\hat T}^{\lambda}_{\lambda}(x)\rangle\right).
\label{EFEVE}
\end{equation}
This is what is 
known as a semiclassical approximation to quantum gravity.

In order for the field equations to be integrable, the expectation value 
$\langle {\hat T}^{\mu\nu}(x)\rangle$ must be divergence free \cite{Steph}, that 
is
\begin{equation}
\langle {\hat T}^{\mu\nu}(x)\rangle_{;\nu}=0.
\label{DF}
\end{equation}
It should be noted that eq.(\ref{DF}) is not a consequence of the equations governing 
the quantum fields but rather a constraint on these quantities \cite{Steph} 
for example, the states which are used to form the expectation values 
(just as we saw happen in chapter \ref{ChapTmunu}). 
\section{Far fields}
In the last chapter we derived explicit equations for the expectation value of the 
string energy-momentum tensor. Therefore, we may now proceed to solve 
eq.(\ref{EFE}).

As we have already shown in chapter \ref{ChapTmunu}, 
$\langle {\hat T}^{\lambda}_{\lambda}(x)\rangle=0$. Thus eq.(\ref{EFEVE}) 
simplifies to
\begin{equation}
\Box h^{\mu\nu}(x)=-16\pi G\;\langle{\hat T}^{\mu\nu}(x)\rangle.
\label{EJE}
\end{equation}
This differential equation can be solved with the help of the retarded 
potential Green's function \cite{jac} leading to
\begin{equation}
h^{\mu\nu}=4G\int d^3 \vec{x}'
\frac{\langle {\hat T}^{\mu\nu}(\vec{x}',t_{ret})\rangle}{|\vec{x}-\vec{x}'|},
\label{EQSOLV}
\end{equation}
where
$$t_{ret}=t-\frac{|\vec{x}-\vec{x}'|}{c}.$$
Of course, to this solution we can always add the solution of the  homogeneous
part:
$$\Box h^{\mu\nu}=0.$$
So we can interpret the two solutions as follows: the solution (\ref{EQSOLV}) 
is the gravitational radiation produced by the source, in our case a 
quantum bosonic string; the solution of the homogeneous equation represents the 
gravitational radiation coming in from infinity \cite{Wei}. 

The retarded time $t_{ret}$ that appears in the argument of 
the energy-momentum tensor in 
eq.(\ref{EQSOLV}) tells us that gravitational effects due to this 
source propagate at the speed of light.

\begin{figure}
\centerline{\epsfxsize=15cm\epsfbox{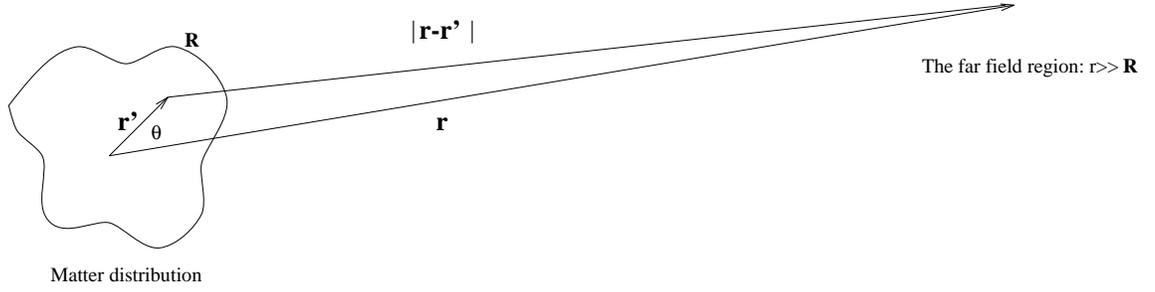}}
\caption{The far field approximation. The source is assumed to be contained 
within a region of radius $R$. We compute the weak-field metric in a region 
at a distance $r$ far from the source.}
\label{ffield}
\end{figure}

At the quantum level the string energy-momentum tensor does not behave 
like a localised source, we cannot blindly use the retarded potential technique 
to solve Einstein's field equations; therefore, in order to solve 
eq.(\ref{EQSOLV}), we need to make some reasonable assumptions about the 
behaviour of our source. We have learnt for example in the previous 
chapter that the string energy-momentum tensor vanishes asymptotically for 
large values of $r$ and $t$. In this case, a far field approximation seems to be appropriate 
(see fig.(\ref{ffield})), where we would be interested in computing the 
gravitational field in a region at a distance $r$ far from the matter 
distribution. That is, in the region where $r\gg R$, where $R$ acts like a 
cut-off, in that most of the matter would be concentrated within a region of 
radius $R$ and we will be neglecting any other contributions from 
outside this region. We may visualise this scenario as if looking 
at the gravitational field exerted by a planetary system or a galaxy 
viewed from a location far away from them and neglecting all interstellar 
matter in between them and our location. Notice that our calculation for 
$\langle{\hat T}^{\mu\nu}(\vec{x},t)\rangle$ was carried out using 
Gaussian-wave packets peaked around $E=0$. Since $E=\sqrt{\vec{p}^2+m^2}$ 
(because we are always working on-shell) we have that for the massless case we 
are considering here $E=\sqrt{\vec{p}^2}$ implies that we will be considering 
the weak gravitational field produced by a source moving with 
centre of mass velocities 
small compared with the velocity of light. This kind of approximation is 
the one presented in most general relativity textbooks (e.g. \cite{Steph}, 
\cite{Meis}, \cite{Wei}, \cite{sch}.) 
\section{The quantum bosonic string gravitational field}
From chapter \ref{ChapTmunu} eqs.(\ref{rotaT}) and (\ref{forfac}) 
we learnt that the quantum string energy-momentum tensor components 
may be written as:
\begin{eqnarray}
\langle\hat{T}^{00}(\vec{x},t)\rangle &=&
\frac{1}{r}\left[F(t+r)-F(t-r)\right],\nonumber\\
 \langle\hat{T}^{11}(\vec{x},t)\rangle &=&
-\frac{1}{r^2}\left[H(t+r)+H(t-r)\right]+\frac{1}{r^3}
\left[E(t+r)-E(t-r)\right]+
\nonumber\\ & & \cos^2\phi\;\sin^2\gamma\;\left\{
\frac{1}{r}\left[F(t+r)-F(t-r)\right]+\frac{3}{r^2}\left[H(t+r)+H(t-r)\right]
\right.\nonumber\\ & & \left. -\frac{3}{r^3}\left[E(t+r)-E(t-r)\right]\right\},
\nonumber\\
 \langle\hat{T}^{22}(\vec{x},t)\rangle &=&
\frac{1}{r^2}\left[H(t+r)+H(t-r)\right]+\frac{1}{r^3}
\left[E(t+r)-E(t-r)\right]+
\nonumber\\ & & \sin^2\phi\;\sin^2\gamma\;\left\{
\frac{1}{r}\left[F(t+r)-F(t-r)\right]+\frac{3}{r^2}\left[H(t+r)+H(t-r)\right]
\right.\nonumber\\ & & \left. -\frac{3}{r^3}\left[E(t+r)-E(t-r)\right]\right\},
\nonumber\\
 \langle\hat{T}^{33}(\vec{x},t)\rangle &=&
\frac{1}{r^2}\left[H(t+r)+H(t-r)\right]+\frac{1}{r^3}
\left[E(t+r)-E(t-r)\right]+
\nonumber\\ & & \cos^2\gamma\;\left\{
\frac{1}{r}\left[F(t+r)-F(t-r)\right]+\frac{3}{r^2}\left[H(t+r)+H(t-r)\right]
\right.\nonumber\\ & & \left. -\frac{3}{r^3}\left[E(t+r)-E(t-r)\right]\right\},
\nonumber\\
 \langle\hat{T}^{01}(\vec{x},t)\rangle &=&
-\cos\phi\;\sin\gamma\;\left\{\frac{1}{r}\left[F(t+r)+F(t-r)\right]
+\frac{1}{r^2}\left[H(t+r)-H(t-r)\right]\right\},\nonumber\\
 \langle\hat{T}^{02}(\vec{x},t)\rangle &=&
-\sin\phi\;\sin\gamma\;\left\{\frac{1}{r}\left[F(t+r)+F(t-r)\right]
+\frac{1}{r^2}\left[H(t+r)-H(t-r)\right]\right\},\nonumber\\
 \langle\hat{T}^{03}(\vec{x},t)\rangle &=&
 -\cos\gamma\;\left\{\frac{1}{r}\left[F(t+r)+F(t-r)\right]
+\frac{1}{r^2}\left[H(t+r)-H(t-r)\right]\right\}.
\label{Texp}
\end{eqnarray}
We will insert expressions (\ref{Texp}) into eq.(\ref{EQSOLV}) shortly. 
First, in order to work out the far field of our quantum string, 
we need to expand $|\vec{x}-\vec{x}'|$ as a power series 
\cite{Steph,jac,Landa,Gold}:
\begin{eqnarray}
|\vec{x}-\vec{x}'|&=&r-\frac{x^{i}x'^{i}}{r}-
\frac{1}{2}\frac{x^{i}x^{j}}
{r^3}\left(x'^{i}x'^{j}-r'^2\delta^{ij}\right)+
\dots,\nonumber\\
\frac{1}{|\vec{x}-\vec{x}'|}&=&\sum^{\infty}_{l=0}
\frac{r'^{l}}{r^{l+1}}P_{l}(cos\theta)=
\frac{1}{r}+\frac{x^{i}x'^{i}}{r^3}+
\frac{1}{2}\frac{x^{i}x^{j}}
{r^5}\left(3x'^{i}x'^{j}-r'^2\delta^{ij}\right)+
\dots
\label{expan}
\end{eqnarray}
where $i,j=1,2,3$. 
Substituting this expression into the argument $t-|\vec{x}-\vec{x}'|$ 
of the energy-momentum 
tensor and expanding the components of $\langle {\hat T}^{\mu\nu}\rangle$, 
we obtain:
\begin{eqnarray}
\int d^3 \vec{x}'
\frac{\langle {\hat T}^{\mu\nu}(\vec{x}',t)\rangle}{|\vec{x}-\vec{x}'|}&=&
\int d^3 \vec{x}'\langle {\hat T}^{\mu\nu}(\vec{x}',t-r)\rangle
\left\{\frac{1}{r}+\frac{x^{i}x'^{i}}{r^3}+
\frac{1}{2}\frac{x^{i}x^{j}}
{r^5}\left(3x'^{i}x'^{j}-r'^2\delta^{ij}\right)+
\dots\right\}\nonumber\\
& &+\int d^3 \vec{x}'\langle \dot{{\hat T}}^{\mu\nu}(\vec{x}',t-r)\rangle
\left\{\frac{x^{i}x'^{i}}{r^2}+
\frac{1}{2}\frac{x^{i}x^{j}}
{r^4}\left(3x'^{i}x'^{j}-r'^2\delta^{ij}\right)+
\dots\right\}\nonumber\\
& &+\int d^3 \vec{x}'\frac{1}{2}\langle 
\ddot{{\hat T}}^{\mu\nu}(\vec{x}',t-r)\rangle\frac{x^{i}x^{j}
x'^{i}x'^{j}}{r^3}+\dots,
\label{expan2}
\end{eqnarray}
where the dot means differentiation with respect to time $t$. 
The first three terms of this series are the analogue to 
the `charge', `dipole' and `quadrupole' terms of the electro-magnetic field 
multipole expansion \cite{jac}. 

We will work only with the terms we have displayed above 
since they are the ones that will dominate the expansion in the region of 
large $r$. 

Inserting eq.(\ref{expan2}) into expression (\ref{EQSOLV}), we 
obtain:
\begin{eqnarray}
h^{\mu\nu}(\vec{x},t)&=& 
\left[h^{\mu\nu}_{a}(\vec{x},t)+h^{\mu\nu}_{b}(\vec{x},t)+
\frac{1}{2}h^{\mu\nu}_{c}(\vec{x},t)+\right.\nonumber\\ & & \left.
r\dot{h}^{\mu\nu}_{b}(\vec{x},t)+\frac{r}{2}\dot{h}^{\mu\nu}_{c}(\vec{x},t)
+\frac{1}{2}\ddot{h}^{\mu\nu}_{d}(\vec{x},t)\right].
\label{MetComp}
\end{eqnarray}
where $i=1,2,3$.
Here, $h^{\mu\nu}_{a}(\vec{x},t)$, $h^{\mu\nu}_{b}(\vec{x},t)$, 
$h^{\mu\nu}_{c}(\vec{x},t)$ and $h^{\mu\nu}_{d}(\vec{x},t)$ are given by:
\begin{equation}
h^{\mu\nu}_{a}(\vec{x},t)=\frac{4G}{r}\int d^3 \vec{x}'\langle 
{\hat T}^{\mu\nu}(\vec{x}',t-r)\rangle,
\label{huno}
\end{equation}
\begin{equation}
h^{\mu\nu}_{b}(\vec{x},t)=4G\frac{x^{i}}{r^3}
\int d^3 \vec{x}'\,x'^{i}\langle 
{\hat T}^{\mu\nu}(\vec{x}',t-r)\rangle,
\label{hdos}
\end{equation}
\begin{equation}
h^{\mu\nu}_{c}(\vec{x},t)=
4G\frac{x^{i}x^{j}}{r^5}\int d^3\;\vec{x}' \;(3\;x'^{i}x'^{j}-
r'^2\delta^{ij})\;\langle 
{\hat T}^{\mu\nu}(\vec{x}',t-r)\rangle
\label{htresa}
\end{equation}
and
\begin{equation}
h^{\mu\nu}_{d}(\vec{x},t)=
4G\frac{x^{i}x^{j}}{r^3}\int d^3 \vec{x}'\;x'^{i}x'^{j}\langle 
{\hat T}^{\mu\nu}(\vec{x}',t-r)\rangle.
\label{htresb}
\end{equation}
Now we can use the explicit expressions eq.(\ref{Texp}) given in chapter 
\ref{Alpha} by eqs.(\ref{FZA})-(\ref{EZA}) for 
the string energy-momentum tensor in order to compute the integrals in
eqs.(\ref{huno})-(\ref{htresb}). 
Here we have chosen 
$$x^i=r(\cos\alpha_0\sin\gamma_0,\sin\alpha_0\sin\gamma_0,\cos\gamma_0)$$ 
and 
$$x'^i=r(\cos\phi\sin\gamma,\sin\phi\sin\gamma,\cos\gamma).$$
After some work (and using the table of integrals in appendix \ref{APEXPRIM}) 
we find the following relations between the different 
components of the metric tensor $h^{\mu\nu}(\vec{x},t)$ (the explicit results 
for all of the metric components are given in appendix \ref{GRAVFOR}):
\begin{equation}
\frac{1}{3}h^{00}_{a}(\vec{x},t)=h^{ii}_{a}(\vec{x},t),
\label{empezar}
\end{equation}
\begin{equation}
h^{0i}_{a}(\vec{x},t)=0,
\end{equation}
\begin{equation}
h^{00}_{b}(\vec{x},t)=h^{ii}_{b}(\vec{x},t)=0,
\label{segundo}
\end{equation}
\begin{equation}
h^{0i}_{b}(\vec{x},t)={\cal W}_i(\alpha_0,\gamma_0){\cal G}_1(\vec{x},t),
\label{tercero}
\end{equation}
\begin{equation}
h^{00}_{c}(\vec{x},t)=h^{0i}_{c}(\vec{x},t)=0,
\label{cuarto}
\end{equation}
\begin{equation}
h^{ii}_{c}(\vec{x},t)={\cal Z}_i(\alpha_0,\gamma_0){\cal F}_1(\vec{x},t),
\label{quinto}
\end{equation}
\begin{equation}
h^{ii}_{d}(\vec{x},t)=r^2\;{\cal S}_i(\alpha_0,\gamma_0){\cal F}_1(\vec{x},t)
+\frac{h^{00}_{d}(\vec{x},t)}{3},
\end{equation}
\begin{equation}
h^{0i}_{d}(\vec{x},t)=0.
\end{equation}
With
\begin{eqnarray}
h^{00}_{a}(\vec{x},t)&=&\frac{1}{4\alpha^2\pi\; r}
\;\sqrt{\frac{2\alpha}{\pi}}\;
\left\{\left[\Phi(\frac{t-r-R}{\sqrt{2\alpha}})-
\Phi(\frac{t-r+R}{\sqrt{2\alpha}})\right]\right.\nonumber\\ & &
\left. \left((t-r)^2-R^2+\frac{2\alpha}{3}
\right)\right.\nonumber\\ & & \left.
+\sqrt{\frac{2\alpha}{\pi}}\left[e^{-\frac{(t-r+R)^2}{2\alpha}}+
e^{-\frac{(t-r-R)^2}{2\alpha}}\right]\left(\frac{2R}{3}-t+r\right)\right\},
\label{h00uno}
\end{eqnarray}
\begin{eqnarray}
h^{00}_{d}(\vec{x},t)&=&\frac{2}{3\alpha^2(2\pi)^2\; r}\;
\sqrt{\frac{2\alpha}{\pi}}\;
\left\{\frac{\pi}{4}\left[\Phi(\frac{t-r+R}{\sqrt{2\alpha}})-
\Phi(\frac{t-r-R}{\sqrt{2\alpha}})\right]\right.\nonumber\\ & & \left. 
\left(R^4-(t-r)^4-\alpha^2-4\alpha
(t-r)^2\right)+\frac{\alpha(t-r)}{4}\sqrt{\frac{2\pi}{\alpha}}
\left[e^{-\frac{(t-r+R)^2}{2\alpha}}-e^{-\frac{(t-r-R)^2}{2\alpha}}\right]
\right.\nonumber\\ & & \left.
\left((t-r)^2-R^2-3\alpha\right)
+\frac{\alpha R}{4}\sqrt{\frac{2\pi}{\alpha}}
\left[e^{-\frac{(t-r+R)^2}{2\alpha}}+e^{-\frac{(t-r-R)^2}{2\alpha}}\right]
\right.\nonumber\\ & & \left.
\left((t-r)^2+\frac{R^2}{3}+\alpha\right)\right\},
\label{h00cuatro}
\end{eqnarray}
\begin{eqnarray}
{\cal F}_1(\vec{x},t)&=&\frac{\pi}{8\alpha^2(2\pi)^3\;r^3}\;
\sqrt{\frac{2\alpha}{\pi}}\;
\left\{\left[\Phi(\frac{t-r+R}{\sqrt{2\alpha}})-
\Phi(\frac{t-r-R}{\sqrt{2\alpha}})\right]\times\right.\nonumber\\ & & \left.
\left(6(t-r)^2R^2-R^4-5(t-r)^4-5\alpha^2-20\alpha(t-r)^2-4\alpha R^2\right)+
\right.\nonumber\\ & & \left. (t-r)\sqrt{\frac{2\alpha}{\pi}}\left(
3(t-r)^2+R^2-3\alpha\right)
\left[e^{-\frac{(t-r+R)^2}{2\alpha}}-e^{-\frac{(t-r-R)^2}{2\alpha}}\right]
+\right.\nonumber\\ & & \left. R\sqrt{\frac{2\alpha}{\pi}}\left(9(t-r)^2+
\frac{11}{3}R^2+5\alpha\right)
\left[e^{-\frac{(t-r+R)^2}{2\alpha}}+e^{-\frac{(t-r-R)^2}{2\alpha}}\right]
\right\},
\label{Fprim}
\end{eqnarray}
\begin{eqnarray}
{\cal G}_1(\vec{x},t)&=&\frac{\pi}{2\alpha^2(2\pi)^3\;r^2}\;
\sqrt{\frac{2\alpha}{\pi}}\;
\left\{\left[\Phi(\frac{t-r+R}{\sqrt{2\alpha}})-
\Phi(\frac{t-r-R}{\sqrt{2\alpha}})\right]\times\right.\nonumber\\ & & \left.
\left((t-r)R^2-(t-r)^3-2\alpha(t-r)\right)-\alpha\sqrt{\frac{2\alpha}{\pi}}
\left[e^{-\frac{(t-r+R)^2}{2\alpha}}-e^{-\frac{(t-r-R)^2}{2\alpha}}\right]
+\right.\nonumber\\ & & \left. R(t-r)\sqrt{\frac{2\alpha}{\pi}}
\left[e^{-\frac{(t-r+R)^2}{2\alpha}}+e^{-\frac{(t-r-R)^2}{2\alpha}}\right]
\right\},
\label{Gprim}
\end{eqnarray}
and
\begin{equation}
{\cal Z}_{1}(\alpha_0,\gamma_0)=
\frac{8\pi}{15}\left(3\cos^2\alpha_0\sin^2\gamma_0-1\right),
\label{Z1}
\end{equation}
\begin{equation}
{\cal Z}_{2}(\alpha_0,\gamma_0)=
\frac{8\pi}{15}\left(3\sin^2\alpha_0\sin^2\gamma_0-1\right),
\end{equation}
\begin{equation}
{\cal Z}_{3}(\gamma_0)=
\frac{8\pi}{15}\left(2-3\sin^2\gamma_0\right),
\label{Z3}
\end{equation}
\begin{equation}
{\cal S}_{1}(\alpha_0,\gamma_0)=
\frac{8\pi}{15}\left(\cos^2\alpha_0\sin^2\gamma_0-1\right),
\label{S1}
\end{equation}
\begin{equation}
{\cal S}_{2}(\alpha_0,\gamma_0)=
\frac{8\pi}{15}\left(\sin^2\alpha_0\sin^2\gamma_0-1\right),
\label{S2}
\end{equation}
\begin{equation}
{\cal S}_{3}(\gamma_0)=
\frac{16\pi}{45}\left(1-\frac{3}{2}\sin^2\gamma_0\right),
\label{S3}
\end{equation}
\begin{equation}
{\cal W}_{1}(\alpha_0,\gamma_0)=
\frac{4\pi}{3}\cos\alpha_0\sin\gamma_0,
\label{W1}
\end{equation}
\begin{equation}
{\cal W}_{2}(\alpha_0,\gamma_0)=
\frac{4\pi}{3}\sin\alpha_0\sin\gamma_0,
\end{equation}
\begin{equation}
{\cal W}_{3}(\gamma_0)=
\frac{4\pi}{3}\cos\gamma_0,
\label{Compfin}
\end{equation}

We can see from eqs.(\ref{empezar})-(\ref{Compfin}), 
that all the metric components are written in terms of Gaussian 
exponentials and error functions. As we will see in the next chapter, 
all of these terms tend to zero as $r\rightarrow\infty$ ($t_{ret}$ fixed). 
Thus, all the metric components presented here approach those 
of a flat Minkowski metric at large distances from the matter distribution, 
as we should have expected. 

We must not forget that in this chapter we have 
stated that most of the mass distribution is localised inside a region of 
radius $R$ and we are neglecting all the contribution from matter outside this 
region. The calculations in this chapter make sense as long as we are far away 
from this distribution; therefore, in principle our calculations 
do not depend on 
how large $R$ is as long as $r\gg R$. 

With the results we have just obtained, we can now write the metric for 
a massless quantum bosonic string as:

\begin{eqnarray}
dS^2 &=&
\left(1+h^{00}_a+\frac{1}{2}\ddot{h}^{00}_d\right)dt^2\nonumber\\ & & 
-\left(1-\frac{1}{3}h^{00}_a-\frac{1}{2}{\cal Z}_1(\alpha_0,\gamma_0)
{\cal F}_1(\vec{x},t)-\frac{r}{2}{\cal Z}_1(\alpha_0,\gamma_0)
\dot{{\cal F}}_1(\vec{x},t)\right.\nonumber\\ & & \left.
-\frac{r^2}{2}{\cal S}_1(\alpha_0,\gamma_0)\ddot{{\cal F}}_1(\vec{x},t)
-\frac{1}{6}\ddot{h}^{00}_d\right) dx^2\nonumber\\ & &
-\left(1-\frac{1}{3}h^{00}_a-\frac{1}{2}{\cal Z}_2(\alpha_0,\gamma_0)
{\cal F}_1(\vec{x},t)-\frac{r}{2}{\cal Z}_2(\alpha_0,\gamma_0)
\dot{{\cal F}}_1(\vec{x},t)\right.\nonumber\\ & & \left.
-\frac{r^2}{2}{\cal S}_2(\alpha_0,\gamma_0)\ddot{{\cal F}}_1(\vec{x},t)
-\frac{1}{6}\ddot{h}^{00}_d\right)dy^2\nonumber\\ & &
-\left(1-\frac{1}{3}h^{00}_a-\frac{1}{2}{\cal Z}_3(\gamma_0)
{\cal F}_1(\vec{x},t)-\frac{r}{2}{\cal Z}_3(\gamma_0)
\dot{{\cal F}}_1(\vec{x},t)\right.\nonumber\\ & & \left.
-\frac{r^2}{2}{\cal S}_3(\gamma_0)\ddot{{\cal F}}_1(\vec{x},t)
-\frac{1}{6}\ddot{h}^{00}_d\right)dz^2+\nonumber\\ & &
\left({\cal W}_1(\alpha_0,\gamma_0){\cal G}_1(\vec{x},t)+
r{\cal W}_1(\alpha_0,\gamma_0)\dot{{\cal G}}_1(\vec{x},t)\right)dxdt+
\nonumber\\ & &
\left({\cal W}_2(\alpha_0,\gamma_0){\cal G}_1(\vec{x},t)+
r{\cal W}_2(\alpha_0,\gamma_0)\dot{{\cal G}}_1(\vec{x},t)\right)dydt+
\nonumber\\ & & 
\left({\cal W}_3(\gamma_0){\cal G}_1(\vec{x},t)+
r{\cal W}_1(\gamma_0)\dot{{\cal G}}_1(\vec{x},t)\right)dzdt.
\label{Fullmetric}
\end{eqnarray}
We can see that this metric presents anisotropies introduced by the functions 
${\cal Z}_i(\alpha_0,\gamma_0)$, ${\cal S}_i(\alpha_0,\gamma_0)$, 
${\cal W}_i(\alpha_0,\gamma_0)$.

In the following chapter we will discuss in a more `graphical' way the 
properties of the far field approximation we have developed here and 
we will mention some results of the gravitational field of 
classical strings (cosmic strings).

\chapter{The properties of the far field of a quantum bosonic string}
\label{Back2}
In the previous chapter we dealt with most of the technical issues involved in 
computing 
the weak-field metric produced by a quantum bosonic string. 
Here we will focus on the 
analysis of the results found there. First we must outline the main differences 
between our present computations and those that can be found in the literature 
regarding cosmic strings (classical strings).

As we have said elsewhere, cosmic strings are, from the mathematical point of 
view, nothing more than very long classical strings (provided we ignore their 
microstructure).
Saying that cosmic strings are classical strings means that in such a 
theory we will not have 
to worry about anomalies since these only emerge when we quantise the 
theory.  

It is important to notice that we cannot make a direct comparison between 
the results presented in this chapter and 
the results for cosmic strings. Physically, quantum fundamental strings are very 
different objects from cosmic strings. In addition we have that for example 
the solutions found by Vilenkin \cite{Alex} 
for static cosmic strings, are not compatible with our computation since
fundamental strings are not static 
(in general they move at relativistic speeds). Therefore, it should not surprise 
us that our results 
present a different type of behaviour for the $h^{\mu\nu}(\vec{x},t)$ 
metric from those found for cosmic strings. 
\section{The quantum string dipole and quadrupole radiation}
As was seen in chapter~\ref{Back}, the far field approximation 
consists of a multipole 
expansion in terms of the `field' $h^{\mu\nu}$ and its derivatives 
with respect to time. This expansion is in clear analogy with the one studied in 
the electro-magnetic theory \cite{Steph, jac}. In both cases the main contributions to the 
multipole expansion come from the first three terms of the series; 
that is: the `charge' term, the `dipole moment' term and the `quadrupole 
moment' term. 
\subsection{The dipole radiation}
First of all let us check the well established result that gravitational 
dipole radiation is zero when the energy-momentum tensor in the source is 
a conserved quantity \cite{Steph,Meis,Wei}. That is, 
we have to check that there is no radiation 
emerging from the {\em angular momentum}. The angular momentum can be 
defined as follows \cite{Steph}:
\begin{equation}
\Omega^{i}=
\varepsilon^{ij}_{k}\;\int d^3\vec{x}'x'^{k}
\langle {\hat T}_{0j}(\vec{x}',t-r)\rangle,
\label{ang}
\end{equation}
where $\varepsilon^{ij}_{k}$ is the 3-dimensional antisymmetric tensor. Thus, 
from eq.(\ref{ang}) we find:
\begin{eqnarray}
\Omega^{1}&=&\varepsilon^{1j}_{k}\;
\int d^3\vec{x}'x'^{k}\langle {\hat T}_{0j}(\vec{x}',t-r)\rangle,\nonumber\\
\Omega^{1}&=&\varepsilon^{12}_{k}\;
\int d^3\vec{x}'x'^{k}\langle {\hat T}_{02}(\vec{x}',t-r)\rangle+
\varepsilon^{13}_{k}\;
\int d^3\vec{x}x'^{k}\langle {\hat T}_{03}(\vec{x}',t-r)\rangle,\nonumber\\
\Omega^{1}&=&\varepsilon^{12}_{3}\;
\int d^3\vec{x}x'^{3}\langle {\hat T}_{02}(\vec{x}',t-r)\rangle+
\varepsilon^{13}_{2}\;
\int d^3\vec{x}x'^{2}\langle {\hat T}_{03}(\vec{x}',t-r)\rangle.\nonumber
\end{eqnarray}
Now, substituting for $\langle {\hat T}_{02}\rangle$ and 
$\langle {\hat T}_{03}\rangle$ from eq.(\ref{Texp}):
\begin{eqnarray}
\Omega^{1}&=&-\varepsilon^{12}_{3}\;
\int dr'r'^{3}\cos\gamma\sin\phi\sin^2\gamma\;d\gamma\;d\phi
\left\{\frac{1}{r'}\left[F(t-r+r')+F(t-r-r')\right]+\nonumber\right.
\\ & & \left.
\frac{1}{r'^3}\left[H(t-r+r')-H(t-r-r')\right]\right\}-\nonumber\\
& & \varepsilon^{13}_{2}\;
\int dr'r'^{3}\cos\gamma\sin\phi\sin^2\gamma\;d\gamma\;d\phi
\left\{\frac{1}{r'}\left[F(t-r+r')+F(t-r-r')\right]+\nonumber\right.
\\ & & \left.
\frac{1}{r'^3}\left[H(t-r+r')-H(t-r-r')\right]\right\},
\end{eqnarray}
hence,
\begin{eqnarray}
\Omega^{1}&=& 0.
\label{angres}
\end{eqnarray}
Similarly we find that $\Omega^2=\Omega^3=0$. Thus, since there is 
no angular momentum here, we may conclude that there is no gravitational 
radiation radiated in the form of dipole radiation (we recall that 
radiation is associated with the change in momentum with respect to time). 
Indeed the dipole radiation must be zero 
because, in general relativity the 
angular momentum of the system is a constant of motion the general result being:
$\Omega^i=constant$, thus there cannot possibly be any radiation emerging from 
the angular momentum.
\subsection{The quadrupole radiation}
Let us investigate now the behaviour of the quadrupole radiation 
of the quantum bosonic string. The following calculation is an easy way to 
obtain an accurate estimate of the quadrupole gravitational radiation without 
going into long tensor calculus \cite{Meis}. 

The quadrupole moment can be defined as \cite{Steph,jac,Gold,sch}:
\begin{equation}
Q^{ij}=\int d^3 x'\; T^{00}(\vec{x}',t-r)\;x'^{i}x'^{j}
\label{quadru}
\end{equation}
or as
\begin{equation}
Q^{ij}_{red}=\int d^3 x'\; T^{00}(\vec{x}',t-r)\;
 \left(3x'^{i}x'^{j}-r'^2\delta^{ij}\right)
\label{quadru2}
\end{equation}
In our calculation $T^{00}(\vec{x}',t-r)$ must be replaced of course by 
$\langle {\hat T}^{00}(\vec{x}',t-r)\rangle $. Eq.(\ref{quadru2}) 
is known as the {\em reduced} quadrupole moment and it is 
useful when one chooses to work in the transverse-traceless gauge for the 
weak-field metric since $Q^{ij}_{red}$ is a traceless tensor. An
examination of eq.(\ref{quadru2}) shows that this tensor is of little 
use to us in 
order to determine the quadrupole radiation since it identically vanishes. In 
any case since we have not chosen to work in the transverse-traceless gauge 
we must work with eq.(\ref{quadru}) in order to compute the radiation associated 
with it.

Let us consider the following two expressions arising from 
energy-momentum conservation. 
\begin{equation}
\frac{\partial\langle{\hat T}_{i0}\rangle}{\partial t}-
\frac{\partial\langle{\hat T}_{ij}\rangle}{\partial x^j}=0,
\label{CONS1}
\end{equation}
\begin{equation}
\frac{\partial\langle{\hat T}_{00}\rangle}{\partial t}-
\frac{\partial\langle{\hat T}_{0j}\rangle}{\partial x^j}=0.
\label{CONS2}
\end{equation}
Let us multiply eq.(\ref{CONS1}) by $x^k$:
\begin{equation}
\frac{\partial\langle{\hat T}_{i0}\rangle}{\partial t}\;x^k-
\frac{\partial\langle{\hat T}_{ij}\rangle}{\partial x^j}\;x^k=0,
\label{CONS1a}
\end{equation}
then integrating this expression over a volume we obtain:
\begin{equation}
\int dV\;\frac{\partial\langle{\hat T}_{i0}\rangle}{\partial t}\;x^k=
\int dV\;\frac{\partial(\langle{\hat T}_{ij}\rangle\;x^k)}{\partial x^j}-
\int dV\;\langle{\hat T}^k_{i}\rangle.
\label{CONS1b}
\end{equation}
Since $\langle{\hat T}^{ij}\rangle\rightarrow 0$ at infinity we have that 
the first integral on the RHS of eq.(\ref{CONS1b}) can be dropped (after 
using Gauss theorem), thus:
\begin{eqnarray}
\int dV\;\frac{\partial\langle{\hat T}^{i0}\rangle}{\partial t}\;x^k &=&
-\int dV\;\langle{\hat T}^{ik}\rangle\nonumber\\ 
\int dV\;\langle{\hat T}^{ik}\rangle &=&
-\frac{1}{2}\frac{\partial}{\partial t}
\int dV\;\left(\langle{\hat T}^{i0}\rangle\;x^k+
\langle{\hat T}^{k0}\rangle\;x^i\right).
\label{CONS1c}
\end{eqnarray}
Now, let us consider eq.(\ref{CONS2}). Multiplying this expression by $x^ix^k$ 
we obtain:
\begin{equation}
\frac{\partial\langle{\hat T}_{00}\rangle}{\partial t}\;x^ix^k-
\frac{\partial\langle{\hat T}_{0j}\rangle}{\partial x^j}
\;x^ix^k=0.
\label{CONS2a}
\end{equation}
Integrating this expression over a volume we find
\begin{eqnarray}
\int dV\;\frac{\partial\langle{\hat T}^{00}\rangle}{\partial t}\;x^ix^k&=&
-\int dV\;\frac{\partial}{\partial x^j}\left(\langle{\hat T}^{0j}\rangle
\;x^ix^k\right)\nonumber\\ & &
-\int dV\;(\langle{\hat T}^{0i}\rangle\;x^k+\langle{\hat T}^{0k}\rangle\;x^i),
\label{CONS2b}
\end{eqnarray}
Using Gauss theorem again we can drop the first of the integrals in the 
RHS of eq.(\ref{CONS2b}):
\begin{eqnarray}
\int dV\;\frac{\partial\langle{\hat T}^{00}\rangle}{\partial t}\;x^ix^k &=&
-\int dV\;(\langle{\hat T}^{0i}\rangle\;x^k+\langle{\hat T}^{0k}\rangle\;x^i).
\label{CONS2c}
\end{eqnarray}
Comparing eq.(\ref{CONS2c}) with eq.(\ref{CONS1c}) we can re-write 
eq.(\ref{CONS2c}) as
\begin{equation}
\frac{\partial^2}{\partial t^2}\int dV\;
\langle{\hat T}^{00}\rangle\;x^ix^k=
2\int dV\;\langle{\hat T}^{ik}\rangle
\label{CONS2d}
\end{equation}
thus, we have that 
\begin{eqnarray} 
\int dV\;\langle{\hat T}^{ik}\rangle &=& \frac{1}{2}\ddot{Q}^{ik}.
\label{CONSQ}
\end{eqnarray}
At this point it may be worth to remind the reader what the different 
components of the energy-momentum tensor actually represent \cite{Landa,sch}:

\begin{tabular}{rl}
 & \\
$T^{00}=$ & energy density,\\
$T^{0i}=$ & energy flux across $x^i$ surface,\\
$T^{ij}=$ & flux of $i$ momentum across $j$ surface.\\
 &
\end{tabular}

Thus, the LHS of eq.(\ref{CONSQ}) represents the $i$-momentum which has crossed 
a surface perpendicular to the $k$-axis. The radiation is given by the change of 
momentum crossing the surface with respect to time:
\begin{equation}
{\cal P}^{ik}\sim\;\stackrel{...}{Q}^{\;ik}
\label{CONSQ1}
\end{equation}
This is the power flowing from one side of our system to another. The power 
output ({\em luminosity}) is given approximately by the square of ${\cal P}^{ik}$ 
\cite{Meis}:
\begin{equation}
{\cal I}\sim\;
\stackrel{...}{Q}^{\;ik}\stackrel{...}{Q}_{\;ik}
\label{CONSQ2}
\end{equation}
this expression as it stands is a good estimate for the order of magnitude 
of the {\em luminosity} of the gravitational radiation. 

\begin{figure}
\centerline{\rotatebox{-90}{\resizebox{10cm}{!}
{\includegraphics{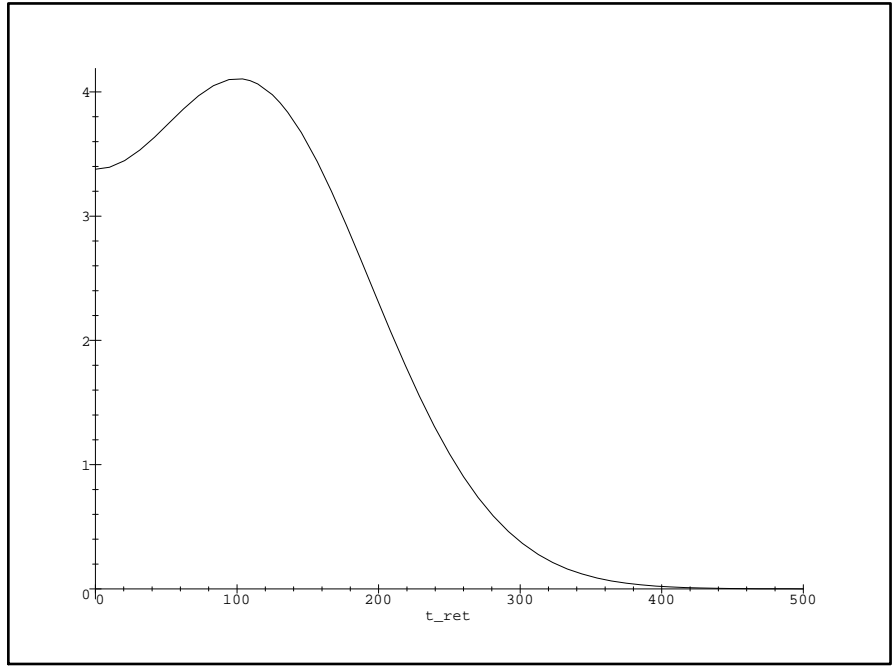}}}}
\caption{The $\stackrel{...}{Q}^{\;ik}$ quadrupole power flux of radiation for 
a quantum bosonic 
string in the far field region, $\alpha=10000$. ($\times 10^{-6}$)}
\label{quadii}
\end{figure}

\begin{figure}
\centerline{\rotatebox{-90}{\resizebox{10cm}{!}
{\includegraphics{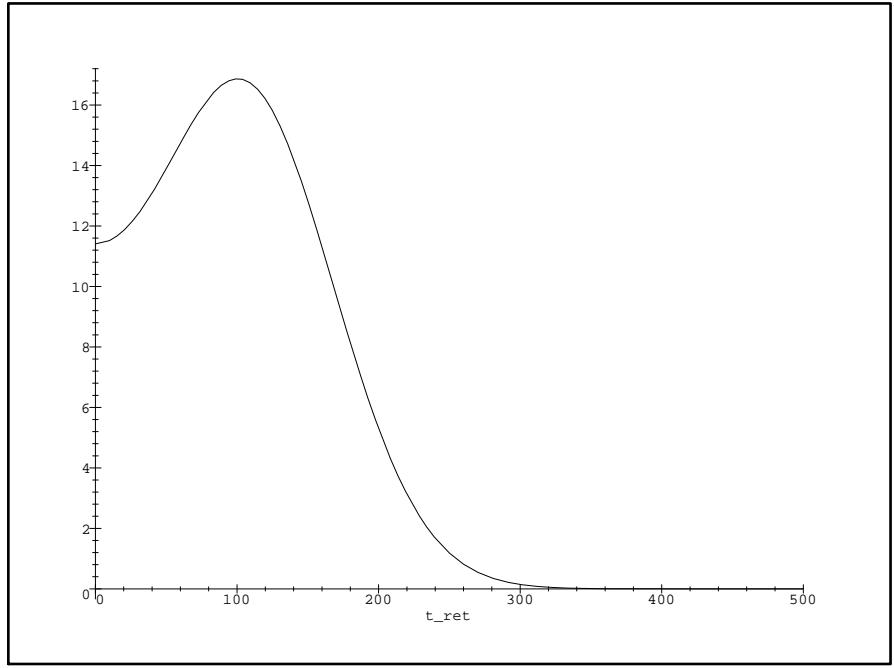}}}}
\caption{The $\stackrel{...}{Q}^{\;ik}\;\stackrel{...}{Q}_{\;ik}$ 
quadrupole power output of radiation for a quantum bosonic 
string in the far field region, $\alpha=10000$. ($\times 10^{-12}$)}
\label{quadii2}
\end{figure}

We can now compute $\stackrel{...}{Q}^{\;ik}$ from the explicit expressions 
(\ref{Texp}) for the string energy-momentum tensor obtaining 
the following results:
\begin{equation}
\stackrel{...}{Q}^{\;jl}=0,\;\;\;j\neq l
\label{Qradz}
\end{equation}
\begin{eqnarray}
\stackrel{...}{Q}^{\;ii}(\vec{x},t)&=&\frac{1}{6\alpha^2\pi^2}
\;\left\{\left[e^{-\frac{(t-r-R)^2}{2\alpha}}-
e^{-\frac{(t-r+R)^2}{2\alpha}}\right]\times\right.\nonumber\\ & & \left.
\left(\frac{2\alpha}{3}-\frac{R^2}{3}+(t-r)^2-R(t-r)\right)+\right.\nonumber
\\ & & \left. \left[e^{-\frac{(t-r+R)^2}{2\alpha}}+
e^{-\frac{(t-r-R)^2}{2\alpha}}\right]\left((t-r)^2-\frac{2R}{3}(t-r)+\alpha
\right)+\right.\nonumber\\ & & \left.
2(t-r)\left[\Phi(\frac{t-r-R}{\sqrt{2\alpha}})-
\Phi(\frac{t-r+R}{\sqrt{2\alpha}})\right]\right\}.
\label{Quadradz2}
\end{eqnarray}
The $\stackrel{...}{Q}^{\;ik}$ and 
$\stackrel{...}{Q}^{\;ik}\;\stackrel{...}{Q}_{\;ik}$ terms 
are shown in figures (\ref{quadii})-(\ref{quadii2}). 
We notice that all the $\stackrel{...}{Q}^{\;ik}$ components 
as well as $\stackrel{...}{Q}^{\;ik}\;\stackrel{...}{Q}_{\;ik}$ are 
finite in the far field region and that they vanish asymptotically as 
$t_{ret}\rightarrow\infty$. The peak in these figures occurs at the elapsed 
time for the gravitational radiation to damp if, for example, turbulence, 
heat conduction or any other effect do not damp it sooner \cite{Meis}.

In summary we have learnt in this section that   
quantum bosonic strings produce no dipole radiation, but do radiate 
gravitationally in the form of 
quadrupole radiation. This general picture 
agrees with standard results of general relativity 
regarding the nature of gravitational radiation. 
 
\section{The quantum string gravitational field in the far field approximation}
So far we have been discussing the gravitational radiation properties 
of a quantum string. Now let us discuss another important aspect: the 
space-time geometry induced by a quantum 
string (the weak field metric components are fully presented in 
appendix \ref{GRAVFOR}). One of the first important things we should 
notice is that 
our weak-field approximation does not breakdown in the region under study, 
which is the region $r\gg R$. 
From here onwards the results stated in the remainder of 
the chapter 
are valid only in this far field region. 
From these results and our discussion in 
the previous sections, it is expected that individual massless 
quantum strings produce 
very small disturbances in the space-time. 

We can see from the expression for the first term of our multipole 
expansion (eq.(\ref{huno}))
that this term is finite in the 
region where $r$ is large. This term is the actual gravitational 
field produced by the string in the absence of higher order terms in the 
multipole expansion of the gravitational field worked out in the 
previous chapter.

\begin{figure}
\centerline{\rotatebox{-90}{\resizebox{10cm}{!}
{\includegraphics{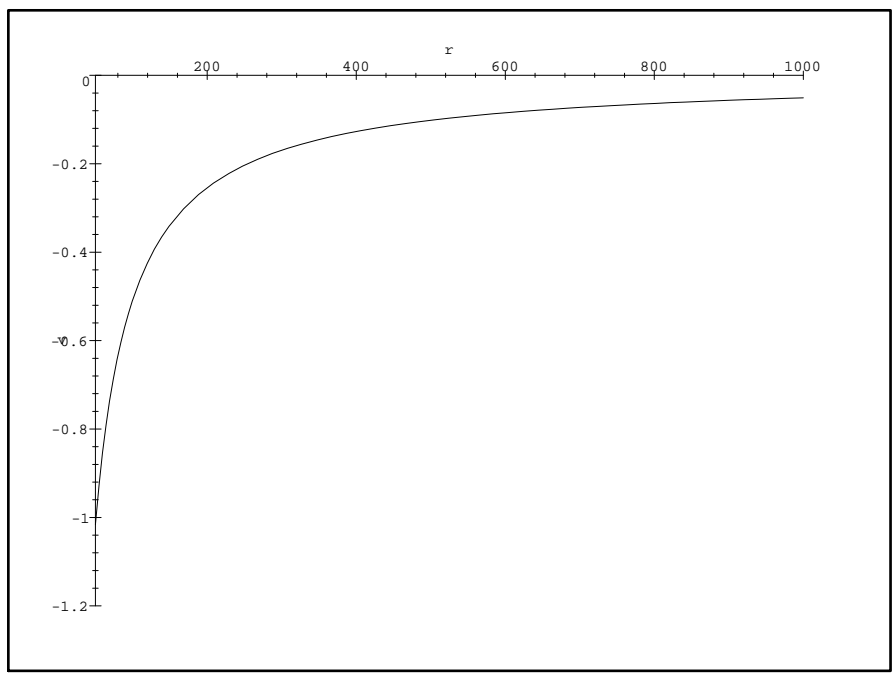}}}}
\caption{A plot of $h^{00}_{a}(r)$. 
This term represents the gravitational field induced 
by the matter source. ($r\gg R$, $t_{ret}$ fixed and $\alpha=10000$. 
In units of $G \times 10^{-6}$.)}
\label{ploth001}
\end{figure}

Let us start by plotting this term, $h^{00}_{a}(\vec{x},t)$ 
fixing the time $t_{ret}$
(see fig.(\ref{ploth001})). 
\begin{figure}
\centerline{\rotatebox{-90}{\resizebox{10cm}{!}
{\includegraphics{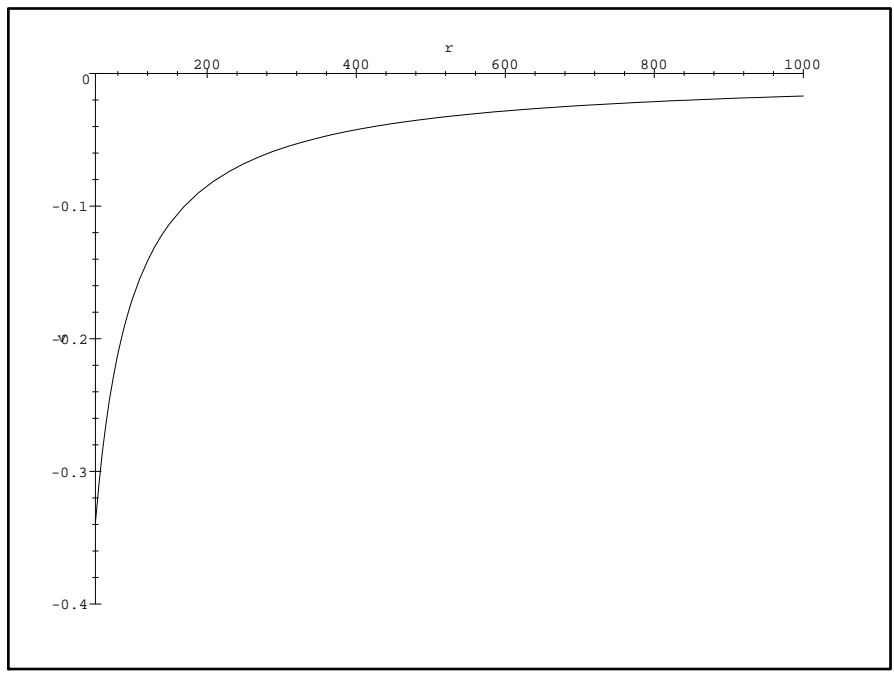}}}}
\caption{A plot of $h^{ii}_{a}(r)$.
 ($r\gg R$, $t_{ret}$ fixed and $\alpha=10000$. 
In units of $G \times 10^{-6}$.)}
\label{plothii1}
\end{figure}
We see that the gravitational field of a quantum bosonic string without the 
contribution from the gravitational radiation is asymptotically flat. 
The behaviour of the 
space elements $h^{ii}_{a}(r)$ of the metric is completely similar to 
the one for $h^{00}_{a}(r)$ as we can see from eq.(\ref{empezar}) and 
figure (\ref{plothii1}). In these plots we have used the following set 
of values for $\alpha$, $R$ and $t_{ret}$: ($\alpha=10^4$, $R=0.5$,$t_{ret}=1$.)

\begin{figure}
\centerline{\rotatebox{-90}{\resizebox{10cm}{!}
{\includegraphics{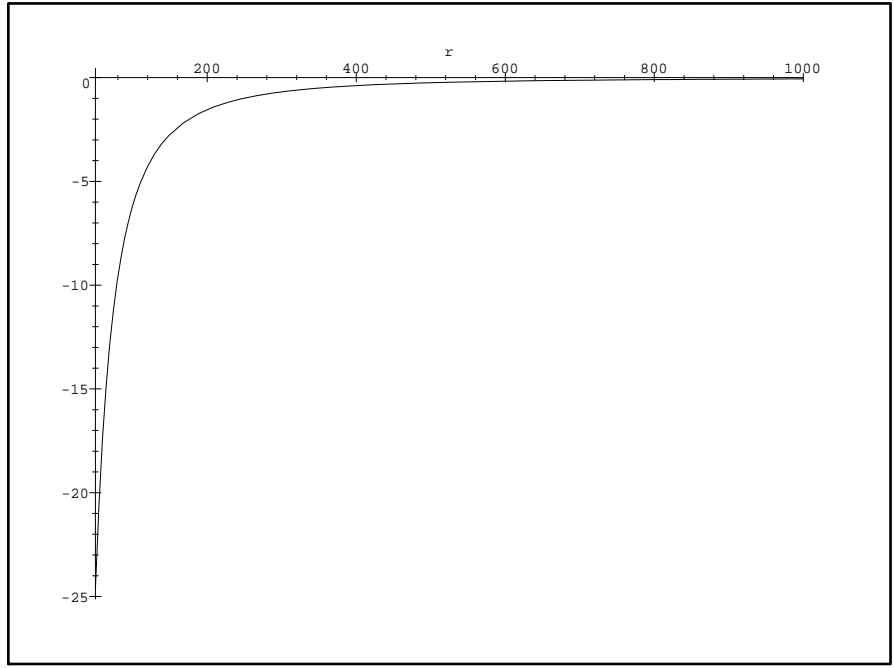}}}}
\caption{Behaviour of the term $h^{0i}_{b}(r)$
of the metric. ($r\gg R$, $t_{ret}$ fixed and $\alpha=10000$. 
In units of $G \times 10^{-14}$.) }
\label{ploth0i2}
\end{figure}

Now, let us plot the second order term components of the metric eq.(\ref{hdos})
(see fig.(\ref{ploth0i2})). As we can see from eqs.(\ref{segundo})-(\ref{tercero}), 
only the $h^{0i}_{b}$ component of the metric contributes 
at second order to $h^{\mu\nu}$.

For a fixed $t_{ret}$ we can estimate how strongly the higher order terms 
contribute to the metric by estimating the ratio 
$h^{\mu\nu}_a/h^{\mu\nu}_{h.o.}\sim h^{00}_a/h^{\mu\nu}_{h.o.}$ where by 
$h^{\mu\nu}_{h.o.}$ we mean the higher order terms of the metric in 
eq.(\ref{MetComp}). From eqs.(\ref{empezar})-(\ref{Compfin}) we obtain:
\begin{equation}
\frac{h^{00}_a}{h^{0i}_{b}}\sim\;\frac{\alpha\; r}{R\;t_{ret}^2}.
\label{estimate1}
\end{equation}
We can see from this expression that the contribution of $h^{0i}_{b}$ to 
the metric is negligible compared to the contribution given by the 
`monopole' or `charge' terms whenever $r$ or $\alpha$ are large 
($r$ large is always the case 
in this far field approximation) and gets 
smaller as $r\rightarrow\infty$ or $\alpha\rightarrow\infty$. 
We can see in fig.(\ref{ploth0i2}) that 
for the given 
values of $\alpha$, $R$, and $t_{ret}$ we have used to plot the metric 
components the contribution of $h^{0i}_{b}$ to 
the metric is 
about 8 orders of magnitude smaller than the  `monopole' terms 
shown in figures (\ref{ploth001}) and (\ref{plothii1}). 
Notice that if this term or any of the other higher order terms in our multipole 
expansion of the gravitational field were to make the space-time metric 
deviate significantly from that of a Minkowski space-time, we would need to take into 
account the curvature on the metric induced by this term when we 
derive the string equations of motion in the first place. 
We would expect in that hypothetical case
that the behaviour of the string equations of motion would change drastically 
from the one we have been using in this thesis and therefore the string 
solutions we have used in this computation would no longer be valid. 
The validity of our results therefore, would have been encompassed 
in a region $r_0 > r \gg R$, where $r_0$ would be the critical point 
in which our original quantum string solutions need to be modified in order to 
take into account the curvature of the space-time. 

\begin{figure}
\centerline{\rotatebox{-90}{\resizebox{10cm}{!}
{\includegraphics{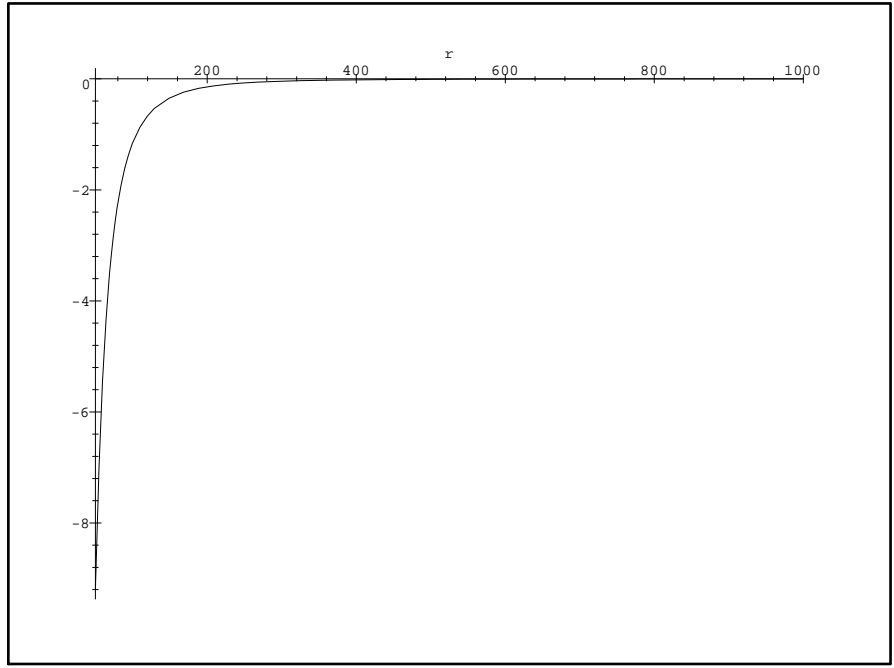}}}}
\caption{The quadrupole term $h^{ii}_{c}(r)$ 
of the metric. ($r\gg R$, $t_{ret}$ fixed and $\alpha=10000$. 
In units of $G \times 10^{-11}$.) }
\label{plothii3}
\end{figure}

\begin{figure}
\centerline{\rotatebox{-90}{\resizebox{10cm}{!}
{\includegraphics{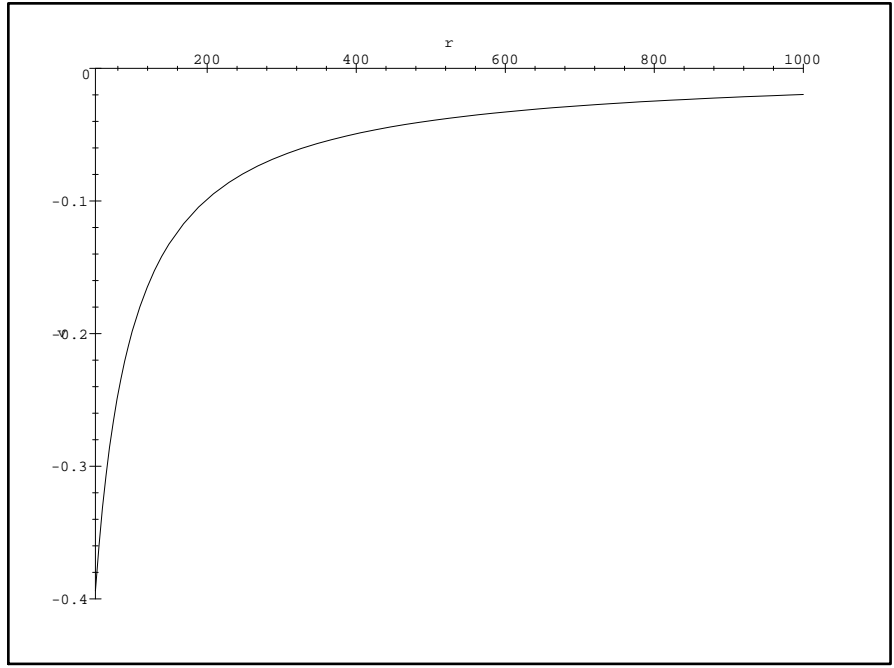}}}}
\caption{The $\dot{h}^{0i}_{b}(r)$ component 
of the metric. ($r\gg R$, $t_{ret}$ fixed and $\alpha=10000$. 
In units of $G \times 10^{-10}$.) }
\label{dh0i2t}
\end{figure}

\begin{figure}
\centerline{\rotatebox{-90}{\resizebox{10cm}{!}
{\includegraphics{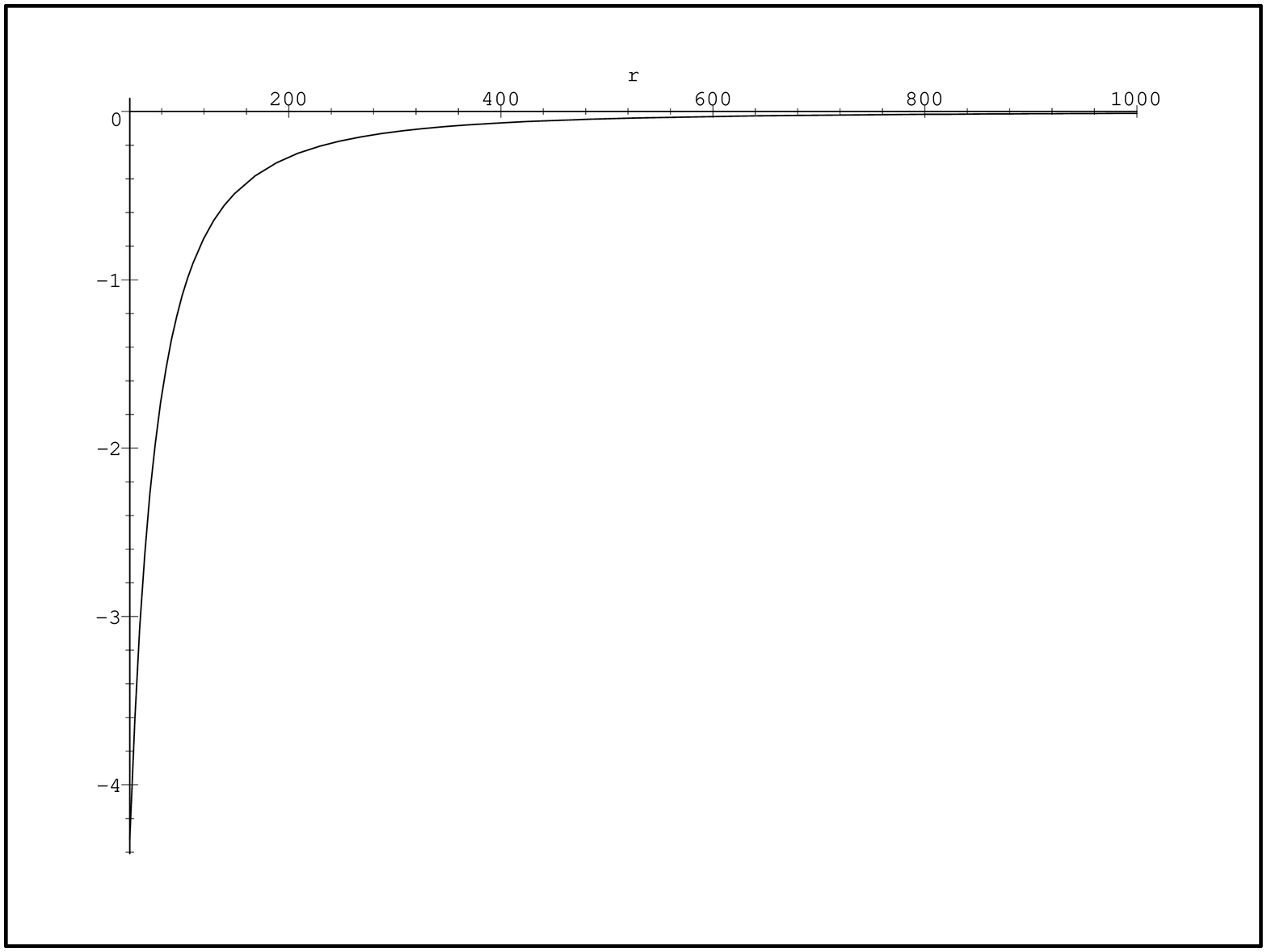}}}}
\caption{The $\dot{h}^{ii}_{c}(r)$ component  
of the metric. ($r\gg R$, $t_{ret}$ fixed and $\alpha=10000$. 
In units of $G \times 10^{-9}$.)}
\label{dhii3t}
\end{figure}

\begin{figure}
\centerline{\rotatebox{-90}{\resizebox{10cm}{!}
{\includegraphics{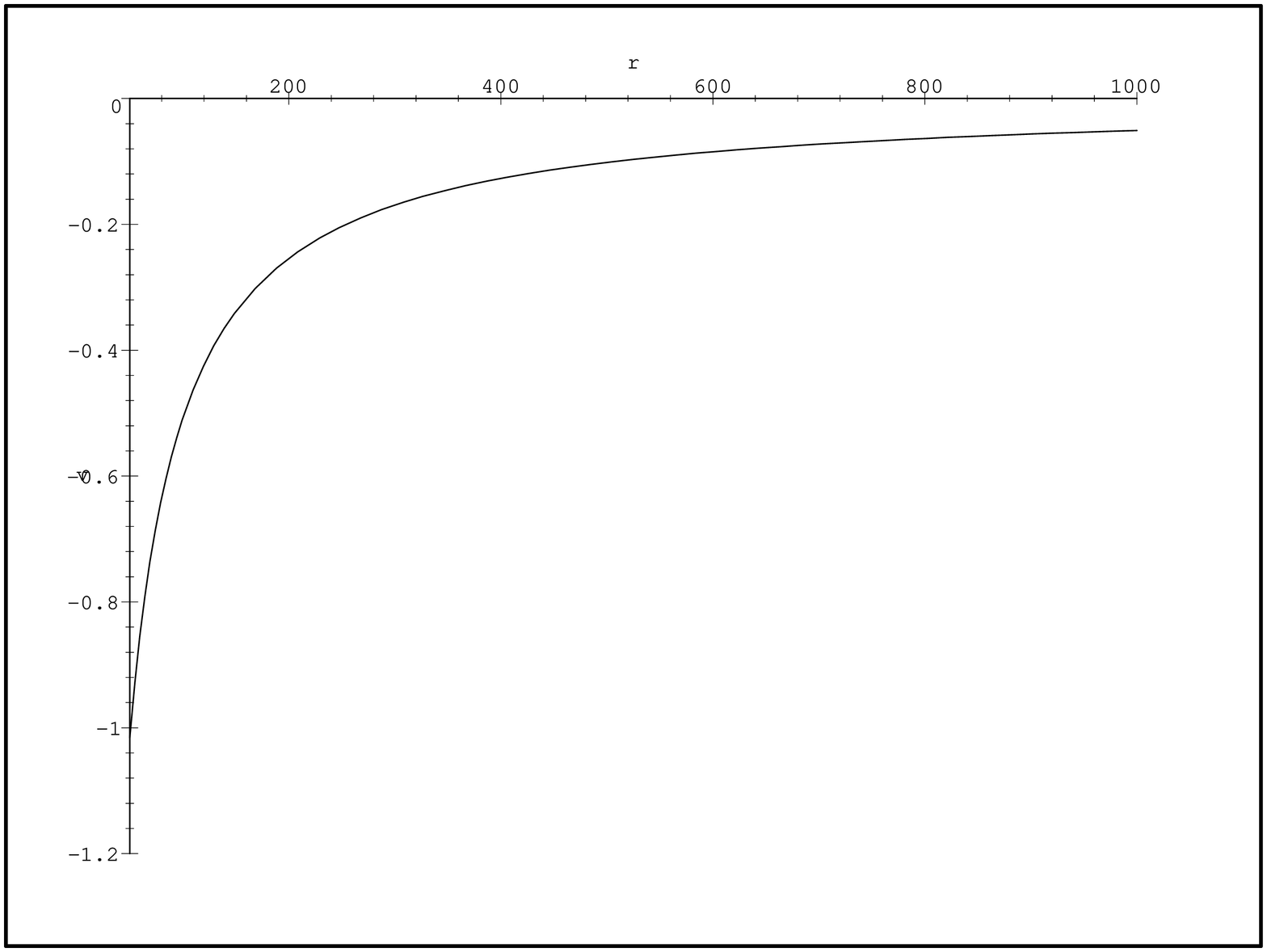}}}}
\caption{The $\ddot{h}^{00}_{d}(r)$ component  
of the metric. ($r\gg R$, $t_{ret}$ fixed and $\alpha=10000$. 
In units of $G \times 10^{-10}$.)}
\label{ddh003}
\end{figure}

\begin{figure}
\centerline{\rotatebox{-90}{\resizebox{10cm}{!}
{\includegraphics{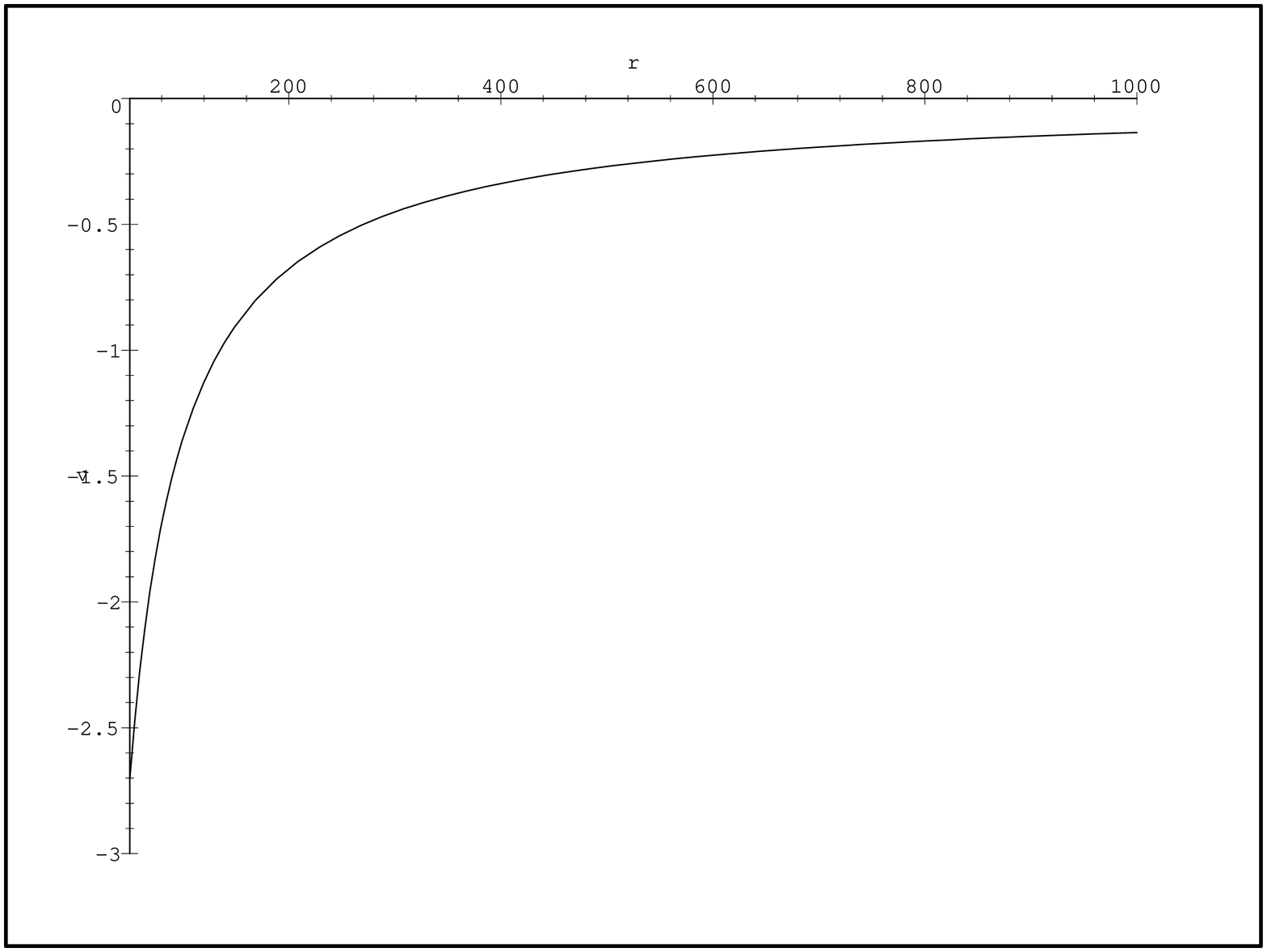}}}}
\caption{The $\ddot{h}^{ii}_{d}(r)$ component  
of the metric. ($r\gg R$, $t_{ret}$ fixed and $\alpha=10000$. 
In units of $G \times 10^{-11}$.)}
\label{ddhii4t}
\end{figure}

We now turn our attention to the contribution from the $h^{\mu\nu}_{c}$ 
term eq.(\ref{htresa}). We have seen in eq.(\ref{cuarto}) 
that there is no 'mass'-quadrupole term $h^{00}_{c}$ 
contributing to the gravitational field. In fact from the same eq.(\ref{cuarto}) 
we can see that the only contributions 
to the metric given by the $h^{\mu\nu}_{c}$ term are through its space  
components. This term is plotted in fig.(\ref{plothii3}). The ratio 
$h^{00}_a/h^{\mu\nu}_{h.o.}$ for this term is given by
\begin{equation}
\frac{h^{00}_a}{h^{ii}_{c}}\sim\;\frac{r^2}{R\;t_{ret}}.
\label{estimate2}
\end{equation}
We can see that $h^{ii}_{c}$ makes a 
minimal contribution to the metric (as compared to $h^{00}_a$) 
when $r$ is large. We can see in  fig.(\ref{plothii3}) that for the given 
values of $\alpha$, $R$, and $t_{ret}$ we have used to plot the metric 
components its contribution is about 5 orders of magnitude smaller 
than the 'monopole' term. 
Similarly, we can estimate the ratio 
$h^{00}_a/h^{\mu\nu}_{h.o.}$ for all the other terms that contribute 
to the metric produced by a massless quantum 
bosonic string. these terms   
involve the time derivatives of $h^{0i}_{b}(\vec{x},t)$, 
$h^{ii}_{c}(\vec{x},t)$,  
$h^{00}_{d}(\vec{x},t)$ and $h^{ii}_{d}(\vec{x},t)$ in eq.(\ref{MetComp}). 
Their ratios $h^{00}_a/h^{\mu\nu}_{h.o.}$ are given by
\begin{equation}
\frac{h^{00}_a}{r\dot{h}^{0i}_{b}}\sim\;\frac{\alpha}{R\;t_{ret}},
\label{estimate3}
\end{equation}
\begin{equation}
\frac{h^{00}_a}{r/2\dot{h}^{ii}_{c}}\sim\;\frac{r}{R},
\label{estimate4}
\end{equation}
\begin{equation}
\frac{h^{00}_a}{1/2\ddot{h}^{00}_{d}}\sim\;\frac{\alpha}{R\;t_{ret}}
\label{estimate5}
\end{equation}
and
\begin{equation}
\frac{h^{00}_a}{1/2\ddot{h}^{ii}_{d}}\sim\;\frac{\alpha}{R\;t_{ret}}.
\label{estimate6}
\end{equation}
We can see from expressions (\ref{estimate3})-(\ref{estimate6}) that 
the contributions from these higher order terms in the metric are also 
small compared to the `monopole' term whenever $r$ or $\alpha$ are large. 
For the set of values we are using to plot our results we find that
the $\dot{h}^{0i}_{b}$ terms 
give a contribution about 5  
orders of magnitude smaller than the 'monopole' term and 
$\dot{h}^{ii}_{c}$ a contribution 3 orders of magnitude smaller 
than the 'monopole' term. The contribution 
from the $\ddot{h}^{00}_{d}$ is also small, 
their contribution to the metric 
being about 4 orders of magnitude smaller than the 'monopole' term. Finally, 
the $\ddot{h}^{ii}_{d}$ 
term is also four orders of magnitude smaller than its corresponding 'monopole' 
term (see figs.(\ref{dh0i2t})-(\ref{ddhii4t}).) 
\section{The gravitational force exerted on non-relativistic particles by a 
quantum string}
We will proceed now to study the properties of the 
gravitational field 
$h^{00}(\vec{x},t)$ derived in the previous chapter. Let us start by considering 
the force non-relativistic particles would experience if they were 
to move in such a gravitational field.

The force exerted on a non-relativistic particle by $h^{00}(\vec{x},t)$ is
given by \cite{Meis, Wei}
\begin{equation}
\frac{d^2 x^{i}}{dt^2}+\Gamma^{i}_{00}=0,
\label{gravfor1}
\end{equation}
where
\begin{eqnarray}
\Gamma^{i}_{00}&=&-\frac{1}{2}(2h_{0i,0}-h_{00,i})
\label{gravfor}
\end{eqnarray}
and $h^{\mu\nu}(\vec{x},t)$ is given by eq.(\ref{MetComp}) and 
eqs.(\ref{empezar})-(\ref{Compfin}).
From these expressions we see that non-relativistic particles experience a 
force due to the mass term $h^{00}_{a}(\vec{x},t)$, the terms $h^{0i}_{b}(\vec{x},t)$, 
$\dot{h}^{0i}_{b}(\vec{x},t)$ and the term $\ddot{h}^{00}_{d}(\vec{x},t)$. 
Thus, eq.(\ref{gravfor1}) can be written in the following way:
\begin{eqnarray}
{\cal F}(\vec{x},t)\hat{x}^i=\frac{d^2 x^{i}}{dt^2}&=&
2G\left[\frac{8\pi}{3}\hat{x}^i\left(\frac{d{\cal G}_1}{dt}+
r\frac{d^2{\cal G}_1}{dt^2}\right)
\right.\nonumber\\ & & \left. -\frac{dh^{00}_{a}}{dr}\;\hat{x}^i-
\frac{1}{2}\frac{d\ddot{h}^{00}_{d}}{dr}\;\hat{x}^i\right].
\label{gravfor2}
\end{eqnarray}
From the discussion in the last section we know that the dominant term in the above 
expression is 
$$\frac{dh^{00}_{a}}{dr}\;\hat{x}^i$$ 
but let us keep all the terms in eq.(\ref{gravfor2}) just to be sure we do not miss 
any additional information in its behaviour. 
The exact expressions from these contributions are not difficult to obtain, 
however, these expressions are exceedingly lengthy, for this reason we have 
not included them here (the technical details of this computation and 
the corresponding expressions for 
the gravitational force exerted by the string on non-relativistic particles 
can be found in appendix \ref{GRAVFOR}.) Here we discuss their behaviour 
in a schematic way.  

\begin{figure}
\centerline{\rotatebox{-90}{\resizebox{10cm}{!}
{\includegraphics{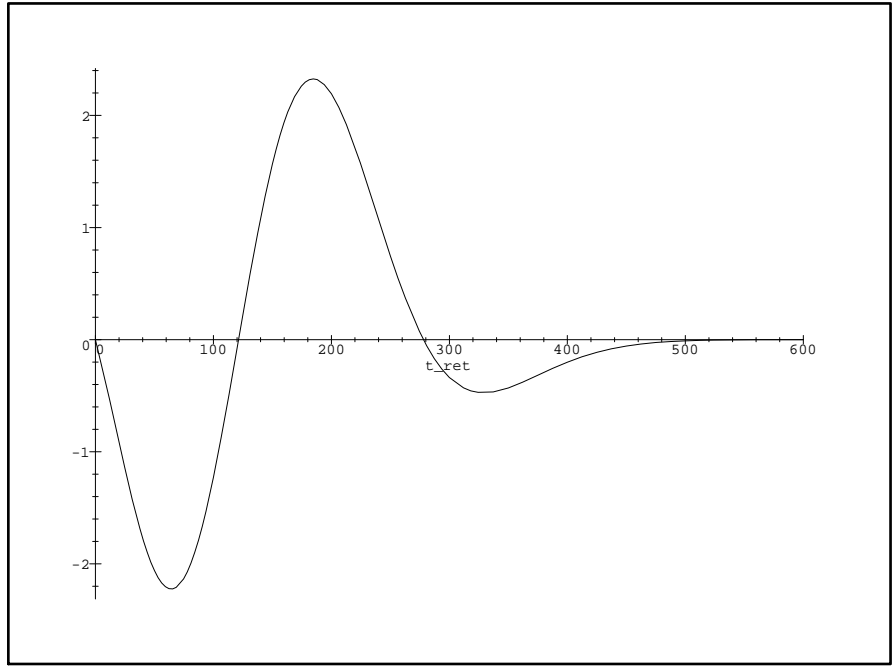}}}}
\caption{The gravitational force exerted by a quantum string on 
non-relativistic particles on the $\hat{x}^i$ direction.  
($r$ fixed and equal to 100 and $\alpha=10000$. 
In units of $G \times 10^{-9}$.)}
\label{F123rfixed}
\end{figure}

\begin{figure}
\centerline{\rotatebox{-90}{\resizebox{10cm}{!}
{\includegraphics{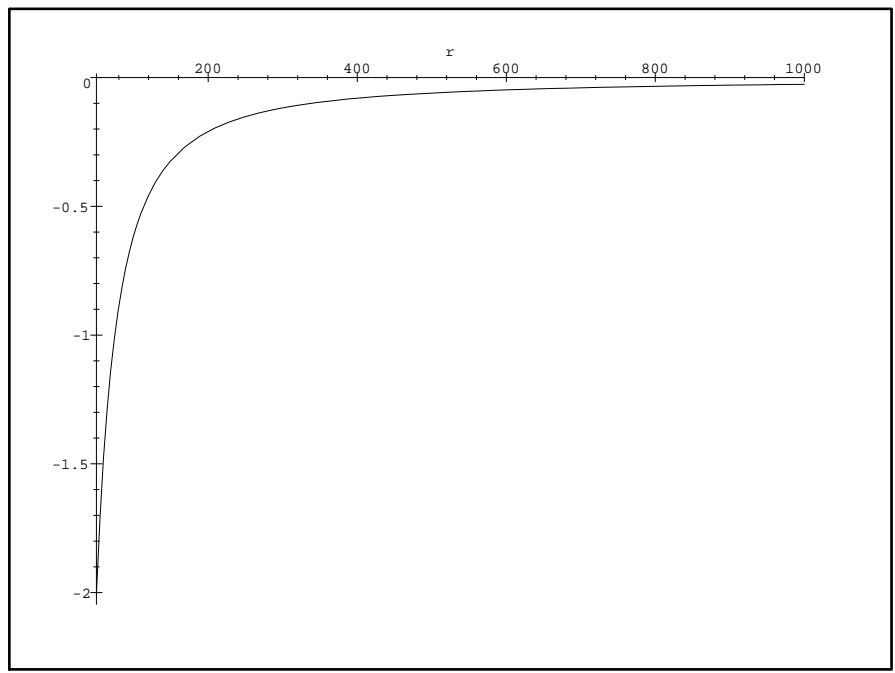}}}}
\caption{The averaged gravitational force exerted by a quantum string on 
non-relativistic particles on the $\hat{x}^i$ direction.  
($J=1000$ and $\alpha=10000$. 
In units of $G \times 10^{-14}$.)}
\label{Favg}
\end{figure}

\begin{figure}
\centerline{\rotatebox{-90}{\resizebox{10cm}{!}
{\includegraphics{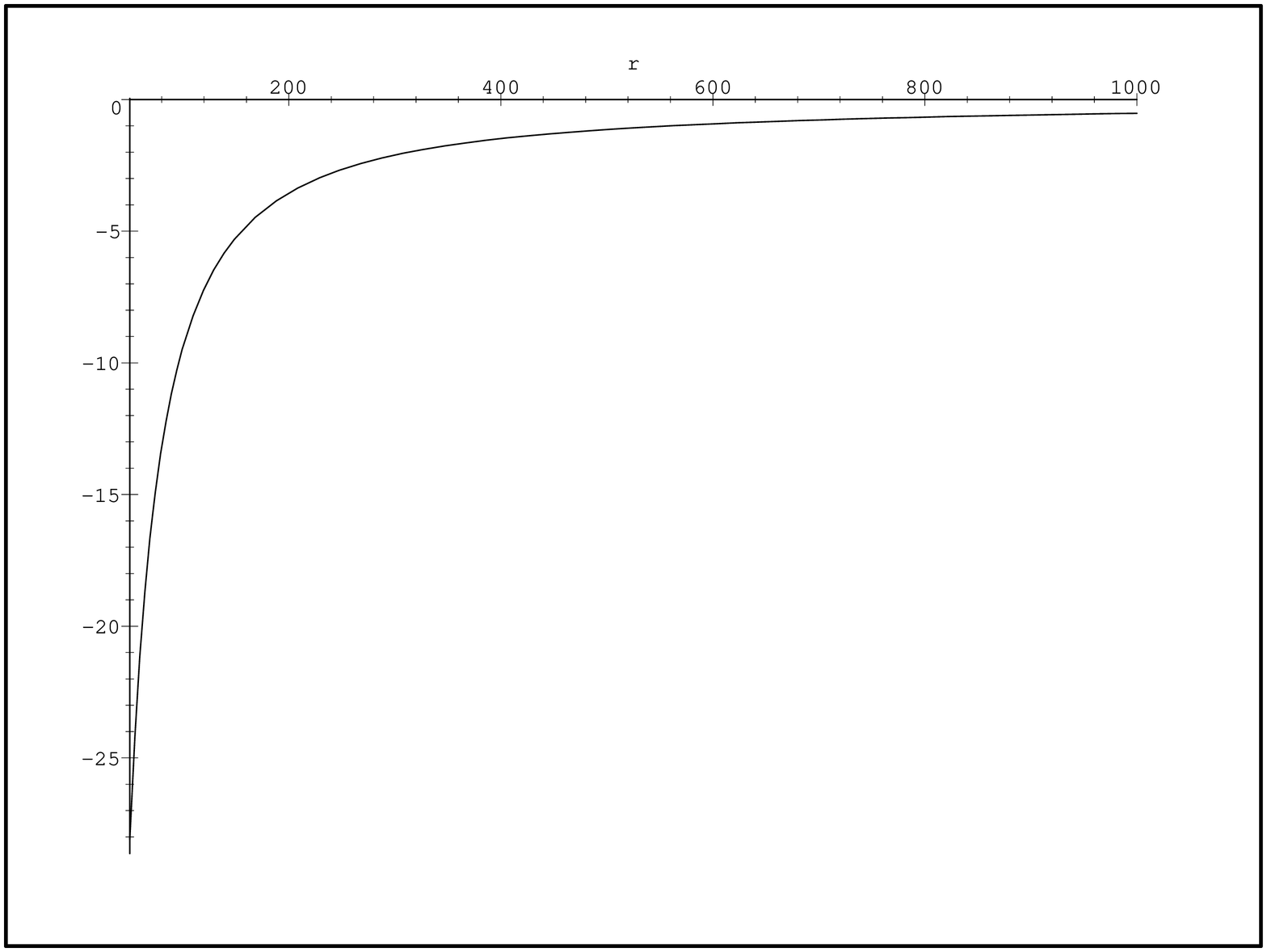}}}}
\caption{The averaged resultant gravitational force exerted by a quantum string on 
non-relativistic particles. Here  $J=500$, 
$\alpha=10000$ (In units of $G \times 10^{-13}$ ) }
\label{Fresavg}
\end{figure}

As we can see from figure (\ref{F123rfixed}), if we fix the position 
and plot the gravitational force as time evolves, 
we find that non-relativistic particles will experience small attractive and 
repulsive forces. Therefore we must take in such a case a time average of the 
force exerted on the particles to see how the particles behave overall:
\begin{equation}
{\cal F}_{avg}(\vec{x})\hat{x}^i=\overline{{\cal F}(\vec{x})}\hat{x}^i=
\frac{\hat{x}^i}{J}\int^{J}_{0}dt_{ret}\;{\cal F}(\vec{x},t_{ret})
\end{equation}
where $J$ is the time elapsed in a few periods of oscillation. From figure 
(\ref{F123rfixed}) we might want to take a value for $J$ equal to $J=500$.  
We can see now in figure (\ref{Favg}) 
that non-relativistic particles experience a small attractive force.
The resultant force being:
\begin{eqnarray}
F_{res}(\vec{x})&=& {\cal F}_{avg}(\vec{x})\hat{x}^1+
{\cal F}_{avg}(\vec{x})\hat{x}^2
+{\cal F}_{avg}(\vec{x})\hat{x}^3\nonumber\\ 
 &=&-\sqrt{3}\left|\overline{{\cal F}(\vec{x})}\right|\,\hat{r}
 \end{eqnarray}
Its plot is given in fig.(\ref{Fresavg}).


\section{Remarks}
We have seen that considering the quantum nature of massless bosonic 
strings provides us with 
several new and interesting features in respect to  
cosmic strings. Some of these features are:
\begin{enumerate}
\item{Our weak-field approximation for the metric derived from the low energy excitation 
of a quantum bosonic string holds 
in the region under 
study namely the far field limit ($r\gg R$). The metric computed in chapter \ref{Back} approaches that 
of flat Minkowski space-time as we go to distances far from the source. 
The divergent logarithmic behaviour of cosmic strings 
\cite{Alex} is not present here.} 
\item{The metric presented in chapter \ref{Back} 
and studied in this chapter, does not present an equivalent `deficit angle' as 
the metric for cosmic string does (at least not in any obvious way).} 
Since quantum strings are fundamentally very different from cosmic strings, 
we should not be expecting this to be the case.
\item{The space-time geometry generated by a quantum bosonic string 
presents small anisotropies introduced by the higher order terms in the 
multipole expansion of our metric namely: $h^{0i}_{b}$, $h^{ii}_{c}$
$\dot{h}^{0i}_{b}$, $\dot{h}^{ii}_{c}$ and $\ddot{h}^{ii}_{d}$. These 
terms may be of interest from a Cosmological point of view.}
\end{enumerate}
We found that the behaviour of the gravitational 
force as described here is not too different from some of the behaviour 
that can be found for cosmic strings 
(see for example \cite{Tur}), which considered classical non-static strings. 
In \cite{Tur} it was found that the gravitational
field of a non-static string is the one given by its total mass. That would have 
been the case here as well if we neglected all the terms in eq.(\ref{gravfor2}) 
which are subdominant in our slow motion approximation which is 
what is done in \cite{Tur}, however we have chosen 
to keep all the relevant terms in eq.(\ref{gravfor2}). We can see this more 
clearly by looking at eqs.(\ref{huno})-(\ref{htresb}). The subdominant terms 
are given by eqs.(\ref{hdos})-(\ref{htresb}) and to compute the resultant 
gravitational force we have performed an average over the time; thus, integrating 
eq.(\ref{huno}):
\begin{equation}
h^{\mu\nu}_{a}(\vec{x},t)=\frac{4G}{r}\int d^3 \vec{x}'\langle 
{\hat T}^{\mu\nu}(\vec{x}',t_{ret})\rangle,
\end{equation}
over $t_{ret}$ we find that the RHS of this expression is just $constant/r$ 
which can be interpreted as usual \cite{Steph, Meis} as $M/R$, $M$ being the 
total mass of the source. Hence to leading order, the gravitational force 
associated with this term is the one given by 
the total mass of the source.

To close the analysis of these results, let us recall that the linearised 
Einstein's equations are based upon the assumption that we possess sufficient 
knowledge of the behaviour of our source. Furthermore, we have made also 
the assumption that most of the matter is concentrated within a certain 
region of radius $R$ and we have neglected contribution from matter outside 
this region. These assumptions, may or may not be true. Further 
study of the behaviour of quantum 
strings in the space-time geometry presented in this chapter and the previous 
one will throw more light on the gravitational properties 
of quantum strings. 
However, the intention of this work has been mainly to outline the differences 
between the behaviour of classical strings and their quantum counterpart and 
to present a starting point for future developments in this direction.

\chapter{Quantum bosonic strings in shock-wave space-times}
In this chapter we will present some results on 
quantum bosonic strings in a shock-wave background. Strings in this 
type of space-time configuration have been studied in detail by 
a number of authors \cite{HJV,dan}, \cite{Cos}-\cite{LoustN}. 
Shock-wave space-times are important to consider 
because we may interpret this geometry as the one produced by an 
ultra-relativistic particle; therefore, they present an opportunity 
to study the scattering between strings and ultra-relativistic particles 
close to the Planck energy scale, where the gravitational interaction 
is the one that dominates the scattering process. 
These geometries are also important because they can give us a 
description of the geometry upon which a string moves when it is
in the presence of other strings \cite{HJV} (in the previous chapters we 
considered only isolated strings.) 

In this chapter we will try to show in a schematic way that at 
the quantum level the string energy-momentum tensor in a shock-wave 
space-time configuration  
leads to extra terms which result from the excitation of the oscillation modes 
of the string when the string collides with the shock-wave. We will also 
show some basic differences between the results computed in \cite{HJV} 
and ours, which 
maintain the string nature of the energy-momentum tensor. (In \cite{HJV} 
the authors 
integrated the string energy-momentum tensor over a volume completely enclosing 
the string; thus, loosing its stringy features.) As we will see, these 
extra terms are important to consider since their contributions to the 
string energy-momentum tensor seem to come from all the possible 
interactions of the oscillation modes of the string which are being 
excited by the shock-wave.
\section{The Aichelburg-Sexl geometry}
Let us very briefly summarise some of the results of the 
Aichelburg-Sexl geometry. 
In \cite{dray} the Aichelburg-Sexl geometry was interpreted as that of a 
shock-wave produced by a neutral and spinless ultra-relativistic particle. 
This geometry has been generalised by a number of authors 
so as to include ultra-relativistic particles with charge and spin. 
(See for example \cite{Loust,LoustN}).

The Aichelburg-Sexl metric can be made to take the following form:
\begin{equation}
dS^2=dUdV-(dX^i)^2+f(\rho)\delta(U)dU^2.
\label{Sexl}
\end{equation}
(The derivation of this metric can be found in \cite{Sexl} and an alternative 
computation is presented also in \cite{dray}.)
Here $U$ and $V$ are given by \cite{HJV}:
$$U=X^0-X^1$$ and
$$V=X^0+X^1.$$ In addition we have that
$\rho=|X^i|$, $2\leq i\leq D-2$, and $f(\rho)$ satisfies the following expression:
$$\nabla^2_{\perp}f=16\pi G\rho(X^i)$$
where $\nabla_{\perp}$ refers to the transverse part of $\nabla$. 
As we can see, the space-time is everywhere flat except at the location of 
the shock-wave $U=0$.
Some of the main features of this type of metric are:
\begin{enumerate}
\item{The Aichelburg-Sexl metric is basically that of a boosted 
black-hole metric where we demand that the momentum of the ultra-relativistic 
particle remains 
constant as it approaches the speed of light. That is, we demand 
that its mass $m\rightarrow 0$ as its velocity $v\rightarrow c$.}
\item{Geodesics in this metric have a discontinuity at the 
intersection with the shock-wave.}
\item{There exists a `time shift' owing to the discontinuity mentioned above. 
Clocks slow down as the shock-wave passes through them.}
\item{There exists a `spatial refraction' as well. Objects are pushed towards 
the trajectory of the ultra-relativistic particle. Objects are moved also 
towards the general direction of the shock-wave.} 
\end{enumerate}
\section{The string non-linear transformations}
In this section we will present the main results found in the literature 
on bosonic strings in shock-wave 
space-time configurations \cite{dan}, \cite{Cos}-\cite{LoustN}. 
The equations of motion 
for the strings are 
highly non-linear owing to the curvature of the space-time.
In the light-cone gauge $U=2\alpha' p^{u}\tau $ one finds that the string 
equations of  motion are the same as those for a flat space-time when $
\tau>0$ and $\tau<0$, $\tau=0$ being the moment when the string interacts with 
the shock-wave. Thus, the string coordinates are given by:
\begin{equation}\label{sseven}
    X^{\mu}(\sigma,\tau)=q_{<}^{\mu}+2\alpha'p_{<}^{\mu}\tau
    +i\sqrt{\alpha'}\sum_{n\neq
    0}\frac{1}{n}[\alpha^{\mu}_{n<}\;e^{-in(\tau-\sigma)}
    +\tilde{\alpha}^{\mu}_{n<}\;e^{-in(\tau+\sigma)}]
\end{equation}
and by:
\begin{equation}\label{sseven2}
    X^{\mu}(\sigma,\tau)=q_{>}^{\mu}+2\alpha'p_{>}^{\mu}\tau
    +i\sqrt{\alpha'}\sum_{n\neq
    0}\frac{1}{n}[\alpha^{\mu}_{n>}\;e^{-in(\tau-\sigma)}
    +\tilde{\alpha}^{\mu}_{n>}\;e^{-in(\tau+\sigma)}].
\end{equation}
The transformations that take us from the $<$ region to the $>$ are given 
by \cite{HJV}, \cite{dray}:
\begin{equation}
\label{primera}
q^{i}_{>}-q^{i}_{<}=0,
\end{equation}
\begin{equation}
p^{i}_{>}-p^{i}_{<}=i\frac{p^{u}}{4\pi}\int^{2\pi}_{0}d\sigma\int
d^{D-2}p p^{i}\varphi(\vec{p}):e^{i\vec{p}\cdot C(\sigma)}:,
\end{equation}
\begin{equation}
\alpha^{i}_{n>}-\alpha^{i}_{n<}=i\frac{\sqrt{\alpha'}p^{u}}{4\pi}
\int^{2\pi}_{0}d\sigma\int
d^{D-2}p p^{i}\varphi(\vec{p}):e^{i\vec{p}\cdot C(\sigma)}: e^{in\sigma},
\end{equation}
\begin{equation}
\tilde{\alpha}^{i}_{n>}-\tilde{\alpha}^{i}_{n<}
=i\frac{\sqrt{\alpha'}p^{u}}{4\pi}\int^{2\pi}_{0}d\sigma\int
d^{D-2}p p^{i}\varphi(\vec{p}):e^{i\vec{p}\cdot C(\sigma)}: e^{-in\sigma}.
\label{ultima}
\end{equation}
Here, $C(\sigma)$ is given by:
$$C^{i}(\sigma)=X^{i}_{<}(\sigma,\tau=0)=X^{i}_{>}(\sigma,\tau=0)=
q^{i}+i\sqrt{\alpha'}\sum_{n\neq 0}\frac{e^{in\sigma}}{n}[\alpha^{i}_{n<}-
\tilde{\alpha}^{i}_{n<}],$$
and $\varphi(\vec{p})$ is related to the matter-density of the source by 
means of a Fourier transformation:
$$\varphi(\vec{p})=-\frac{16\pi G}{p^2}\int \frac{d^{D-2}X}{(2\pi)^{D-2}}
\rho(X)e^{-i\vec{p}\cdot X}.$$
\section{The string energy-momentum tensor in a shock-wave space-time}
In this section we want to outline some of the main differences between the 
results we computed in chapter \ref{ChapTmunu} for strings in a Minkowski 
space-time metric and those for strings in a shock-wave space-time metric. 
We will also see that keeping the stringy nature of the energy-momentum tensor 
will lead to very different results from those found in \cite{HJV}, 
where the authors 
integrated the string energy-momentum tensor over a volume completely enclosing 
the string.

Consider the string energy-momentum tensor:
\begin{equation}
\label{TensorS}
T^{\mu\nu}(x)=\frac{1}{2\pi\alpha'}\int d\sigma d\tau   \;
(\dot{X}^{\mu}\dot{X}^{\nu}-X'^{\mu}X'^{\nu})\delta(x-X(\sigma,\tau)).
\end{equation}
We are now interested in computing the quantum expectation 
value of $T^{\mu\nu}(x)$ once the string has collided with the shock-wave with respect to 
the ground state. That is, we want to compute the expectation value
$\langle 0_{<}|{\hat T}_{>}^{\mu\nu}(x)|0_{<}\rangle$.
\subsection{Keeping the string nature of $\langle {\hat T}^{\mu\nu}(x)\rangle $}
To start, let us examine the delta function appearing in eq.(\ref{TensorS}). 
We can write it as:
\begin{equation}
\label{deltaT}
\delta(x-X_{>}(\sigma,\tau))=\frac{1}{(2\pi)^{D}}\int d^{D} \lambda 
e^{i\lambda\cdot (x-X_{>}(\sigma,\tau))}
\end{equation}
just as we did in chapter \ref{ChapTmunu}. We can see that 
expression (\ref{deltaT}) is 
not a trivial one: we can write $X_{>}(\sigma,\tau)$ as:
$$X_{>}(\sigma,\tau)=X_{>cm}+X_{>-}+X_{>+},$$
where $-$ and $+$ refer to the creation and 
annihilation parts of $X_{>}(\sigma,\tau)$ and $cm$ to the centre of 
mass coordinates: 
$q_{>}+2\alpha'p_{>}\tau $. Expressing all the $>$ operators in terms of 
the $<$, which are those we know how to work with, we find that in 
addition to the usual terms we presented in chapter 
\ref{ChapTmunu} we also have 
terms which represent the interaction of the centre of mass coordinates 
and the oscillation 
modes of the string with the shock-wave of an ultra-relativistic particle.

In chapter \ref{ChapTmunu} we found that normal ordering eq.(\ref{deltaT}), 
only the centre of mass coordinates of this operator contributes to the 
energy-momentum expectation value. Here, however, we can see that is not 
the case. The additional terms representing the interaction of the string with 
the shock-wave do not vanish and will give extra contributions to the 
expectation value. Furthermore, the centre of mass coordinates will also present 
extra terms from the string interaction with the shock-wave.
Notice that none of these additional terms would be present if we integrated 
over a spatial volume enclosing the string. Therefore, these `additional' 
terms are effects arising from the extended nature of strings. 

After some work we find that the terms in brackets in eq.(\ref{TensorS}) can 
be written as:
\begin{eqnarray}
\dot{X}^{\mu}_{>}(\sigma,\tau)\dot{X}^{\nu}_{>}(\sigma,\tau)-
X'^{\mu}_{>}(\sigma,\tau)X'^{\nu}_{>}(\sigma,\tau)= 
4\alpha'^2p^{\mu}_{>}p^{\nu}_{>}+2(\alpha')^{3/2}p^{\mu}_{>}
\times\nonumber\\  \sum_{m\neq 0}
e^{-im\tau}[\alpha^{\nu}_{m>}e^{im\sigma}+
\tilde{\alpha}^{\nu}_{m>}e^{-im\sigma}]+
2(\alpha')^{3/2}\sum_{n\neq 0}
e^{-in\tau}[\alpha^{\mu}_{n>}e^{in\sigma}+
\tilde{\alpha}^{\mu}_{n>}e^{-in\sigma}]p^{\nu}_{>}+\nonumber\\
\alpha'\sum_{n\neq 0}\sum_{m\neq 0} e^{-i(n+m)\tau}[2\alpha^{\mu}_{n>}
\tilde{\alpha}^{\nu}_{m>}e^{i(n-m)\sigma}+2\tilde{\alpha}^{\mu}_{n>}
\alpha^{\nu}_{m>}e^{-i(n-m)\sigma}] 
\label{shock1}
\end{eqnarray}
Of course, in order to take the expectation value we need to normal 
order the above expression.

In order to obtain a better insight into the situation we are working with, 
let us just look at the first term of eq.(\ref{shock1}). If we express this 
term in terms of the $<$ operators, we obtain (setting $\alpha'=1/2$):

\begin{eqnarray}
p^{\mu}_{>}p^{\nu}_{>}&=&
p^{\mu}_{<}p^{\nu}_{<}+
\left[-\frac{p^{u}}{16\pi^2}\int^{2\pi}_{0}d\sigma
\int d^{D-2}p p^{i}\varphi(\vec{p}):
e^{i\vec{p}\cdot C(\sigma)}:\delta_{\mu,i}\right]\times\nonumber\\ & & 
\left[p^{u}\int^{2\pi}_{0}d\sigma'
\int d^{D-2}p' p^{j}\varphi(\vec{p}'):
e^{i\vec{p}'\cdot C(\sigma')}:\delta_{\nu,j}\right]+\nonumber\\ & & 
\left[i\frac{p^{u}}{4\pi}\int^{2\pi}_{0}d\sigma
\int d^{D-2}p p^{i}\varphi(\vec{p}):
e^{i\vec{p}\cdot C(\sigma)}:\delta_{\mu,i}\right]p^{\nu}_{<}+\nonumber\\ & &
i\frac{p^{\mu}_{<}p^{u}}{4\pi}\int^{2\pi}_{0}d\sigma'
\int d^{D-2}p' p^{j}\varphi(\vec{p}'):e^{i\vec{p}'\cdot C(\sigma')}:
\delta_{\nu,j}.
\label{Shock2}
\end{eqnarray}
It is to be noticed that upon integration of 
this expression (with the delta function factor included as well) 
over a spatial volume we would obtain the same results presented 
in \cite{HJV}.
\section{A brief analysis of the string energy-momentum tensor expectation value in 
a shock-wave space-time configuration}
It has not been the intention in this thesis to compute or present 
explicit results for the expectation value of the string 
energy-momentum tensor in shock-wave space-times 
since the highly non-linear transformations make such 
a task extremely difficult to achieve. Instead, 
from here on, we will proceed very schematically. We will try to 
show what additional terms emerge as a consequence of keeping the extended 
nature of the string and examine their relevance.

From eq.(\ref{Shock2}) we see that the first term of this expression is the 
term we worked with in the previous chapters of the thesis. At this point, 
we may be tempted 
to think that we will obtain our results obtained in chapter \ref{ChapTmunu} 
and that the other terms in eq.(\ref{Shock2}) are just additional terms 
arising from the fact that we are now in a non-trivial background. 
However, this is not entirely true. Certainly, the extra terms 
in eq.(\ref{Shock2}) are additional to the expectation value 
of the energy-momentum tensor we computed in chapter 
\ref{ChapTmunu}  but even if these terms were not present 
and we concentrated 
only on the first one, we would still obtain different results here. 
Such a term would give a completely different contribution from the one we 
found previously, the reason 
being that the delta function present in the energy-momentum tensor 
turns out to be a very complex quantum operator. 

Let us consider, just as an example, the first term in expression (\ref{Shock2}). 
Its expectation value is given by:
\begin{equation}
\langle p^{\mu}_{<}p^{\nu}_{<}\rangle=\langle 0_<|e^{-i\lambda\cdot X_{-}}
p^{\mu}_{<}p^{\nu}_{<}e^{i\lambda\cdot\vec{x}}e^{-i\lambda\cdot X_{cm}}
e^{-i\lambda\cdot X_{+}}
|0_<\rangle.
\label{term1}
\end{equation}
Here $e^{-i\lambda\cdot X_{cm}}$ is given by:
\begin{eqnarray}
e^{-i\lambda\cdot X_{cm}}&=&
e^{-i\lambda\cdot( q^{\beta}_{<}+p^0\tau\delta_{\beta,0}
+p^1\tau\delta_{\beta,1}+p^i_{<}\tau\delta_{\beta,i}+
i\frac{\tau p^{u}}{8\pi}\int^{2\pi}_{0}d\sigma\int d^{D-2}p p^{i}
\varphi(\vec{p}):e^{i\vec{p}\cdot C(\sigma)}:
\delta_{\beta,i})},\nonumber\\ & &
\label{Vert1}
\end{eqnarray}
where the repeated indexes do not represent implicit sums. 
The $e^{-i\lambda\cdot X_{\pm}}$ are given by:
\begin{eqnarray}
& & e^{-i\lambda\cdot X_{\pm}}=e^{-i\lambda\cdot (X_{R\pm}+X_{L\pm})}
\nonumber\\
& &e^{\pm\frac{
\lambda}{\sqrt{2}}\cdot(\sum_{s=0}\frac{e^{\mp is\tau}}{s}
(\alpha_{\pm s<}^0\delta_{\beta,0}+
\alpha_{\pm s<}^1\delta_{\beta,1}+[\alpha_{\pm s<}^i
+i\frac{p^{u}}{4\sqrt{2}\pi}\int^{2\pi}_{0}d\sigma\int d^{D-2}p p^{i}
\varphi(\vec{p}):e^{i\vec{p}\cdot C(\sigma)}:e^{\pm is\sigma}])
\delta_{\beta,i}\,e^{in\sigma})+\dots},\nonumber\\ & & 
\label{Vert2}
\end{eqnarray}
and ($\dots$) is a similar expression involving the left-moving 
operators of the string. From this expression we can clearly see that the 
expectation value of a quantum bosonic string in a curved space-time background, 
in particular in a shock-wave space-time, will contain very complex terms arising 
from the interaction of the string with the shock-wave. And 
remembering from chapter 
\ref{ChapTmunu} that the delta function in the string energy-momentum tensor, 
when considered a quantum operator, is basically of the same form of 
a vertex operator, we see that it seems we have got here 
an expansion in terms of vertex operators. 
When the string collides with the shock-wave all the 
oscillation modes of the string become excited. In other words, when we 
keep the stringy features of the string energy-momentum value, $\langle 
{\hat T}^{\mu\nu}(x)\rangle$ takes into account all possible interactions between
the string and the shock-wave plus all the interactions emerging from the 
oscillation modes excited by the shock-wave:
\begin{eqnarray}
\delta(x-X(\sigma,\tau))&=&\frac{1}{(2\pi)^D}
\int d^{D}\lambda 
e^{i\lambda\cdot\vec{x}}e^{-i\lambda\cdot(X^{\mu}+
V_{\beta})}\nonumber\\ &=&\frac{1}{(2\pi)^D}
\int d^{D}\lambda 
e^{i\lambda\cdot\vec{x}}e^{-i\lambda\cdot X^{\mu}}
\left(1-i\lambda\cdot V_{\beta}+\frac{(i\lambda\cdot V_{\beta})^2}{2!}-
\frac{(i\lambda\cdot V_{\beta})^3}{3!}+\dots\right)\times\nonumber\\ & &
e^{-\lambda^2[X,V_{\beta}]}.
\label{Vertex}
\end{eqnarray}
Here $V_{\beta}$ is the nonlinear transformation between the $<$ and $>$ 
operators of the string. This transformation is nothing other than the 
``\dots transverse part of a vertex operator 
$e^{i\vec{p}\cdot X(\sigma)}$ integrated 
over $p$ and over the world-sheet at $\tau=0\;$'' \cite{HJV}. 
The importance of not losing the 
stringy nature of $\langle{\hat T}^{\mu\nu}(x)
\rangle$ now seems to be clear. With this expression we can now see that 
the expectation value given by eq.(\ref{term1}) will take the following 
form:
\begin{eqnarray}
\langle p^{\mu}_{<}p^{\nu}_{<}\rangle&=&\frac{1}{(2\pi)^D}
\int d^{D}\lambda \langle 0_{<}|e^{-i\lambda\cdot(X^{\mu}_{-}+
V_{\beta -})}p^{\mu}_{<}p^{\nu}_{<}
e^{i\lambda\cdot\vec{x}}e^{-i\lambda\cdot(X^{\mu}_{+}+
V_{\beta +})}|0_{<}\rangle\nonumber\\ &=&\frac{1}{(2\pi)^D}
\int d^{D}\lambda \langle 0_{<}|
e^{i\lambda\cdot\vec{x}}e^{-i\lambda\cdot X^{\mu}_{-}}
\left(1-i\lambda\cdot V_{\beta -}+\frac{(i\lambda\cdot V_{\beta -})^2}{2!}-
\frac{(i\lambda\cdot V_{\beta -})^3}{3!}+\dots\right)\times\nonumber\\ & &
e^{-\lambda^2[X_{-},V_{\beta -}]}\,
p^{\mu}_{<}p^{\nu}_{<}\,e^{-i\lambda\cdot X^{\mu}_{cm}}
e^{-i\lambda\cdot X^{\mu}_{+}}\times\nonumber\\ & &
\left(1-i\lambda\cdot V_{\beta cm}+\frac{(i\lambda\cdot V_{\beta cm})^2}{2!}-
\frac{(i\lambda\cdot V_{\beta cm})^3}{3!}+\dots\right)\times\nonumber\\ & &
\left(1-i\lambda\cdot V_{\beta +}+\frac{(i\lambda\cdot V_{\beta +})^2}{2!}-
\frac{(i\lambda\cdot V_{\beta +})^3}{3!}+\dots\right)
e^{-\lambda^2[X_{+},V_{\beta +}]}|0_{<}\rangle.\nonumber\\ & &
\label{Vertex2}
\end{eqnarray}
It is important to notice that the result we have just presented above 
would not have arisen if we had integrated over a 
spatial volume totally enclosing the string.
\section{Remarks}
In this chapter we have tried to present very briefly some of the main differences 
between the computation for the string energy-momentum tensor presented in 
\cite{HJV} and ours. Whilst in \cite{HJV} the string energy-momentum tensor 
was integrated over a spatial volume and therefore we no longer had all the 
string features of $\langle{\hat T}^{\mu\nu}(x)
\rangle$, here, we have kept all the string features of it. In doing this 
we find that the expectation value presents contributions from 
all the interactions take place when the string collides with a shock-wave. 
The `price' we have to pay for this is that explicit calculations 
are exceedingly difficult to perform. 

As a final note, it has to be remembered that shock-wave space-times 
are not candidates for the 
string vacua. The shock-wave metric does not satisfy the conditions  
required by conformal invariance \cite{HJV}. 

\chapter{Conclusions}

In this work we have attempted to provide the preliminaries for a more
sophisticated study of the gravitational properties of fundamental
strings. We have shown in this thesis that the behaviour of the
energy-momentum tensor when treated as a quantum operator presents
features which are different from its classical counterpart. 

In this thesis, we wanted to
outline the 
quantum results of a bosonic string in the context of the gravitational 
properties they present. We asked the question: 
Is the behaviour of fundamental 
strings at the quantum level similar to that of a classical 
string (e.g. a cosmic 
string)? As we have seen throughout the chapters of this thesis, the answer is 
no,  
there are a number of differences that certainly deserve 
further study since 
their interpretation is still far from obvious. Let us recall what we 
hope we have learnt from the chapters of this thesis. 

The main differences obtained with respect to the classical theory of
strings are:
\begin{enumerate}
\item{As shown in chapter \ref{ChapTmunu}, 
the energy-momentum tensor loses its locality. This happens because
quantum fluctuations smear the string position. This is a purely quantum
mechanics effect and it is in some sense a reflection of 
Heisenberg's principle of uncertainty.}
\item{The energy-momentum tensor at the quantum level behaves roughly
like a string vertex operator. In this thesis we have concentrated mainly 
on the string massless states;
that is, photons, gravitons and dilatons, although we have written down solutions 
for massive states.}
\item{The string energy-density for massless string states decays as $1/r$ when
$r\rightarrow\infty$ and $t$ fixed. We showed in chapters 
\ref{ChapTmunu} and \ref{Alpha} that  
the total energy formally diverges in this regime 
only if we work in a temporal box 
which is very small compared to the radius of the volume of space, an acausal 
situation. On the other hand in the causal limit $r<t$, the energy remained 
finite.}
\end{enumerate}
After analysing in detail the string energy-momentum tensor, we
proceeded to study the gravitational properties of quantum bosonic 
strings in a
first order approximation to Einstein's field equations. Because of the
difficulties presented by the very complex structure of the quantum
string energy-momentum tensor, we studied a multipole expansion of the
weak-field metric $h^{\mu\nu}(x)$ in the far field limit, 
in a way analogous to that which can
be performed in the electro-magnetic theory.  We showed 
in chapter \ref{Back2} that these  
properties are different in form and in content 
from those found for classical strings (cosmic strings):
\begin{enumerate} 
\item{We found that in the quantum counterpart there exists 
no obvious relation between 
potential divergences that may have emerged there and the position 
of the string since the string
position has been smeared out by quantum fluctuations. 
This is a purely quantum effect.}

\item{We also found that because fundamental strings are of a 
quantum nature and move at relativistic speeds, static solutions
like those found by Vilenkin \cite{Alex} are no longer relevant.
We found that the metric obtained for a quantum bosonic string does not
resemble in any obvious way the metric for a static classical string (cosmic
string) found by Vilenkin. We showed that $h^{\mu\nu}(x)$ falls away at large 
distances a result consistent with the weak field assumptions made.}

\item{We studied the gravitational radiation produced by a quantum string
and found that the string radiates in the form of
quadrupole radiation and that there is no dipole radiation. 
This is in total agreement with standard results in 
general relativity.}

\item{We compute the gravitational force exerted by a massless bosonic string 
on non-relativistic particles. We found that when the force is integrated over 
periods of time the force is that given by the total mass of the source. This 
results is similar to the one found by \cite{Tur}.}

\item{Our results represent a 
semi-classical approximation in that the energy-momentum 
source in Einstein's equations is of a quantum nature, entering 
via its expectation value. The expectation value 
was taken with respect to the low excitations of a quantum bosonic string; 
that is, 
the states in the expectation value represent massless string states 
such as gravitons,
photons and dilatons.}
\end{enumerate}

From the points mentioned above, it is clear that much more work 
is needed in order to assess
the relevance of quantum strings in Cosmology. As we have said
elsewhere, this work has been developed as an 
approach to understanding the gravitational properties of quantum strings. 

The study of the gravitational properties of quantum
bosonic strings in non-trivial space-times has already been contemplated. 
In particular we are looking at bosonic strings in a shock-wave
background (for which some ideas were presented in the previous chapter). 
In this case the situation is expected to be much more complicated
since there are non-trivial transformations between the string
oscillation modes before the string collides with the shock-wave and
after the string has collided. We expect to report on that case also at some
time in the future.


\pagebreak
\vspace*{9cm}
\begin{center}\addcontentsline{toc}{chapter}{APPENDICES}
APPENDICES
\end{center}
\appendix

\chapter{THE WEAK-FIELD METRIC $h^{\mu\nu}$ FROM A QUANTUM BOSONIC STRING}
\label{GRAVFOR}
\input{mapleenv.sty}
\section{The metric components from the quantum bosonic string energy-momentum 
tensor in Minkowski space-time}
In this section we will present the explicit results for the weak-field metric 
components of a quantum bosonic string. Most of the results presented in this 
appendix were performed using Maple V Release 3.
From eqs.(\ref{huno})-(\ref{htresb}) and eqs.(\ref{Texp}), 
(\ref{FZA})-(\ref{EZA}) we find:
\begin{equation}
h^{00}_{a}(\vec{x},t)=\frac{4G}{r}\int d^3\vec{x}'\;\langle{\hat T}^{00}
(r',t_{ret})\rangle
\end{equation}
\begin{equation}
h^{00}_{a}(\vec{x},t)=\frac{16\pi\;G}{r}\int^{R}_{0}dr'\;r'\left[
F(t_{ret}-r')-F(t_{ret}-r')\right]
\label{Aini}
\end{equation}
we can now solve the remaining integral over $r'$\footnote{the 
final results in this appendix were performed using MapleV v.3, the output style 
in latex is that of maple.sty} with the help of appendix 
\ref{APEXPRIM} 
\begin{maplelatex}
\begin{eqnarray}
\lefteqn{h^{00}_{a}(\vec{x},t)= {G}\,\;\sqrt{\frac{2\alpha}{\pi}}\; 
\left( {\vrule 
height1.03em width0em depth1.03em} \right. \! \!  \left( \! \,
\Phi \left( \! \,{\displaystyle \frac {1}{2}}\,
{\displaystyle \frac {(\,{(t-r)} - {R}\,)\,\sqrt {2}}{\sqrt {{\alpha}}}}\,
 \!  \right)  - \Phi \left( \! \,{\displaystyle \frac {1}{2
}}\,
{\displaystyle \frac {(\,{(t-r)} + {R}\,)\,\sqrt {2}}{\sqrt {{\alpha}}
}}\, \!  \right) \, \!  \right)\times}\nonumber\\ & & \mbox{}
 \left( \! \,{(t-r)}^{2} - {R}^{2}
 + {\displaystyle \frac {2}{3}}\,{\alpha}\, \!  \right)  
    +\left( \! \,{\rm e}^{ \left( \! \, - \,1/2\,
\frac {(\,{(t-r)} + {R}\,)^{2}}{{\alpha}}\, \!  \right) } + {\rm e}^{
 \left( \! \, - \,1/2\,\frac {(\,{(t-r)} - {R}\,)^{2}}{{\alpha}}\, \! 
 \right) }\, \!  \right) \times\nonumber\\ & & \mbox{}
 \sqrt {2}\,\sqrt {{\displaystyle 
\frac {{\alpha}}{{ \pi}}}}\, \left( \! \,{\displaystyle \frac {2}{3}}
\,{R} - {(t-r)}\, \!  \right)  \! \! \left. {\vrule 
height1.03em width0em depth1.03em} \right)  \left/ {\vrule 
height0.37em width0em depth0.37em} \right. \! \! (\,{ \pi}\,{\alpha}^{
2}\,{r}\,)\mbox{\hspace{23pt}}
\end{eqnarray}
\end{maplelatex}
\vspace*{1cm}
\begin{equation}
h^{11}_{a}(\vec{x},t)=\frac{4G}{r}\int d^3\vec{x}'\;\langle{\hat T}^{11}
(r',t_{ret})\rangle
\end{equation}
\begin{eqnarray}
h^{11}_{a}(\vec{x},t)&=&\frac{16\pi\;G}{r}\int^{R}_{0}dr'\;r'^2
\left(-\frac{1}{r'^2}\left[
H(t_{ret}-r')+H(t_{ret}-r')\right]+\right.\nonumber\\ & & \left.
\frac{1}{r'^3}\left[E(t_{ret}-r')-E(t_{ret}-r')\right]\right)+
\frac{4G}{r}\int^{R}_{0}dr'\int^{2\pi}_0d\phi\int^{\pi}_0 d\gamma\;
r'^2\cos^2\phi\sin^3\gamma\times\nonumber\\ & & \left(\frac{1}{r'}\left[
F(t_{ret}-r')-F(t_{ret}-r')\right]+
\frac{3}{r'^2}\left[
H(t_{ret}-r')+H(t_{ret}-r')\right]\right.\nonumber\\ & & \left.-\frac{1}{r'^3}
\left[E(t_{ret}-r')-E(t_{ret}-r')\right]\right)
\end{eqnarray}
performing the integrals over $\alpha$ and $\gamma$ we see that
\begin{equation}
h^{11}_{a}(\vec{x},t)=\frac{1}{3}h^{00}_{a}(\vec{x},t)
\end{equation}
similarly one finds 
$$
h^{22}_{a}(\vec{x},t)=h^{22}_{a}(\vec{x},t)=\frac{1}{3}h^{00}_{a}(\vec{x},t)
$$
Thus, we obtain:
\begin{maplelatex}
\begin{eqnarray}
\lefteqn{h^{ii}_{a}(\vec{x},t)= \frac{G}{3}\,\;\sqrt{\frac{2\alpha}{\pi}}\; 
\left( {\vrule 
height1.03em width0em depth1.03em} \right. \! \!  \left( \! \,
\Phi \left( \! \,{\displaystyle \frac {1}{2}}\,
{\displaystyle \frac {(\,{(t-r)} - {R}\,)\,\sqrt {2}}{\sqrt {{\alpha}}}}\,
 \!  \right)  - \Phi \left( \! \,{\displaystyle \frac {1}{2
}}\,
{\displaystyle \frac {(\,{(t-r)} + {R}\,)\,\sqrt {2}}{\sqrt {{\alpha}}
}}\, \!  \right) \, \!  \right)\times}\nonumber\\ & & \mbox{}
 \left( \! \,{(t-r)}^{2} - {R}^{2}
 + {\displaystyle \frac {2}{3}}\,{\alpha}\, \!  \right)  
    +\left( \! \,{\rm e}^{ \left( \! \, - \,1/2\,
\frac {(\,{(t-r)} + {R}\,)^{2}}{{\alpha}}\, \!  \right) } + {\rm e}^{
 \left( \! \, - \,1/2\,\frac {(\,{(t-r)} - {R}\,)^{2}}{{\alpha}}\, \! 
 \right) }\, \!  \right) \times\nonumber\\ & & \mbox{}
 \sqrt {2}\,\sqrt {{\displaystyle 
\frac {{\alpha}}{{ \pi}}}}\, \left( \! \,{\displaystyle \frac {2}{3}}
\,{R} - {(t-r)}\, \!  \right)  \! \! \left. {\vrule 
height1.03em width0em depth1.03em} \right)  \left/ {\vrule 
height0.37em width0em depth0.37em} \right. \! \! (\,{ \pi}\,{\alpha}^{
2}\,{r}\,)\mbox{\hspace{23pt}}
\end{eqnarray}
\end{maplelatex}
\vspace*{1cm}
\begin{equation}
h^{01}_{a}(\vec{x},t)=\frac{4G}{r}\int d^3\vec{x}'\;
\langle{\hat T}^{01}(r',t_{ret})\rangle
\end{equation}
\begin{equation}
h^{01}_{a}(\vec{x},t)=\frac{4G}{r}\int^{R}_0 dr'\int^{2\pi}_0 d\phi
\int^{\pi}_0 d\gamma\;r'^2\;\cos\phi\sin^2\gamma\;C(t_{ret},r')
\end{equation}
\begin{equation}
h^{01}_{a}(\vec{x},t)=0
\end{equation}
Similarly one finds that
$$h^{03}_{a}(\vec{x},t)=h^{03}_{a}(\vec{x},t)=0.$$
For $h^{\mu\nu}_{b}(\vec{x},t)$ we have the following:
\begin{equation}
h^{00}_{b}(\vec{x},t)=\frac{4G\;x^i}{r^3}\int d^3\vec{x}'\;
x'^i\;\langle{\hat T}^{00}(r',t_{ret})\rangle
\end{equation}
\begin{eqnarray}
h^{00}_{b}(\vec{x},t)&=&\frac{4G}{r^2}\int^{R}_0 dr'\int^{2\pi}_0 d\phi
\int^{\pi}_0 d\gamma\;r'^3\;
\langle{\hat T}^{00}(r',t_{ret})\rangle\times\nonumber\\ & &
\left[\cos\phi\sin^2\gamma\cos\alpha_0\sin\gamma_0+
\sin\phi\sin\gamma\sin\alpha_0\sin\gamma_0+\sin\gamma\cos\gamma\cos\gamma_0
\right]\nonumber\\ & &
\end{eqnarray}
thus, $h^{00}_{b}(\vec{x},t)=0$ similarly one finds $h^{ii}_{b}(\vec{x},t)=0$. 
For $h^{0i}_{b}(\vec{x},t)$ we have:
\begin{equation}
h^{0i}_{b}(\vec{x},t)=\frac{4G\;x^i}{r^3}\int d^3\vec{x}'\;
x'^i\;\langle{\hat T}^{0i}(r',t_{ret})\rangle
\end{equation}
\begin{equation}
h^{01}_{b}(\vec{x},t)=\frac{4G}{r^2}\cos\alpha_0\sin\gamma_0
\int^R_0 dr'\int^{2\pi}_0 d\phi
\int^{\pi}_0 d\gamma\;r'^3\;cos^2\alpha\sin^3\gamma\;C(t_{ret},r')
\end{equation}
\begin{equation}
h^{01}_{b}(\vec{x},t)=\frac{16\pi\;G}{3r^2}\cos\alpha_0\sin\gamma_0\int^R_0
dr'\;r'^3\;C(t_{ret},r')
\end{equation}
substituting the expression for $C(t_{ret},r')$ and performing the $r'$ 
integration we obtain:
\begin{maplelatex}
\begin{eqnarray}
\lefteqn{h^{01}_{b}(\vec{x},t)= {\displaystyle \frac {1}{3}}{G}\,\;
\sqrt{\frac{2\alpha}{\pi}}\;
 \left( \right.\left( \! \,\Phi \left( \! \,{\displaystyle \frac {1}{
2}}\,{\displaystyle \frac {(\,{(t-r)} + {R}\,)\,\sqrt {2}}{\sqrt {{\alpha}
}}}\, \!  \right)  - \Phi \left( \! \,{\displaystyle \frac {
1}{2}}\,{\displaystyle \frac {(\,{(t-r)} - {R}\,)\,\sqrt {2}}{\sqrt {
{\alpha}}}}\, \!  \right) \, \!  \right) }\times\nonumber\\
 & &\left.
(\,{(t-r)}\,{R}^{2} - {(t-r)}^{3}
 - 2\,{\alpha}\,{(t-r)}\,) 
 - {\alpha}\, \left( \! \,{\rm e}^{ \left( \! \, - \,1/2\,
\frac {(\,{(t-r)} + {R}\,)^{2}}{{\alpha}}\, \!  \right) } - {\rm e}^{
 \left( \! \, - \,1/2\,\frac {(\,{(t-r)} - {R}\,)^{2}}{{\alpha}}\, \! 
 \right) }\, \!  \right) \times\right.\nonumber\\ & &
 \sqrt {2}\,\sqrt {{\displaystyle 
\frac {{\alpha}}{{ \pi}}}} + {R}\,{(t-r)}\, \left( \! \,{\rm e}^{ \left( \! \, - \,
1/2\,\frac {(\,{(t-r)} + {R}\,)^{2}}{{\alpha}}\, \!  \right) } + {\rm e}^{
 \left( \! \, - \,1/2\,\frac {(\,{(t-r)} - {R}\,)^{2}}{{\alpha}}\, \! 
 \right) }\, \!  \right) \,\sqrt {2}\,\sqrt {{\displaystyle 
\frac {{\alpha}}{{ \pi}}}} \! \! \left. {\vrule 
height1.03em width0em depth1.03em} \right)\times \nonumber\\ & & 
{\rm cos}(\,{ \alpha_0)}
\,)\,{\rm sin}(\,{ \gamma_0}\,) \left/ {\vrule 
height0.37em width0em depth0.37em} \right. \! \! (\,{ \pi}\,
{\alpha}^{2}{r}^{2}\, )\\ & & \nonumber
\end{eqnarray}
\end{maplelatex}
Similarly we obtain:
\begin{maplelatex}
\begin{eqnarray}
\lefteqn{h^{02}_{b}(\vec{x},t)= {\displaystyle 
\frac {1}{3}}{G}\,\;\sqrt{\frac{2\alpha}{\pi}}\;
 \left( {\vrule height1.03em width0em depth1.03em} \right. \! \! 
} \nonumber\\
 & &  \left( \! \,\Phi \left( \! \,{\displaystyle \frac {1}{
2}}\,{\displaystyle \frac {(\,{(t-r)} + {R}\,)\,\sqrt {2}}{\sqrt {{\alpha}
}}}\, \!  \right)  - \Phi \left( \! \,{\displaystyle \frac {
1}{2}}\,{\displaystyle \frac {(\,{(t-r)} - {R}\,)\,\sqrt {2}}{\sqrt {
{\alpha}}}}\, \!  \right) \, \!  \right) \,(\,{(t-r)}\,{R}^{2} - {(t-r)}^{3}
 - 2\,{\alpha}\,{(t-r)}\,) \nonumber\\
 & & \mbox{} - {\alpha}\, \left( \! \,{\rm e}^{ \left( \! \, - \,1/2\,
\frac {(\,{(t-r)} + {R}\,)^{2}}{{\alpha}}\, \!  \right) } - {\rm e}^{
 \left( \! \, - \,1/2\,\frac {(\,{(t-r)} - {R}\,)^{2}}{{\alpha}}\, \! 
 \right) }\, \!  \right) \,\sqrt {2}\,\sqrt {{\displaystyle 
\frac {{\alpha}}{{ \pi}}}} \nonumber\\
 & & \mbox{} + {R}\,{(t-r)}\, \left( \! \,{\rm e}^{ \left( \! \, - \,
1/2\,\frac {(\,{(t-r)} + {R}\,)^{2}}{{\alpha}}\, \!  \right) } + {\rm e}^{
 \left( \! \, - \,1/2\,\frac {(\,{(t-r)} - {R}\,)^{2}}{{\alpha}}\, \! 
 \right) }\, \!  \right) \,\sqrt {2}\,\sqrt {{\displaystyle 
\frac {{\alpha}}{{ \pi}}}} \! \! \left. {\vrule 
height1.03em width0em depth1.03em} \right) {\rm sin}(\,{ \alpha_0}
\,)\,{\rm sin}(\,{ \gamma_0}\,) \left/ {\vrule 
height0.37em width0em depth0.37em} \right. \! \! (\,{ \pi}\,{\alpha}^{
2}{r}^{2}\,)\nonumber\\ & &
\end{eqnarray}
\end{maplelatex}
\begin{maplelatex}
\begin{eqnarray}
\lefteqn{h^{03}_{b}(\vec{x},t)= {\displaystyle \frac {1}{3}}{G}\,\;\sqrt{\frac{2\alpha}{\pi}}\;
 \left( {\vrule height1.03em width0em depth1.03em} \right. \! \! 
} \nonumber\\
 & &  \left( \! \,\Phi \left( \! \,{\displaystyle \frac {1}{
2}}\,{\displaystyle \frac {(\,{(t-r)} + {R}\,)\,\sqrt {2}}{\sqrt {{\alpha}
}}}\, \!  \right)  - \Phi \left( \! \,{\displaystyle \frac {
1}{2}}\,{\displaystyle \frac {(\,{(t-r)} - {R}\,)\,\sqrt {2}}{\sqrt {
{\alpha}}}}\, \!  \right) \, \!  \right) \,(\,{(t-r)}\,{R}^{2} - {(t-r)}^{3}
 - 2\,{\alpha}\,{(t-r)}\,) \nonumber\\
 & & \mbox{} - {\alpha}\, \left( \! \,{\rm e}^{ \left( \! \, - \,1/2\,
\frac {(\,{(t-r)} + {R}\,)^{2}}{{\alpha}}\, \!  \right) } - {\rm e}^{
 \left( \! \, - \,1/2\,\frac {(\,{(t-r)} - {R}\,)^{2}}{{\alpha}}\, \! 
 \right) }\, \!  \right) \,\sqrt {2}\,\sqrt {{\displaystyle 
\frac {{\alpha}}{{ \pi}}}} \nonumber\\
 & & \mbox{} + {R}\,{(t-r)}\, \left( \! \,{\rm e}^{ \left( \! \, - \,
1/2\,\frac {(\,{(t-r)} + {R}\,)^{2}}{{\alpha}}\, \!  \right) } + {\rm e}^{
 \left( \! \, - \,1/2\,\frac {(\,{(t-r)} - {R}\,)^{2}}{{\alpha}}\, \! 
 \right) }\, \!  \right) \,\sqrt {2}\,\sqrt {{\displaystyle 
\frac {{\alpha}}{{ \pi}}}} \! \! \left. {\vrule 
height1.03em width0em depth1.03em} \right) {\rm cos}(\,{ \gamma_0}
\,) \left/ {\vrule height0.37em width0em depth0.37em} \right. \! 
\! (\,{ \pi}\,{\alpha}^{2}\,{r}^{2}\,)\\ & &\nonumber
\end{eqnarray}
\end{maplelatex}
For $h^{\mu\nu}_3(\vec{x},t)$ we have the following:
\begin{equation}
h^{00}_{c}(\vec{x},t)=\frac{4G\;x^ix^j}{r^5}\int d^3\vec{x}'\;
(3x'^ix'^j-r'^2\delta^{ij})\;\langle{\hat T}^{00}(r',t_{ret})\rangle
\end{equation}
\begin{eqnarray}
h^{00}_{c}(\vec{x},t)&=&\frac{4G}{r^3}\int^R_0 dr'\int^{2\pi}_0 d\phi
\int^{\pi}_0 d\gamma\;r'^4\;\sin\gamma\;
\langle{\hat T}^{00}(r',t_{ret})\rangle\times
\nonumber\\ & & \left[3(\cos^2\phi\sin^2\gamma\cos^\alpha_0\sin^2\gamma_0+
\sin^2\phi\sin^2\gamma\sin^2\alpha_0\sin^2\gamma_0+\cos^2\gamma\cos^2\gamma_0)
-1\right]\nonumber\\ & &
\end{eqnarray}
performing the remaining $\phi$ and $\gamma$ integrals we find that
\begin{equation}
h^{00}_{c}(\vec{x},t)=0.
\end{equation}
\begin{equation}
h^{11}_{c}(\vec{x},t)=\frac{4G\;x^ix^j}{r^5}\int d^3\vec{x}'\;
(3x'^ix'^j-r'^2\delta^{ij})\;\langle{\hat T}^{11}(r',t_{ret})\rangle
\end{equation}
\begin{eqnarray}
h^{11}_{c}(\vec{x},t)&=&\frac{4G}{r^3}\int^R_0 dr'\int^{2\pi}_0 d\phi
\int^{\pi}_0 d\gamma\;r'^4\;\cos^2\phi\sin^3\gamma\;
\langle{\hat T}^{00}(r',t_{ret})\rangle\times
\nonumber\\ & & \left[3(\cos^2\phi\sin^2\gamma\cos^\alpha_0\sin^2\gamma_0+
\sin^2\phi\sin^2\gamma\sin^2\alpha_0\sin^2\gamma_0+\cos^2\gamma\cos^2\gamma_0)
-1\right]\times\nonumber\\ & &
\left(\frac{1}{r'}\left[F(t_{ret}+r')-F(t_{ret}-r')\right]+\frac{3}{r'^2}
\left[H(t_{ret}+r')+H(t_{ret}-r')\right]\right.\nonumber\\ & & \left. 
-\frac{3}{r'^3}\left[E(t_{ret}+r')-E(t_{ret}-r')\right]\right)
\end{eqnarray}
performing the remaining $\phi$ and $\gamma$ integrals we find:
\begin{eqnarray}
h^{11}_{c}(\vec{x},t)&=&\frac{32\pi\;G}{15r^3}
\left[3\cos^2\alpha_0\sin^2\gamma_0-1\right]\int^R_0 dr'\;r'^4\left(
\frac{1}{r'}\left[F(t_{ret}+r')-F(t_{ret}-r')\right]+\right.\nonumber\\ 
& & \left. \frac{3}{r'^2}\left[H(t_{ret}+r')+H(t_{ret}-r')\right]-
\frac{3}{r'^3}\left[E(t_{ret}+r')-E(t_{ret}-r')\right]\right)
\end{eqnarray}
performing the $r'$ integration we obtain:
\begin{maplelatex}
\begin{eqnarray}
\lefteqn{h^{11}_{c}(\vec{x},t) = {\displaystyle \frac {1}{30}} \left( \! 
\,3\,{\rm cos}(\,{ \alpha_0}\,)^{2}\,{\rm sin}(\,{ \gamma_0}\,)^{2}
 - 1\, \!  \right) \,{G}\,\;\sqrt{\frac{2\alpha}{\pi}}\; \left( {\vrule 
height1.03em width0em depth1.03em} \right. \! \! } \nonumber\\
 & &  \left( \! \,\Phi \left( \! \,{\displaystyle \frac {1}{
2}}\,{\displaystyle \frac {(\,{(t-r)} + {R}\,)\,\sqrt {2}}{\sqrt {{\alpha}
}}}\, \!  \right)  - \Phi \left( \! \,{\displaystyle \frac {
1}{2}}\,{\displaystyle \frac {(\,{(t-r)} - {R}\,)\,\sqrt {2}}{\sqrt {
{\alpha}}}}\, \!  \right) \, \!  \right)  \nonumber\\
 & & (\, - {R}^{4} - 5\,{(t-r)}^{4} - 5\,{\alpha}^{2} - 20\,{\alpha}\,{(t-r)}^{2}
 + 6\,{(t-r)}^{2}\,{R}^{2} - 4\,{\alpha}\,{R}^{2}\,) \nonumber\\
 & & \mbox{} + {(t-r)}\,\sqrt {2}\,\sqrt {{\displaystyle \frac {{\alpha}}{
{ \pi}}}}\,(\,3\,{(t-r)}^{2} + {R}^{2} - 3\,{\alpha}\,)\, \left( \! \,
{\rm e}^{ \left( \! \, - \,1/2\,\frac {(\,{(t-r)} + {R}\,)^{2}}{{\alpha}}
\, \!  \right) } - {\rm e}^{ \left( \! \, - \,1/2\,\frac {(\,{(t-r)}
 - {R}\,)^{2}}{{\alpha}}\, \!  \right) }\, \!  \right)  \nonumber\\
 & & \mbox{} + {R}\,\sqrt {2}\,\sqrt {{\displaystyle \frac {{\alpha}}{
{ \pi}}}}\, \left( \! \,9\,{(t-r)}^{2} + {\displaystyle \frac {11}{3
}}\,{R}^{2} + 5\,{\alpha}\, \!  \right) \, \left( \! \,{\rm e}^{
 \left( \! \, - \,1/2\,\frac {(\,{(t-r)} + {R}\,)^{2}}{{\alpha}}\, \! 
 \right) } + {\rm e}^{ \left( \! \, - \,1/2\,\frac {(\,{(t-r)} - {R}
\,)^{2}}{{\alpha}}\, \!  \right) }\, \!  \right)  \! \! \left. 
{\vrule height1.03em width0em depth1.03em} \right)  \left/ 
{\vrule height0.37em width0em depth0.37em} \right. \! \! ( \nonumber\\
 & & { \pi}\,{\alpha}^{2}\,{r}^{3}\,)\\ & & \nonumber
\end{eqnarray}
\end{maplelatex}
Similarly we find:
\begin{maplelatex}
\begin{eqnarray}
\lefteqn{h^{22}_{c}(\vec{x},t) = {\displaystyle \frac {1}{30}} \left( \! 
\,3\,{\rm sin}(\,{ \alpha_0}\,)^{2}\,{\rm sin}(\,{ \gamma_0}\,)^{2}
 - 1\, \!  \right) \,{G}\,\;\sqrt{\frac{2\alpha}{\pi}}\; \left( {\vrule 
height1.03em width0em depth1.03em} \right. \! \! } \nonumber\\
 & &  \left( \! \,\Phi \left( \! \,{\displaystyle \frac {1}{
2}}\,{\displaystyle \frac {(\,{(t-r)} + {R}\,)\,\sqrt {2}}{\sqrt {{\alpha}
}}}\, \!  \right)  - \Phi \left( \! \,{\displaystyle \frac {
1}{2}}\,{\displaystyle \frac {(\,{(t-r)} - {R}\,)\,\sqrt {2}}{\sqrt {
{\alpha}}}}\, \!  \right) \, \!  \right)  \nonumber\\
 & & (\, - {R}^{4} - 5\,{(t-r)}^{4} - 5\,{\alpha}^{2} - 20\,{\alpha}\,{(t-r)}^{2}
 + 6\,{(t-r)}^{2}\,{R}^{2} - 4\,{\alpha}\,{R}^{2}\,) \nonumber\\
 & & \mbox{} + {(t-r)}\,\sqrt {2}\,\sqrt {{\displaystyle \frac {{\alpha}}{
{ \pi}}}}\,(\,3\,{(t-r)}^{2} + {R}^{2} - 3\,{\alpha}\,)\, \left( \! \,
{\rm e}^{ \left( \! \, - \,1/2\,\frac {(\,{(t-r)} + {R}\,)^{2}}{{\alpha}}
\, \!  \right) } - {\rm e}^{ \left( \! \, - \,1/2\,\frac {(\,{(t-r)}
 - {R}\,)^{2}}{{\alpha}}\, \!  \right) }\, \!  \right)  \nonumber\\
 & & \mbox{} + {R}\,\sqrt {2}\,\sqrt {{\displaystyle \frac {{\alpha}}{
{ \pi}}}}\, \left( \! \,9\,{(t-r)}^{2} + {\displaystyle \frac {11}{3
}}\,{R}^{2} + 5\,{\alpha}\, \!  \right) \, \left( \! \,{\rm e}^{
 \left( \! \, - \,1/2\,\frac {(\,{(t-r)} + {R}\,)^{2}}{{\alpha}}\, \! 
 \right) } + {\rm e}^{ \left( \! \, - \,1/2\,\frac {(\,{(t-r)} - {R}
\,)^{2}}{{\alpha}}\, \!  \right) }\, \!  \right)  \! \! \left. 
{\vrule height1.03em width0em depth1.03em} \right)  \left/ 
{\vrule height0.37em width0em depth0.37em} \right. \! \! ( \nonumber\\
 & & { \pi}\,{\alpha}^{2}\,{r}^{3}\,)
\end{eqnarray}
\end{maplelatex}
\vspace*{1cm}
\begin{maplelatex}
\begin{eqnarray}
\lefteqn{h^{33}_{c}(\vec{x},t)= {\displaystyle \frac {1}{30}} \left( \! 
\,2 - 3\,{\rm sin}(\,{ \gamma_0}\,)^{2}\, \!  \right) \,{G}\,\;\sqrt{\frac{2\alpha}{\pi}}\;
 \left( {\vrule height1.03em width0em depth1.03em} \right. \! \! 
} \nonumber\\
 & &  \left( \! \,\Phi \left( \! \,{\displaystyle \frac {1}{
2}}\,{\displaystyle \frac {(\,{(t-r)} + {R}\,)\,\sqrt {2}}{\sqrt {{\alpha}
}}}\, \!  \right)  - \Phi \left( \! \,{\displaystyle \frac {
1}{2}}\,{\displaystyle \frac {(\,{(t-r)} - {R}\,)\,\sqrt {2}}{\sqrt {
{\alpha}}}}\, \!  \right) \, \!  \right)  \nonumber\\
 & & (\, - {R}^{4} - 5\,{(t-r)}^{4} - 5\,{\alpha}^{2} - 20\,{\alpha}\,{(t-r)}^{2}
 + 6\,{(t-r)}^{2}\,{R}^{2} - 4\,{\alpha}\,{R}^{2}\,) \nonumber\\
 & & \mbox{} + {(t-r)}\,\sqrt {2}\,\sqrt {{\displaystyle \frac {{\alpha}}{
{ \pi}}}}\,(\,3\,{(t-r)}^{2} + {R}^{2} - 3\,{\alpha}\,)\, \left( \! \,
{\rm e}^{ \left( \! \, - \,1/2\,\frac {(\,{(t-r)} + {R}\,)^{2}}{{\alpha}}
\, \!  \right) } - {\rm e}^{ \left( \! \, - \,1/2\,\frac {(\,{(t-r)}
 - {R}\,)^{2}}{{\alpha}}\, \!  \right) }\, \!  \right)  \nonumber\\
 & & \mbox{} + {R}\,\sqrt {2}\,\sqrt {{\displaystyle \frac {{\alpha}}{
{ \pi}}}}\, \left( \! \,9\,{(t-r)}^{2} + {\displaystyle \frac {11}{3
}}\,{R}^{2} + 5\,{\alpha}\, \!  \right) \, \left( \! \,{\rm e}^{
 \left( \! \, - \,1/2\,\frac {(\,{(t-r)} + {R}\,)^{2}}{{\alpha}}\, \! 
 \right) } + {\rm e}^{ \left( \! \, - \,1/2\,\frac {(\,{(t-r)} - {R}
\,)^{2}}{{\alpha}}\, \!  \right) }\, \!  \right)  \! \! \left. 
{\vrule height1.03em width0em depth1.03em} \right)  \left/ 
{\vrule height0.37em width0em depth0.37em} \right. \! \! ( \nonumber\\
 & & { \pi}\,{\alpha}^{2}\,{r}^{3}\,)
\end{eqnarray}
\end{maplelatex}
\vspace*{1cm}
\begin{equation}
h^{01}_{c}(\vec{x},t)=\frac{4G\;x^ix^j}{r^5}\int d^3\vec{x}'\;
(3x'^ix'^j-r'^2\delta^{ij})\;\langle{\hat T}^{01}(r',t_{ret})\rangle
\end{equation}
\begin{eqnarray}
h^{01}_{c}(\vec{x},t)&=&\frac{4G}{r^3}\int^R_0 dr'\int^{2\pi}_0 d\phi
\int^{\pi}_0 d\gamma\;r'^4\cos\phi\sin^2\gamma\;
C(r',t_{ret})\times\nonumber\\ & & 
\left[3(\cos^2\phi\sin^2\gamma\cos^\alpha_0\sin^2\gamma_0+
\sin^2\phi\sin^2\gamma\sin^2\alpha_0\sin^2\gamma_0+\cos^2\gamma\cos^2\gamma_0)
-1\right]\nonumber\\ & &
\end{eqnarray}
performing the $\phi$ and $\gamma$ integrals we find:
\begin{equation}
h^{01}_{c}(\vec{x},t)=0
\end{equation}
similarly one finds that 
$$h^{02}_{c}(\vec{x},t)=h^{03}_{c}(\vec{x},t)=0$$
\begin{equation}
h^{00}_{d}(\vec{x},t)=\frac{4G\;x^ix^j}{r^3}\int d^3\vec{x}'\;
x'^ix'^j\;\langle{\hat T}^{00}(r',t_{ret})\rangle
\end{equation}
\begin{eqnarray}
h^{00}_{d}(\vec{x},t)&=&\frac{4G}{r}\int^{R}_0 dr'\int^{2\pi}_0 d\phi
\int^{\pi}_0 d\gamma\;r'^4\;\sin\gamma\;\langle{\hat T}^{00}(r',t_{ret})\rangle
\times\nonumber\\ & &
\left[\cos^2\phi\sin^2\gamma\cos\alpha_0\sin^2\gamma_0+
\sin^2\phi\sin^2\gamma\sin\alpha_0\sin^2\gamma_0+\cos^2\gamma\cos^2\gamma_0
\right]\nonumber\\ & &
\end{eqnarray}
performing the $\phi$ and $\gamma$ integrals we obtain:
\begin{equation}
h^{00}_{d}(\vec{x},t)=\frac{16\pi\;G}{3r}\int^R_0 dr'\;r'^3\left[
F(t_{ret}+r')-F(t_{ret}-r')\right]
\end{equation}
\begin{equation}
h^{11}_{d}(\vec{x},t)=\frac{4G\;x^ix^j}{r^3}\int d^3\vec{x}'\;
x'^ix'^j\;\langle{\hat T}^{11}(r',t_{ret})\rangle
\end{equation}
\begin{eqnarray}
h^{11}_{d}(\vec{x},t)&=&\frac{4G}{r}\int^{R}_0 dr'\int^{2\pi}_0 d\phi
\int^{\pi}_0 d\gamma\;r'^4\;\cos^2\phi\sin^3\gamma\times\nonumber\\ & &
\left[
\cos^2\phi\sin^2\gamma\cos\alpha_0\sin^2\gamma_0+
\sin^2\phi\sin^2\gamma\sin\alpha_0\sin^2\gamma_0+\cos^2\gamma\cos^2\gamma_0
\right]\times\nonumber\\ & & \left(
\frac{1}{r'}\left[F(t_{ret}+r')-F(t_{ret}+r')\right]+
\frac{3}{r'^2}\left[H(t_{ret}+r')+H(t_{ret}+r')\right]\right.\nonumber\\ & &
\left. -\frac{3}{r'^3}\left[E(t_{ret}+r')-E(t_{ret}+r')\right]\right)+\nonumber
\\ & & 
\frac{16\pi\;G}{3r}\int^{R}_0 dr'\;r'^4
\left(-\frac{1}{r'^2}\left[H(t_{ret}+r')+H(t_{ret}+r')\right]+\right.
\nonumber\\ & &\left.
\frac{1}{r'^3}\left[E(t_{ret}+r')-E(t_{ret}+r')\right]\right)\\ & & \nonumber
\end{eqnarray}
performing the remaining $\phi$ and $\gamma$ integrals
\begin{eqnarray}
h^{11}_{d}(\vec{x},t)&=&\frac{16\pi\;G}{15r}\left[2\cos^2\alpha_0\sin^2\gamma_0
+1\right]\int^{R}_0 dr'\;\left(\frac{1}{r'}\left[F(t_{ret}+r')-F(t_{ret}+r')\right]
+\right.\nonumber\\ & & \left. 
\frac{3}{r'^2}\left[H(t_{ret}+r')+H(t_{ret}+r')\right]
-\frac{3}{r'^3}\left[E(t_{ret}+r')-E(t_{ret}+r')\right]\right)+
\nonumber\\ & & \frac{16\pi\;G}{3r}\int^{R}_0 dr'\;r'^4
\left(-\frac{1}{r'^2}\left[H(t_{ret}+r')+H(t_{ret}+r')\right]+\right.
\nonumber\\ & &\left.
\frac{1}{r'^3}\left[E(t_{ret}+r')-E(t_{ret}+r')\right]\right)
\\ & & \nonumber
\end{eqnarray}
Similarly one finds:
\begin{eqnarray}
h^{22}_{d}(\vec{x},t)&=&\frac{16\pi\;G}{15r}\left[2\sin^2\alpha_0\sin^2\gamma_0
+1\right]\int^{R}_0 dr'\;\left(\frac{1}{r'}\left[F(t_{ret}+r')-F(t_{ret}+r')\right]
+\right.\nonumber\\ & & \left. 
\frac{3}{r'^2}\left[H(t_{ret}+r')+H(t_{ret}+r')\right]
-\frac{3}{r'^3}\left[E(t_{ret}+r')-E(t_{ret}+r')\right]\right)+
\nonumber\\ & & \frac{16\pi\;G}{3r}\int^{R}_0 dr'\;r'^4
\left(-\frac{1}{r'^2}\left[H(t_{ret}+r')+H(t_{ret}+r')\right]+\right.\nonumber
\\ & & \left.
\frac{1}{r'^3}\left[E(t_{ret}+r')-E(t_{ret}+r')\right]\right)\nonumber\\ & &
\end{eqnarray}
\begin{eqnarray}
h^{22}_{d}(\vec{x},t)&=&\frac{16\pi\;G}{5r}\left[1-\frac{2}{3}
\sin^2\gamma_0\right]\int^{R}_0 dr'\;\left(\frac{1}{r'}
\left[F(t_{ret}+r')-F(t_{ret}+r')\right]
+\right.\nonumber\\ & & \left. 
\frac{3}{r'^2}\left[H(t_{ret}+r')+H(t_{ret}+r')\right]
-\frac{3}{r'^3}\left[E(t_{ret}+r')-E(t_{ret}+r')\right]\right)+
\nonumber\\ & & \frac{16\pi\;G}{3r}\int^{R}_0 dr'\;r'^4
\left(-\frac{1}{r'^2}\left[H(t_{ret}+r')+H(t_{ret}+r')\right]+
\right.\nonumber\\ & &\left.
\frac{1}{r'^3}\left[E(t_{ret}+r')-E(t_{ret}+r')\right]\right)
\end{eqnarray}
\begin{equation}
h^{01}_{d}(\vec{x},t)=\frac{4G\;x^ix^j}{r^3}\int d^3\vec{x}'\;
x'^ix'^j\;\langle{\hat T}^{01}(r',t_{ret})\rangle
\end{equation}
\begin{eqnarray}
h^{01}_{d}(\vec{x},t)&=&\frac{4G}{r}\int^{R}_0 dr'\int^{2\pi}_0 d\phi
\int^{\pi}_0 d\gamma\;r'^4\;\cos\phi\sin^2\gamma\;C(r',t_{ret})\times
\nonumber\\ & & \left[\cos^2\phi\sin^2\gamma\cos\alpha_0\sin^2\gamma_0+
2\cos\phi\sin\phi\sin^2\gamma\cos\alpha_0\sin\alpha_0\sin^2\gamma_0+
\right.\nonumber\\ & & \left. 
\sin^2\phi\sin^2\gamma\sin\alpha_0\sin^2\gamma_0+\cos^2\gamma\cos^2\gamma_0
+\right.\nonumber\\ & & \left.
2\cos\phi\sin\gamma\cos\gamma\cos\alpha_0\sin\gamma_0\cos\gamma_0+
2\sin\phi\sin\gamma\cos\gamma\sin\alpha_0\sin\gamma_0\cos\gamma_0
\right]\nonumber\\ & &
\end{eqnarray}
performing the $\phi$ and $\gamma$ integrals we find that
\begin{equation}
h^{01}_{d}(\vec{x},t)=0
\label{Afin}
\end{equation}
similarly one finds:
$$h^{02}_{d}(\vec{x},t)=h^{03}_{d}(\vec{x},t)=0.$$
The other terms in the weak-field metric can be obtained from 
eqs.(\ref{Aini})-(\ref{Afin}) with the results:
\begin{maplelatex}
\begin{eqnarray}
\lefteqn{\dot{h}^{01}_{b}(\vec{x},t) = {\displaystyle \frac {1}{3}}{G}\,\;\sqrt{\frac{2\alpha}{\pi}}\;
 \left( {\vrule height0.89em width0em depth0.89em} \right. \! \! 
 \left( \! \,{\displaystyle \frac {{\rm \%2}\,\sqrt {2}}{\sqrt {{
 \pi}}\,\sqrt {{\alpha}}}} - {\displaystyle \frac {{\rm \%1}\,\sqrt {2
}}{\sqrt {{ \pi}}\,\sqrt {{\alpha}}}}\, \!  \right) \,(\,{(t-r)}\,{R}^{2}
 - {(t-r)}^{3} - 2\,{\alpha}\,{(t-r)}\,)} \nonumber\\
 & & \mbox{} +  \left( \! \,\Phi \left( \! \,{\displaystyle 
\frac {1}{2}}\,{\displaystyle \frac {(\,{(t-r)} + {R}\,)\,\sqrt {2}}{
\sqrt {{\alpha}}}}\, \!  \right)  - \Phi \left( \! \,
{\displaystyle \frac {1}{2}}\,{\displaystyle \frac {(\,{(t-r)} - {R}
\,)\,\sqrt {2}}{\sqrt {{\alpha}}}}\, \!  \right) \, \!  \right) \,(\,{
R}^{2} - 3\,{(t-r)}^{2} - 2\,{\alpha}\,) \nonumber\\
 & & \mbox{} - {\alpha}\, \left( \! \, - \,{\displaystyle \frac {(\,{t
} + {R}\,)\,{\rm \%2}}{{\alpha}}} + {\displaystyle \frac {(\,{(t-r)} - {R}
\,)\,{\rm \%1}}{{\alpha}}}\, \!  \right) \,\sqrt {2}\,\sqrt {
{\displaystyle \frac {{\alpha}}{{ \pi}}}} + {R}\,(\,{\rm \%2} + {\rm 
\%1}\,)\,\sqrt {2}\,\sqrt {{\displaystyle \frac {{\alpha}}{{ \pi}}}}
 \nonumber\\
 & & \mbox{} + {R}\,{(t-r)}\, \left( \! \, - \,{\displaystyle \frac {
(\,{(t-r)} + {R}\,)\,{\rm \%2}}{{\alpha}}} - {\displaystyle \frac {(\,{(t-r)}
 - {R}\,)\,{\rm \%1}}{{\alpha}}}\, \!  \right) \,\sqrt {2}\,\sqrt {
{\displaystyle \frac {{\alpha}}{{ \pi}}}} \! \! \left. {\vrule 
height0.89em width0em depth0.89em} \right) {\rm cos}(\,{ \alpha_0}
\,)\,{\rm sin}(\,{ \gamma_0}\,) \left/ {\vrule 
height0.37em width0em depth0.37em} \right. \! \! (\,{r}\,{ \pi}
 \nonumber\\
 & & {\alpha}^{2}\,) 
\end{eqnarray}
\end{maplelatex}
\begin{maplelatex}
\begin{eqnarray}
\lefteqn{\dot{h}^{02}_{b}(\vec{x},t) = {\displaystyle \frac {1}{3}}{G}\,\;\sqrt{\frac{2\alpha}{\pi}}\;
 \left( {\vrule height0.89em width0em depth0.89em} \right. \! \! 
 \left( \! \,{\displaystyle \frac {{\rm \%2}\,\sqrt {2}}{\sqrt {{
 \pi}}\,\sqrt {{\alpha}}}} - {\displaystyle \frac {{\rm \%1}\,\sqrt {2
}}{\sqrt {{ \pi}}\,\sqrt {{\alpha}}}}\, \!  \right) \,(\,{(t-r)}\,{R}^{2}
 - {(t-r)}^{3} - 2\,{\alpha}\,{(t-r)}\,)} \nonumber\\
 & & \mbox{} +  \left( \! \,\Phi \left( \! \,{\displaystyle 
\frac {1}{2}}\,{\displaystyle \frac {(\,{(t-r)} + {R}\,)\,\sqrt {2}}{
\sqrt {{\alpha}}}}\, \!  \right)  - \Phi \left( \! \,
{\displaystyle \frac {1}{2}}\,{\displaystyle \frac {(\,{(t-r)} - {R}
\,)\,\sqrt {2}}{\sqrt {{\alpha}}}}\, \!  \right) \, \!  \right) \,(\,{
R}^{2} - 3\,{(t-r)}^{2} - 2\,{\alpha}\,) \nonumber\\
 & & \mbox{} - {\alpha}\, \left( \! \, - \,{\displaystyle \frac {(\,{t
} + {R}\,)\,{\rm \%2}}{{\alpha}}} + {\displaystyle \frac {(\,{(t-r)} - {R}
\,)\,{\rm \%1}}{{\alpha}}}\, \!  \right) \,\sqrt {2}\,\sqrt {
{\displaystyle \frac {{\alpha}}{{ \pi}}}} + {R}\,(\,{\rm \%2} + {\rm 
\%1}\,)\,\sqrt {2}\,\sqrt {{\displaystyle \frac {{\alpha}}{{ \pi}}}}
 \nonumber\\
 & & \mbox{} + {R}\,{(t-r)}\, \left( \! \, - \,{\displaystyle \frac {
(\,{(t-r)} + {R}\,)\,{\rm \%2}}{{\alpha}}} - {\displaystyle \frac {(\,{(t-r)}
 - {R}\,)\,{\rm \%1}}{{\alpha}}}\, \!  \right) \,\sqrt {2}\,\sqrt {
{\displaystyle \frac {{\alpha}}{{ \pi}}}} \! \! \left. {\vrule 
height0.89em width0em depth0.89em} \right) {\rm sin}(\,{ \alpha_0}
\,)\,{\rm sin}(\,{ \gamma_0}\,) \left/ {\vrule 
height0.37em width0em depth0.37em} \right. \! \! (\,{r}\,{ \pi}
 \nonumber\\
 & & {\alpha}^{2}\,) 
\end{eqnarray}
\end{maplelatex}
\begin{maplelatex}
\begin{eqnarray}
\lefteqn{\dot{h}^{03}_{b}(\vec{x},t)= {\displaystyle \frac {1}{3}}{G}\,\;\sqrt{\frac{2\alpha}{\pi}}\;
 \left( {\vrule height0.89em width0em depth0.89em} \right. \! \! 
 \left( \! \,{\displaystyle \frac {{\rm \%2}\,\sqrt {2}}{\sqrt {{
 \pi}}\,\sqrt {{\alpha}}}} - {\displaystyle \frac {{\rm \%1}\,\sqrt {2
}}{\sqrt {{ \pi}}\,\sqrt {{\alpha}}}}\, \!  \right) \,(\,{(t-r)}\,{R}^{2}
 - {(t-r)}^{3} - 2\,{\alpha}\,{(t-r)}\,)} \nonumber\\
 & & \mbox{} +  \left( \! \,\Phi \left( \! \,{\displaystyle 
\frac {1}{2}}\,{\displaystyle \frac {(\,{(t-r)} + {R}\,)\,\sqrt {2}}{
\sqrt {{\alpha}}}}\, \!  \right)  - \Phi \left( \! \,
{\displaystyle \frac {1}{2}}\,{\displaystyle \frac {(\,{(t-r)} - {R}
\,)\,\sqrt {2}}{\sqrt {{\alpha}}}}\, \!  \right) \, \!  \right) \,(\,{
R}^{2} - 3\,{(t-r)}^{2} - 2\,{\alpha}\,) \nonumber\\
 & & \mbox{} - {\alpha}\, \left( \! \, - \,{\displaystyle \frac {(\,{t
} + {R}\,)\,{\rm \%2}}{{\alpha}}} + {\displaystyle \frac {(\,{(t-r)} - {R}
\,)\,{\rm \%1}}{{\alpha}}}\, \!  \right) \,\sqrt {2}\,\sqrt {
{\displaystyle \frac {{\alpha}}{{ \pi}}}} + {R}\,(\,{\rm \%2} + {\rm 
\%1}\,)\,\sqrt {2}\,\sqrt {{\displaystyle \frac {{\alpha}}{{ \pi}}}}
 \nonumber\\
 & & \mbox{} + {R}\,{(t-r)}\, \left( \! \, - \,{\displaystyle \frac {
(\,{(t-r)} + {R}\,)\,{\rm \%2}}{{\alpha}}} - {\displaystyle \frac {(\,{(t-r)}
 - {R}\,)\,{\rm \%1}}{{\alpha}}}\, \!  \right) \,\sqrt {2}\,\sqrt {
{\displaystyle \frac {{\alpha}}{{ \pi}}}} \! \! \left. {\vrule 
height0.89em width0em depth0.89em} \right) {\rm cos}(\,{ \gamma_0}
\,) \left/ {\vrule height0.37em width0em depth0.37em} \right. \! 
\! (\,{r}\,{ \pi}\,{\alpha}^{2}\,) \nonumber\\ & &
\end{eqnarray}
\end{maplelatex}
\begin{maplelatex}
\begin{eqnarray}
\lefteqn{\dot{h}^{11}_{c}(\vec{x},t)= {\displaystyle \frac {1}{60}} \left( 
\! \,3\,{\rm cos}(\,{ \alpha_0}\,)^{2}\,{\rm sin}(\,{ \gamma_0}\,)^{2
} - 1\, \!  \right) \,{G}\,\;\sqrt{\frac{2\alpha}{\pi}}\; \left( {\vrule 
height0.89em width0em depth0.89em} \right. \! \! } \nonumber\\
 & &  \left( \! \,{\displaystyle \frac {{\rm \%2}\,\sqrt {2}}{
\sqrt {{ \pi}}\,\sqrt {{\alpha}}}} - {\displaystyle \frac {{\rm \%1}\,
\sqrt {2}}{\sqrt {{ \pi}}\,\sqrt {{\alpha}}}}\, \!  \right) \,(\, - {R
}^{4} - 5\,{(t-r)}^{4} - 5\,{\alpha}^{2} - 20\,{\alpha}\,{(t-r)}^{2} + 6\,{(t-r)}^{2}\,
{R}^{2} - 4\,{\alpha}\,{R}^{2}\,) +  \nonumber\\
 & &  \left( \! \,\Phi \left( \! \,{\displaystyle \frac {1}{
2}}\,{\displaystyle \frac {(\,{(t-r)} + {R}\,)\,\sqrt {2}}{\sqrt {{\alpha}
}}}\, \!  \right)  - \Phi \left( \! \,{\displaystyle \frac {
1}{2}}\,{\displaystyle \frac {(\,{(t-r)} - {R}\,)\,\sqrt {2}}{\sqrt {
{\alpha}}}}\, \!  \right) \, \!  \right)  \nonumber\\
 & & (\, - 20\,{(t-r)}^{3} - 40\,{\alpha}\,{(t-r)} + 12\,{(t-r)}\,{R}^{2}\,)
\mbox{} + \sqrt {2}\,\sqrt {{\displaystyle \frac {{\alpha}}{{ \pi}}}}
\,(\,3\,{(t-r)}^{2} + {R}^{2} - 3\,{\alpha}\,)\,(\,{\rm \%2} - {\rm \%1}\,
) \nonumber\\
 & & \mbox{} + 6\,{(t-r)}^{2}\,\sqrt {2}\,\sqrt {{\displaystyle 
\frac {{\alpha}}{{ \pi}}}}\,(\,{\rm \%2} - {\rm \%1}\,) \nonumber\\
 & & \mbox{} + {(t-r)}\,\sqrt {2}\,\sqrt {{\displaystyle \frac {{\alpha}}{
{ \pi}}}}\,(\,3\,{(t-r)}^{2} + {R}^{2} - 3\,{\alpha}\,)\, \left( \! \, - 
\,{\displaystyle \frac {(\,{(t-r)} + {R}\,)\,{\rm \%2}}{{\alpha}}} + 
{\displaystyle \frac {(\,{(t-r)} - {R}\,)\,{\rm \%1}}{{\alpha}}}\, \! 
 \right)  \nonumber\\
 & & \mbox{} + 18\,{R}\,{(t-r)}\,(\,{\rm \%2} + {\rm \%1}\,)\,\sqrt {
2}\,\sqrt {{\displaystyle \frac {{\alpha}}{{ \pi}}}} \nonumber\\
 & & \mbox{} + {R}\,\sqrt {2}\,\sqrt {{\displaystyle \frac {{\alpha}}{
{ \pi}}}}\, \left( \! \,9\,{(t-r)}^{2} + {\displaystyle \frac {11}{3
}}\,{R}^{2} + 5\,{\alpha}\, \!  \right) \, \left( \! \, - \,
{\displaystyle \frac {(\,{(t-r)} + {R}\,)\,{\rm \%2}}{{\alpha}}} - 
{\displaystyle \frac {(\,{(t-r)} - {R}\,)\,{\rm \%1}}{{\alpha}}}\, \! 
 \right)  \! \! \left. {\vrule height0.89em width0em depth0.89em}
 \right)  \left/ {\vrule height0.37em width0em depth0.37em}
 \right. \! \! ( \nonumber\\
 & & {r}^{2}\,{ \pi}\,{\alpha}^{2}\,) 
\end{eqnarray}
\end{maplelatex}
\begin{maplelatex}
\begin{eqnarray}
\lefteqn{\dot{h}^{22}_{c}(\vec{x},t)= {\displaystyle \frac {1}{60}} \left( 
\! \,3\,{\rm sin}(\,{ \alpha_0}\,)^{2}\,{\rm sin}(\,{ \gamma_0}\,)^{2
} - 1\, \!  \right) \,{G}\,\;\sqrt{\frac{2\alpha}{\pi}}\; \left( {\vrule 
height0.89em width0em depth0.89em} \right. \! \! } \nonumber\\
 & &  \left( \! \,{\displaystyle \frac {{\rm \%2}\,\sqrt {2}}{
\sqrt {{ \pi}}\,\sqrt {{\alpha}}}} - {\displaystyle \frac {{\rm \%1}\,
\sqrt {2}}{\sqrt {{ \pi}}\,\sqrt {{\alpha}}}}\, \!  \right) \,(\, - {R
}^{4} - 5\,{(t-r)}^{4} - 5\,{\alpha}^{2} - 20\,{\alpha}\,{(t-r)}^{2} + 6\,{(t-r)}^{2}\,
{R}^{2} - 4\,{\alpha}\,{R}^{2}\,) +  \nonumber\\
 & &  \left( \! \,\Phi \left( \! \,{\displaystyle \frac {1}{
2}}\,{\displaystyle \frac {(\,{(t-r)} + {R}\,)\,\sqrt {2}}{\sqrt {{\alpha}
}}}\, \!  \right)  - \Phi \left( \! \,{\displaystyle \frac {
1}{2}}\,{\displaystyle \frac {(\,{(t-r)} - {R}\,)\,\sqrt {2}}{\sqrt {
{\alpha}}}}\, \!  \right) \, \!  \right)  \nonumber\\
 & & (\, - 20\,{(t-r)}^{3} - 40\,{\alpha}\,{(t-r)} + 12\,{(t-r)}\,{R}^{2}\,)
\mbox{} + \sqrt {2}\,\sqrt {{\displaystyle \frac {{\alpha}}{{ \pi}}}}
\,(\,3\,{(t-r)}^{2} + {R}^{2} - 3\,{\alpha}\,)\,(\,{\rm \%2} - {\rm \%1}\,
) \nonumber\\
 & & \mbox{} + 6\,{(t-r)}^{2}\,\sqrt {2}\,\sqrt {{\displaystyle 
\frac {{\alpha}}{{ \pi}}}}\,(\,{\rm \%2} - {\rm \%1}\,) \nonumber\\
 & & \mbox{} + {(t-r)}\,\sqrt {2}\,\sqrt {{\displaystyle \frac {{\alpha}}{
{ \pi}}}}\,(\,3\,{(t-r)}^{2} + {R}^{2} - 3\,{\alpha}\,)\, \left( \! \, - 
\,{\displaystyle \frac {(\,{(t-r)} + {R}\,)\,{\rm \%2}}{{\alpha}}} + 
{\displaystyle \frac {(\,{(t-r)} - {R}\,)\,{\rm \%1}}{{\alpha}}}\, \! 
 \right)  \nonumber\\
 & & \mbox{} + 18\,{R}\,{(t-r)}\,(\,{\rm \%2} + {\rm \%1}\,)\,\sqrt {
2}\,\sqrt {{\displaystyle \frac {{\alpha}}{{ \pi}}}} \nonumber\\
 & & \mbox{} + {R}\,\sqrt {2}\,\sqrt {{\displaystyle \frac {{\alpha}}{
{ \pi}}}}\, \left( \! \,9\,{(t-r)}^{2} + {\displaystyle \frac {11}{3
}}\,{R}^{2} + 5\,{\alpha}\, \!  \right) \, \left( \! \, - \,
{\displaystyle \frac {(\,{(t-r)} + {R}\,)\,{\rm \%2}}{{\alpha}}} - 
{\displaystyle \frac {(\,{(t-r)} - {R}\,)\,{\rm \%1}}{{\alpha}}}\, \! 
 \right)  \! \! \left. {\vrule height0.89em width0em depth0.89em}
 \right)  \left/ {\vrule height0.37em width0em depth0.37em}
 \right. \! \! ( \nonumber\\
 & & {r}^{2}\,{ \pi}\,{\alpha}^{2}\,)
\end{eqnarray}
\end{maplelatex}
\begin{maplelatex}
\begin{eqnarray}
\lefteqn{\dot{h}^{33}_{c}(\vec{x},t)= {\displaystyle \frac {1}{60}} \left( 
\! \,2 - 3\,{\rm sin}(\,{ \gamma_0}\,)^{2}\, \!  \right) \,{G}\,\;\sqrt{\frac{2\alpha}{\pi}}\;
 \left( {\vrule height0.89em width0em depth0.89em} \right. \! \! 
} \nonumber\\
 & &  \left( \! \,{\displaystyle \frac {{\rm \%2}\,\sqrt {2}}{
\sqrt {{ \pi}}\,\sqrt {{\alpha}}}} - {\displaystyle \frac {{\rm \%1}\,
\sqrt {2}}{\sqrt {{ \pi}}\,\sqrt {{\alpha}}}}\, \!  \right) \,(\, - {R
}^{4} - 5\,{(t-r)}^{4} - 5\,{\alpha}^{2} - 20\,{\alpha}\,{(t-r)}^{2} + 6\,{(t-r)}^{2}\,
{R}^{2} - 4\,{\alpha}\,{R}^{2}\,) +  \nonumber\\
 & &  \left( \! \,\Phi \left( \! \,{\displaystyle \frac {1}{
2}}\,{\displaystyle \frac {(\,{(t-r)} + {R}\,)\,\sqrt {2}}{\sqrt {{\alpha}
}}}\, \!  \right)  - \Phi \left( \! \,{\displaystyle \frac {
1}{2}}\,{\displaystyle \frac {(\,{(t-r)} - {R}\,)\,\sqrt {2}}{\sqrt {
{\alpha}}}}\, \!  \right) \, \!  \right)  \nonumber\\
 & & (\, - 20\,{(t-r)}^{3} - 40\,{\alpha}\,{(t-r)} + 12\,{(t-r)}\,{R}^{2}\,)
\mbox{} + \sqrt {2}\,\sqrt {{\displaystyle \frac {{\alpha}}{{ \pi}}}}
\,(\,3\,{(t-r)}^{2} + {R}^{2} - 3\,{\alpha}\,)\,(\,{\rm \%2} - {\rm \%1}\,
) \nonumber\\
 & & \mbox{} + 6\,{(t-r)}^{2}\,\sqrt {2}\,\sqrt {{\displaystyle 
\frac {{\alpha}}{{ \pi}}}}\,(\,{\rm \%2} - {\rm \%1}\,) \nonumber\\
 & & \mbox{} + {(t-r)}\,\sqrt {2}\,\sqrt {{\displaystyle \frac {{\alpha}}{
{ \pi}}}}\,(\,3\,{(t-r)}^{2} + {R}^{2} - 3\,{\alpha}\,)\, \left( \! \, - 
\,{\displaystyle \frac {(\,{(t-r)} + {R}\,)\,{\rm \%2}}{{\alpha}}} + 
{\displaystyle \frac {(\,{(t-r)} - {R}\,)\,{\rm \%1}}{{\alpha}}}\, \! 
 \right)  \nonumber\\
 & & \mbox{} + 18\,{R}\,{(t-r)}\,(\,{\rm \%2} + {\rm \%1}\,)\,\sqrt {
2}\,\sqrt {{\displaystyle \frac {{\alpha}}{{ \pi}}}} \nonumber\\
 & & \mbox{} + {R}\,\sqrt {2}\,\sqrt {{\displaystyle \frac {{\alpha}}{
{ \pi}}}}\, \left( \! \,9\,{(t-r)}^{2} + {\displaystyle \frac {11}{3
}}\,{R}^{2} + 5\,{\alpha}\, \!  \right) \, \left( \! \, - \,
{\displaystyle \frac {(\,{(t-r)} + {R}\,)\,{\rm \%2}}{{\alpha}}} - 
{\displaystyle \frac {(\,{(t-r)} - {R}\,)\,{\rm \%1}}{{\alpha}}}\, \! 
 \right)  \! \! \left. {\vrule height0.89em width0em depth0.89em}
 \right)  \left/ {\vrule height0.37em width0em depth0.37em}
 \right. \! \! ( \nonumber\\
 & & {r}^{2}\,{ \pi}\,{\alpha}^{2}\,) 
\end{eqnarray}
\end{maplelatex}
\begin{maplelatex}
\begin{eqnarray}
\lefteqn{\ddot{h}^{00}_{d}(\vec{x},t)= {\displaystyle \frac {1}{12}}{G}\,\;\sqrt{\frac{2\alpha}{\pi}}\;
 \left( {\vrule height0.92em width0em depth0.92em} \right. \! \! 
} \nonumber\\
 & & { \pi}\, \left( \! \, - \,{\displaystyle \frac {(\,{(t-r)} + {R}
\,)\,{\rm \%2}\,\sqrt {2}}{\sqrt {{ \pi}}\,{\alpha}^{3/2}}} + 
{\displaystyle \frac {(\,{(t-r)} - {R}\,)\,{\rm \%1}\,\sqrt {2}}{
\sqrt {{ \pi}}\,{\alpha}^{3/2}}}\, \!  \right) \,(\,{R}^{4} - {(t-r)}^{4}
 - {\alpha}^{2} - 4\,{\alpha}\,{(t-r)}^{2}\,) \nonumber\\
 & & \mbox{} + 2\,{ \pi}\, \left( \! \,{\displaystyle \frac {
{\rm \%2}\,\sqrt {2}}{\sqrt {{ \pi}}\,\sqrt {{\alpha}}}} - 
{\displaystyle \frac {{\rm \%1}\,\sqrt {2}}{\sqrt {{ \pi}}\,
\sqrt {{\alpha}}}}\, \!  \right) \,(\, - 4\,{(t-r)}^{3} - 8\,{\alpha}\,{(t-r)}\,)
 \nonumber\\
 & & \mbox{} + { \pi}\, \left( \! \,\Phi \left( \! \,
{\displaystyle \frac {1}{2}}\,{\displaystyle \frac {(\,{(t-r)} + {R}
\,)\,\sqrt {2}}{\sqrt {{\alpha}}}}\, \!  \right)  - \Phi \left( 
\! \,{\displaystyle \frac {1}{2}}\,{\displaystyle \frac {(\,{(t-r)}
 - {R}\,)\,\sqrt {2}}{\sqrt {{\alpha}}}}\, \!  \right) \, \!  \right) 
\,(\, - 12\,{(t-r)}^{2} - 8\,{\alpha}\,) \nonumber\\
 & & \mbox{} + 2\,{\alpha}\, \left( \! \, - \,{\displaystyle \frac {(
\,{(t-r)} + {R}\,)\,{\rm \%2}}{{\alpha}}} + {\displaystyle \frac {(\,{(t-r)}
 - {R}\,)\,{\rm \%1}}{{\alpha}}}\, \!  \right) \,\sqrt {2}\,\sqrt {
{\displaystyle \frac {{ \pi}}{{\alpha}}}}\,(\,{(t-r)}^{2} - {R}^{2} - 3\,{
\alpha}\,) \nonumber\\
 & & \mbox{} + 6\,{\alpha}\,(\,{\rm \%2} - {\rm \%1}\,)\,\sqrt {2}\,
\sqrt {{\displaystyle \frac {{ \pi}}{{\alpha}}}}\,{(t-r)} + {\alpha}\,{(t-r)} \nonumber\\
 & &  \left( \! \, - \,{\displaystyle \frac {{\rm \%2}}{{\alpha}}} + 
{\displaystyle \frac {(\,{(t-r)} + {R}\,)^{2}\,{\rm \%2}}{{\alpha}^{2}}}
 + {\displaystyle \frac {{\rm \%1}}{{\alpha}}} - {\displaystyle 
\frac {(\,{(t-r)} - {R}\,)^{2}\,{\rm \%1}}{{\alpha}^{2}}}\, \!  \right) \,
\sqrt {2}\,\sqrt {{\displaystyle \frac {{ \pi}}{{\alpha}}}}\,(\,{(t-r)}^{2
} - {R}^{2} - 3\,{\alpha}\,) \nonumber\\
 & & \mbox{} + 4\,{\alpha}\,{(t-r)}^{2}\, \left( \! \, - \,{\displaystyle 
\frac {(\,{(t-r)} + {R}\,)\,{\rm \%2}}{{\alpha}}} + {\displaystyle \frac {
(\,{(t-r)} - {R}\,)\,{\rm \%1}}{{\alpha}}}\, \!  \right) \,\sqrt {2}\,
\sqrt {{\displaystyle \frac {{ \pi}}{{\alpha}}}} + {\alpha}\,{R} \nonumber\\
 & &  \left( \! \, - \,{\displaystyle \frac {{\rm \%2}}{{\alpha}}} + 
{\displaystyle \frac {(\,{(t-r)} + {R}\,)^{2}\,{\rm \%2}}{{\alpha}^{2}}}
 - {\displaystyle \frac {{\rm \%1}}{{\alpha}}} + {\displaystyle 
\frac {(\,{(t-r)} - {R}\,)^{2}\,{\rm \%1}}{{\alpha}^{2}}}\, \!  \right) \,
\sqrt {2}\,\sqrt {{\displaystyle \frac {{ \pi}}{{\alpha}}}} \nonumber\\
 & &  \left( \! \,{(t-r)}^{2} + {\displaystyle \frac {1}{3}}\,{R}^{2}
 + {\alpha}\, \!  \right) \mbox{} + 4\,{\alpha}\,{R}\, \left( \! \, - \,
{\displaystyle \frac {(\,{(t-r)} + {R}\,)\,{\rm \%2}}{{\alpha}}} - 
{\displaystyle \frac {(\,{(t-r)} - {R}\,)\,{\rm \%1}}{{\alpha}}}\, \! 
 \right) \,\sqrt {2}\,\sqrt {{\displaystyle \frac {{ \pi}}{{\alpha}}}}
\,{(t-r)} \nonumber\\
 & & \mbox{} + 2\,{\alpha}\,{R}\,(\,{\rm \%2} + {\rm \%1}\,)\,\sqrt {2
}\,\sqrt {{\displaystyle \frac {{ \pi}}{{\alpha}}}} \! \! \left. 
{\vrule height0.92em width0em depth0.92em} \right)  \left/ 
{\vrule height0.43em width0em depth0.43em} \right. \! \!  \left( 
\! \,{ \pi}^{2}\,{\alpha}^{2}\,{r}\, \!  \right) 
\end{eqnarray}
\end{maplelatex}
\begin{maplelatex}
\begin{eqnarray}
\lefteqn{\ddot{h}^{11}_{d}(\vec{x},t)= {\displaystyle \frac {1}{60}}{G}\,{\sqrt{\frac{2\alpha}{\pi}}}
 \left( {\vrule height0.92em width0em depth0.92em} \right. \! \! 
 \left( \! \, - \,{\displaystyle \frac {(\,{(t-r)} + {R}\,)\,{\rm \%2
}\,\sqrt {2}}{\sqrt {{ \pi}}\,{\alpha}^{3/2}}} + {\displaystyle 
\frac {(\,{(t-r)} - {R}\,)\,{\rm \%1}\,\sqrt {2}}{\sqrt {{ \pi}}\,{\alpha}
^{3/2}}}\, \!  \right) }\nonumber \\
 & & (\, - {R}^{4} - 5\,{(t-r)}^{4} - 5\,{\alpha}^{2} - 20\,{\alpha}\,{(t-r)}^{2}
 + 6\,{(t-r)}^{2}\,{R}^{2} - 4\,{\alpha}\,{R}^{2}\,) \nonumber\\
 & & \mbox{} + 2\, \left( \! \,{\displaystyle \frac {{\rm \%2}\,
\sqrt {2}}{\sqrt {{ \pi}}\,\sqrt {{\alpha}}}} - {\displaystyle \frac {
{\rm \%1}\,\sqrt {2}}{\sqrt {{ \pi}}\,\sqrt {{\alpha}}}}\, \! 
 \right) \,(\, - 20\,{(t-r)}^{3} - 40\,{\alpha}\,{(t-r)} + 12\,{(t-r)}\,{R}^{2}\,)
 + \nonumber \\
 & &  \left( \! \,\Phi \left( \! \,{\displaystyle \frac {1}{
2}}\,{\displaystyle \frac {(\,{(t-r)} + {R}\,)\,\sqrt {2}}{\sqrt {{\alpha}
}}}\, \!  \right)  - \Phi \left( \! \,{\displaystyle \frac {
1}{2}}\,{\displaystyle \frac {(\,{(t-r)} - {R}\,)\,\sqrt {2}}{\sqrt {
{\alpha}}}}\, \!  \right) \, \!  \right) \,(\, - 60\,{(t-r)}^{2} - 40\,{\alpha}
 + 12\,{R}^{2}\,) \nonumber\\
 & & \mbox{} + 18\,\sqrt {2}\,\sqrt {{\displaystyle \frac {{\alpha}}{{
 \pi}}}}\,{(t-r)}\,(\,{\rm \%2} - {\rm \%1}\,) + 2\,\sqrt {2}\,
\sqrt {{\displaystyle \frac {{\alpha}}{{ \pi}}}}\,(\,3\,{(t-r)}^{2} + {R}
^{2} - 3\,{\alpha}\,)\,{\rm \%3} \nonumber\\
 & & \mbox{} + 12\,{(t-r)}^{2}\,\sqrt {2}\,\sqrt {{\displaystyle 
\frac {{\alpha}}{{ \pi}}}}\,{\rm \%3} + {(t-r)}\,\sqrt {2}\,\sqrt {
{\displaystyle \frac {{\alpha}}{{ \pi}}}}\,(\,3\,{(t-r)}^{2} + {R}^{2} - 3
\,{\alpha}\,) \nonumber\\
 & &  \left( \! \, - \,{\displaystyle \frac {{\rm \%2}}{{\alpha}}} + 
{\displaystyle \frac {(\,{(t-r)} + {R}\,)^{2}\,{\rm \%2}}{{\alpha}^{2}}}
 + {\displaystyle \frac {{\rm \%1}}{{\alpha}}} - {\displaystyle 
\frac {(\,{(t-r)} - {R}\,)^{2}\,{\rm \%1}}{{\alpha}^{2}}}\, \!  \right) 
 \nonumber\\
 & & \mbox{} + 18\,{R}\,(\,{\rm \%2} + {\rm \%1}\,)\,\sqrt {2}\,
\sqrt {{\displaystyle \frac {{\alpha}}{{ \pi}}}} \nonumber\\
 & & \mbox{} + 36\,{R}\,{(t-r)}\, \left( \! \, - \,{\displaystyle 
\frac {(\,{(t-r)} + {R}\,)\,{\rm \%2}}{{\alpha}}} - {\displaystyle \frac {
(\,{(t-r)} - {R}\,)\,{\rm \%1}}{{\alpha}}}\, \!  \right) \,\sqrt {2}\,
\sqrt {{\displaystyle \frac {{\alpha}}{{ \pi}}}} + {R}\,\sqrt {2}\,
\sqrt {{\displaystyle \frac {{\alpha}}{{ \pi}}}} \nonumber\\
 & &  \left( \! \,9\,{(t-r)}^{2} + {\displaystyle \frac {11}{3}}\,{R}
^{2} + 5\,{\alpha}\, \!  \right) \, \left( \! \, - \,{\displaystyle 
\frac {{\rm \%2}}{{\alpha}}} + {\displaystyle \frac {(\,{(t-r)} + {R}\,)^{
2}\,{\rm \%2}}{{\alpha}^{2}}} - {\displaystyle \frac {{\rm \%1}}{{\alpha}}}
 + {\displaystyle \frac {(\,{(t-r)} - {R}\,)^{2}\,{\rm \%1}}{{\alpha}^{2}
}}\, \!  \right)  \! \! \left. {\vrule 
height0.92em width0em depth0.92em} \right)  \nonumber\\
 & &  \left( \! \,{\rm cos}(\,{ \alpha_0}\,)^{2}\,{\rm sin}(\,{ 
\gamma_0}\,)^{2} - 1\, \!  \right)  \left/ {\vrule 
height0.37em width0em depth0.37em} \right. \! \! (\,{r}\,{ \pi}\,
{\alpha}^{2}\,)\mbox{} + {\displaystyle \frac {1}{36}}{G}\,{\sqrt{\frac{2\alpha}{\pi}}} \left( 
{\vrule height0.92em width0em depth0.92em} \right. \! \!  \nonumber\\
 & & { \pi}\, \left( \! \, - \,{\displaystyle \frac {(\,{(t-r)} + {R}
\,)\,{\rm \%2}\,\sqrt {2}}{\sqrt {{ \pi}}\,{\alpha}^{3/2}}} + 
{\displaystyle \frac {(\,{(t-r)} - {R}\,)\,{\rm \%1}\,\sqrt {2}}{
\sqrt {{ \pi}}\,{\alpha}^{3/2}}}\, \!  \right) \,(\,{R}^{4} - {(t-r)}^{4}
 - {\alpha}^{2} - 4\,{\alpha}\,{(t-r)}^{2}\,) \nonumber\\
 & & \mbox{} + 2\,{ \pi}\, \left( \! \,{\displaystyle \frac {
{\rm \%2}\,\sqrt {2}}{\sqrt {{ \pi}}\,\sqrt {{\alpha}}}} - 
{\displaystyle \frac {{\rm \%1}\,\sqrt {2}}{\sqrt {{ \pi}}\,
\sqrt {{\alpha}}}}\, \!  \right) \,(\, - 4\,{(t-r)}^{3} - 8\,{\alpha}\,{(t-r)}\,)
\nonumber \\
 & & \mbox{} + { \pi}\, \left( \! \,\Phi \left( \! \,
{\displaystyle \frac {1}{2}}\,{\displaystyle \frac {(\,{(t-r)} + {R}
\,)\,\sqrt {2}}{\sqrt {{\alpha}}}}\, \!  \right)  - \Phi \left( 
\! \,{\displaystyle \frac {1}{2}}\,{\displaystyle \frac {(\,{(t-r)}
 - {R}\,)\,\sqrt {2}}{\sqrt {{\alpha}}}}\, \!  \right) \, \!  \right) 
\,(\, - 12\,{(t-r)}^{2} - 8\,{\alpha}\,) \nonumber\\
 & & \mbox{} + 2\,{\alpha}\,{\rm \%3}\,\sqrt {2}\,\sqrt {
{\displaystyle \frac {{ \pi}}{{\alpha}}}}\,(\,{(t-r)}^{2} - {R}^{2} - 3\,{
a}\,) + 6\,{\alpha}\,(\,{\rm \%2} - {\rm \%1}\,)\,\sqrt {2}\,\sqrt {
{\displaystyle \frac {{ \pi}}{{\alpha}}}}\,{(t-r)} + {\alpha}\,{(t-r)} 
\nonumber\\
 & &  \left( \! \, - \,{\displaystyle \frac {{\rm \%2}}{{\alpha}}} + 
{\displaystyle \frac {(\,{(t-r)} + {R}\,)^{2}\,{\rm \%2}}{{\alpha}^{2}}}
 + {\displaystyle \frac {{\rm \%1}}{{\alpha}}} - {\displaystyle 
\frac {(\,{(t-r)} - {R}\,)^{2}\,{\rm \%1}}{{\alpha}^{2}}}\, \!  \right) \,
\sqrt {2}\,\sqrt {{\displaystyle \frac {{ \pi}}{{\alpha}}}}\,(\,{(t-r)}^{2
} - {R}^{2} - 3\,{\alpha}\,) \nonumber\\
 & & \mbox{} + 4\,{\alpha}\,{(t-r)}^{2}\,{\rm \%3}\,\sqrt {2}\,\sqrt {
{\displaystyle \frac {{ \pi}}{{\alpha}}}} + {\alpha}\,{R} \nonumber\\
 & &  \left( \! \, - \,{\displaystyle \frac {{\rm \%2}}{{\alpha}}} + 
{\displaystyle \frac {(\,{(t-r)} + {R}\,)^{2}\,{\rm \%2}}{{\alpha}^{2}}}
 - {\displaystyle \frac {{\rm \%1}}{{\alpha}}} + {\displaystyle 
\frac {(\,{(t-r)} - {R}\,)^{2}\,{\rm \%1}}{{\alpha}^{2}}}\, \!  \right) \,
\sqrt {2}\,\sqrt {{\displaystyle \frac {{ \pi}}{{\alpha}}}} \nonumber\\
 & &  \left( \! \,{(t-r)}^{2} + {\displaystyle \frac {1}{3}}\,{R}^{2}
 + {\alpha}\, \!  \right) \mbox{} + 4\,{\alpha}\,{R}\, \left( \! \, - \,
{\displaystyle \frac {(\,{(t-r)} + {R}\,)\,{\rm \%2}}{{\alpha}}} - 
{\displaystyle \frac {(\,{(t-r)} - {R}\,)\,{\rm \%1}}{{\alpha}}}\, \! 
 \right) \,\sqrt {2}\,\sqrt {{\displaystyle \frac {{ \pi}}{{\alpha}}}}
\,{(t-r)} \nonumber\\
 & & \mbox{} + 2\,{\alpha}\,{R}\,(\,{\rm \%2} + {\rm \%1}\,)\,\sqrt {2
}\,\sqrt {{\displaystyle \frac {{ \pi}}{{\alpha}}}} \! \! \left. 
{\vrule height0.92em width0em depth0.92em} \right)  \left/ 
{\vrule height0.43em width0em depth0.43em} \right. \! \!  \left( 
\! \,{ \pi}^{2}\,{\alpha}^{2}\,{r}\, \!  \right)  
\end{eqnarray}
\end{maplelatex}
\begin{maplelatex}
\begin{eqnarray}
\lefteqn{\ddot{h}^{22}_{d}(\vec{x},t) = {\displaystyle \frac {1}{60}}{G}\,{\sqrt{\frac{2\alpha}{\pi}}}
 \left( {\vrule height0.92em width0em depth0.92em} \right. \! \! 
 \left( \! \, - \,{\displaystyle \frac {(\,{(t-r)} + {R}\,)\,{\rm \%2
}\,\sqrt {2}}{\sqrt {{ \pi}}\,{\alpha}^{3/2}}} + {\displaystyle 
\frac {(\,{(t-r)} - {R}\,)\,{\rm \%1}\,\sqrt {2}}{\sqrt {{ \pi}}\,{\alpha}
^{3/2}}}\, \!  \right) }\nonumber \\
 & & (\, - {R}^{4} - 5\,{(t-r)}^{4} - 5\,{\alpha}^{2} - 20\,{\alpha}\,{(t-r)}^{2}
 + 6\,{(t-r)}^{2}\,{R}^{2} - 4\,{\alpha}\,{R}^{2}\,) \nonumber\\
 & & \mbox{} + 2\, \left( \! \,{\displaystyle \frac {{\rm \%2}\,
\sqrt {2}}{\sqrt {{ \pi}}\,\sqrt {{\alpha}}}} - {\displaystyle \frac {
{\rm \%1}\,\sqrt {2}}{\sqrt {{ \pi}}\,\sqrt {{\alpha}}}}\, \! 
 \right) \,(\, - 20\,{(t-r)}^{3} - 40\,{\alpha}\,{(t-r)} + 12\,{(t-r)}\,{R}^{2}\,)
 +  \nonumber\\
 & &  \left( \! \,\Phi \left( \! \,{\displaystyle \frac {1}{
2}}\,{\displaystyle \frac {(\,{(t-r)} + {R}\,)\,\sqrt {2}}{\sqrt {{\alpha}
}}}\, \!  \right)  - \Phi \left( \! \,{\displaystyle \frac {
1}{2}}\,{\displaystyle \frac {(\,{(t-r)} - {R}\,)\,\sqrt {2}}{\sqrt {
{\alpha}}}}\, \!  \right) \, \!  \right) \,(\, - 60\,{(t-r)}^{2} - 40\,{\alpha}
 + 12\,{R}^{2}\,) \nonumber\\
 & & \mbox{} + 18\,\sqrt {2}\,\sqrt {{\displaystyle \frac {{\alpha}}{{
 \pi}}}}\,{(t-r)}\,(\,{\rm \%2} - {\rm \%1}\,) + 2\,\sqrt {2}\,
\sqrt {{\displaystyle \frac {{\alpha}}{{ \pi}}}}\,(\,3\,{(t-r)}^{2} + {R}
^{2} - 3\,{\alpha}\,)\,{\rm \%3} \nonumber\\
 & & \mbox{} + 12\,{(t-r)}^{2}\,\sqrt {2}\,\sqrt {{\displaystyle 
\frac {{\alpha}}{{ \pi}}}}\,{\rm \%3} + {(t-r)}\,\sqrt {2}\,\sqrt {
{\displaystyle \frac {{\alpha}}{{ \pi}}}}\,(\,3\,{(t-r)}^{2} + {R}^{2} - 3
\,{\alpha}\,) \nonumber\\
 & &  \left( \! \, - \,{\displaystyle \frac {{\rm \%2}}{{\alpha}}} + 
{\displaystyle \frac {(\,{(t-r)} + {R}\,)^{2}\,{\rm \%2}}{{\alpha}^{2}}}
 + {\displaystyle \frac {{\rm \%1}}{{\alpha}}} - {\displaystyle 
\frac {(\,{(t-r)} - {R}\,)^{2}\,{\rm \%1}}{{\alpha}^{2}}}\, \!  \right) 
 \nonumber\\
 & & \mbox{} + 18\,{R}\,(\,{\rm \%2} + {\rm \%1}\,)\,\sqrt {2}\,
\sqrt {{\displaystyle \frac {{\alpha}}{{ \pi}}}} \nonumber\\
 & & \mbox{} + 36\,{R}\,{(t-r)}\, \left( \! \, - \,{\displaystyle 
\frac {(\,{(t-r)} + {R}\,)\,{\rm \%2}}{{\alpha}}} - {\displaystyle \frac {
(\,{(t-r)} - {R}\,)\,{\rm \%1}}{{\alpha}}}\, \!  \right) \,\sqrt {2}\,
\sqrt {{\displaystyle \frac {{\alpha}}{{ \pi}}}} + {R}\,\sqrt {2}\,
\sqrt {{\displaystyle \frac {{\alpha}}{{ \pi}}}} \nonumber\\
 & &  \left( \! \,9\,{(t-r)}^{2} + {\displaystyle \frac {11}{3}}\,{R}
^{2} + 5\,{\alpha}\, \!  \right) \, \left( \! \, - \,{\displaystyle 
\frac {{\rm \%2}}{{\alpha}}} + {\displaystyle \frac {(\,{(t-r)} + {R}\,)^{
2}\,{\rm \%2}}{{\alpha}^{2}}} - {\displaystyle \frac {{\rm \%1}}{{\alpha}}}
 + {\displaystyle \frac {(\,{(t-r)} - {R}\,)^{2}\,{\rm \%1}}{{\alpha}^{2}
}}\, \!  \right)  \! \! \left. {\vrule 
height0.92em width0em depth0.92em} \right) \nonumber \\
 & &  \left( \! \,{\rm sin}(\,{ \alpha_0}\,)^{2}\,{\rm sin}(\,{ 
\gamma_0}\,)^{2} - 1\, \!  \right)  \left/ {\vrule 
height0.37em width0em depth0.37em} \right. \! \! (\,{r}\,{ \pi}\,
{\alpha}^{2}\,)\mbox{} + {\displaystyle \frac {1}{36}}{G}\,{\sqrt{\frac{2\alpha}{\pi}}} \left( 
{\vrule height0.92em width0em depth0.92em} \right. \! \!  \nonumber\\
 & & { \pi}\, \left( \! \, - \,{\displaystyle \frac {(\,{(t-r)} + {R}
\,)\,{\rm \%2}\,\sqrt {2}}{\sqrt {{ \pi}}\,{\alpha}^{3/2}}} + 
{\displaystyle \frac {(\,{(t-r)} - {R}\,)\,{\rm \%1}\,\sqrt {2}}{
\sqrt {{ \pi}}\,{\alpha}^{3/2}}}\, \!  \right) \,(\,{R}^{4} - {(t-r)}^{4}
 - {\alpha}^{2} - 4\,{\alpha}\,{(t-r)}^{2}\,) \nonumber\\
 & & \mbox{} + 2\,{ \pi}\, \left( \! \,{\displaystyle \frac {
{\rm \%2}\,\sqrt {2}}{\sqrt {{ \pi}}\,\sqrt {{\alpha}}}} - 
{\displaystyle \frac {{\rm \%1}\,\sqrt {2}}{\sqrt {{ \pi}}\,
\sqrt {{\alpha}}}}\, \!  \right) \,(\, - 4\,{(t-r)}^{3} - 8\,{\alpha}\,{(t-r)}\,)
 \nonumber\\
 & & \mbox{} + { \pi}\, \left( \! \,\Phi \left( \! \,
{\displaystyle \frac {1}{2}}\,{\displaystyle \frac {(\,{(t-r)} + {R}
\,)\,\sqrt {2}}{\sqrt {{\alpha}}}}\, \!  \right)  - \Phi \left( 
\! \,{\displaystyle \frac {1}{2}}\,{\displaystyle \frac {(\,{(t-r)}
 - {R}\,)\,\sqrt {2}}{\sqrt {{\alpha}}}}\, \!  \right) \, \!  \right) 
\,(\, - 12\,{(t-r)}^{2} - 8\,{\alpha}\,) \nonumber\\
 & & \mbox{} + 2\,{\alpha}\,{\rm \%3}\,\sqrt {2}\,\sqrt {
{\displaystyle \frac {{ \pi}}{{\alpha}}}}\,(\,{(t-r)}^{2} - {R}^{2} - 3\,{
a}\,) + 6\,{\alpha}\,(\,{\rm \%2} - {\rm \%1}\,)\,\sqrt {2}\,\sqrt {
{\displaystyle \frac {{ \pi}}{{\alpha}}}}\,{(t-r)} + {\alpha}\,{(t-r)} \nonumber\\
 & &  \left( \! \, - \,{\displaystyle \frac {{\rm \%2}}{{\alpha}}} + 
{\displaystyle \frac {(\,{(t-r)} + {R}\,)^{2}\,{\rm \%2}}{{\alpha}^{2}}}
 + {\displaystyle \frac {{\rm \%1}}{{\alpha}}} - {\displaystyle 
\frac {(\,{(t-r)} - {R}\,)^{2}\,{\rm \%1}}{{\alpha}^{2}}}\, \!  \right) \,
\sqrt {2}\,\sqrt {{\displaystyle \frac {{ \pi}}{{\alpha}}}}\,(\,{(t-r)}^{2
} - {R}^{2} - 3\,{\alpha}\,) \nonumber\\
 & & \mbox{} + 4\,{\alpha}\,{(t-r)}^{2}\,{\rm \%3}\,\sqrt {2}\,\sqrt {
{\displaystyle \frac {{ \pi}}{{\alpha}}}} + {\alpha}\,{R} \nonumber\\
 & &  \left( \! \, - \,{\displaystyle \frac {{\rm \%2}}{{\alpha}}} + 
{\displaystyle \frac {(\,{(t-r)} + {R}\,)^{2}\,{\rm \%2}}{{\alpha}^{2}}}
 - {\displaystyle \frac {{\rm \%1}}{{\alpha}}} + {\displaystyle 
\frac {(\,{(t-r)} - {R}\,)^{2}\,{\rm \%1}}{{\alpha}^{2}}}\, \!  \right) \,
\sqrt {2}\,\sqrt {{\displaystyle \frac {{ \pi}}{{\alpha}}}} \nonumber\\
 & &  \left( \! \,{(t-r)}^{2} + {\displaystyle \frac {1}{3}}\,{R}^{2}
 + {\alpha}\, \!  \right) \mbox{} + 4\,{\alpha}\,{R}\, \left( \! \, - \,
{\displaystyle \frac {(\,{(t-r)} + {R}\,)\,{\rm \%2}}{{\alpha}}} - 
{\displaystyle \frac {(\,{(t-r)} - {R}\,)\,{\rm \%1}}{{\alpha}}}\, \! 
 \right) \,\sqrt {2}\,\sqrt {{\displaystyle \frac {{ \pi}}{{\alpha}}}}
\,{(t-r)}\nonumber \\
 & & \mbox{} + 2\,{\alpha}\,{R}\,(\,{\rm \%2} + {\rm \%1}\,)\,\sqrt {2
}\,\sqrt {{\displaystyle \frac {{ \pi}}{{\alpha}}}} \! \! \left. 
{\vrule height0.92em width0em depth0.92em} \right)  \left/ 
{\vrule height0.43em width0em depth0.43em} \right. \! \!  \left( 
\! \,{ \pi}^{2}\,{\alpha}^{2}\,{r}\, \!  \right)  
\end{eqnarray}
\end{maplelatex}
\begin{maplelatex}
\begin{eqnarray}
\lefteqn{\ddot{h}^{33}_{d}(\vec{x},t) = {\displaystyle \frac {1}{90}}{G}\,{\sqrt{\frac{2\alpha}{\pi}}}
 \left( {\vrule height0.92em width0em depth0.92em} \right. \! \! 
 \left( \! \, - \,{\displaystyle \frac {(\,{(t-r)} + {R}\,)\,{\rm \%2
}\,\sqrt {2}}{\sqrt {{ \pi}}\,{\alpha}^{3/2}}} + {\displaystyle 
\frac {(\,{(t-r)} - {R}\,)\,{\rm \%1}\,\sqrt {2}}{\sqrt {{ \pi}}\,{\alpha}
^{3/2}}}\, \!  \right) } \nonumber\\
 & & (\, - {R}^{4} - 5\,{(t-r)}^{4} - 5\,{\alpha}^{2} - 20\,{\alpha}\,{(t-r)}^{2}
 + 6\,{(t-r)}^{2}\,{R}^{2} - 4\,{\alpha}\,{R}^{2}\,) \nonumber\\
 & & \mbox{} + 2\, \left( \! \,{\displaystyle \frac {{\rm \%2}\,
\sqrt {2}}{\sqrt {{ \pi}}\,\sqrt {{\alpha}}}} - {\displaystyle \frac {
{\rm \%1}\,\sqrt {2}}{\sqrt {{ \pi}}\,\sqrt {{\alpha}}}}\, \! 
 \right) \,(\, - 20\,{(t-r)}^{3} - 40\,{\alpha}\,{(t-r)} + 12\,{(t-r)}\,{R}^{2}\,)
 +  \nonumber\\
 & &  \left( \! \,\Phi \left( \! \,{\displaystyle \frac {1}{
2}}\,{\displaystyle \frac {(\,{(t-r)} + {R}\,)\,\sqrt {2}}{\sqrt {{\alpha}
}}}\, \!  \right)  - \Phi \left( \! \,{\displaystyle \frac {
1}{2}}\,{\displaystyle \frac {(\,{(t-r)} - {R}\,)\,\sqrt {2}}{\sqrt {
{\alpha}}}}\, \!  \right) \, \!  \right) \,(\, - 60\,{(t-r)}^{2} - 40\,{\alpha}
 + 12\,{R}^{2}\,)\nonumber \\
 & & \mbox{} + 18\,\sqrt {2}\,\sqrt {{\displaystyle \frac {{\alpha}}{{
 \pi}}}}\,{(t-r)}\,(\,{\rm \%2} - {\rm \%1}\,) + 2\,\sqrt {2}\,
\sqrt {{\displaystyle \frac {{\alpha}}{{ \pi}}}}\,(\,3\,{(t-r)}^{2} + {R}
^{2} - 3\,{\alpha}\,)\,{\rm \%3} \nonumber\\
 & & \mbox{} + 12\,{(t-r)}^{2}\,\sqrt {2}\,\sqrt {{\displaystyle 
\frac {{\alpha}}{{ \pi}}}}\,{\rm \%3} + {(t-r)}\,\sqrt {2}\,\sqrt {
{\displaystyle \frac {{\alpha}}{{ \pi}}}}\,(\,3\,{(t-r)}^{2} + {R}^{2} - 3
\,{\alpha}\,) \nonumber\\
 & &  \left( \! \, - \,{\displaystyle \frac {{\rm \%2}}{{\alpha}}} + 
{\displaystyle \frac {(\,{(t-r)} + {R}\,)^{2}\,{\rm \%2}}{{\alpha}^{2}}}
 + {\displaystyle \frac {{\rm \%1}}{{\alpha}}} - {\displaystyle 
\frac {(\,{(t-r)} - {R}\,)^{2}\,{\rm \%1}}{{\alpha}^{2}}}\, \!  \right) 
 \nonumber\\
 & & \mbox{} + 18\,{R}\,(\,{\rm \%2} + {\rm \%1}\,)\,\sqrt {2}\,
\sqrt {{\displaystyle \frac {{\alpha}}{{ \pi}}}} \nonumber\\
 & & \mbox{} + 36\,{R}\,{(t-r)}\, \left( \! \, - \,{\displaystyle 
\frac {(\,{(t-r)} + {R}\,)\,{\rm \%2}}{{\alpha}}} - {\displaystyle \frac {
(\,{(t-r)} - {R}\,)\,{\rm \%1}}{{\alpha}}}\, \!  \right) \,\sqrt {2}\,
\sqrt {{\displaystyle \frac {{\alpha}}{{ \pi}}}} + {R}\,\sqrt {2}\,
\sqrt {{\displaystyle \frac {{\alpha}}{{ \pi}}}} \nonumber\\
 & &  \left( \! \,9\,{(t-r)}^{2} + {\displaystyle \frac {11}{3}}\,{R}
^{2} + 5\,{\alpha}\, \!  \right) \, \left( \! \, - \,{\displaystyle 
\frac {{\rm \%2}}{{\alpha}}} + {\displaystyle \frac {(\,{(t-r)} + {R}\,)^{
2}\,{\rm \%2}}{{\alpha}^{2}}} - {\displaystyle \frac {{\rm \%1}}{{\alpha}}}
 + {\displaystyle \frac {(\,{(t-r)} - {R}\,)^{2}\,{\rm \%1}}{{\alpha}^{2}
}}\, \!  \right)  \! \! \left. {\vrule 
height0.92em width0em depth0.92em} \right)  \nonumber\\
 & &  \left( \! \,1 - {\displaystyle \frac {3}{2}}\,{\rm sin}(\,{
 \gamma_0}\,)^{2}\, \!  \right)  \left/ {\vrule 
height0.37em width0em depth0.37em} \right. \! \! (\,{r}\,{ \pi}\,
{\alpha}^{2}\,)\mbox{} + {\displaystyle \frac {1}{36}}{G}\,{\sqrt{\frac{2\alpha}{\pi}}} \left( 
{\vrule height0.92em width0em depth0.92em} \right. \! \!  \nonumber\\
 & & { \pi}\, \left( \! \, - \,{\displaystyle \frac {(\,{(t-r)} + {R}
\,)\,{\rm \%2}\,\sqrt {2}}{\sqrt {{ \pi}}\,{\alpha}^{3/2}}} + 
{\displaystyle \frac {(\,{(t-r)} - {R}\,)\,{\rm \%1}\,\sqrt {2}}{
\sqrt {{ \pi}}\,{\alpha}^{3/2}}}\, \!  \right) \,(\,{R}^{4} - {(t-r)}^{4}
 - {\alpha}^{2} - 4\,{\alpha}\,{(t-r)}^{2}\,) \nonumber\\
 & & \mbox{} + 2\,{ \pi}\, \left( \! \,{\displaystyle \frac {
{\rm \%2}\,\sqrt {2}}{\sqrt {{ \pi}}\,\sqrt {{\alpha}}}} - 
{\displaystyle \frac {{\rm \%1}\,\sqrt {2}}{\sqrt {{ \pi}}\,
\sqrt {{\alpha}}}}\, \!  \right) \,(\, - 4\,{(t-r)}^{3} - 8\,{\alpha}\,{(t-r)}\,)
 \nonumber\\
 & & \mbox{} + { \pi}\, \left( \! \,\Phi \left( \! \,
{\displaystyle \frac {1}{2}}\,{\displaystyle \frac {(\,{(t-r)} + {R}
\,)\,\sqrt {2}}{\sqrt {{\alpha}}}}\, \!  \right)  - \Phi \left( 
\! \,{\displaystyle \frac {1}{2}}\,{\displaystyle \frac {(\,{(t-r)}
 - {R}\,)\,\sqrt {2}}{\sqrt {{\alpha}}}}\, \!  \right) \, \!  \right) 
\,(\, - 12\,{(t-r)}^{2} - 8\,{\alpha}\,) \nonumber\\
 & & \mbox{} + 2\,{\alpha}\,{\rm \%3}\,\sqrt {2}\,\sqrt {
{\displaystyle \frac {{ \pi}}{{\alpha}}}}\,(\,{(t-r)}^{2} - {R}^{2} - 3\,{
a}\,) + 6\,{\alpha}\,(\,{\rm \%2} - {\rm \%1}\,)\,\sqrt {2}\,\sqrt {
{\displaystyle \frac {{ \pi}}{{\alpha}}}}\,{(t-r)} + {\alpha}\,{(t-r)} 
\nonumber\\
 & &  \left( \! \, - \,{\displaystyle \frac {{\rm \%2}}{{\alpha}}} + 
{\displaystyle \frac {(\,{(t-r)} + {R}\,)^{2}\,{\rm \%2}}{{\alpha}^{2}}}
 + {\displaystyle \frac {{\rm \%1}}{{\alpha}}} - {\displaystyle 
\frac {(\,{(t-r)} - {R}\,)^{2}\,{\rm \%1}}{{\alpha}^{2}}}\, \!  \right) \,
\sqrt {2}\,\sqrt {{\displaystyle \frac {{ \pi}}{{\alpha}}}}\,(\,{(t-r)}^{2
} - {R}^{2} - 3\,{\alpha}\,) \nonumber\\
 & & \mbox{} + 4\,{\alpha}\,{(t-r)}^{2}\,{\rm \%3}\,\sqrt {2}\,\sqrt {
{\displaystyle \frac {{ \pi}}{{\alpha}}}} + {\alpha}\,{R} \nonumber\\
 & &  \left( \! \, - \,{\displaystyle \frac {{\rm \%2}}{{\alpha}}} + 
{\displaystyle \frac {(\,{(t-r)} + {R}\,)^{2}\,{\rm \%2}}{{\alpha}^{2}}}
 - {\displaystyle \frac {{\rm \%1}}{{\alpha}}} + {\displaystyle 
\frac {(\,{(t-r)} - {R}\,)^{2}\,{\rm \%1}}{{\alpha}^{2}}}\, \!  \right) \,
\sqrt {2}\,\sqrt {{\displaystyle \frac {{ \pi}}{{\alpha}}}}\nonumber \\
 & &  \left( \! \,{(t-r)}^{2} + {\displaystyle \frac {1}{3}}\,{R}^{2}
 + {\alpha}\, \!  \right) \mbox{} + 4\,{\alpha}\,{R}\, \left( \! \, - \,
{\displaystyle \frac {(\,{(t-r)} + {R}\,)\,{\rm \%2}}{{\alpha}}} - 
{\displaystyle \frac {(\,{(t-r)} - {R}\,)\,{\rm \%1}}{{\alpha}}}\, \! 
 \right) \,\sqrt {2}\,\sqrt {{\displaystyle \frac {{ \pi}}{{\alpha}}}}
\,{(t-r)} \nonumber\\
 & & \mbox{} + 2\,{\alpha}\,{R}\,(\,{\rm \%2} + {\rm \%1}\,)\,\sqrt {2
}\,\sqrt {{\displaystyle \frac {{ \pi}}{{\alpha}}}} \! \! \left. 
{\vrule height0.92em width0em depth0.92em} \right)  \left/ 
{\vrule height0.43em width0em depth0.43em} \right. \! \!  \left( 
\! \,{ \pi}^{2}\,{\alpha}^{2}\,{r}\, \!  \right)  \nonumber\\
 & & {\rm \%1} := {\rm e}^{ \left( \! \, - \,1/2\,\frac {(\,{(t-r)}
 - {R}\,)^{2}}{{\alpha}}\, \!  \right) } \nonumber\\
 & & {\rm \%2} := {\rm e}^{ \left( \! \, - \,1/2\,\frac {(\,{(t-r)}
 + {R}\,)^{2}}{{\alpha}}\, \!  \right) } \nonumber\\
 & & {\rm \%3} :=  - \,{\displaystyle \frac {(\,{(t-r)} + {R}\,)\,
{\rm \%2}}{{\alpha}}} + {\displaystyle \frac {(\,{(t-r)} - {R}\,)\,{\rm 
\%1}}{{\alpha}}}
\end{eqnarray}
\end{maplelatex}
\section{
The gravitational force exerted on non-relativistic 
particles by a quantum string}
In this section we will compute the gravitational force exerted on non-relativistic 
particles as given by eq.(\ref{gravfor2}):
\begin{eqnarray}
{\cal F}(\vec{x},t)\hat{x}^i=\frac{d^2 x^{i}}{dt^2}&=&
2G\left[\frac{8\pi}{3}\hat{x}^i\left(\frac{d{\cal G}_1}{dt}+
r\frac{d^2{\cal G}_1}{dt^2}\right)
\right.\nonumber\\ & & \left. -\frac{dh^{00}_{a}}{dr}\;\hat{x}^i-
\frac{1}{2}\frac{d\ddot{h}^{00}_{d}}{dr}\;\hat{x}^i\right].
\end{eqnarray}
$h^{\mu\nu}(\vec{x},t)$ is given by eq.(\ref{MetComp}) and 
eqs.(\ref{empezar})-(\ref{Compfin}).
\begin{maplelatex}
\begin{eqnarray}
\lefteqn{{\cal F}(\vec{x},t)\hat{x}^i=\frac{d^2 x^{i}}{dt^2} 
= {\displaystyle \frac {8}{3}}{G}\,{ \pi}\,{x}
^{{i}} \left( {\vrule height0.92em width0em depth0.92em}
 \right. \! \! {\displaystyle \frac {1}{8}}\sqrt {2}\,\sqrt {
{\displaystyle \frac {{\alpha}}{{ \pi}}}} \left( {\vrule 
height0.89em width0em depth0.89em} \right. \! \!  \left( \! \,
{\displaystyle \frac {{\rm \%2}\,\sqrt {2}}{\sqrt {{ \pi}}\,
\sqrt {{\alpha}}}} - {\displaystyle \frac {{\rm \%1}\,\sqrt {2}}{
\sqrt {{ \pi}}\,\sqrt {{\alpha}}}}\, \!  \right) \,(\,{t_{ret}}\,{R}^{2} - {
t_{ret}}^{3} - 2\,{\alpha}\,{t_{ret}}\,)} \nonumber\\
 & & \mbox{} + {\rm \%3}\,(\,{R}^{2} - 3\,{t_{ret}}^{2} - 2\,{\alpha}\,) - {
\alpha}\, \left( \! \, - \,{\displaystyle \frac {(\,{t_{ret}} + {R}\,)\,
{\rm \%2}}{{\alpha}}} + {\displaystyle \frac {(\,{t_{ret}} - {R}\,)\,{\rm 
\%1}}{{\alpha}}}\, \!  \right) \,\sqrt {2}\,\sqrt {{\displaystyle 
\frac {{\alpha}}{{ \pi}}}} \nonumber\\
 & & \mbox{} + {R}\,(\,{\rm \%2} + {\rm \%1}\,)\,\sqrt {2}\,
\sqrt {{\displaystyle \frac {{\alpha}}{{ \pi}}}} \nonumber\\
 & & \mbox{} + {R}\,{t_{ret}}\, \left( \! \, - \,{\displaystyle \frac {
(\,{t_{ret}} + {R}\,)\,{\rm \%2}}{{\alpha}}} - {\displaystyle \frac {(\,{t_{ret}}
 - {R}\,)\,{\rm \%1}}{{\alpha}}}\, \!  \right) \,\sqrt {2}\,\sqrt {
{\displaystyle \frac {{\alpha}}{{ \pi}}}} \! \! \left. {\vrule 
height0.89em width0em depth0.89em} \right)  \left/ {\vrule 
height0.43em width0em depth0.43em} \right. \! \!  \left( \! \,{ 
\pi}^{2}\,{\alpha}^{2}\,{r}^{2}\, \!  \right) \mbox{} + 
{\displaystyle \frac {1}{8}}\sqrt {2} \nonumber\\
 & & \sqrt {{\displaystyle \frac {{\alpha}}{{ \pi}}}} \left( {\vrule 
height0.92em width0em depth0.92em} \right. \! \!  \left( \! \, - 
\,{\displaystyle \frac {(\,{t_{ret}} + {R}\,)\,{\rm \%2}\,\sqrt {2}}{
\sqrt {{ \pi}}\,{\alpha}^{3/2}}} + {\displaystyle \frac {(\,{t_{ret}} - {R}
\,)\,{\rm \%1}\,\sqrt {2}}{\sqrt {{ \pi}}\,{\alpha}^{3/2}}}\, \! 
 \right) \,(\,{t_{ret}}\,{R}^{2} - {t_{ret}}^{3} - 2\,{\alpha}\,{t_{ret}}\,) 
 \nonumber\\
 & & \mbox{} + 2\, \left( \! \,{\displaystyle \frac {{\rm \%2}\,
\sqrt {2}}{\sqrt {{ \pi}}\,\sqrt {{\alpha}}}} - {\displaystyle \frac {
{\rm \%1}\,\sqrt {2}}{\sqrt {{ \pi}}\,\sqrt {{\alpha}}}}\, \! 
 \right) \,(\,{R}^{2} - 3\,{t_{ret}}^{2} - 2\,{\alpha}\,) - 6\,{\rm \%3}\,{t
} \nonumber\\
 & & \mbox{} - {\alpha}\, \left( \! \, - \,{\displaystyle \frac {{\rm 
\%2}}{{\alpha}}} + {\displaystyle \frac {(\,{t_{ret}} + {R}\,)^{2}\,{\rm \%2
}}{{\alpha}^{2}}} + {\displaystyle \frac {{\rm \%1}}{{\alpha}}} - 
{\displaystyle \frac {(\,{t_{ret}} - {R}\,)^{2}\,{\rm \%1}}{{\alpha}^{2}}}\,
 \!  \right) \,\sqrt {2}\,\sqrt {{\displaystyle \frac {{\alpha}}{{ \pi
}}}}\nonumber \\
 & & \mbox{} + 2\,{R}\, \left( \! \, - \,{\displaystyle \frac {(
\,{t_{ret}} + {R}\,)\,{\rm \%2}}{{\alpha}}} - {\displaystyle \frac {(\,{t_{ret}}
 - {R}\,)\,{\rm \%1}}{{\alpha}}}\, \!  \right) \,\sqrt {2}\,\sqrt {
{\displaystyle \frac {{\alpha}}{{ \pi}}}}\nonumber \\
 & & \mbox{} + {R}\,{t_{ret}}\, \left( \! \, - \,{\displaystyle \frac {
{\rm \%2}}{{\alpha}}} + {\displaystyle \frac {(\,{t_{ret}} + {R}\,)^{2}\,
{\rm \%2}}{{\alpha}^{2}}} - {\displaystyle \frac {{\rm \%1}}{{\alpha}}} + 
{\displaystyle \frac {(\,{t_{ret}} - {R}\,)^{2}\,{\rm \%1}}{{\alpha}^{2}}}\,
 \!  \right) \,\sqrt {2}\,\sqrt {{\displaystyle \frac {{\alpha}}{{ \pi
}}}} \! \! \left. {\vrule height0.92em width0em depth0.92em}
 \right)  \left/ {\vrule height0.43em width0em depth0.43em}
 \right. \! \!  \left( {\vrule height0.43em width0em depth0.43em}
 \right. \! \! \,{r}\,{ \pi}^{2} \nonumber\\
 & & {\alpha}^{2}\, \! \! \left. {\vrule 
height0.43em width0em depth0.43em} \right)  \! \! \left. {\vrule 
height0.92em width0em depth0.92em} \right) \mbox{} + 
{\displaystyle \frac {1}{24}}{G}\,{x}^{{i}}\,\sqrt {2}\,\sqrt {
{\displaystyle \frac {{\alpha}}{{ \pi}}}} \left( {\vrule 
height0.92em width0em depth0.92em} \right. \! \! \nonumber \\
 & & { \pi}\, \left( \! \, - \,{\displaystyle \frac {(\,{t_{ret}} + {R}
\,)\,{\rm \%2}\,\sqrt {2}}{\sqrt {{ \pi}}\,{\alpha}^{3/2}}} + 
{\displaystyle \frac {(\,{t_{ret}} - {R}\,)\,{\rm \%1}\,\sqrt {2}}{
\sqrt {{ \pi}}\,{\alpha}^{3/2}}}\, \!  \right) \,(\,{R}^{4} - {t_{ret}}^{4}
 - {\alpha}^{2} - 4\,{\alpha}\,{t_{ret}}^{2}\,) \nonumber\\
 & & \mbox{} + 2\,{ \pi}\, \left( \! \,{\displaystyle \frac {
{\rm \%2}\,\sqrt {2}}{\sqrt {{ \pi}}\,\sqrt {{\alpha}}}} - 
{\displaystyle \frac {{\rm \%1}\,\sqrt {2}}{\sqrt {{ \pi}}\,
\sqrt {{\alpha}}}}\, \!  \right) \,(\, - 4\,{t_{ret}}^{3} - 8\,{\alpha}\,{t_{ret}}\,)
 + { \pi}\,{\rm \%3}\,(\, - 12\,{t_{ret}}^{2} - 8\,{\alpha}\,) \nonumber\\
 & & \mbox{} + 2\,{\alpha}\, \left( \! \, - \,{\displaystyle \frac {(
\,{t_{ret}} + {R}\,)\,{\rm \%2}}{{\alpha}}} + {\displaystyle \frac {(\,{t_{ret}}
 - {R}\,)\,{\rm \%1}}{{\alpha}}}\, \!  \right) \,\sqrt {2}\,\sqrt {
{\displaystyle \frac {{ \pi}}{{\alpha}}}}\,(\,{t_{ret}}^{2} - {R}^{2} - 3\,{
\alpha}\,) \nonumber\\
 & & \mbox{} + 6\,{\alpha}\,(\,{\rm \%2} - {\rm \%1}\,)\,\sqrt {2}\,
\sqrt {{\displaystyle \frac {{ \pi}}{{\alpha}}}}\,{t_{ret}} + {\alpha}\,{t_{ret}} 
\nonumber\\
 & &  \left( \! \, - \,{\displaystyle \frac {{\rm \%2}}{{\alpha}}} + 
{\displaystyle \frac {(\,{t_{ret}} + {R}\,)^{2}\,{\rm \%2}}{{\alpha}^{2}}}
 + {\displaystyle \frac {{\rm \%1}}{{\alpha}}} - {\displaystyle 
\frac {(\,{t_{ret}} - {R}\,)^{2}\,{\rm \%1}}{{\alpha}^{2}}}\, \!  \right) \,
\sqrt {2}\,\sqrt {{\displaystyle \frac {{ \pi}}{{\alpha}}}}\,(\,{t_{ret}}^{2
} - {R}^{2} - 3\,{\alpha}\,) \nonumber\\
 & & \mbox{} + 4\,{\alpha}\,{t_{ret}}^{2}\, \left( \! \, - \,{\displaystyle 
\frac {(\,{t_{ret}} + {R}\,)\,{\rm \%2}}{{\alpha}}} + {\displaystyle \frac {
(\,{t_{ret}} - {R}\,)\,{\rm \%1}}{{\alpha}}}\, \!  \right) \,\sqrt {2}\,
\sqrt {{\displaystyle \frac {{ \pi}}{{\alpha}}}} + {\alpha}\,{R} \nonumber\\
 & &  \left( \! \, - \,{\displaystyle \frac {{\rm \%2}}{{\alpha}}} + 
{\displaystyle \frac {(\,{t_{ret}} + {R}\,)^{2}\,{\rm \%2}}{{\alpha}^{2}}}
 - {\displaystyle \frac {{\rm \%1}}{{\alpha}}} + {\displaystyle 
\frac {(\,{t_{ret}} - {R}\,)^{2}\,{\rm \%1}}{{\alpha}^{2}}}\, \!  \right) \,
\sqrt {2}\,\sqrt {{\displaystyle \frac {{ \pi}}{{\alpha}}}}\nonumber \\
 & &  \left( \! \,{t_{ret}}^{2} + {\displaystyle \frac {1}{3}}\,{R}^{2}
 + {\alpha}\, \!  \right) \mbox{} + 4\,{\alpha}\,{R}\, \left( \! \, - \,
{\displaystyle \frac {(\,{t_{ret}} + {R}\,)\,{\rm \%2}}{{\alpha}}} - 
{\displaystyle \frac {(\,{t_{ret}} - {R}\,)\,{\rm \%1}}{{\alpha}}}\, \! 
 \right) \,\sqrt {2}\,\sqrt {{\displaystyle \frac {{ \pi}}{{\alpha}}}}
\,{t_{ret}} \nonumber\\
 & & \mbox{} + 2\,{\alpha}\,{R}\,(\,{\rm \%2} + {\rm \%1}\,)\,\sqrt {2
}\,\sqrt {{\displaystyle \frac {{ \pi}}{{\alpha}}}} \! \! \left. 
{\vrule height0.92em width0em depth0.92em} \right)  \left/ 
{\vrule height0.43em width0em depth0.43em} \right. \! \!  \left( 
\! \,{ \pi}^{2}\,{\alpha}^{2}\,{r}^{2}\, \!  \right)  \nonumber\\
 & & {\rm \%1} = {\rm e}^{ \left( \! \, - \,1/2\,\frac {(\,{t_{ret}}
 - {R}\,)^{2}}{{\alpha}}\, \!  \right) }\nonumber \\
 & & {\rm \%2} = {\rm e}^{ \left( \! \, - \,1/2\,\frac {(\,{t_{ret}}
 + {R}\,)^{2}}{{\alpha}}\, \!  \right) } \nonumber\\
 & & {\rm \%3} = \Phi \left( \! \,{\displaystyle \frac {1}{
2}}\,{\displaystyle \frac {(\,{t_{ret}} + {R}\,)\,\sqrt {2}}{\sqrt {{\alpha}
}}}\, \!  \right)  - \Phi \left( \! \,{\displaystyle \frac {
1}{2}}\,{\displaystyle \frac {(\,{t_{ret}} - {R}\,)\,\sqrt {2}}{\sqrt {
{\alpha}}}}\, \!  \right) \nonumber\\ & &
\end{eqnarray}
\end{maplelatex}

\chapter{TABLE OF INTEGRALS AND FUNCTIONS MOST USED IN THIS WORK}
\label{APEXPRIM}
This appendix is intended to provide the reader with some of the useful
technical content used in this work in order to avoid otherwise necessary 
recourse to the  
appropriate references \cite{Tablas,Pock} or to consulting special 
packages such as Maple.

\section{Integrals}
\begin{eqnarray}
\int^{\pi}_0 \sin\gamma\; e^{-iEr\cos\gamma}\;d\gamma &=& 
\frac{2}{Er}\sin Er
\label{T1}
\end{eqnarray}

\begin{eqnarray}
\int^{\pi}_0 \sin\gamma\; e^{iEr\cos\gamma}\;d\gamma &=& 
\frac{2}{Er}\sin Er
\label{T2}
\end{eqnarray}

\begin{eqnarray}
\int^{\pi}_0 \sin^3\gamma\; e^{-iEr\cos\gamma}\;d\gamma &=& 
-\frac{4}{E^2r^2}\left(\cos Er-\frac{\sin Er}{Er}\right)
\label{T3}
\end{eqnarray}

\begin{eqnarray}
\int^{\pi}_0 \cos\gamma\sin\gamma\; e^{-iEr\cos\gamma}\;d\gamma &=& 
\frac{2i}{E^2r^2}\left(Er\cos Er-\sin Er\right)
\label{T7}
\end{eqnarray}

\begin{eqnarray}
\int^{\pi}_0 \cos^2\gamma\sin\gamma\; 
e^{-iEr\cos\gamma}\;d\gamma &=& 2\frac{\sin Er}{Er}
+\frac{4}{E^2r^2}\left(\cos Er-\frac{\sin Er}{Er}\right)
\label{T9}
\end{eqnarray}

\begin{eqnarray}
\int^{2\pi}_0  e^{-iE\rho\cos\alpha\sin\gamma}\;d\alpha &=& 
2\pi J_0( E\rho\sin\gamma)
\label{T11}
\end{eqnarray}

\begin{eqnarray}
\int^{2\pi}_0  e^{iE\rho\cos\alpha\sin\gamma}\;d\alpha &=& 
2\pi J_0( E\rho\sin\gamma)
\label{T12}
\end{eqnarray}

\begin{eqnarray}
\int^{2\pi}_0 \cos\alpha\; e^{-iE\rho\cos\alpha\sin\gamma}\;d\alpha &=& 
-2\pi  
i J_1( E\rho\sin\gamma)
\label{T13}
\end{eqnarray}

\begin{eqnarray}
\int^{2\pi}_0 \cos\alpha\; e^{iE\rho\cos\alpha\sin\gamma}\;d\alpha &=& 2\pi  
i J_1( E\rho\sin\gamma)
\label{T14}
\end{eqnarray}

\begin{eqnarray}
\int^{2\pi}_0 \cos^2\alpha\; e^{-iE\rho\cos\alpha\sin\gamma}\;d\alpha &=& 
\pi\left( J_0( E\rho\sin\gamma)-J_2( E\rho\sin\gamma)\right)
\label{T15}
\end{eqnarray}

\begin{eqnarray}
\int^{2\pi}_0 \cos^2\alpha\; e^{iE\rho\cos\alpha\sin\gamma}\;d\alpha &=& 
\pi\left( J_0( E\rho\sin\gamma)-J_2( E\rho\sin\gamma)\right)
\label{T16}
\end{eqnarray}

\begin{eqnarray}
\int^{2\pi}_0 \sin\alpha\; e^{-iE\rho\cos\alpha\sin\gamma}\;d\alpha 
=\int^{2\pi}_0 \sin\alpha\; e^{iE\rho\cos\alpha\sin\gamma}\;d\alpha &=& 0
\label{T17}
\end{eqnarray}

\begin{eqnarray}
\int^{2\pi}_0 \sin\alpha\cos\alpha\; 
e^{-iE\rho\cos\alpha\sin\gamma}\;d\alpha 
=\int^{2\pi}_0 \sin\alpha\cos\alpha\; 
e^{iE\rho\cos\alpha\sin\gamma}\;d\alpha &=& 0
\label{T18}
\end{eqnarray}

\begin{eqnarray}
\int^{2\pi}_0 \sin^2\alpha\; e^{-iE\rho\cos\alpha\sin\gamma}\;d\alpha &=& 
\pi\left( J_0( E\rho\sin\gamma)+J_2( E\rho\sin\gamma)\right)
\label{T19}
\end{eqnarray}

\begin{eqnarray}
\int^{2\pi}_0 \sin^2\alpha\; e^{iE\rho\cos\alpha\sin\gamma}\;d\alpha &=& 
\pi\left( J_0( E\rho\sin\gamma)+J_2( E\rho\sin\gamma)\right)
\label{T20}
\end{eqnarray}

\begin{eqnarray}
\int e^{-(ax^2+2bx)}\;
dx &=&\frac{e^{\frac{b^2}{a}}}{2}
\sqrt{\frac{\pi}{a}}\Phi(\sqrt{a}x+\frac{b}{\sqrt{a}})
\label{I1}
\end{eqnarray}

\begin{eqnarray}
\int \Phi(x)\; dx=x\Phi(x)+\frac{e^{-x^2}}{\sqrt{\pi}}
\label{I2}
\end{eqnarray}

\begin{eqnarray}
\int \Phi(\sqrt{a}x+\frac{b}{\sqrt{a}})\; dx =
(x+\frac{b}{a})\Phi(\sqrt{a}x+\frac{b}{\sqrt{a}})
+\frac{e^{-(ax^2+2bx+\frac{b^2}{a})}}{\sqrt{\pi a}}
\label{I3}
\end{eqnarray}

\begin{eqnarray}
\int xe^{-(ax^2+2bx)}\; dx = -\frac{e^{-(ax^2+2bx)}}{2a}-
\frac{e^{\frac{b^2}{a}}}{2}
\sqrt{\frac{\pi}{a}}\frac{b}{a}\Phi(\sqrt{a}x+\frac{b}{\sqrt{a}})
\label{I4}
\end{eqnarray}

\begin{eqnarray}
\int x^2 e^{-(ax^2+2bx)}\;
 dx &=& -\frac{x}{2a}e^{-(ax^2+2bx)}+
\frac{e^{\frac{b^2}{a}}}{4}
\sqrt{\frac{\pi}{a}}\frac{1}{a}\Phi(\sqrt{a}x+\frac{b}{\sqrt{a}})
+\nonumber\\ 
 & & \sqrt{\frac{\pi}{a}}\frac{b^2}{2a^2}e^{\frac{b^2}{a}}
\Phi(\sqrt{a}x+\frac{b}{\sqrt{a}})+\frac{b}{2a^2}e^{-(ax^2+2bx)}
\label{I5}
\end{eqnarray}

\begin{eqnarray}
\int x^3 e^{-(ax^2+2bx)}\;
dx &=&-\frac{x^2}{2a}e^{-(ax^2+2bx)}-
b\frac{e^{\frac{b^2}{a}}}{2a^2}
\sqrt{\frac{\pi}{a}}\Phi(\sqrt{a}x+\frac{b}{\sqrt{a}})-\nonumber\\
& & \sqrt{\frac{\pi}{a}}\frac{b^3}{2a^3}e^{\frac{b^2}{a}}
\Phi(\sqrt{a}x+\frac{b}{\sqrt{a}})-
\sqrt{\frac{\pi}{a}}\frac{b}{4a^2}e^{\frac{b^2}{a}}
\Phi(\sqrt{a}x+\frac{b}{\sqrt{a}})-\nonumber\\
& & \frac{b^2}{2a^3}e^{-(ax^2+2bx)}+
\frac{1}{2a^2}e^{-(ax^2+2bx)}\left(bx-1\right)
\label{I6}
\end{eqnarray}

\begin{eqnarray}
\int x^4 e^{-(ax^2+2bx)}\; dx &=&
-\frac{x^3}{2a}e^{-(ax^2+2bx)}+\frac{bx^2}{2a^2}e^{-(ax^2+2bx)}-
\frac{3x}{4a^2}e^{-(ax^2+2bx)}-\nonumber\\
 & &\frac{b^2 x}{2a^3}e^{-(ax^2+2bx)}+\frac{5b}{4a^3}e^{-(ax^2+2bx)}+
\frac{b^3}{2a^4}e^{-(ax^2+2bx)}+\nonumber\\
& & \sqrt{\frac{\pi}{a}}\frac{3}{8a^2}e^{\frac{b^2}{a}}
\Phi(\sqrt{a}x+\frac{b}{\sqrt{a}})+
\sqrt{\frac{\pi}{a}}\frac{b^4}{2a^4}e^{\frac{b^2}{a}}
\Phi(\sqrt{a}x+\frac{b}{\sqrt{a}})+\nonumber\\
& & \sqrt{\frac{\pi}{a}}\frac{6b^2}{4a^3}e^{\frac{b^2}{a}}
\Phi(\sqrt{a}x+\frac{b}{\sqrt{a}})
\label{I7}
\end{eqnarray}

\begin{eqnarray}
\int x^5 e^{-\frac{x^2}{2\alpha}}\; dx &=&
-e^{-\frac{x^2}{2\alpha}}\alpha\left(x^4+4\alpha x^2+8\alpha^2\right)
\label{I7a}
\end{eqnarray}

\begin{eqnarray}
\int x^6 e^{-\frac{x^2}{2\alpha}}\; dx &=&
\frac{15\alpha^3}{2}\sqrt{2\alpha\pi}\Phi(\frac{x}{\sqrt{2\alpha}})
-e^{-\frac{x^2}{2\alpha}}x\alpha\left(x^4+5\alpha x^2+15\alpha^2 x\right)
\label{I7b}
\end{eqnarray}

\begin{eqnarray}
\int x\Phi(\frac{x}{\sqrt{2\alpha}})\; dx &=&
\frac{x^2}{2}\Phi(\frac{x}{\sqrt{2\alpha}})-\frac{\alpha}{2}
\Phi(\frac{x}{\sqrt{2\alpha}})-\frac{\alpha x}{2}\sqrt{\frac{2}{\alpha\pi}}
e^{-\frac{x^2}{2\alpha}}
\label{I8}
\end{eqnarray}

\begin{eqnarray}
\int x^2\Phi(\frac{x}{\sqrt{2\alpha}})\; dx &=&
\frac{x^3}{3}\Phi(\frac{x}{\sqrt{2\alpha}})+\sqrt{\frac{2\alpha}{\pi}}
e^{-\frac{x^2}{2\alpha}}\left(\frac{x^2}{3}+\frac{2\alpha}{3}\right)
\label{I8a}
\end{eqnarray}

\begin{eqnarray}
\int x^3\Phi(\frac{x}{\sqrt{2\alpha}})\; dx &=&
\frac{1}{4}\Phi(\frac{x}{\sqrt{2\alpha}})\left[x^4-3\alpha^2\right]
+\frac{u}{4}\sqrt{\frac{2\alpha}{\pi}}e^{-\frac{x^2}{2\alpha}}
\left(u^2+3\alpha\right)
\label{I8b}
\end{eqnarray}

\begin{eqnarray}
\int^{\infty}_{0} e^{-\frac{\alpha x^2}{2}}\; dx &=&
\frac{1}{2}\sqrt{\frac{2\pi}{\alpha}}
\label{I9}
\end{eqnarray}

\begin{eqnarray}
\int^{\infty}_{0} x^2 e^{-\frac{\alpha x^2}{2}}\; dx &=&
\frac{3}{2\alpha}\sqrt{\frac{2\pi}{\alpha}}
\label{I10}
\end{eqnarray}

\begin{eqnarray}
\int^{\infty}_{0} x^4 e^{-\frac{\alpha x^2}{2}}\; dx &=&
\frac{3}{2\alpha^2}\sqrt{\frac{2\pi}{\alpha}}
\label{I11}
\end{eqnarray}

\begin{eqnarray}
\int^{\infty}_{-\infty}e^{-\frac{\alpha x^2}{2}} x\sin xy\; dx &=&
\sqrt{\frac{\pi}{2}}\frac{e^{-\frac{y^2}{4\alpha}}}{\alpha}
\left[D_1\left(-\frac{y}{\sqrt{\alpha}}\right)-
D_1\left(\frac{y}{\sqrt{\alpha}}\right)\right]
\label{I12}
\end{eqnarray}

\begin{eqnarray}
\int^{\infty}_{-\infty}e^{-\frac{\alpha x^2}{2}} x^3\sin xy\; dx &=&
\sqrt{\frac{\pi}{2}}\frac{e^{-\frac{y^2}{4\alpha}}}{\alpha^2}
\left[D_3\left(-\frac{y}{\sqrt{\alpha}}\right)-
D_3\left(\frac{y}{\sqrt{\alpha}}\right)\right]
\label{I13}
\end{eqnarray}

\begin{eqnarray}
\int^{\infty}_{-\infty}\frac{e^{-\frac{\alpha x^2}{2}}}{x}\sin xy\; dx &=&
y\sqrt{\frac{2\pi}{\alpha}}e^{-\frac{y^2}{4\alpha}}\;
_{1}F_{1}\left(1;3/2;\frac{y^2}{2\alpha}\right)
\label{I14}
\end{eqnarray}

\begin{eqnarray}
\int^{\infty}_{0}e^{-\alpha x^2}\;\sin xy\; dx &=& 
\frac{y}{2\alpha}\;e^{-\frac{y^2}{4\alpha}}
\; _1F_1(1/2;3/2;\frac{y^2}{4\alpha})
\label{V0}
\end{eqnarray}

\begin{eqnarray}
\int^{\infty}_{0}x\;e^{-\alpha x^2}\;\sin xy\; dx &=& \frac{y}{4\alpha}
\sqrt{\frac{\pi}{\alpha}}\; e^{-\frac{y^2}{4\alpha}}
\label{V1}
\end{eqnarray}

\begin{eqnarray}
\int^{\infty}_{0}e^{-\alpha x^2}\;\frac{\sin xy}{x}\; dx &=& 
\frac{\pi}{2}\Phi(\frac{y}{2\sqrt{\alpha}})
\label{V2}
\end{eqnarray}

\begin{eqnarray}
\int^{\infty}_{0} x^2\;e^{-\alpha x^2}\;
\sin xy\; dx &=& \frac{y}{2\alpha^2}\;e^{-\frac{y^2}{4\alpha}}
\; _1F_1(-1/2;3/2;\frac{y^2}{4\alpha})
\label{V3}
\end{eqnarray}

\begin{eqnarray}
\int^{\infty}_{0}e^{-\alpha x^2}\;\cos xy\; dx &=& \frac{1}{2}
\sqrt{\frac{\pi}{\alpha}}\;e^{-\frac{y^2}{4\alpha}}
\; _1F_1(-1/2;3/2;\frac{y^2}{4\alpha})
\label{V4}
\end{eqnarray}

\begin{eqnarray}
\int^{\infty}_{0}x\;e^{-\alpha x^2}\;\cos xy\; dx &=& \frac{1}{2\alpha}
\; _1F_1(1;1/2;-\frac{y^2}{4\alpha})
\label{V5}
\end{eqnarray}

\begin{eqnarray}
\int^{\infty}_{0}e^{-\alpha x^2}\;\frac{\cos xy}{x}\; 
dx &=& \int du _1F_1(1;3/2;-u)
\label{V6}
\end{eqnarray}

\begin{eqnarray}
\int^{\infty}_{0} x^2\;e^{-\alpha x^2}\;\cos xy\; dx &=& \frac{1}{8\alpha^2}
\sqrt{\frac{\pi}{\alpha}}\;e^{-\frac{y^2}{4\alpha}}\;(2\alpha-y^2)
\label{V7}
\end{eqnarray}

\begin{eqnarray}
\int^{\infty}_{0}\sin(bt)\;t^{z-1}\;dt &=& \frac{\Gamma(z)}{b^z}
\sin(\frac{\pi}{2}z)
\label{V8}
\end{eqnarray}

\begin{eqnarray}
\int^{\infty}_{0}\frac{\sin(bt)}{t}\;dt &=& \frac{\pi}{2}\;sign(b)
\label{V9}
\end{eqnarray}

\section{Special Functions}

\subsection{The Probability function}
The probability function is defined as
\begin{equation}
\Phi(z)\equiv\frac{2}{\sqrt{\pi}}\int^{z}_{0}e^{-t^2}dt
\end{equation}
and it satisfies
\begin{equation}
\Phi(z)\stackrel{z\rightarrow +\infty}{=}1-\frac{e^{-z^2}}{\sqrt{\pi}z}
\end{equation}
\begin{equation}
\Phi(z)\stackrel{z\rightarrow 0}{=}\frac{2 z}{\sqrt{\pi}}
\end{equation}
\begin{equation}
\Phi(-z)=-\Phi(z).
\end{equation}
Its derivative is given by
\begin{equation}
\frac{d\Phi(z)}{dz}=\frac{2}{\sqrt{\pi}}\;e^{-z^2}
\end{equation}
\subsection{The Whittaker functions}
The Whittaker functions are defined as:
\begin{eqnarray}
D_p(z)&=&2^{\frac{p}{2}}e^{-\frac{z^2}{4}}\;
\left\{\frac{\sqrt{\pi}}{\Gamma\left(\frac{1-p}{2}\right)}\;
_1F_1\left(-\frac{p}{2},\frac{1}{2};\frac{z^2}{2}\right)\right.\nonumber\\ & & 
\left.-\frac{\sqrt{2\pi}\;z}{\Gamma\left(-\frac{p}{2}\right)}\;
_1F_1\left(\frac{1-p}{2},\frac{3}{2};\frac{z^2}{2}\right)\right\}.
\end{eqnarray}
or
\begin{eqnarray}
D_{n}(z)&=& 2^{-\frac{n}{2}}e^{-\frac{z^2}{4}}H_{n}(\frac{z}{\sqrt{2}})
\end{eqnarray}
if $n\geq 0$ where $H_{n}(z)$ are the Hermite polynomials of order $n$. The 
assymptotic expansions of the Whittaker functions when $¦z¦\gg 1$, $¦z¦\gg ¦p¦$ 
are:
\begin{eqnarray}
D_p(z)&\sim& e^{-\frac{z^2}{4}}z^p\left(1-\frac{p(p-1)}{2z^2}+
\frac{p(p-1)(p-2)(p-3)}{2\cdot 4z^4}-\dots\right)\nonumber\\ & &
-\frac{\sqrt{2\pi}}{\Gamma(-p)}e^{p\pi i}z^{-p-1}\left(
1+\frac{(p+1)(p+2)}{2z^2}+\frac{(p+1)(p+2)(p+3)(p+4)}{2\cdot 4z^4}+\dots\right)
\nonumber\\ & &
\end{eqnarray}
if $\pi/4<\arg z<5\pi/4$ and
\begin{eqnarray}
D_p(z)&\sim& e^{-\frac{z^2}{4}}z^p\left(1-\frac{p(p-1)}{2z^2}+
\frac{p(p-1)(p-2)(p-3)}{2\cdot 4z^4}-\dots\right)\nonumber\\ & &
-\frac{\sqrt{2\pi}}{\Gamma(-p)}e^{-p\pi i}z^{-p-1}\left(
1+\frac{(p+1)(p+2)}{2z^2}+\frac{(p+1)(p+2)(p+3)(p+4)}{2\cdot 4z^4}+\dots\right)
\nonumber\\ & &
\end{eqnarray}
if $-\pi/4>\arg z>-5\pi/4$.

\subsection{The Confluent Hypergeometric function}
The confluent hypergeometric function satisfies the {\em Kummer's differential 
equation}:
$$z\frac{d^2w}{dz^2}+(b-z)\frac{dw}{dz}-\alpha w=0$$ and it can be written as
\begin{equation}
_{1}F_{1}\left(a;b;z\right)=1+\frac{az}{b}+\frac{(a)_{2}z^2}{(a)_{2}2!}+
\dots+\frac{(a)_{n}z^n}{(a)_{n}n!}+\dots
\end{equation}
where
$$(a)_n=a(a+1)(a+2)\dots (a+n-1),$$
$$(a)_0=1.$$
The asymptotic behaviour of the confluent hypergeometric function is given by 
the following relations:

\begin{equation}
_{1}F_{1}\left(a;b;z\right)=\frac{\Gamma(b)}{\Gamma(a)}e^{z}z^{a-b}\quad 
{\cal R}z>0
\end{equation}

\begin{equation}
_{1}F_{1}\left(a;b;z\right)=\frac{\Gamma(b)}{\Gamma(b-a)}(-z)^{-a}\quad 
{\cal R}z<0.
\end{equation}
And we also have the following relation:
\begin{eqnarray}
_{1}F_{1}\left(1;3/2; z^2\right)&=& e^{z^2}\;_{1}F_{1}\left(1/2;3/2;-z^2
\right).
\end{eqnarray}
The confluent hypergeometric function is also related to 
the probability function in the following way:
\begin{eqnarray}
_{1}F_{1}\left(1/2;3/2;-z^2\right)&=& \frac{\sqrt{\pi}}{2 z}\Phi(z).
\end{eqnarray}



\begin{thebibliography}{99}\addcontentsline{toc}{chapter}{BIBLIOGRAPHY}
\bibitem{SG1}Freedman, von Nieuwenhuizen and Ferrara, {\em Phys. Rev}.
{\bf D13}, 3214 (1976).
\bibitem{SG1a}S. Ferrara, Z. D. Freedman, P. van Nieuwenhuizen, 
P. Breitenlohner, F.
Gliozzi and J. Scherk, {\em Phys. Rev}.
{\bf D15}, 1013 (1977).
\bibitem{SG1aa}S. Ferrara, J. Wess and B. Zumino, {\em Phys. Lett}. {\bf 51 B}, 
239 (1974).
\bibitem{SG1b}S. Ferrara, F. Gliozzi, J. Scherk and 
P. van Nieuwenhuizen, {\em Nucl.
Phys}, {\bf B117}, 333 (1976).
\bibitem{SG2a}S. Desser and B. Zumino, {\em Phys. Rev. Lett}. 
{\bf 38}, 1433 (1977).
\bibitem{SG2}S. Desser and B. Zumino, {\em Phys. Lett}. {\bf 62 B}, 335 (1976).
\bibitem{UFT1}S. Weinberg, {\em Phys. Rev. Lett}. {\bf 19}, 1264 (1967).
\bibitem{UFT2}A. Salam in `Elementary Particle Theory', Proceedings of the 8th Nobel
Symposium, Stockholm 1968.
\bibitem{UFT3}H. Georgi and S. L. Glashow, {\em Phys. Rev. Lett}. {\bf 32}, 438
(1974).
\bibitem{HJV} H.\@ J.\@ de Vega, {\em Int.\ J.\@ Mod.\ Phys}. {\bf A7},
3043 (1992).
\bibitem{HJVQG}H. J. de Vega in Proceedings of the Erice School:
``String Quantum Gravity and Physics at the Planck Energy Scale'',
21-28 June 1992, Edited by N. S\'anchez, World Scientific, 1993. 
Edited by N. S\'anchez and
A. Zichichi, Kluwer Acad. Publ.
\bibitem{Alex2} A. Vilenkin in Proceedings of the Erice School:
``String Gravity and Physics at the Planck Energy Scale'',
8-19 September 1995, Edited by N. S\'anchez, World Scientific, 1996. 
Edited by N. S\'anchez and
A. Zichichi, Kluwer Acad. Publ.
\bibitem{Mark}M. Davis, {\em Preprint astro-ph 9610149v2}, (1996).
\bibitem{piet}L. Pietronero {\em et al.} 
{\em Preprint astro-ph 9611197}, (1996).
\bibitem{GA0}A. Dressler,
{\em Astrophys. J}. {\bf 329}, 519 (1988).
\bibitem{GA3}A. Dressler and S. M. Faber,
{\em Astrophys. J}. {\bf 354}, L45 (1990).
\bibitem{GA4}A. Dressler and S. M. Faber,
{\em Astrophys. J}. {\bf 354}, 13 (1990).
\bibitem{GA5}J. A. Willick,
{\em Astrophys. J}. {\bf 351}, L5 (1990).
\bibitem{GA7}B. I. Hnatyk, V. N. Lukash and B. S. Novosyadlyj,
{\em Astr. Astrophys}. {\bf 300}, 1 (1995).
\bibitem{GA6}J. M. Bardeen, J. R. Bond and G. Efstathion, 
{\em Astrophys. J}. {\bf 321}, 28 (1987).
\bibitem{GA8}D. Burstein, A. Dressler and S. M. Faber, 
{\em Astrophys. J}. {\bf 354}, 18 (1990).
\bibitem{Barys}Yu. V. Baryshev {\em et al.} {\em Preprint astro-ph 9503074}, 
(1995).
\bibitem{Martinez}E. Martinez-Gonzalez and J.L. Sanz, {Astr. Astrophys}. 
{\bf 300}, 346 MS, (1995).
\bibitem{Ellis}R. Marteens, G. F. R. Ellis and W. R. Stoeger, {\em Preprint 
astro-ph 9510126}, (1995).
\bibitem{Inf1}R. Durrer and J. Laukenmann, 
{\em Class. Quant. Grav}. {\bf 13}, 1069 (1996).
\bibitem{Inf3}W. H. Kinney and K. T. Mahanthappa,
{\em Phys.\ Rev}. {\bf D 53}, 5455 (1996).
\bibitem{Inf4}R. F. Langbein, K. Langfeld, H. Reinhardt and L. von Smekal,
{\em Mod. Phys. Lett}. {\bf A11}, 631 (1996).
\bibitem{LKolb}E. W. Kolb and M. S. Turner, {\em The Early Universe}, Addison 
Wesley (1994).
\bibitem{cobe}S. F. Smoot {\em et al.}, 
{\em Astrophys. J}. {\bf 396}, L1 (1992).
\bibitem{cobe2}E. L. Wright {\em et al.}, 
{\em Astrophys. J}. {\bf 396}, L13 (1992).
\bibitem{HH1}J. B. Hartle and S. W. Hawking, 
{\em Phys.\ Rev}. {\bf D 28}, 2960 (1983).
\bibitem{AlexA}A. Vilenkin, 
{\em Phys.\ Rev}. {\bf D 30}, 509 (1984).
\bibitem{Linde}A. D. Linde, {\em Lett. Nuovo Cim}. {\bf 39}, 401 (1984).
\bibitem{Moss1}H. Luckock, I. Moss and D. Toms, 
{\em Nucl.\ Phys}. {\bf B 297}, 748 (1988).
\bibitem{ext1}S. W. Hawking and G. F. R. Ellis, {\em The Large Scale Structure 
of Space-time}, Cambridge University Press (1984).
\bibitem{ext2}R. H. Dicke and P. J. E. Peebles in General Relativity: 
An Einstein Centenary Survey. Edited by S. W. Hawking and W. Israel, 
Cambridge University Press (1979).
\bibitem{Moss4}S. Lansdale and I. Moss, 
{\em Nucl.\ Phys}. {\bf B 298}, 693 (1988).
\bibitem{Car1}S. Carlip, {\em Phys.\ Rev}. {\bf D 46}, 4387 (1992).
\bibitem{Car2}S. Carlip, {\em Class. Quant. Grav}. {\bf 10}, 1057 (1993).
\bibitem{Car3}A. O. Barbinsky, A. Yu. Kamenshchik, 
{\em Class. Quant. Grav}. {\bf 7}, L181 (1990).
\bibitem{GrisA}L. P. Grishchuk and L. V. Rozhansky, 
{\em Phys.\ Lett}. {\bf B 234}, 9 (1990).
\bibitem{AOB}A. O. Barbinsky, A. Yu. Kamenshchik and I. P. Karmazin, 
{\em Anns. Phys}. {\bf 219}, 201 (1992).
\bibitem{Nariai} H.\@ Nariai and K.\@ Tomita, {\em Prog.\ Theor.\
Phys}.\ {\bf 46}, 776 (1971).

\bibitem{Park} L.\@ Parker and S.\@ A.\@ Fulling, {\em Phys. Rev.} {\bf
D7}, 2357 (1973).

\bibitem{Atk} D.\@ Atkatz, and H.\@ Pagels, {\em Phys.\ Rev}.\ {\bf D
25}, 2065 (1982).
\bibitem{Tryon} E.\@ P.\@ Tryon, {\em Nature}, {\bf 246}, 396 (1973).
\bibitem{Nar} R.\@ Brout, {\em et al.}, {\em Nucl.\ Phys}.\
{\bf B170}, 228 (1980).
\bibitem{Gun} R.\@ Brout, F.\@ Englert and E.\@ Gunzig, {\em Ann.\
Phys}.\ {\bf 115}, 78 (1978).
\bibitem{Bai1}P. Candelas, G. T. Horowitz, A. Strominger and E. Witten, 
{\em Nucl.\ Phys}. {\bf B 258}, 46 (1985).
\bibitem{Bai2}D. Bailin, A. Love and S. Thomas, 
{\em Phys.\ Lett}. {\bf B 176}, 81 (1986).
\bibitem{Bai3}D. Bailin, A. Love and S. Thomas, 
{\em Phys.\ Lett}. {\bf B 178}, 15 (1986).
\bibitem{Bai4}D. Bailin, A. Love and S. Thomas, 
{\em Nucl.\ Phys}. {\bf B 273}, 537 (1986).
\bibitem{Bai5}D. Bailin, and A. Love, 
{\em Phys.\ Lett}. {\bf B 157}, 375 (1985).
\bibitem{Bai6}A. S. Majuundar, T. R. Seshadri and S. K. Sethi, 
{\em Phys.\ Lett}. {\bf B 312}, 67 (1993).
\bibitem{Bai7}D. Bailin, and A. Love, 
{\em Phys.\ Lett}. {\bf B 207}, 151 (1988).
\bibitem{Bai8}D.Bailin, D. C. Dunbar and A. Love, 
{\em Int.\ J.\@ Mod.\ Phys}. {\bf A5}, 939 (1990).
\bibitem{Bai9}A. S. Majuundar, A. Mukherjee and R. P. Saxena, 
{\em Mod. Phys. Lett}. {\bf A7}, 3647 (1992).
\bibitem{NuevoBailin}D. Bailin, G. V. Kraniotis and A. Love, {\em Preprint 
hep-th 9705244v2}, (1997).
\bibitem{SSSS}J. H. Schwarz in ``Superstrings and Supergravity'' Proceedings 
of the Twenty-Eighth Scotish Universities Summer School in Physics 1985. 
Edited by A. T. Davies and D. S. Sutherland. 
\bibitem{str1} V.\@ Alessandrini, D.\@ Amati, M.\@ LeBellac, and D.\@
I.\@ Olive, {\em Phys.\ Rep}. {\bf 1C}, 170 (1971).
\bibitem{str2} S.\@ Mandelstam, {\em Phys.\ Rep}. {\bf 13C}, 259 (1974).
\bibitem{str3} G.\@ Veneziano, {\em Phys.\ Rep}. {\bf 9C}, 199 (1974).
\bibitem{str4} J.\@ H.\@ Schwarz, {\em Phys.\ Rep}. {\bf 8C}, 269 (1973).
\bibitem{str5} A.\@ M.\@ Polyakov, 
{\em Phys.\ Lett}. {\bf B 103}, 207, 211 (1981).
\bibitem{str6} P.\@ Ramond, {\em Phys.\ Rev}. {\bf D 3}, 2415 (1971).
\bibitem{str7} A.\@ Neveu and J.\@ H.\@ Schwarz, 
{\em Nucl.\ Phys.} {\bf B 31}, 816 (1971).
\bibitem{str8} A.\@ Neveu and J.\@ H.\@ Schwarz, 
{\em Phys.\ Rev}. {\bf D 4}, 1109 (1971).
\bibitem{str9} K.\@ Kikkawa, B.\@ Sakita, and M.\@ A.\@ Virasoro, 
{\em Phys.\ Rev}. {\bf 184}, 1701 (1969).
\bibitem{str10} M.\@ Kaku and L.\@ P.\@ Yu, 
{\em Phys.\ Lett}. {\bf B 33}, 166 (1970).
\bibitem{str11} M.\@ Kaku and L.\@ P.\@ Yu, 
{\em Phys.\ Rev}. {\bf D 3}, 2992, 3007, 3020 (1971).
\bibitem{str12} M.\@ Kaku and J.\@ Scherk, 
{\em Phys.\ Rev}. {\bf D 3}, 430 (1971).
\bibitem{str13} M.\@ Kaku and J.\@ Scherk, 
{\em Phys.\ Rev}. {\bf D 3}, 2000 (1971).
\bibitem{str14} C.\@ Lovelace, {\em Phys.\ Lett}. {\bf B 34}, 500 (1971).
\bibitem{str15} T.\@ Goto, {\em Phys.\ Theor.\ Phys}. {\bf 46}, 1560 (1971).
\bibitem{str16} J.\@ Scherk, {\em Rev.\ Mod.\ Phys}. {\bf 47}, 123 (1975).
\bibitem{gsw}
M.\@ B.\@ Green, J.\@ H.\@ Schwarz, and E.\@ Witten,
{\em Superstring Theory}, Cambridge University Press, (1985).
\bibitem{kak}M.\ Kaku, {\em Introduction to Superstring Theory}, 
Spring-Verlag, (1988).
\bibitem{strmex} J.\@ M.\ L\'{o}pez R., M.\@ A.\@ Rodr\'{\i}guez S., M.\@ 
Socolovsky and J.\@ L.\@ V\'{a}zquez B, {\em Rev.\ Mex.\ F\'{\i}s}. {\bf 34}, 
452 (1988).
\bibitem{green} B.\@ Green, {\em Basics of CFT, string compactifications, 
mirror symmetry and space-time topology change}, preliminary notes. Summer 
School in High energy Physics and Cosmology, 12 June-28 July 1995, ICTP, Trieste, 
Italy.
\bibitem{shaw} F.\@ Mandel and G.\@ Shaw, {\em Quantum Field theory}, John Wiley 
and Sons, (1991).
\bibitem{MTheory}P. K. Townsend, {\em Preprint hep-th 9612121v2}, (1996).
\bibitem{dvs} H. J. de Vega and N. S\'anchez,
 Phys. Lett. {\bf B 197} (1987), 320;
\bibitem{ERI} Proceedings of the Erice Schools:
``String Quantum Gravity and Physics at the Planck Energy Scale'',
21-28 June 1992, Edited by N. S\'anchez, World Scientific, 1993. 
Edited by N. S\'anchez and
A. Zichichi, Kluwer Acad. Publ.\\
Third D. Chalonge School `Current Topics in Astrofundamental
Physics', 4-16 September 1994,  Edited by N. S\'anchez and
A. Zichichi, NATO ASI Series Vol. 467, Kluwer Acad. Publ.
\bibitem{Leon}L. P. Grishchuk in Proceedings of the Erice School:
``String Gravity and Physics at the Planck Energy Scale'',
8-19 September 1995, Edited by N. S\'anchez, World Scientific, 1996. 
Edited by N. S\'anchez and
A. Zichichi, Kluwer Acad. Publ.

\bibitem{edal}
A.\@ Dabholkar and J.\@ M.\@ Quashnock, {\em Nucl.\ Phys}. 
{\bf B 333}, 815 (1990).
E.\@ J.\@ Copeland, D.\@ Haws, and M.\@ B.\@ Hindmarsh, {\em Phys.\
Rev}.\
 {\bf D 42}, 726 (1990).
\bibitem{dan} D.\@ Amati and C.\@ Klim\v{c}ik, {\em Phys.\ Lett}.\
 {\bf B 210}, 92 (1988).
 H. J. de Vega and N. S\'anchez, {\em Nucl. Phys.} {\bf B 317}, 706
(1989).
\bibitem{shellard} A.\@ Vilenkin and E.\@ P.\@ S.\@ Shellard,
{\em Cosmic Strings and Other Topological Defects} (CUP 1994).

\bibitem{Tablas}I. S. Gradshteyn and I. M. Ryzhik, {\em Table of Integrals, 
Series and Products}. Edited by Alan Jeffrey. Academic Press Inc. (1980).
\bibitem{Pock}M. Abramowitz and I. A. Stegun, {\em Pocketbook of 
Mathematical Functions}, Verlag Harri Deutsch, (1984).

\bibitem{kuo} C. Kuo, {\em Preprint gr-qc 9611064v2}, (1996).

\bibitem{eps} H. Epstein, V. Glaser and A. Jaffe, {\em Nuovo Cimento} 
{\bf 36}, 1016, (1965).

\bibitem{ford} L. H. Ford and T. A. Roman, {\em Preprint gr-qc 9607003v2}, 
(1996).

\bibitem{helf} A. D. Helfer, {\em Preprint gr-qc 9612029}, (1996).

\bibitem{Steph} H. Stephani, {\em General Relativity}, Cambridge University 
Press, (1990).

\bibitem{Meis} C. W. Misner, K. S. Thorne and J. A. Wheeler, {\em Gravitation}, 
W. H. Freeman \& and Co. (1973).

\bibitem{CVV} E.\@ J.\@ Copeland, A.\@ V\'{a}zquez, and H.\@ J.\@ de
Vega.\ {\em Preprint
LPTHE-Paris95-26/SUSX-TH95-31, hep-th 960112}, (1996). (To be revised.)

\bibitem{Alex} A.\@ Vilenkin, {\em Phys.\ Rev}.\ {\bf D 23}, 852 (1981).

\bibitem{Wei} S.\@ Weinberg, {\em Gravitation and Cosmology: Principles
and Applications of the General Theory of Relativity}, John Wiley and
Sons, (1972).
\bibitem{jac} J.\@ D.\@ Jackson, {\em Classical Electrodynamics}, John
Wiley and Sons, (1975).
\bibitem{Landa}L. D. Landau and E. M. Lifshitz, {\em The Classical Theory 
of Fields}, Pergamon Press, (1989).
\bibitem{Gold} H. Goldstein, {\em Classical Mechanics}, Addison Wesley, (1980).
\bibitem{sch} B. F. Schutz, {\em A first course in general relativity}, 
Cambridge University Press, (1990).
\bibitem{KIB}T.W. B. Kibble in `The Formation and Evolution of Cosmic Strings'. 
Proceedings of a Workshop supported by the SERC. 3-7 July 1989. 
Edited by G. Gibbons, S. Hawking and T. Vachaspati. Cambridge University Press.


\bibitem{Tur} N.\@ Turok, {\em Phys.\ Lett}.\ {\bf 126B}, 437 (1983).


\bibitem{Hisc} W. A. Hiscock and J. B. Lail, {\em Phys. Rev.}. {\bf D37}, 
869, (1988).

\bibitem{Sil} J. Silk and A. Vilenkin, {\em Phys. Rev. Lett}. {\bf 53}, 
1700, (1984).

\bibitem{Cos} M.\@ E.\@ V.\@ Costa, and H.\@ J.\@ de Vega, {\em Ann.\
Phys}.\ {\bf 211}, 223 (1991).


\bibitem{dray} T.\@ Dray, and G.\@ 't Hooft, {\em Nucl.\ Phys}.\ {\bf
B253}, 173 (1985).

\bibitem{Loust} C. O. Loust\'{o} and N. S\'{a}nchez, {\em Nucl.\ Phys}.\ {\bf
B355}, 231 (1991).

\bibitem{LoustN} C. O. Loust\'{o} and N. S\'{a}nchez, {\em Phys. Lett.}. 
{\bf B232}, 462 (1989).

\bibitem{Sexl} P. C. Aichelburg and R. U. Sexl, J. Gen. Rel. Grav.2, 303 
(1971).






\end{thebibliography}
\end{document}